\documentclass[final,3p,times]{elsarticle}

\usepackage{amssymb}
\usepackage{amsmath}
\usepackage{hyperref}
\usepackage{subcaption}
\usepackage{lineno}

\journal{Astroparticle Physics}

\begin{document}

\begin{frontmatter}



\title{The source of the cosmic-ray excess in the Centaurus region - constraints on possible candidates, mass composition and cosmic magnetic fields}

\author[label1,label2]{Teresa Bister}
\affiliation[label1]{organization={Nationaal instituut voor subatomaire fysica (NIKHEF)},
            addressline={Science Park 105},
            postcode={1098 XG},
            city={Amsterdam},
            country={The Netherlands}}

\affiliation[label2]{organization={Institute for Mathematics, Astrophysics and Particle Physics, Radboud University Nijmegen},
            addressline={Houtlaan 4},
            postcode={6525 XZ},
            city={Nijmegen},
            country={The Netherlands}}

\begin{abstract}
The most significant excess in the arrival directions of ultra-high-energy cosmic rays with energies $\gtrsim40\,\mathrm{EeV}$ is found in the direction of several interesting source candidates, most prominently the nearby radio galaxy Centaurus A. Naturally, Centaurus A has been suspected to create the anisotropy - but very different scenarios have been proposed. This includes a subdominant source contribution in combination with isotropic background sources, as well as a scenario where Centaurus A supplies the whole cosmic-ray flux above the ankle. Recently, it was suggested that the overdensity could instead consist of strongly deflected events from the Sombrero galaxy. Thanks to the recent development of several models of the Galactic magnetic field, it is now possible to test these proposed scenarios explicitly. 
We find that both sources inside the overdensity region (Centaurus A, NGC 4945, or M83), as well as outside of it (Sombrero galaxy) can in principle reproduce the excess. Leveraging the measured overdensity direction, significance, angular scale, and energy evolution, we place limits on the  allowed signal fraction, the possible ejected charge number and the strength of the extragalactic magnetic field between the respective source and Earth. We find that the scenario of a subdominant source in the overdensity region requires the charge number to be $Z\lesssim6$ and the extragalactic magnetic field quantity $B/\mathrm{nG} \sqrt{L_c/\mathrm{Mpc}}$ to be between $~1$ and $~100$ (depending on the charge and signal fraction). For the Sombrero galaxy to be the source, the dominant charge number has to be around $Z=6$ with $1\lesssim B/\mathrm{nG} \sqrt{L_c/\mathrm{Mpc}}\lesssim20$. We find that a scenario where all the flux above $30\,\mathrm{EeV}$ is supplied by Cen A or M83 is possible for $20\lesssim B/\mathrm{nG} \sqrt{L_c/\mathrm{Mpc}}\lesssim30$ and a mixed composition - explaining both the Centaurus region excess and the distribution of the highest-energy events - however, another contributing source is possibly required in the energy range $<30\,\mathrm{EeV}$.

\end{abstract}




\end{frontmatter}



\section{Introduction}
\begin{figure}[ht]
\centering
\includegraphics[width=0.6\textwidth]{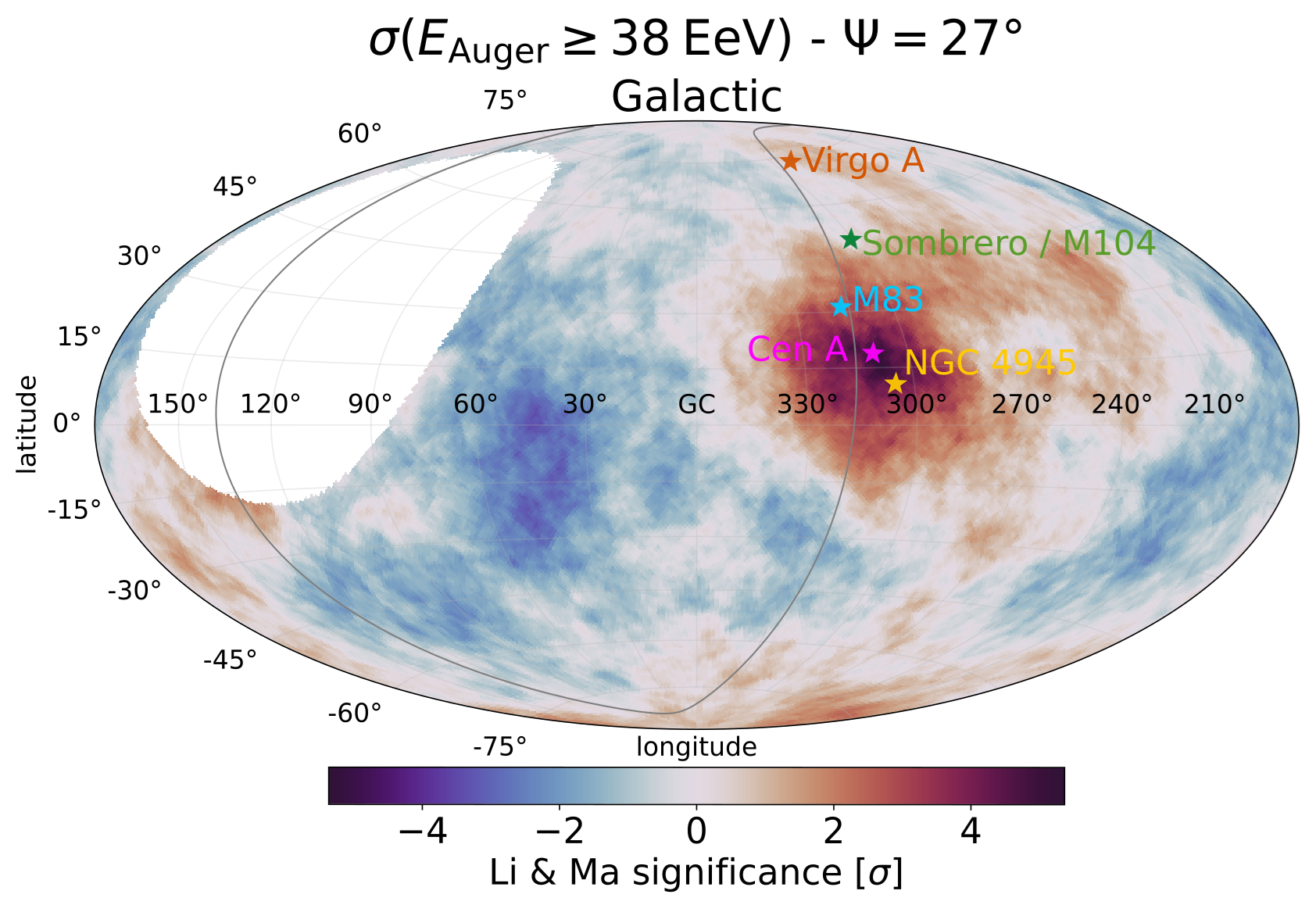}
\caption{Local LiMa significance in Galactic coordinates from a scan of the arrival direction data of the Pierre Auger Observatory~\cite{g_golup_on_behalf_of_the_pierre_auger_collaboration_update_2024}. The direction of several proposed source candidates of the most significant overdensity in the Centaurus region are indicated by the colored star markers.}
\label{fig:ADs_Auger}
\end{figure}

The origin of the highest energy particles reaching Earth - \textit{ultra-high-energy cosmic rays} (UHECRs) with energies $>10^{18}\,\mathrm{eV}$ to beyond $10^{20}\,\mathrm{eV}$ - is still unclear. Because UHECRs are charged nuclei that get deflected by cosmic magnetic fields during their propagation from the sources to Earth, their arrival directions are almost uniformly distributed, making it difficult to identify the directions of the sources. The most significant anisotropy in the arrival directions of UHECRs is a dipole at $E>8\,\mathrm{EeV}$~\cite{the_pierre_auger_collaboration_a_aab_et_al_observation_2017} with a magnitude of $\sim7.3\%$ and a current significance of $6.9\sigma$~\citep{abdul_halim_large-scale_2024}. 
At even higher energies, smaller-scale anisotropies begin to arise. These could indicate contributions of nearby source candidates as the propagation distance of UHECRs is limited to less than a few hundred megaparsec at these energies. 
In a scan over the sky, the Pierre Auger Collaboration identified the most significant excess at an angular scale of $\psi_\mathrm{40\,EeV}\simeq27^\circ$~\cite{g_golup_on_behalf_of_the_pierre_auger_collaboration_update_2024} for $E\gtrsim40\,\mathrm{EeV}$. The LiMa significance~\cite{li_analysis_1983} as a function of the direction is shown in Fig.~\ref{fig:ADs_Auger}. Recently, it was found that the excess also extends to lower energies $>20\,\mathrm{EeV}$, and that the excess direction stays almost constant with the energy threshold~\cite{the_pierre_auger_collaboration_flux_2024}.

Interestingly, the excess direction is very close to the nearby radio galaxy Centaurus A at a distance of only $3.8\,\mathrm{Mpc}$. It is therefore often called the \textit{Centaurus region overdensity}.
Cen A has been proposed as a source of UHECRs for decades~\cite{cavallo_sources_1978, romero_centaurus_1996, farrar_deducing_2000, gorbunov_comment_2007} due to its proximity and powerful radio jets. In~\cite{the_pierre_auger_collaboration_a_abdul_halim_et_al_constraining_2024}, it was found that a model where around $\mathcal{O}(10\%)$ of the UHECR flux at 40\,EeV comes from Cen A describes the UHECR arrival directions well $>16\,\mathrm{EeV}$. In~\cite{berezinskii_predicted_1990, isola_centaurus_2001, mollerach_case_2024}, it was even proposed that 100\% of the UHECR flux above the ankle is supplied by Cen A alone. This scenario is very attractive because the dominance of a single source swiftly explains why the allowed variation between the maximum acceleration energy of the source(s) is as small as found in~\cite{ehlert_curious_2023}, and also why the same  energy spectrum features are present over the whole sky~\cite{collaboration_energy_2025, kim_energy_2025}.
In order to describe the energy spectrum and arrival directions properly, the extragalactic magnetic field (EGMF) between Cen A and Earth has to be strong in that case, but exactly how strong to not produce too large anisotropies was not tested so far.

In addition to Cen A, there are also other nearby source candidates close to the excess direction that could explain the observed overdensity - namely the nearby starburst galaxies NGC 4945 and M83, both at a very similar distance as Cen A of $d\sim4\,\mathrm{Mpc}$~\cite{the_pierre_auger_collaboration_a_aab_et_al_indication_2018, the_pierre_auger_collaboration_a_abdul_halim_et_al_constraining_2024}. Also candidates further outside the overdensity region have been suggested as the source, like the Virgo cluster~\cite{ding_imprint_2021} (here we take the strongest galaxy Virgo A / M87 at $d\sim18.5\,\mathrm{Mpc}$ in the cluster for simplicity) or recently the nearby active galactic nucleus (AGN) M104 (\textit{Sombrero galaxy}) at a distance of $d\sim9\,\mathrm{Mpc}$~\cite{he_evidence_2024}. The directions of these proposed candidates are also indicated in Fig.~\ref{fig:ADs_Auger}.

For a galaxy at a larger angular distance from the excess direction to be the source, there have to be significant displacements by the Galactic magnetic field (GMF) - while for sources very close to the excess direction, coherent displacements have to be small. This implies constraints on the rigidity $R=E/Z$ of the involved particles because magnetic field deflections scale proportionally to $1/R$. Further constraints, also on the EGMF~\cite{Lee_extragalactic_1995} and signal fraction, can be derived from the size of the angular scale, as well as the significance and the parameter's energy evolutions.

In previous works like~\cite{Bray_2018, al-zetoun_constraining_2025}, this has been attempted without taking into account the effect of the GMF. That not only makes it impossible to consider source candidates at larger angular distances to the excess direction, but also leads to imprecise estimates for the allowed EGMF. Note that there are also previous works including older models of the GMF~\cite{vanvliet_extragalactic_2022, neronov_limit_2023}, but \cite{vanvliet_extragalactic_2022} could not yet leverage the energy evolution of the excess that was published this year~\cite{the_pierre_auger_collaboration_flux_2024}, and also explored only one possible model for the mass composition emitted by the sources, while~\cite{neronov_limit_2023} focused on a different, lower-significance anisotropy.
Owing to the recent progress in the development of GMF models~\cite{unger_coherent_2024, korochkin_coherent_2025, unger_galactic_2025, Pelgrims_ICRC_2025}, it is now possible to properly take into account the uncertainty involved with the GMF. That finally allows to determine which source candidates can actually explain the observed Centaurus region excess and its energy evolution, and place comprehensive constraints on the emitted mass composition, signal fraction and the EGMF.

For that purpose, this paper is structured as follows. In sec.~\ref{sec:simulation}, the setup of the simulations is explained. Then, three different scenarios are explored subsequently - \textit{Scenario I} with a subdominant source like Cen A close to the excess region in sec.~\ref{sec:scen1}, \textit{Scenario II} with a subdominant source like the Sombrero galaxy further away from the excess region in sec.~\ref{sec:scen2}, and \textit{Scenario III} where one source supplies the whole flux above the ankle in sec.~\ref{sec:scen3}. A summary of the findings is provided in sec.~\ref{sec:conclusion}. In the appendix, other source candidates and additional figures can be found.

\section{Simulation setup} \label{sec:simulation}
\begin{figure}[ht]
\subfloat[$f=0$]{\label{fig:spec1}\includegraphics[width=0.33\textwidth]{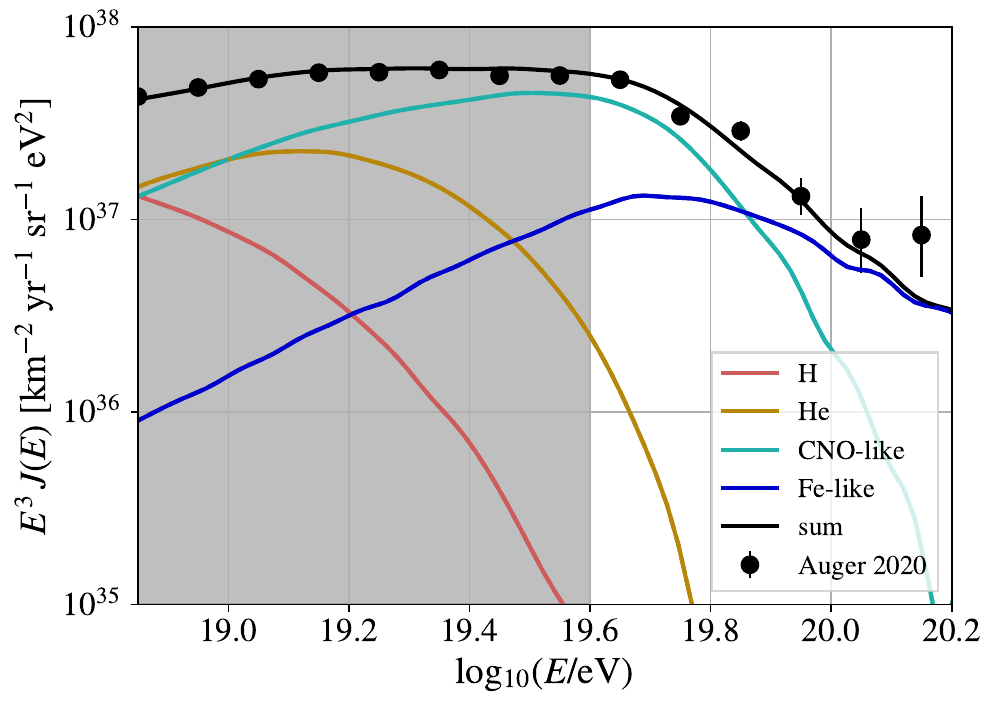}}
\subfloat[$f=0.05$, mixed composition]{\includegraphics[width=0.33\textwidth]{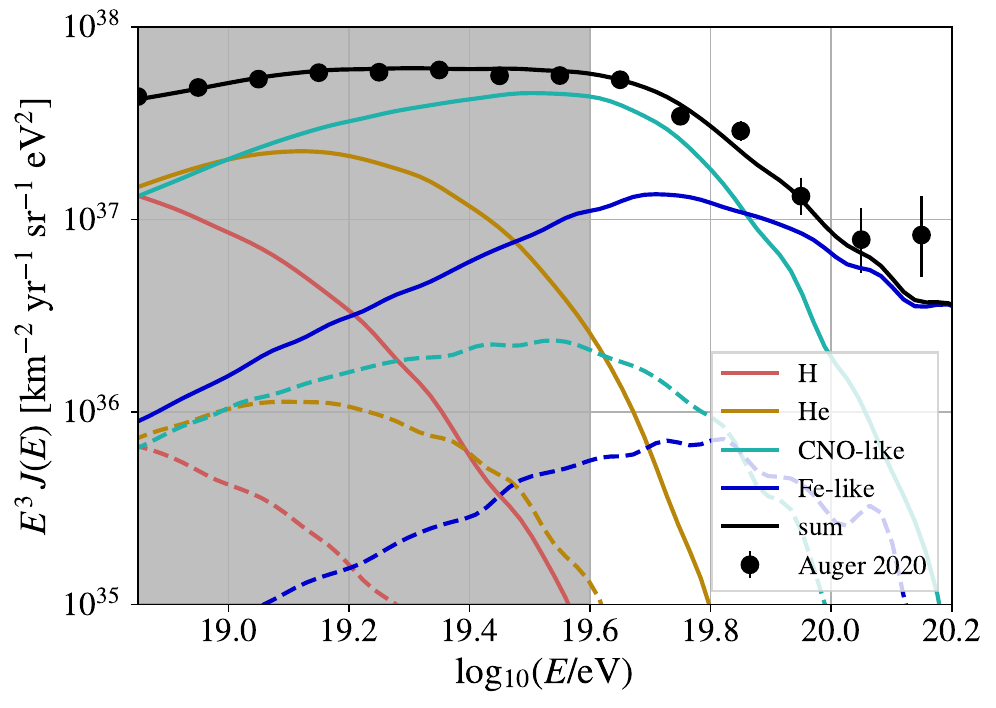}}
\subfloat[$f=0.05$, $Z=2$.]{\includegraphics[width=0.33\textwidth]{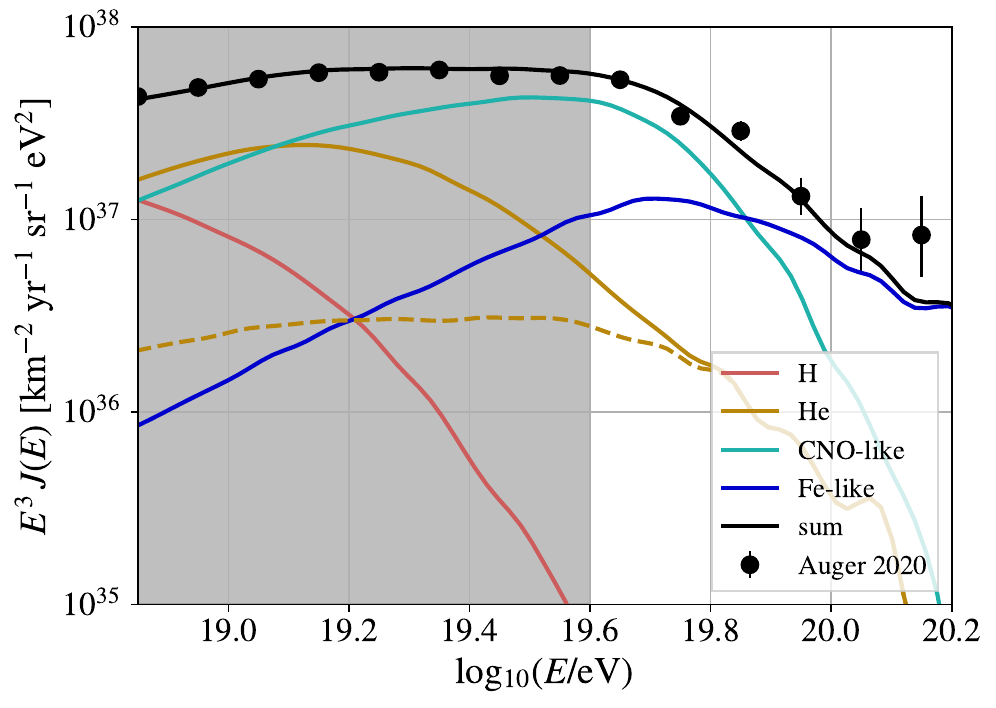}}\\
\subfloat[$f=0$]{\label{fig:comp1}\includegraphics[width=0.33\textwidth]{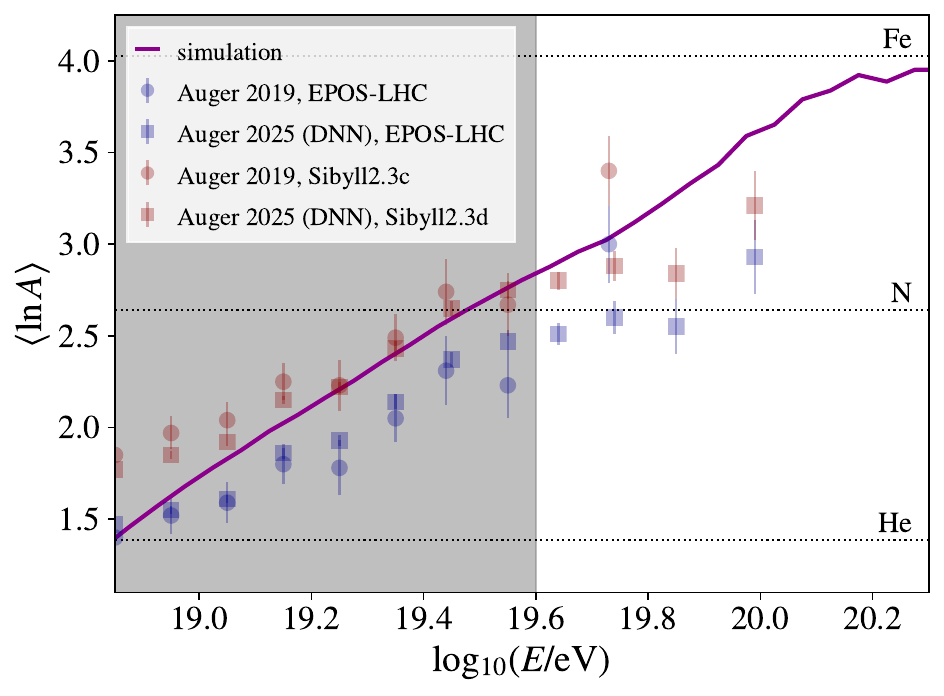}}
\subfloat[$f=0.05$, mixed composition]{\label{fig:comp2}\includegraphics[width=0.33\textwidth]{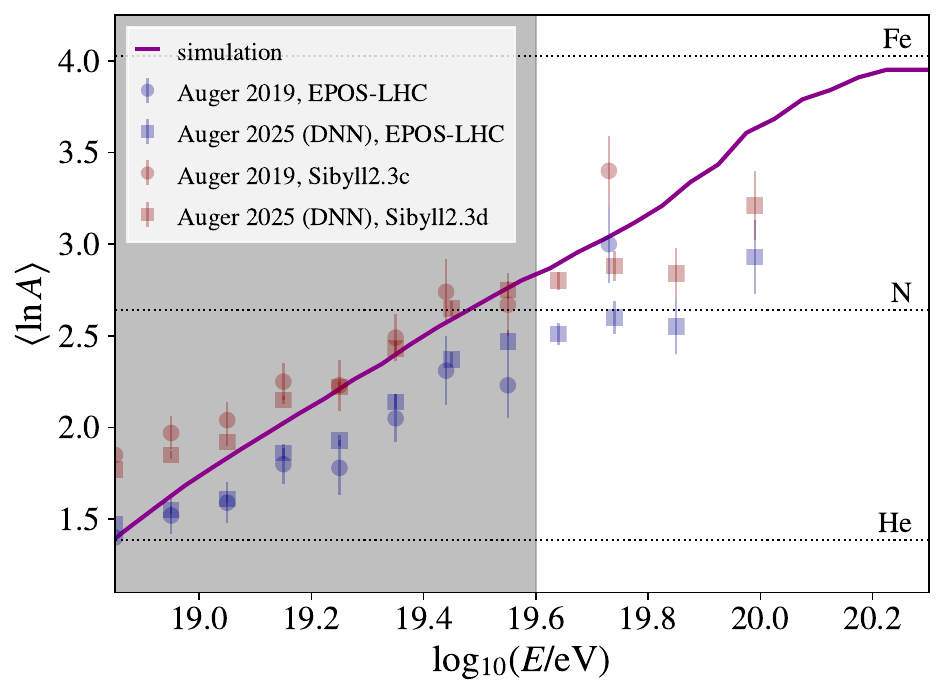}}
\subfloat[$f=0.05$, $Z=2$.]{\includegraphics[width=0.33\textwidth]{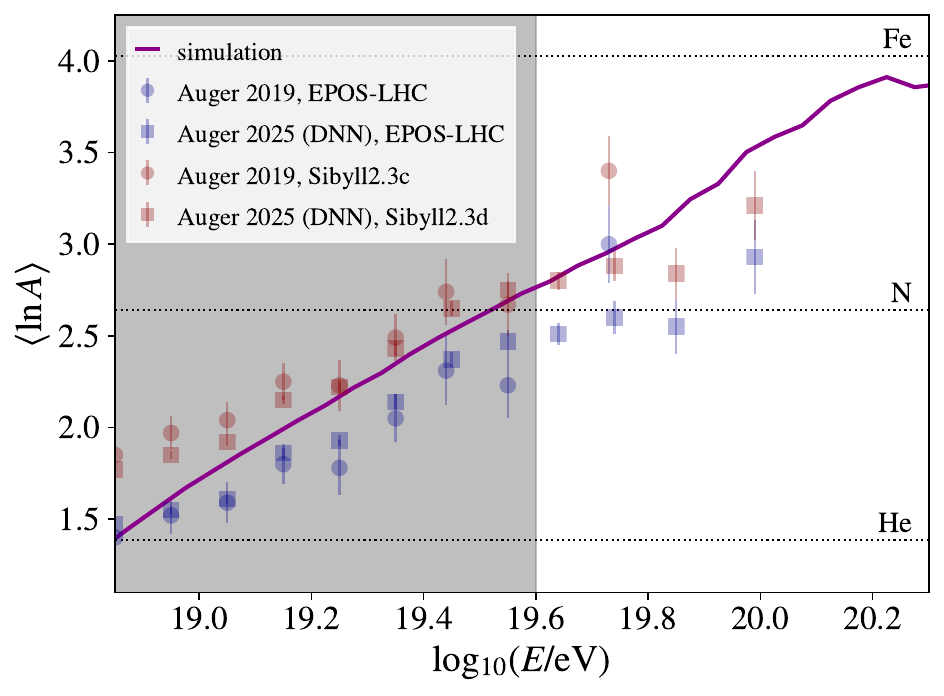}}
\caption{Simulation examples, energy spectrum in the upper row and corresponding mass composition (represented by the mean over the natural logarithm of the mass number $A$) in the lower row. 
The solid lines in the spectra denote the total simulated contribution of a mass group, the dashed lines that of the signal events only. The background always follows a mixed composition while the source composition and signal fraction $f$ can vary, see figure captions. The error bars correspond to Auger data from~\cite{Aab_measurement_2020} for the spectrum and \cite{Yushkov:2019J8} and \cite{Auger_Mass_DNN_2025} for the mass. The gray area denotes energies below $10^{19.6}\,\mathrm{eV}\simeq40\,\mathrm{EeV}$ corresponding to the energy threshold where the Centaurus excess becomes maximally significant.}
\label{fig:spec}
\end{figure}

The simulations consist of a \textit{signal} part from a source candidate, as well as a \textit{background} part. Note that the simulations contain no propagation or direct modeling of source emissions - instead, all event properties are directly drawn from measurements at Earth. The background events are distributed randomly following the directional exposure of the Pierre Auger Observatory. The energies are drawn randomly following the Auger energy spectrum~\cite{Aab_measurement_2020} as shown in Fig.~\ref{fig:spec1}. The charges follow a simple mixed composition model consisting of protons, helium, nitrogen, and iron. The mass composition becomes heavier with increasing energy as expected from a Peters cycle, and is in agreement with mass composition measurements by Auger~\cite{Mayotte_ICRC_2025} as visible in Fig.~\ref{fig:comp1}. Note however that the mass composition of the background has no influence on the figures shown in this paper - it is only relevant for the discussion of directional mass anisotropies or cuts on mass-sensitive variables. 

If the source only contributes a subdominant part of the whole UHECR flux, the overall mass composition is determined by the background events and the source mass composition is not constrained well by measured data (as long as no directionally-dependent mass composition measurements in the Centaurus region become available), see Fig.~\ref{fig:comp2}.
Thus, we test different possibilities for the source mass composition: either a single element (protons, helium, nitrogen, silicon, or iron); or the same mixed composition model as for the background. 
The energy spectrum of the source events at Earth is also not well constrained by current data. From propagation energy losses that have a larger impact on the further away background sources, it is generally expected that the source spectrum at Earth is harder than the background one due to the proximity of the source candidates~\cite{the_pierre_auger_collaboration_a_aab_et_al_features_2020}. The current best available indication comes from the evolution of the number of excess events in the excess region, that follows a power-law $\propto E^{-2.63}$~\cite{the_pierre_auger_collaboration_flux_2024}. But, naturally not all source events have to be inside that region considering magnetic field deflections. Additionally, it becomes problematic to draw source energies from this parameterization when the relative flux contribution of the source becomes large. That is because the parameterized excess region spectrum is harder than the Auger spectrum and does not have a cutoff, so that the simulated spectrum does not reproduce the Auger one anymore when too many energies are drawn from the excess parameterization. Thus, we simply draw the energies of the source from the measured Auger spectrum as well. Note that the chosen source spectrum has only minor effects on the conclusions of this work as discussed below. Three example simulated spectra are displayed in Fig.~\ref{fig:spec}.

For drawing the arrival directions of the events, we consider the effect of the GMF and EGMF. First, the distribution of arriving particles at the edge of the Milky Way is approximated using a Fisher distribution centered on the source direction to mimic the effect of a turbulent EGMF~\cite{Achterberg:1999vr, Harari_2016}:
\begin{equation}
\label{eq:egmf} 
    \vartheta = 29^\circ \frac{\mathrm{EV}}{E/Z} \frac{B}{\mathrm{nG}} \frac{\sqrt{d \ L_{c}}}{\mathrm{Mpc}} := 29^\circ \frac{\mathrm{EV}}{E/Z} \sqrt{\frac{d}{\mathrm{Mpc}}} \beta_{\rm EGMF}   
\end{equation}
Here, the field is characterized by the quantity $\beta_{\rm EGMF} = B/\mathrm{nG} \sqrt{L_c/\mathrm{Mpc}}$ encompassing the field strength $B$ and coherence length $L_c$. The blurring is proportional to the square root of the source distance $d$ and antiproportional to the rigidity $R=E/Z$. 
Note that the parameterization of deflections in a turbulent magnetic field by a Fisher distribution only applies in the diffusive regime where the coherence length $L_c$ is significantly smaller than the distance to the source. If $L_c$ becomes larger than around $10\%$ of the distance (thus around 400\,kpc for Cen A), the EGMF can also lead to coherent displacements and a significantly smaller angular blurring than predicted by eq.~\ref{eq:egmf}~\cite{dolgikh_cen_2025}. The coherence length of the EGMF in the vicinity of the Milky Way is not known, but an estimate could be the coherence lengths observed in galaxy clusters like Coma, which is on the order of $L_c\sim10\,\mathrm{kpc}\ll d$~\cite{bonafede_coma_2010}. Also, the large observed angular scale of the overdensity of $\sim27^\circ$ makes it likely but not guaranteed that diffusion theory applies, in which case the limits set in this work are valid. Another limitation of the Fisher distribution can arise for very small rigidities where CRs can perform several turns and get fully isotropized. Such an additional component can be modeled by an additional isotropic term~\cite{Harari_2016, mollerach_case_2024, harari_cosmic_2021}, but due to the unknown source energy spectrum, it is impossible to determine its contribution in relation to the background events. Thus, we ignore this possible contribution and use only the Fisher distribution to model the EGMF.

After EGMF deflections, the distribution at Earth is calculated using different lenses of the Galactic magnetic field. We use the \texttt{JF12} model~\cite{jansson_galactic_2012} (in combination with no random field, the original random field~\cite{jansson_new_2012}, and the Planck-tuned version~\cite{planck_collaboration_r_adam_et_al_planck_2016} ("Pl")), the recently published \texttt{UF23} suite of 8~\cite{unger_coherent_2024} + 1~\cite{unger_galactic_2025} GMF models (in combination with the Planck-tuned \texttt{JF12} random field model with different coherence lengths of $30\,\mathrm{pc}$ and $60\,\mathrm{pc}$), and lastly the \texttt{KST24}-model (again with the Planck-tuned  \texttt{JF12} random field model as well as without random field). For more information about the models we refer to the original publications as well as other recent works comparing the impact of the models on UHECR arrival direction analyses, e.g.~\cite{bister_large-scale_2024, korochkin_uhecr_2025, unger_where_2024}.
The signal events are drawn from the arrival direction distribution after GMF for their respective rigidity. Lastly, the total signal contribution is weighted with a signal fraction $f$ determined as $f=\frac{N_\mathrm{signal}>40\,\mathrm{EeV}}{(N_\mathrm{signal}+N_\mathrm{background})>40\,\mathrm{EeV}}$. The total number of events in the simulations is the same as in data taken from~\cite{the_pierre_auger_collaboration_flux_2024}.

The free parameters of the simulations and their impact are summarized in the following:
\begin{itemize}
    \item The GMF is a sizeable uncertainty in any study of cosmic-ray arrival directions. The systematic uncertainty of the GMF model is best explored by taking into account multiple GMF models based on different assumptions and nuisance data, such as the 8 models from the UF23~\cite{unger_coherent_2024} model suite.
    For a conservative estimate, we use all GMF models described above to see if a simulation can explain the observations for \textit{any} GMF model. By additionally including different random field models, the involved uncertainty is marginalized over.
    \item The charge, in combination with the GMF, is the only simulation quantity influencing the size of the coherent displacement. For sources inside the excess region, the charge cannot be too big to be able to explain the overdensity direction - while for sources outside of it the charge cannot be too small in order to have large enough coherent displacements. This can be easily seen when looking at deflected events from each source for different charge numbers as shown in Fig.~\ref{fig:ADs} in the appendix. In addition to the coherent deflections, the charge certainly also influences the size of the EGMF blurring and thereby the excess significance and angular scale.
    \item The EGMF parameter $\beta_{\rm EGMF} = B/\mathrm{nG} \sqrt{L_c/\mathrm{Mpc}}$ has an influence on both the angular scale as well as the significance. If there was no GMF, it would directly determine the angular scale through eq.~\ref{eq:egmf}. Also, a larger EGMF blurring leads to larger random variations of the excess direction with the energy.
    \item The signal fraction's main impact is on the significance. Additionally, a smaller signal fraction also leads to larger random variations of the excess direction with the energy.
\end{itemize}
In the following, these simulations are tested to see if they can reproduce the results of the LiMa scan $E>40\,\mathrm{EeV}$, namely the maximum LiMa significance of $\sigma_\mathrm{40\,EeV}\simeq5.1\sigma$~\cite{the_pierre_auger_collaboration_p_abreu_arrival_2022}, plus its direction, and the angular scale $\psi_\mathrm{40\,EeV}=27^\circ$~\cite{the_pierre_auger_collaboration_flux_2024}. From the recent scan of the energy threshold in 6 steps ($>20\,\mathrm{EeV}$, $>25\,\mathrm{EeV}$, $>32\,\mathrm{EeV}$, $>40\,\mathrm{EeV}$, $>50\,\mathrm{EeV}$, $>63\,\mathrm{EeV}$)~\cite{the_pierre_auger_collaboration_flux_2024} it is clear that the excess extends over the whole energy range $>20\,\mathrm{EeV}$. The evolution of the LiMa significance with the energy is directly correlated with the signal contribution (that is directly linked to the unknown source spectrum) and can hence not be used to decide if e.g. the charge or EGMF used in a simulation reproduce the observations or not. The evolution of the significance with the energy is thus only used as a secondary check and not to dismiss a simulation. The angular scale evolution can also not be used as it was kept fixed in the energy dependent scan. The evolution of the direction however is very constraining: it moves by less than a few degrees over the whole energy range which is quite difficult to reproduce as will be discussed below. 
For each combination of GMF, signal charge, $\beta_{\rm EGMF}$ and $f$, we produce 10 random simulations and apply the same scan of the maximum LiMa significance as applied on the data~\cite{the_pierre_auger_collaboration_flux_2024, the_pierre_auger_collaboration_p_abreu_arrival_2022} to each of them.

\section{Scenario I - subdominant source within the Centaurus region} \label{sec:scen1}
\begin{figure}[ht!]
\includegraphics[width=0.49\textwidth]{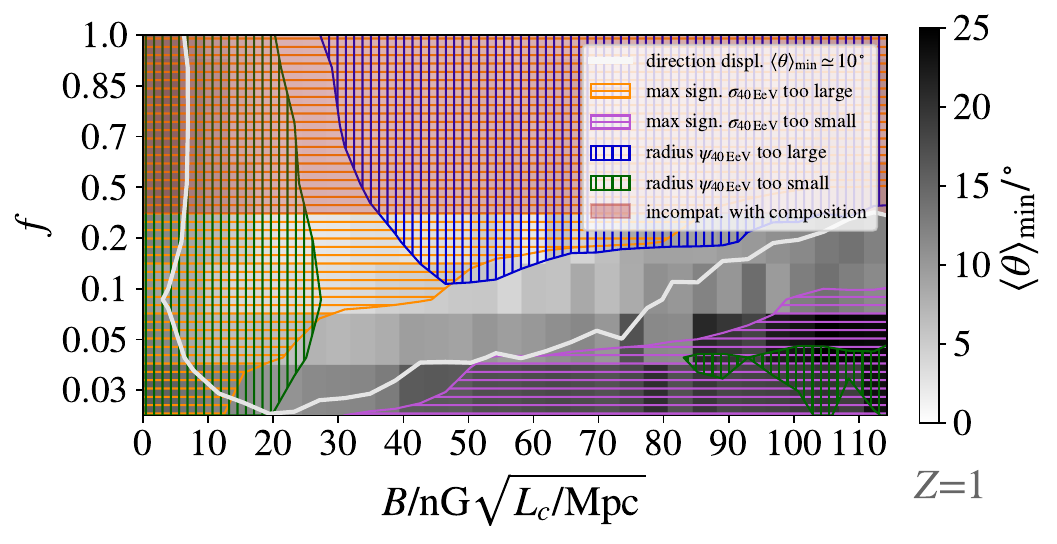}
\includegraphics[width=0.49\textwidth]{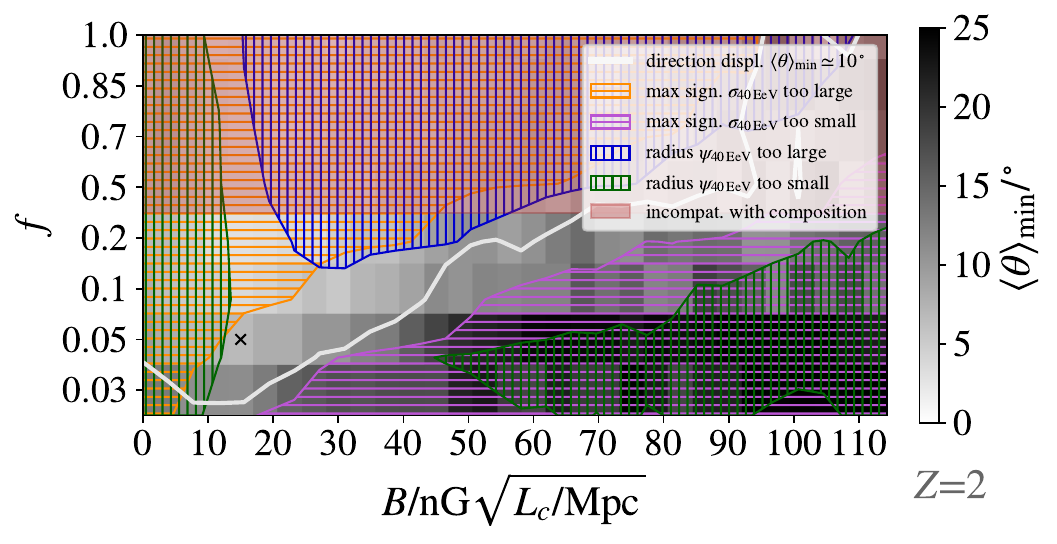}\\
\includegraphics[width=0.49\textwidth]{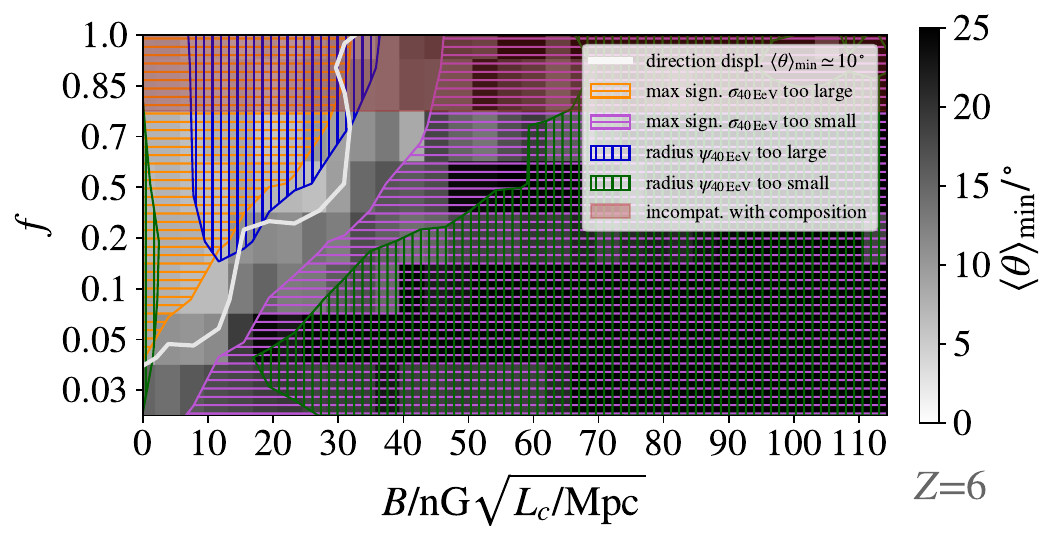}
\includegraphics[width=0.49\textwidth]{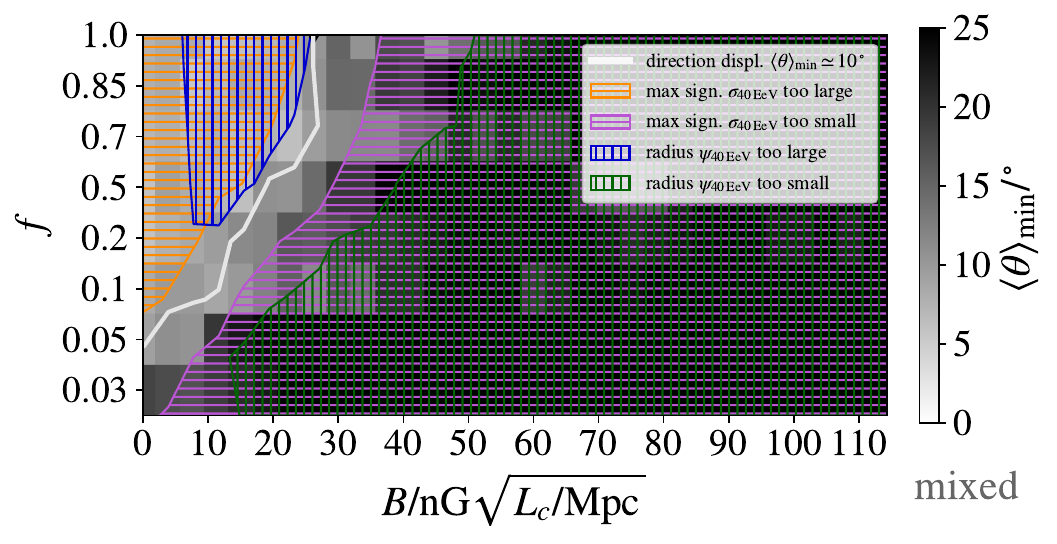}\\
\includegraphics[width=0.49\textwidth]{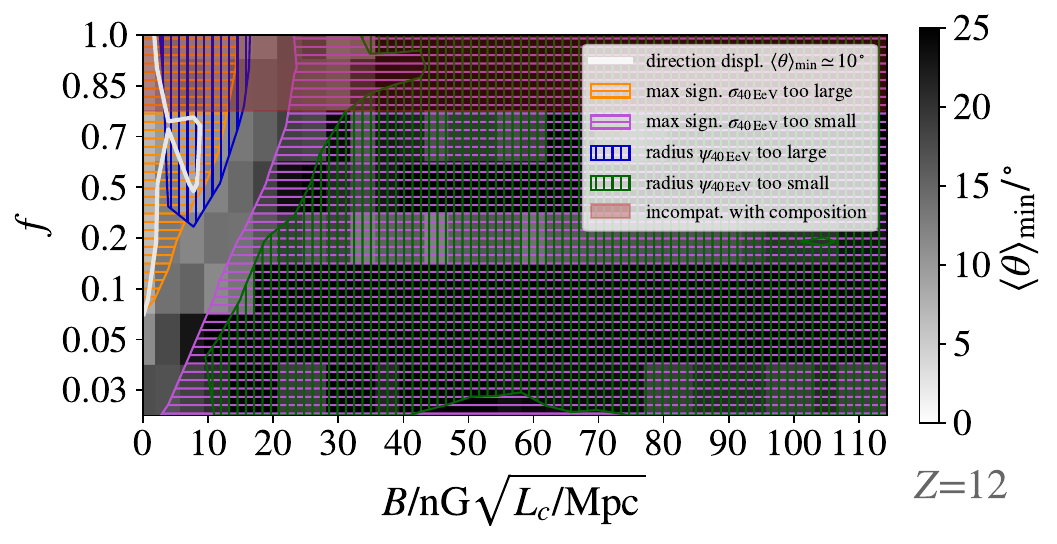}
\includegraphics[width=0.49\textwidth]{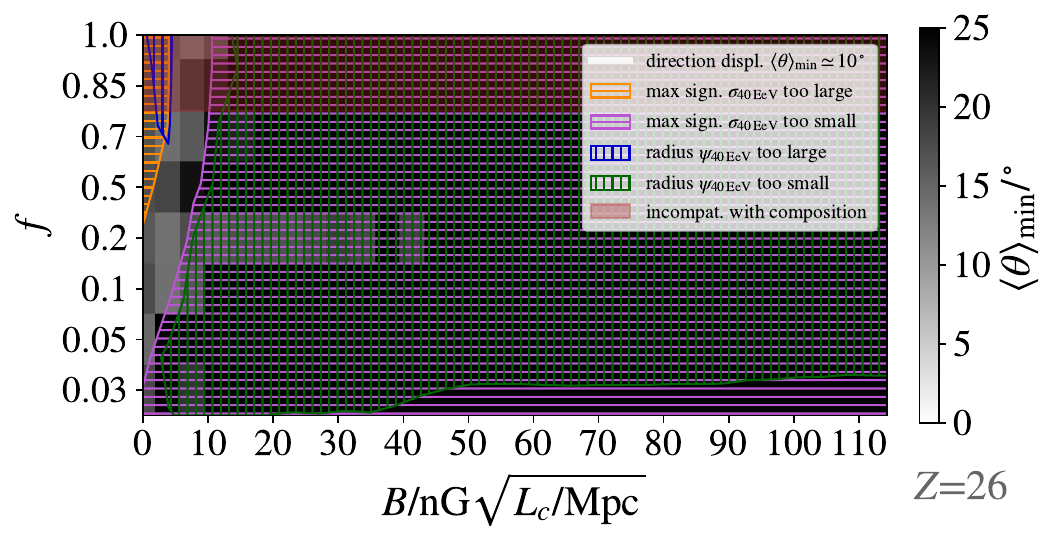}
\caption{Constraints on the parameter space of signal fraction $f$ and EGMF parameter $\beta_\mathrm{EGMF}$ for Cen A as the source of the observed excess, for different charge numbers as indicated in the figures. Parameter combinations in the orange horizontally hatched area lead to too large maximum LiMa significances $\sigma_\mathrm{40\,EeV}\gg5.1\sigma$ not in agreement with the data, while in the pink horizontally hatched area the significance is too small. In the green vertically  hatched area, the angular scale $\psi_\mathrm{40\,EeV}\ll27^\circ$ is too small, in the blue vertically hatched area it is too large. The red shaded region leads to a mass composition not in agreement with Auger mass composition measurements~\cite{Mayotte_ICRC_2025}. The color bar (background gray scale) denotes the minimum (over the 10 simulations per GMF model) of the average (over the scanned energy thresholds~\cite{the_pierre_auger_collaboration_flux_2024}, see also Fig.~\ref{fig:cena_AD}) of the angular difference between the observed excess direction and the simulated ones. Only very few simulations can reproduce the excess direction well over all energy thresholds, so that $\langle\theta\rangle$ is often a lot larger than the observed value of $3.3^\circ$, especially for larger charge numbers. The white contour denotes a value of $\langle\theta\rangle\simeq10^\circ$.
The black cross marker (for $Z=2$) indicates a parameter combination that is discussed more thoroughly below.
For more details see text.}
\label{fig:cena_constraints}
\end{figure}

First, we focus on sources located close to the excess direction. In the following we will discuss the case of Cen A. The corresponding (very similar) results for NGC 4945 and M83 can be found in~\ref{app:ngc4945} and~\ref{app:m83}. 
In Fig.~\ref{fig:cena_constraints}, a summary of the allowed values for the signal fraction $f$ and EGMF parameter $\beta_{\rm EGMF}$ is shown for the different tested mass composition scenarios. The orange horizontally hatched region indicates when $90\%$ of the simulations exhibit a maximum significance larger than the observed value of $\sigma_\mathrm{40\,EeV}\simeq5.1\sigma$. That implies that the simulations are too anisotropic, which can be due to a too large signal fraction $f$ or a too small EGMF blurring. The pink horizontally hatched region on the other hand denotes that 90\% of the simulations have a smaller maximum significance than the observed $\sigma_\mathrm{40\,EeV}\simeq5.1\sigma$, and hence that the simulations are too isotropic compared to the data.
The green vertically hatched region indicates that 90\% of the simulations exhibit an angular scale that is smaller than the observed value of $\psi_\mathrm{40\,EeV}=27^\circ$, which mostly happens when the simulated EGMF blurring is too small. The blue vertically hatched region on the other hand signifies that 90\% of the simulations have an angular scale larger than $\psi_\mathrm{40\,EeV}=27^\circ$, which is mostly related to a too large value of the EGMF for sufficiently anisotropic simulations.

\begin{figure}[ht!]
\includegraphics[trim={13cm 0 0 0}, clip, width=0.24\textwidth]{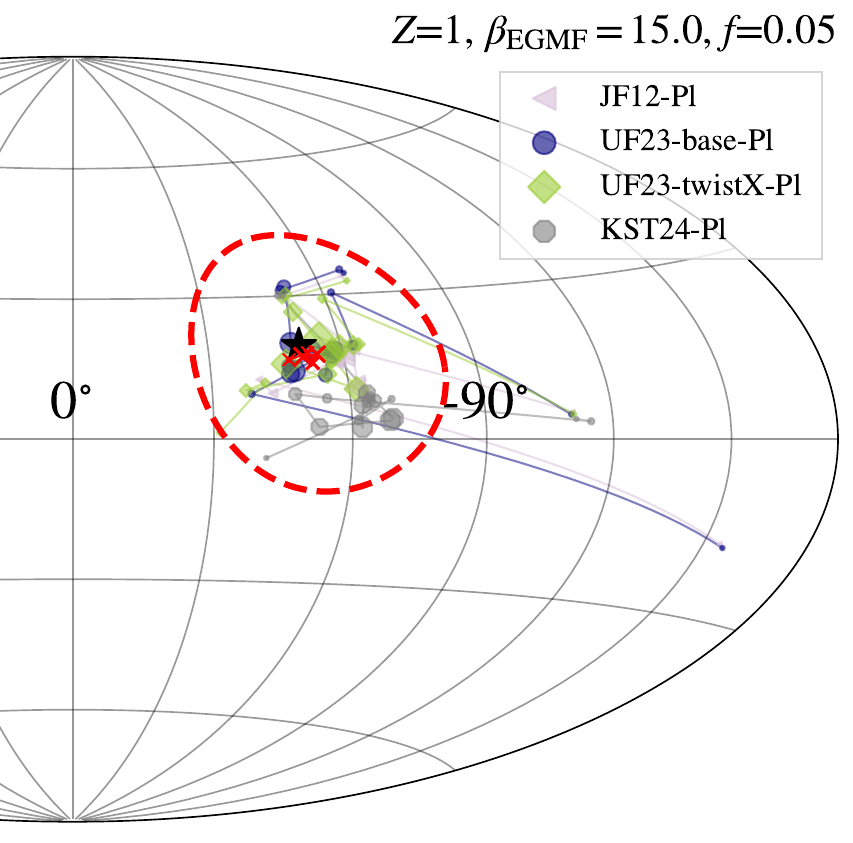}
\includegraphics[trim={13cm 0 0 0}, clip, width=0.24\textwidth]{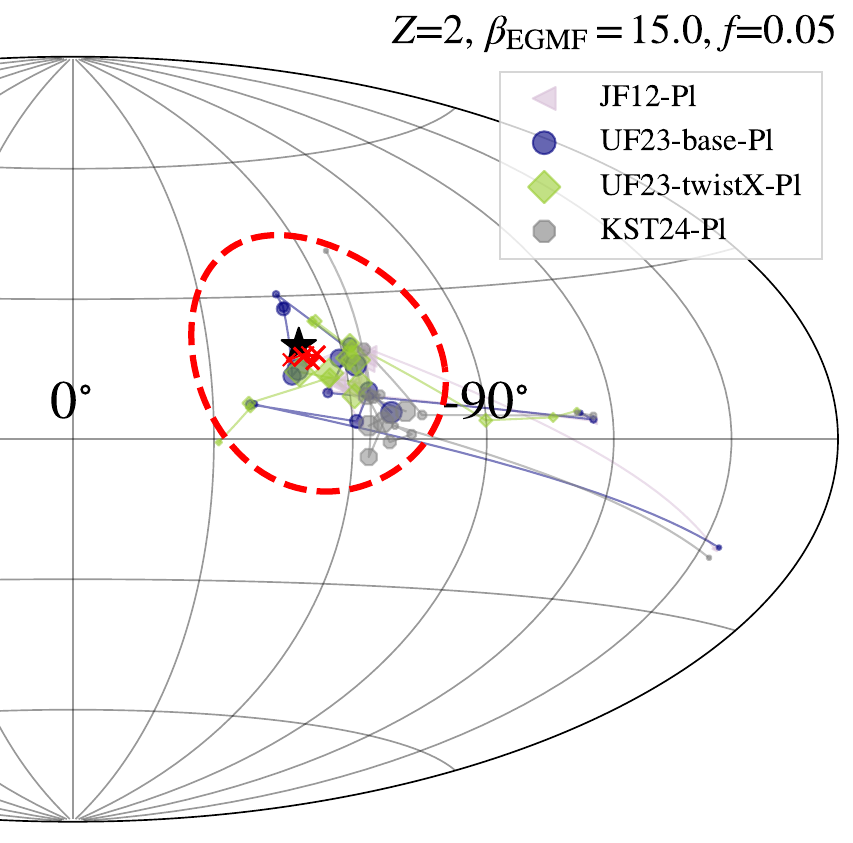}
\includegraphics[trim={13cm 0 0 0}, clip, width=0.24\textwidth]{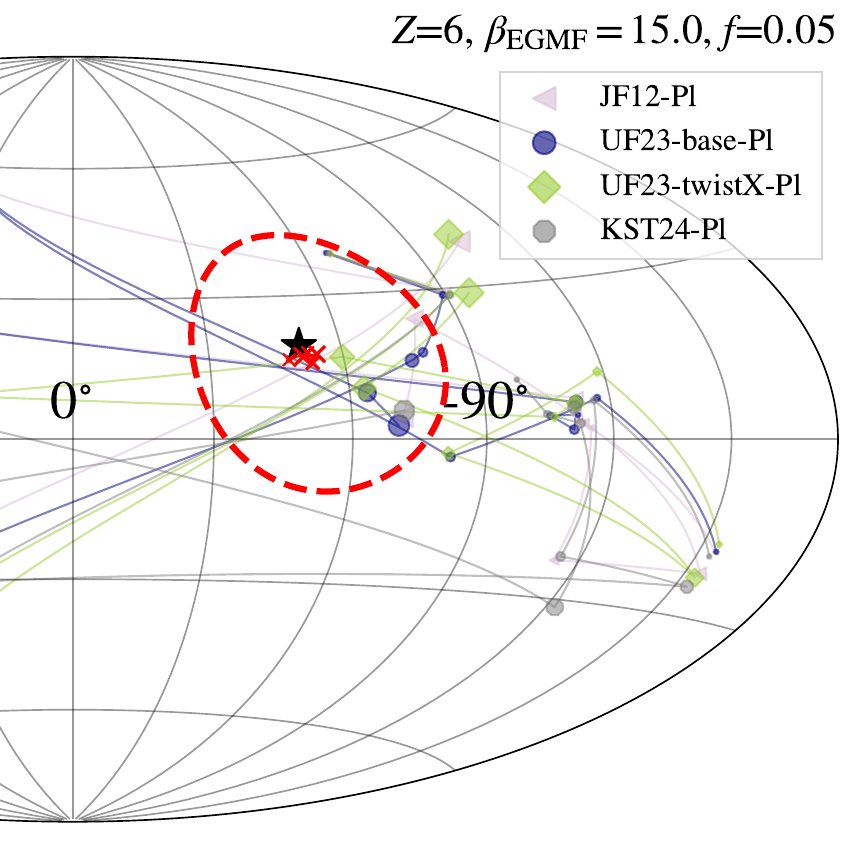}
\includegraphics[trim={13cm 0 0 0}, clip, width=0.24\textwidth]{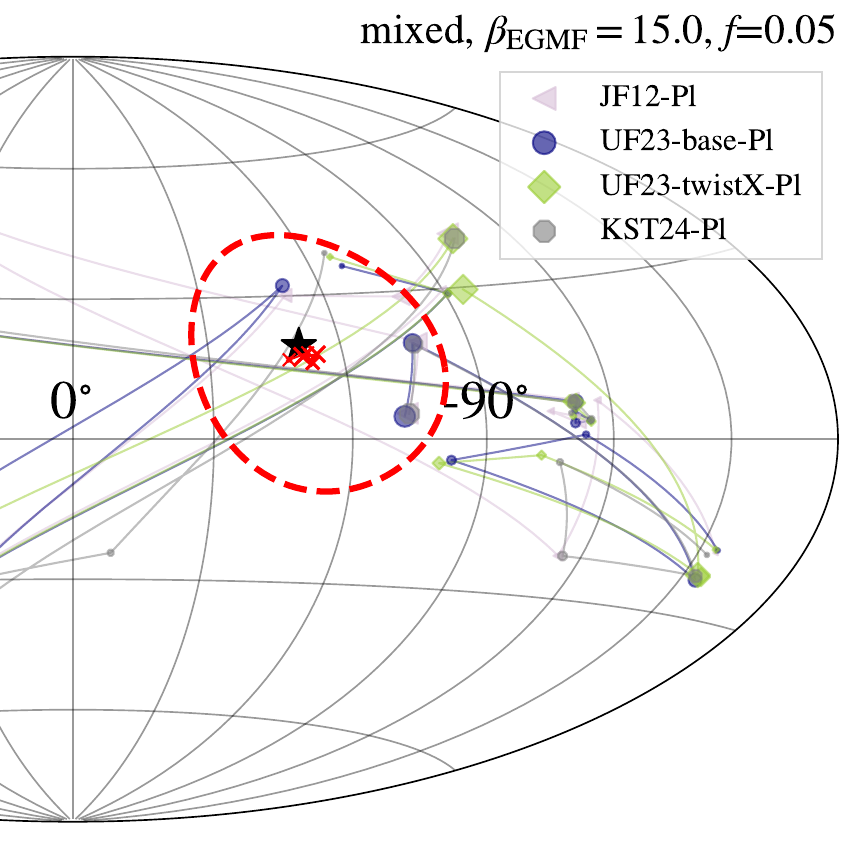}
\caption{Evolution of the excess direction with the energy for example simulations with Cen A (star marker) as the source, $f=0.05$, $\beta_\mathrm{EGMF}=15$, using and different charges (see figure title) and four different GMF models. For each GMF, four randomly drawn simulations are shown. Each simulation's excess directions are indicated by 6 connected markers of decreasing size denoting the energy thresholds $>20\,\mathrm{EeV}$, $>25\,\mathrm{EeV}$, $>32\,\mathrm{EeV}$, $>40\,\mathrm{EeV}$, $>50\,\mathrm{EeV}$, and $>63\,\mathrm{EeV}$ used in~\cite{the_pierre_auger_collaboration_flux_2024}. The red crosses mark the same for the data. The red circle indicates the best angular scale of $\psi_\mathrm{40\,EeV}=27^\circ$~\cite{the_pierre_auger_collaboration_flux_2024} for $>40\,\mathrm{EeV}$. Note that the random seeds of the four simulations are the same for each GMF and charge number, so that the background events are the same. This is why the excess directions are similar for some GMFs, especially for $>63\,\mathrm{EeV}$ where statistics can dominate.}
\label{fig:cena_AD}
\end{figure}

The color bar visualizes the third constraint set by the excess direction. $\langle\theta\rangle$ denotes the angular distance between the observed excess direction for $E_\mathrm{min}=40\,\mathrm{EeV}$~\cite{the_pierre_auger_collaboration_p_abreu_arrival_2022} and the excess direction of the simulations, averaged over the 6 tested energy thresholds~\cite{the_pierre_auger_collaboration_flux_2024}. On data, $\langle\theta\rangle=3.3^\circ$, indicating that the excess is extremely stable with the energy. $\langle\theta\rangle_\mathrm{min}$ represents the smallest value for $\langle\theta\rangle$ reached over all 10 random draws and GMF models. It can be seen that in simulations, the variation is typically a lot larger than the observed value of $\langle\theta\rangle=3.3^\circ$. This can have two reasons. Either, the combination of source direction, GMF, and charge does not work to reproduce the excess direction - or the variations between energy bins are too large. In Fig.~\ref{fig:cena_AD}, the evolution of the excess direction is shown for a few example simulations with $f=0.05$ and $\beta_\mathrm{EGMF}=15$. It is visible that the excess direction is generally well reproduced for $Z\leq2$ (just not for the \texttt{KST24} GMF model). For larger charge numbers, however, the direction jumps significantly with the energy and would then lead to a large $\langle\theta\rangle_\mathrm{min}$ - that combination of parameters is however anyways forbidden by a too small significance as visible in Fig.~\ref{fig:cena_constraints}. 
Because we do not expect that any of the tested GMF models acts exactly as the real one, we set a conservative constraint on $\langle\theta\rangle_\mathrm{min}$ to be smaller than $10^\circ$.
For charge numbers $Z\geq12$, $\langle\theta\rangle_\mathrm{min} \gg 10^\circ$ for all parameter combinations, implying that for sources close to the excess direction like Cen A, M83, or NGC 4945 the charge has to be $Z\lesssim6$. For NGC 4945, the charge even has to be $Z\lesssim2$ (see \ref{app:ngc4945}) because variations of the excess direction get too large even for $Z=6$ due to NGC 4945 being so close to the Galactic plane.

A correlation of the allowed parameter region is clearly visible, indicating that either a combination of small signal fraction and small EGMF, or large signal fraction and large EGMF is possible. It has to be noted, however, that only because a combination of parameters falls into the allowed region, it does not necessarily mean that there are (many) simulations that closely reproduce the data excess direction, significance, and angular scale. The hatched regions indicate that one observational constraint alone is already enough to exclude the parameter combination. Yet, it is usually possible to find a simulation that resembles the data for at least one GMF model in the allowed parameter region. A few example simulations are discussed in the following. First, we will focus on the case of a subdominant source $f\ll50\%$ (\textit{scenario I}). 
Below, in sec.~\ref{sec:scen3}, scenarios where the whole flux is dominated by only one source with $f>50\%$ will be discussed.

\begin{figure}[ht!]
\includegraphics[width=0.24\textwidth]{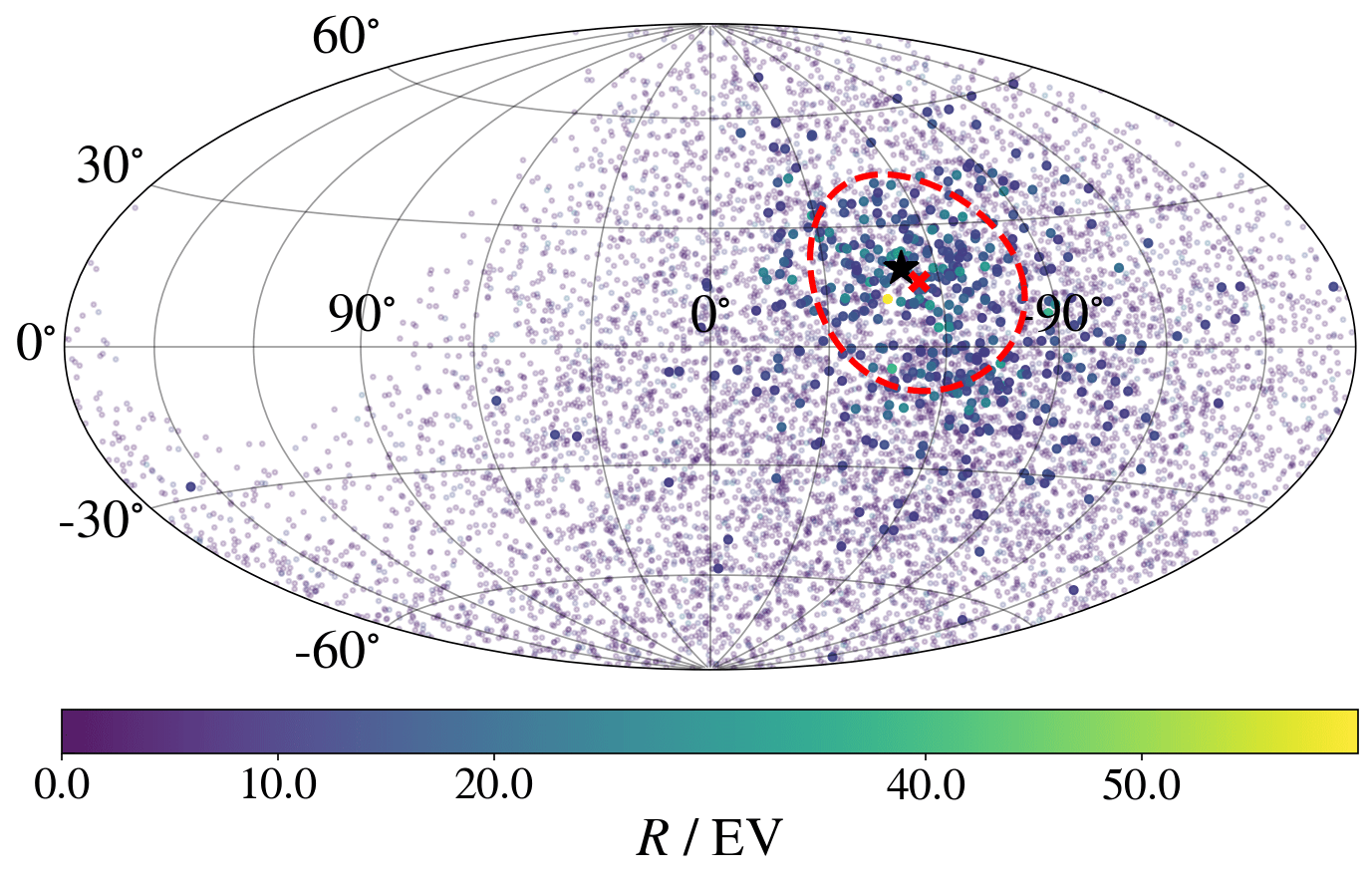}
\includegraphics[width=0.24\textwidth]{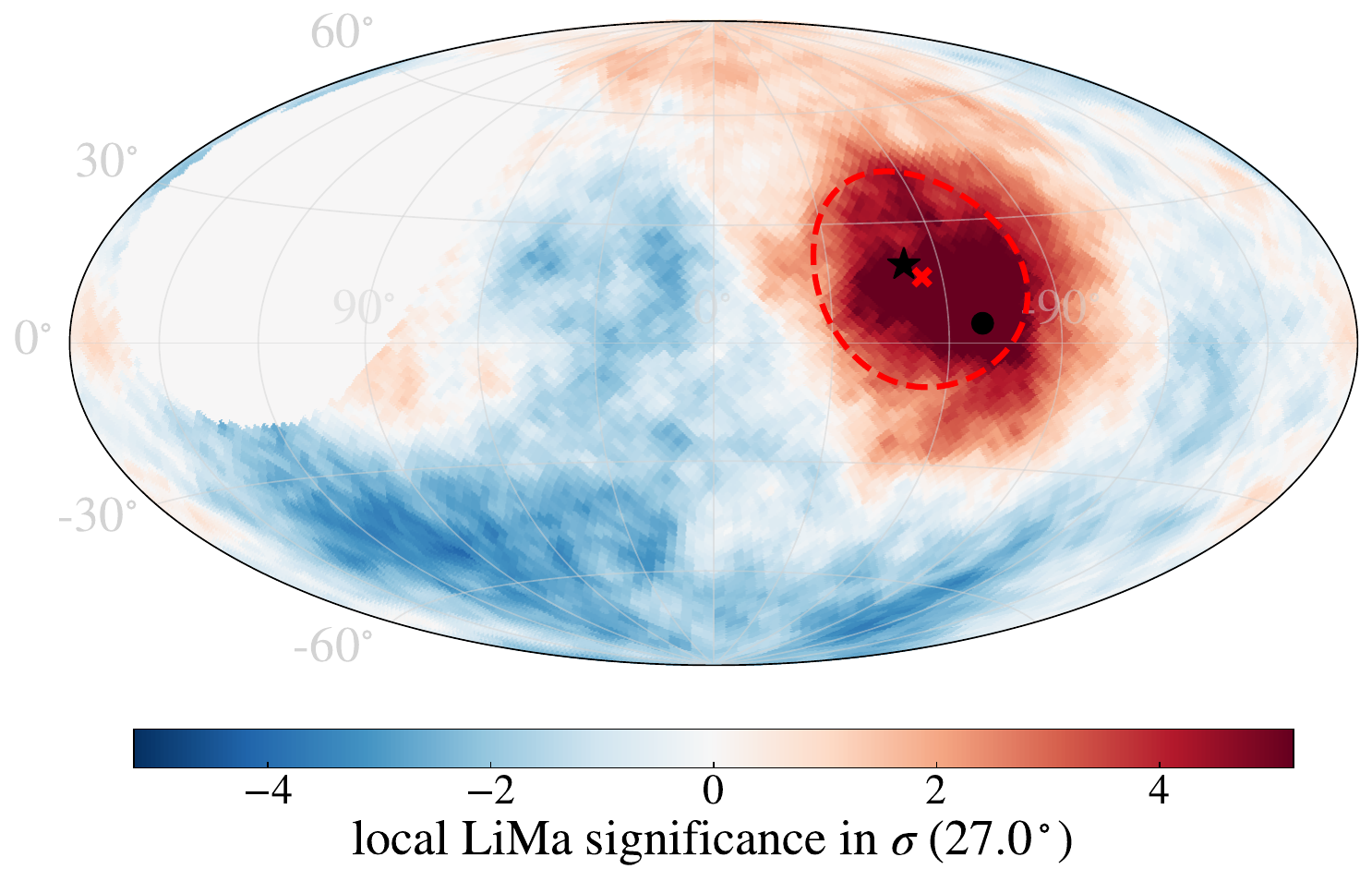}
\includegraphics[width=0.24\textwidth]{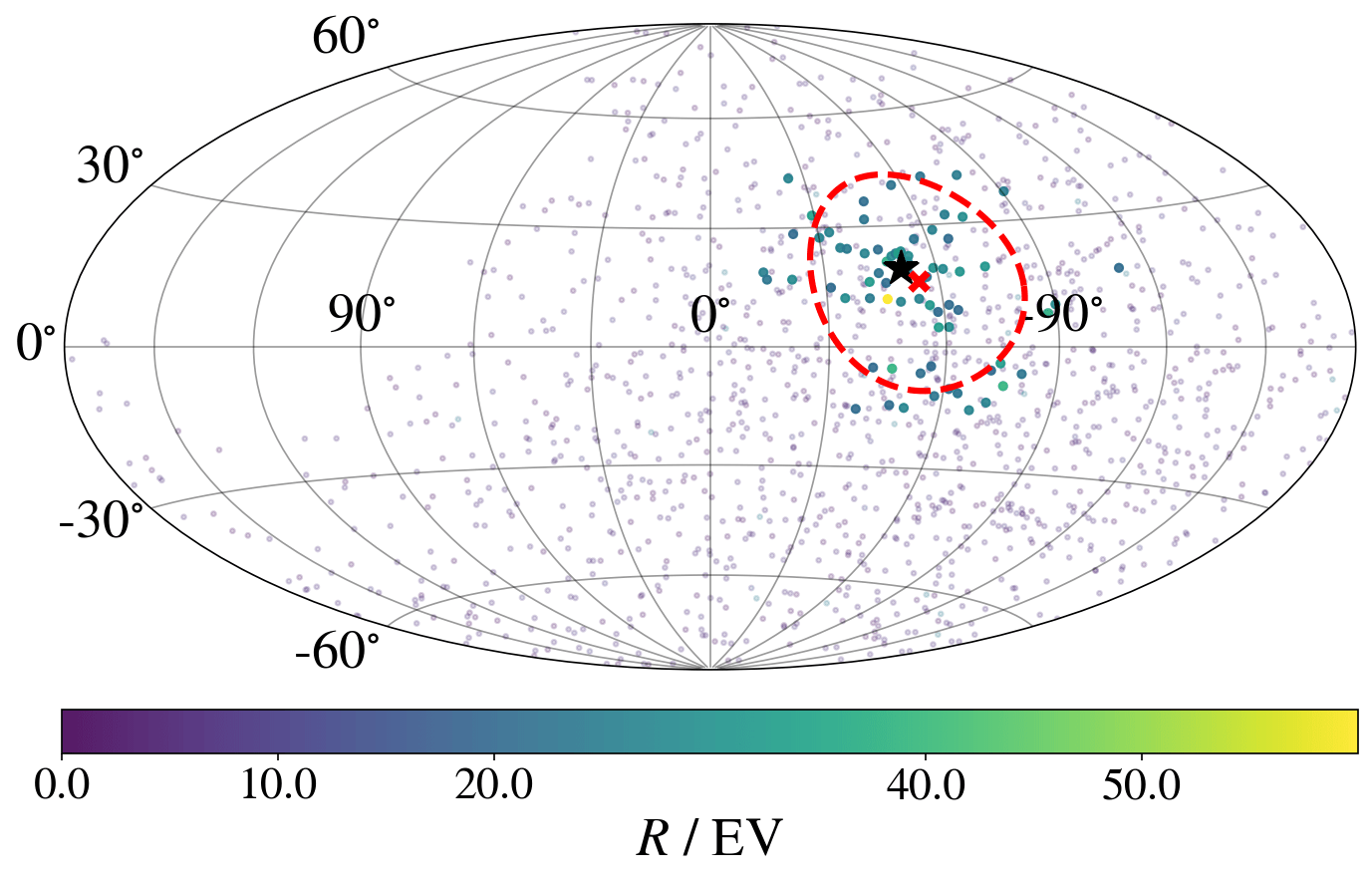}
\includegraphics[width=0.24\textwidth]{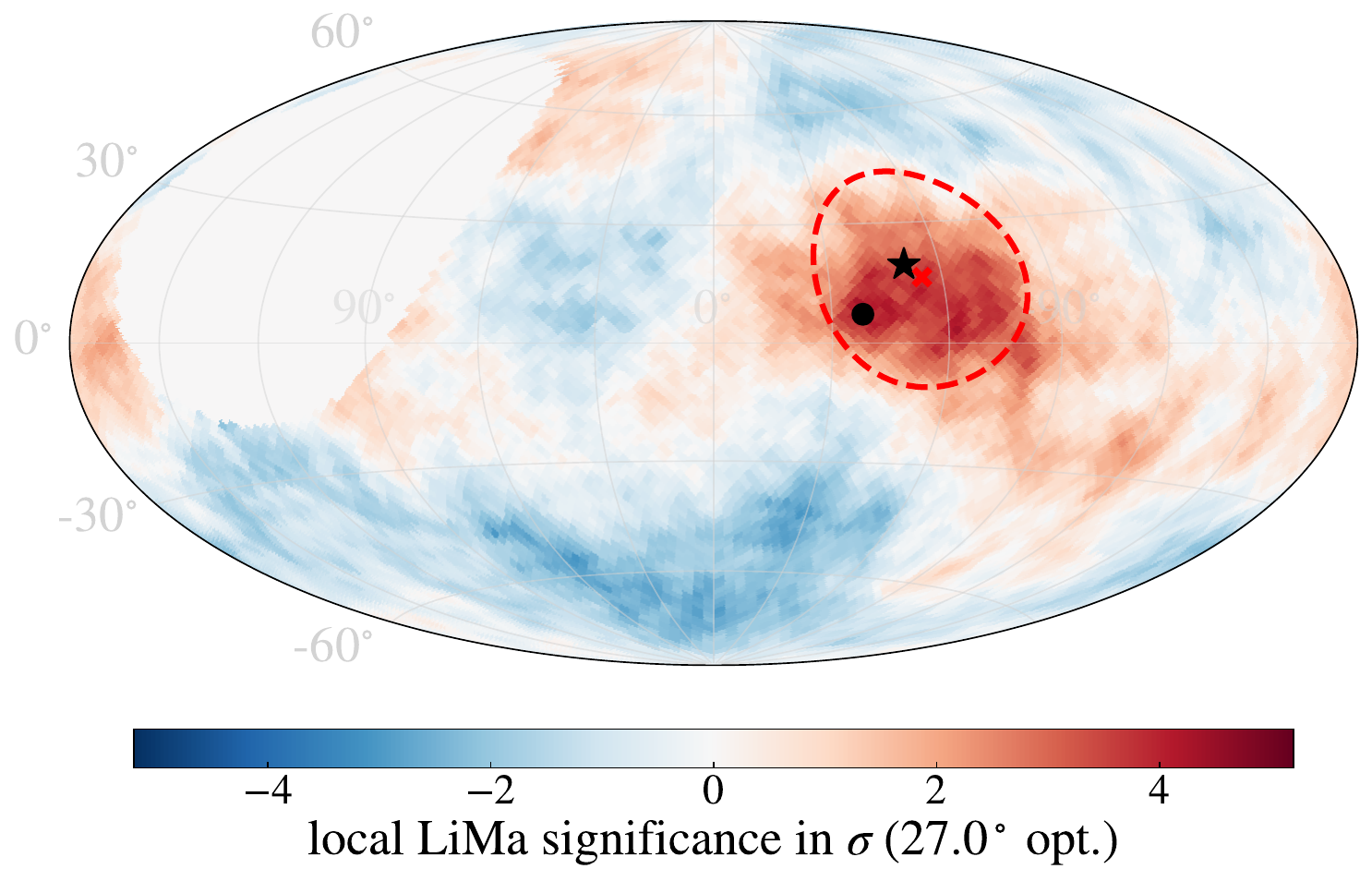}

\includegraphics[width=0.24\textwidth]{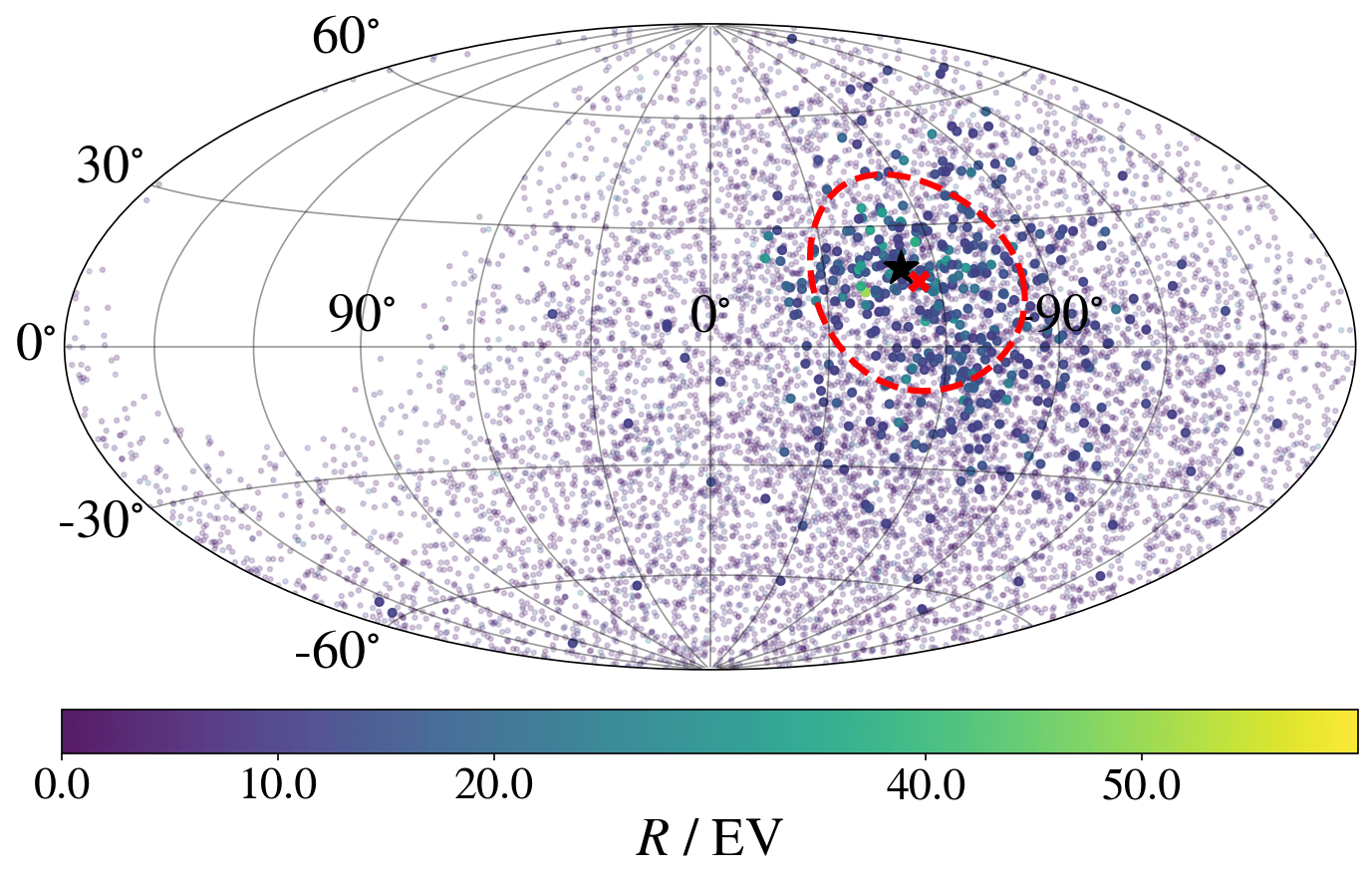}
\includegraphics[width=0.24\textwidth]{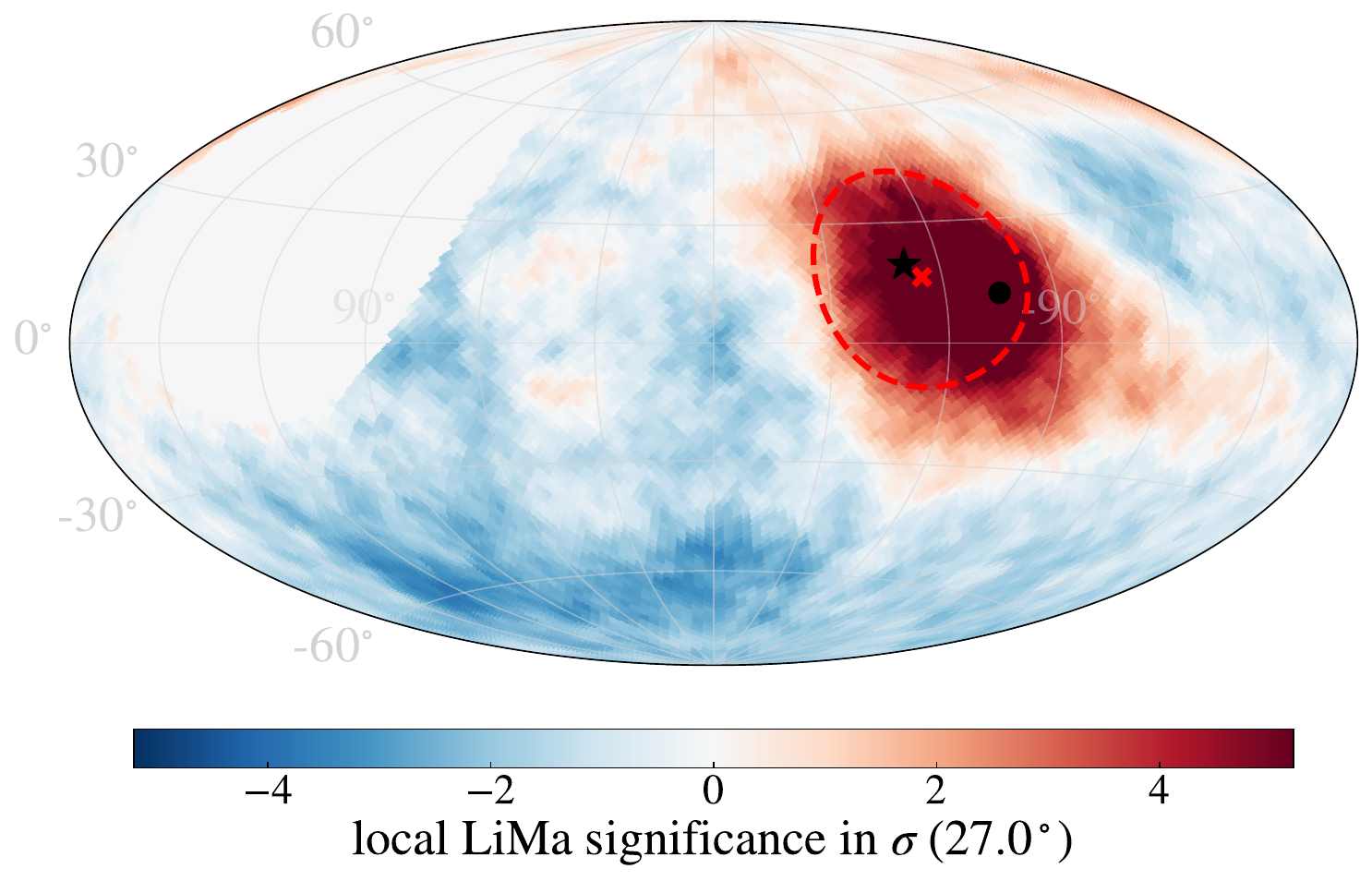}
\includegraphics[width=0.24\textwidth]{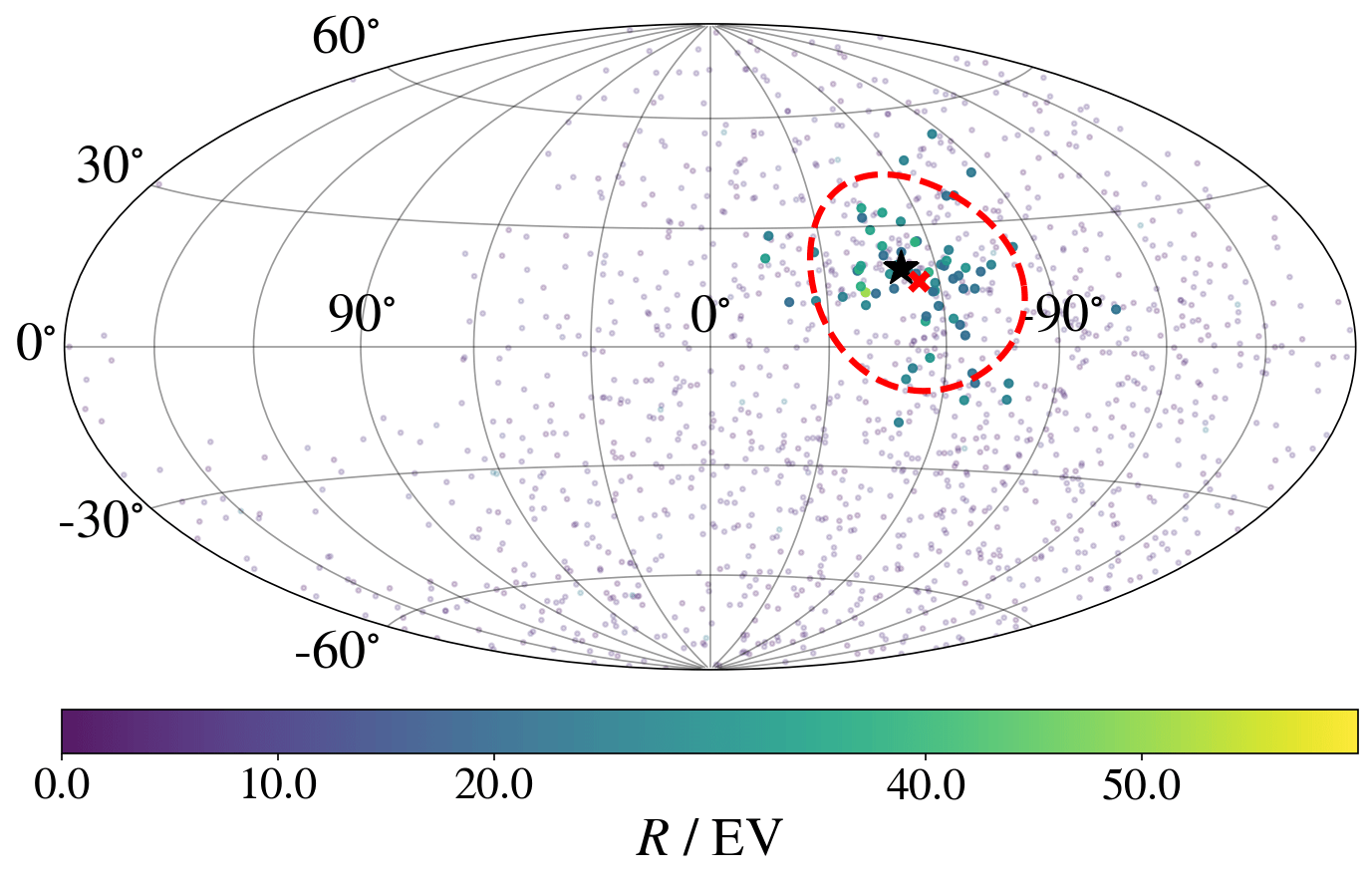}
\includegraphics[width=0.24\textwidth]{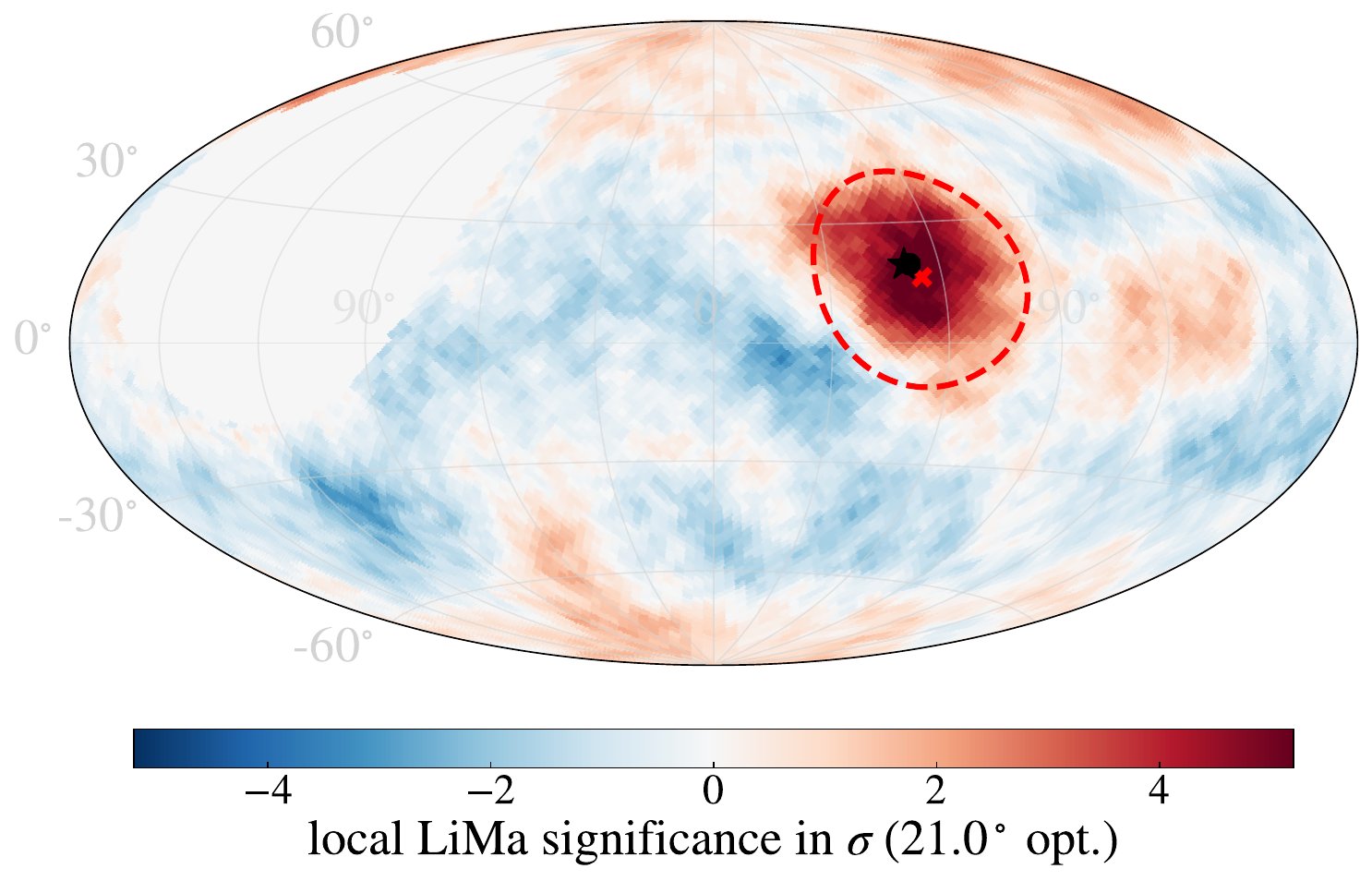}

\includegraphics[width=0.24\textwidth]{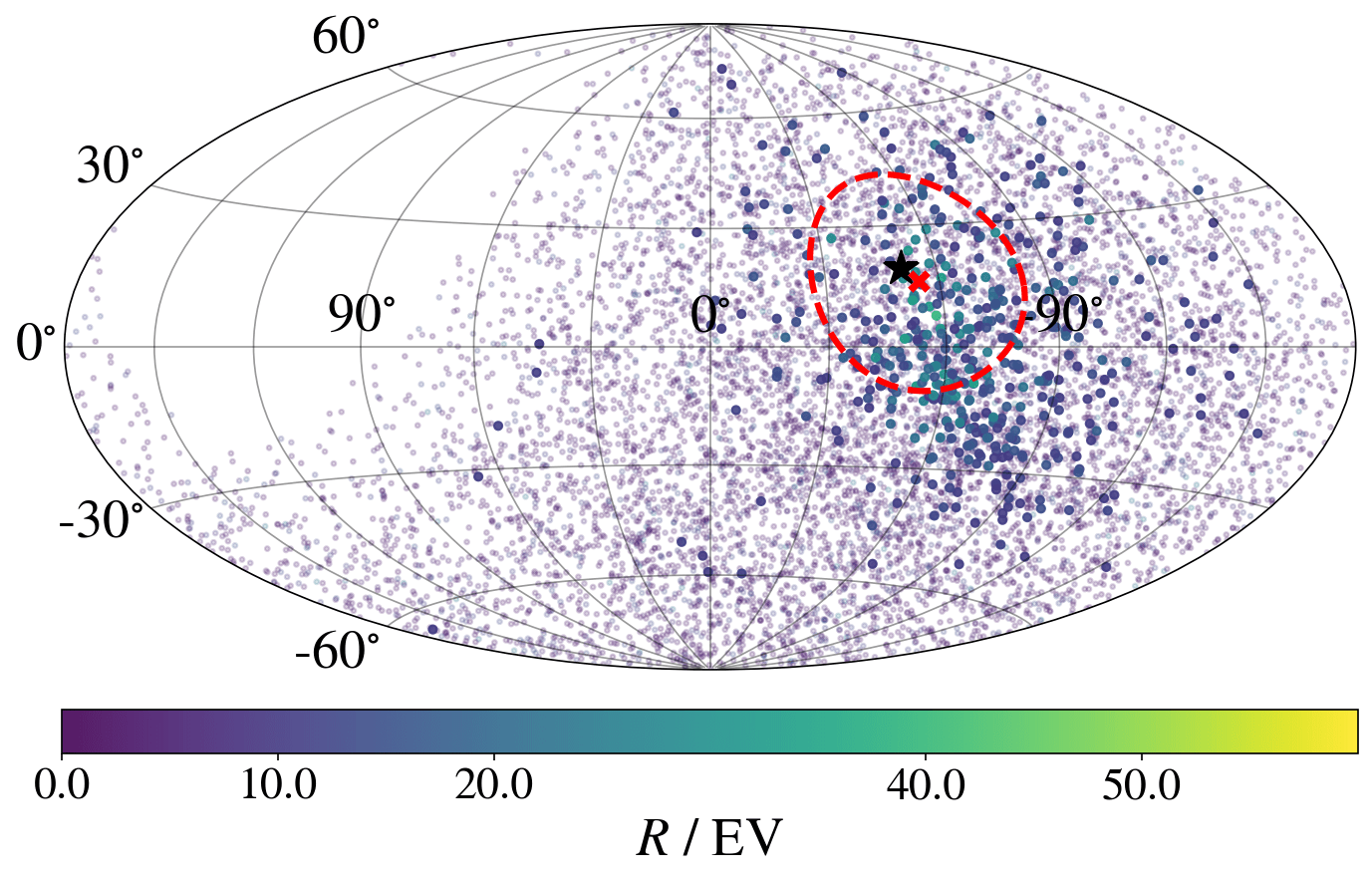}
\includegraphics[width=0.24\textwidth]{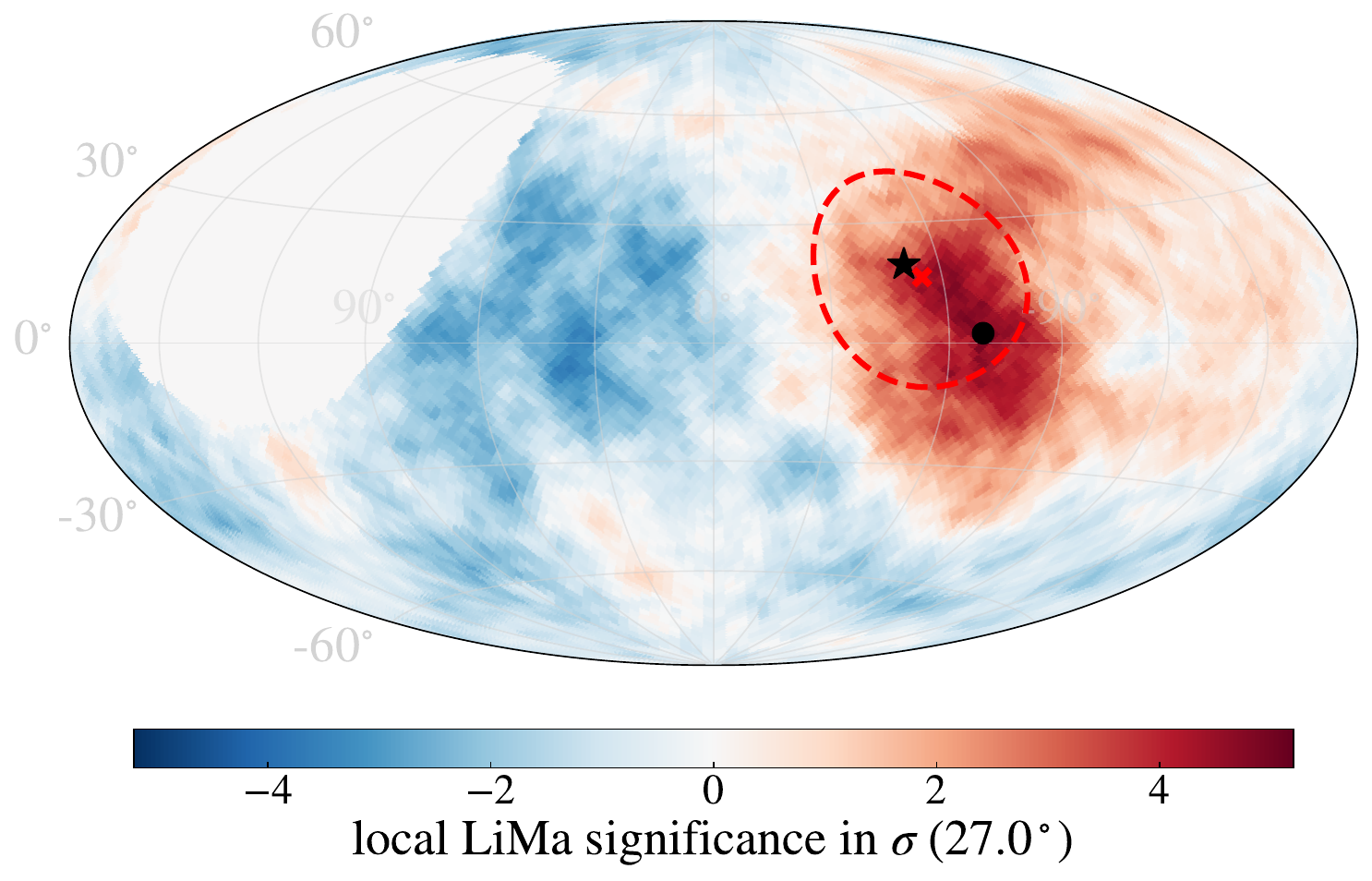}
\includegraphics[width=0.24\textwidth]{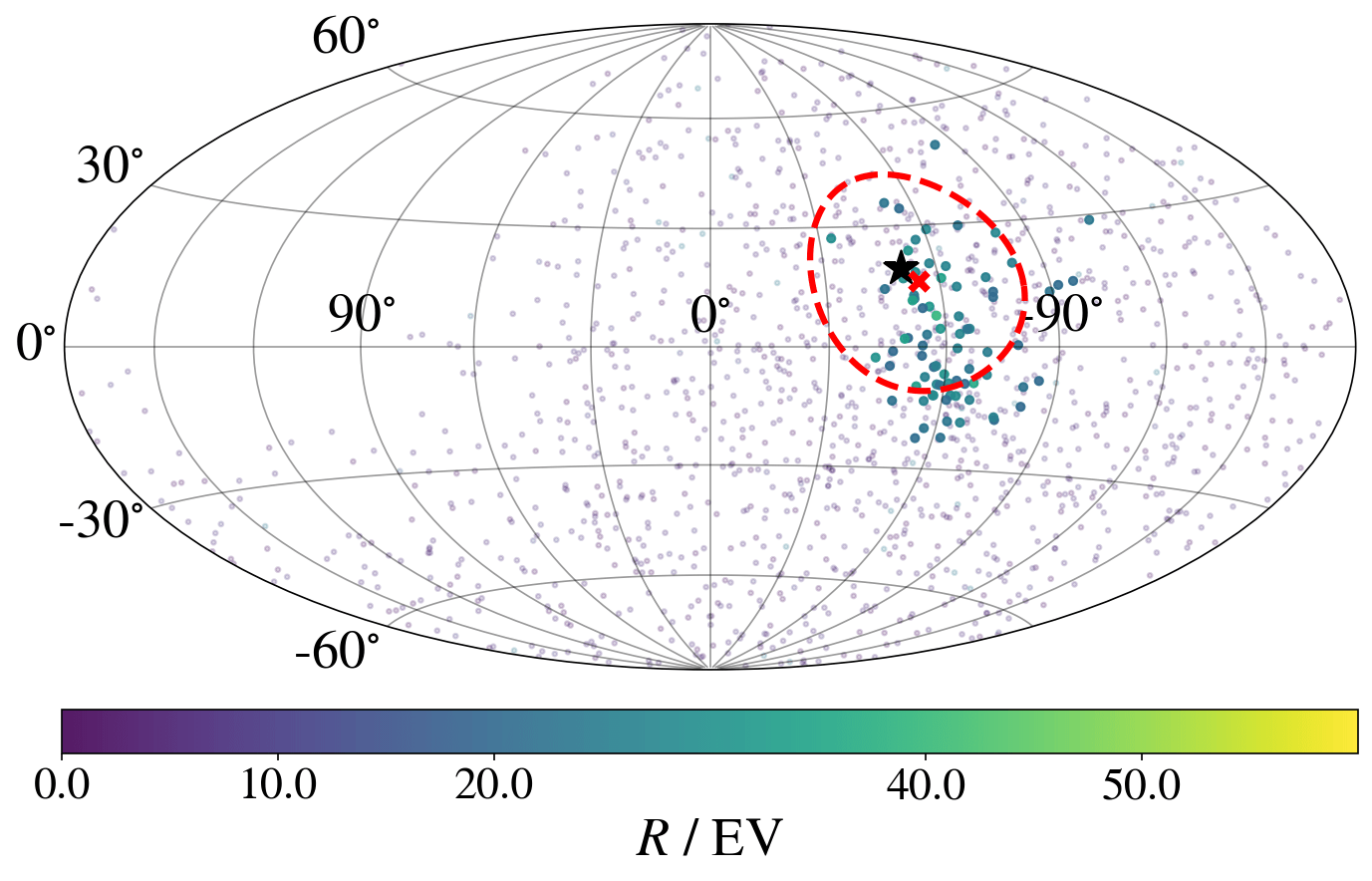}
\includegraphics[width=0.24\textwidth]{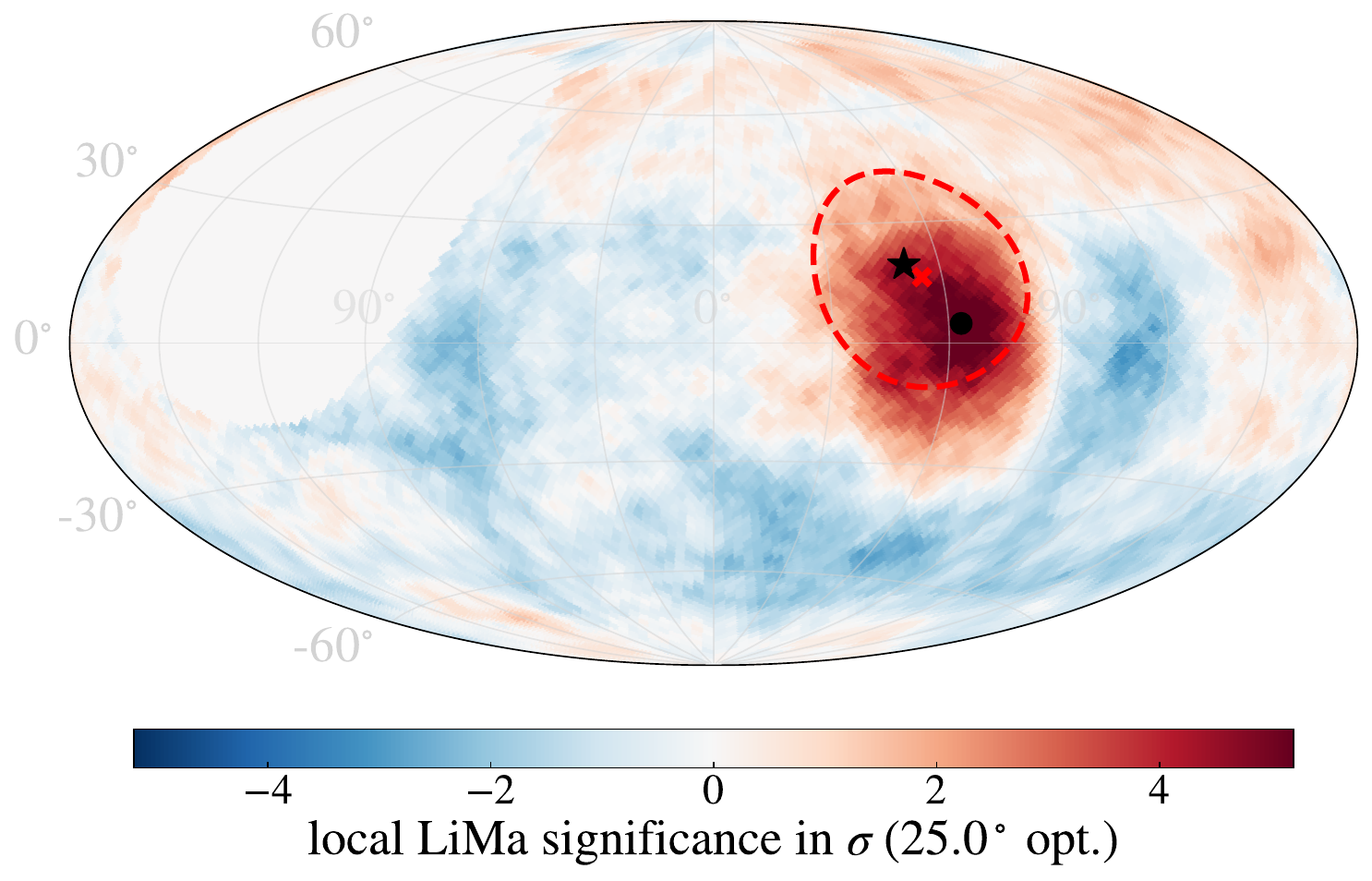}

\includegraphics[width=0.24\textwidth]{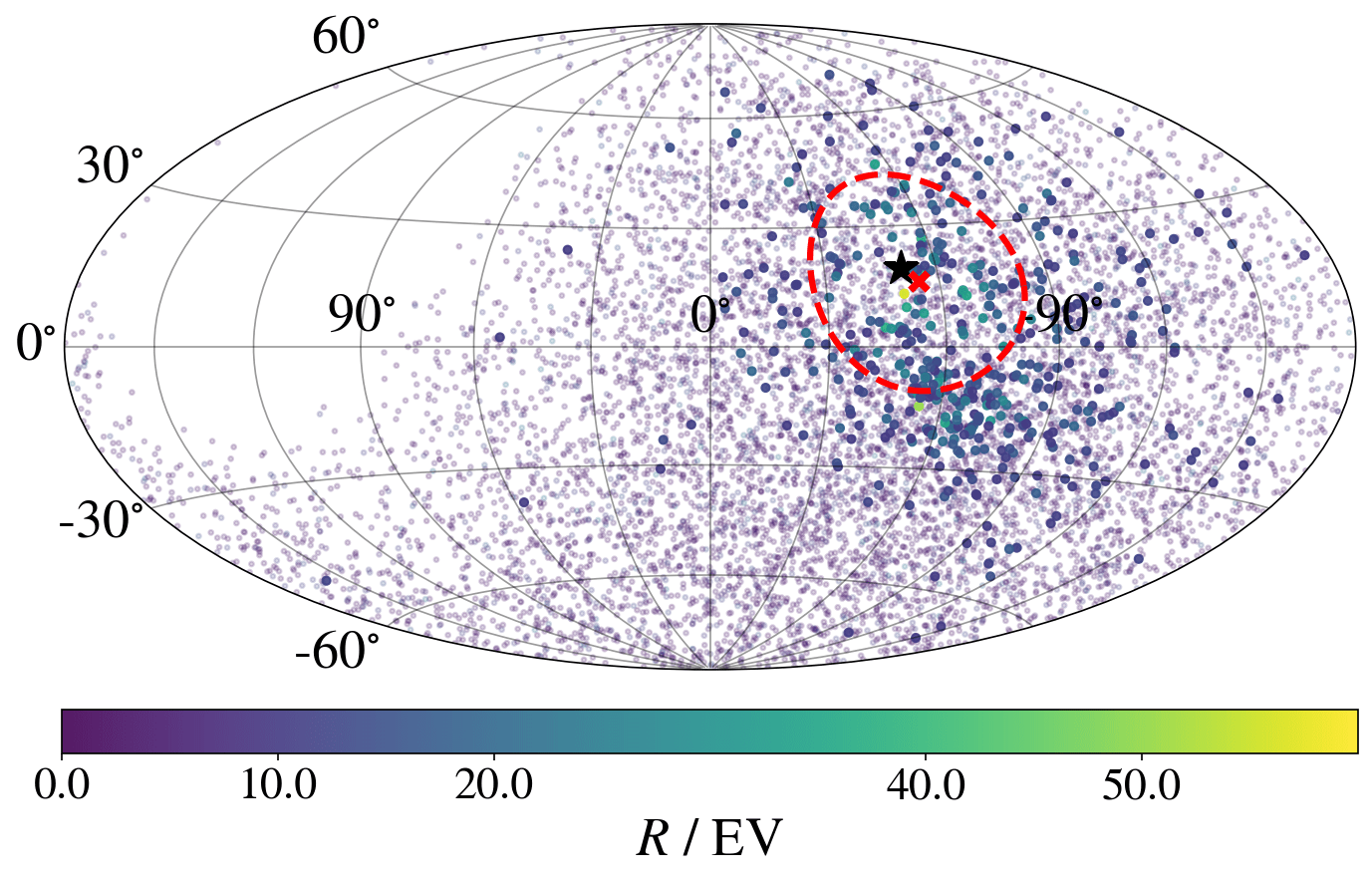}
\includegraphics[width=0.24\textwidth]{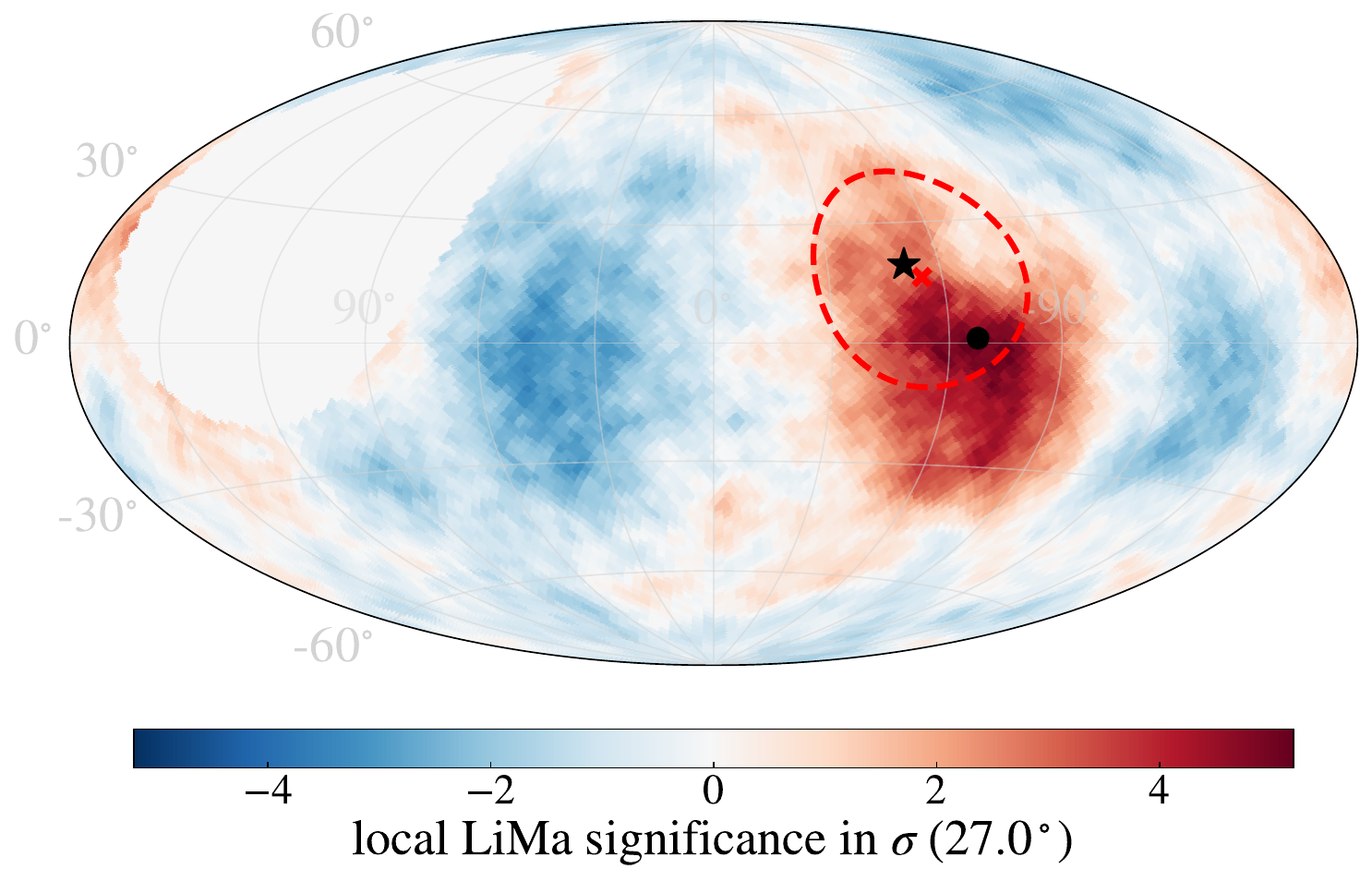}
\includegraphics[width=0.24\textwidth]{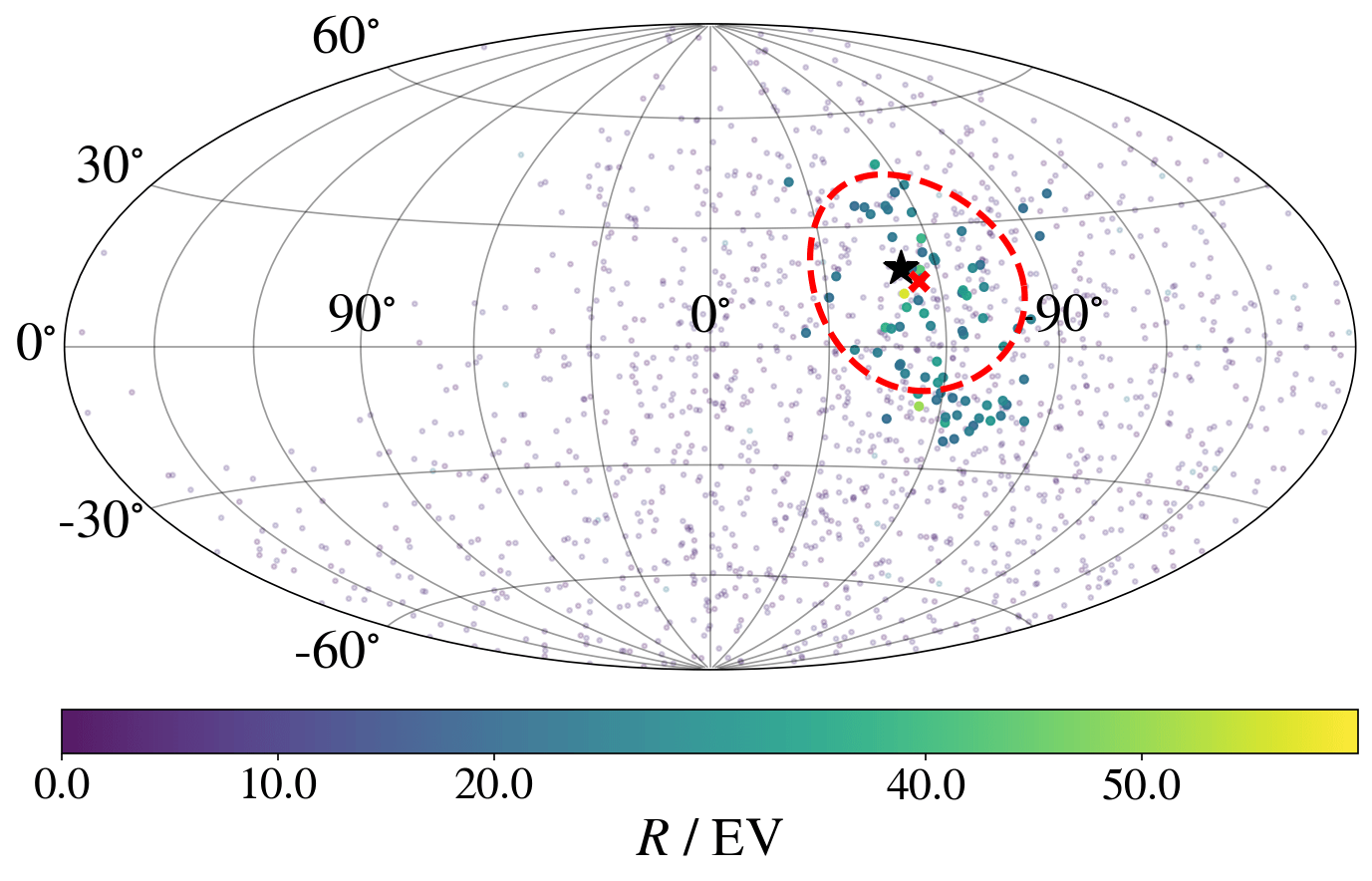}
\includegraphics[width=0.24\textwidth]{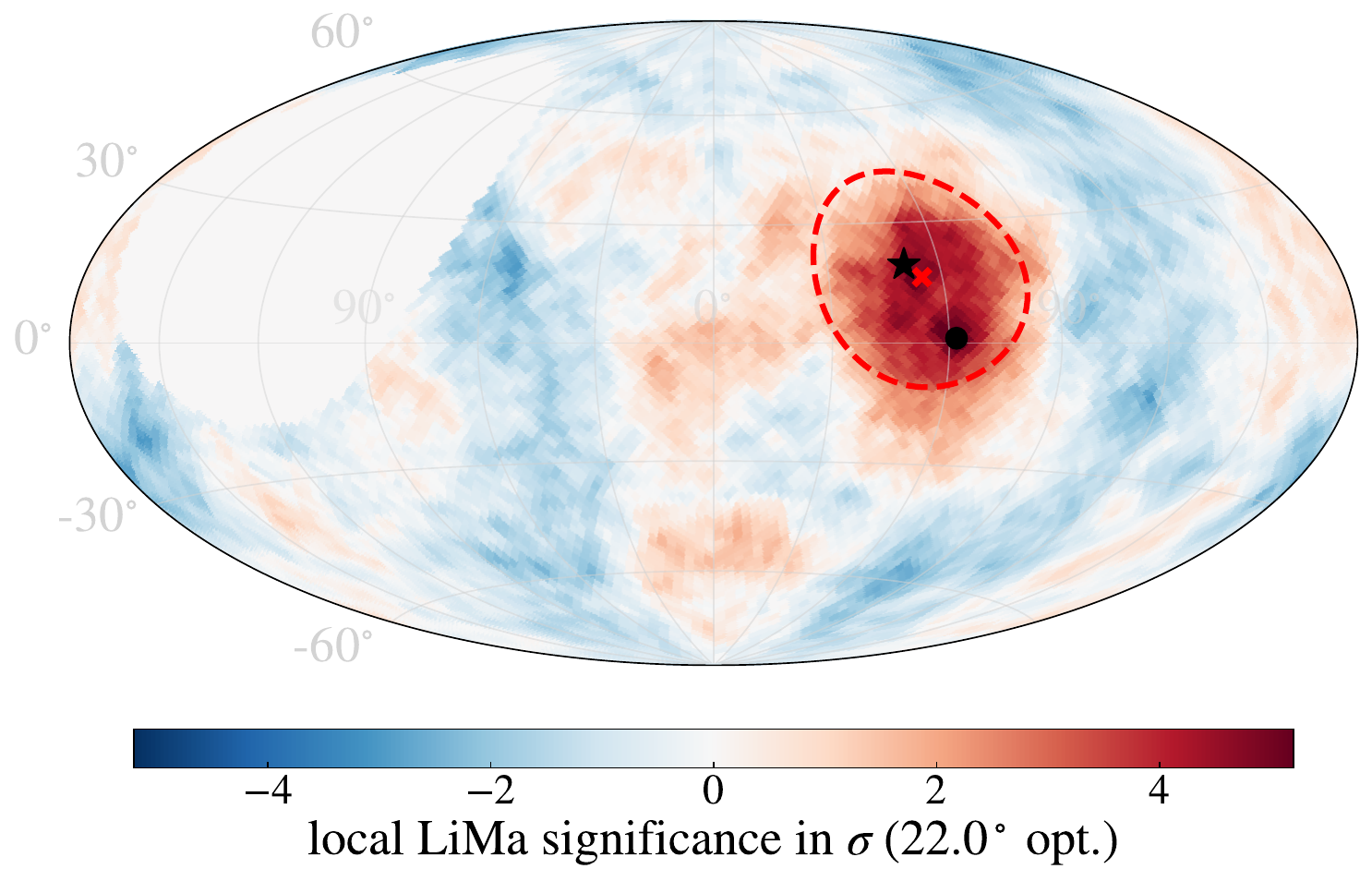}
\caption{Four example simulations with Cen A as the source, with $Z=2$, $f=0.05$, $\beta_\mathrm{EGMF}=15$ (black cross in Fig.~\ref{fig:cena_constraints}). The \textit{upper two rows} are two examples with the \texttt{UF23-base-Pl} GMF model, the \textit{lower two rows} with the \texttt{KST24-Pl} model. The left two figures are for $E_\mathrm{min}=20\,\mathrm{EeV}$, the right two for $E_\mathrm{min}=40\,\mathrm{EeV}$. All skymaps are in Galactic coordinates. The background events are half-transparent. Note that for $E_\mathrm{min}=20\,\mathrm{EeV}$, the angular scale of the LiMa significance scan is not optimized and instead fixed to $27^\circ$ as in~\cite{the_pierre_auger_collaboration_flux_2024}.}
\label{fig:cena_examples}
\end{figure}

In Fig.~\ref{fig:cena_examples}, example simulations are shown for $f=0.05$, $Z=2$, and $\beta_\mathrm{EGMF}=15$. The direction of the overdensity is well reproduced with the \texttt{UF23-base-Pl} GMF model. With the \texttt{KST24-Pl} model, however, coherent deflections are already too large for $Z=2$ (see also Fig.~\ref{fig:ADs}), so that the excess direction is displaced by $\sim25^\circ$ from the observed one for all 20 simulations based on the \texttt{KST24} model. 

\begin{figure}[ht!]
\includegraphics[height=5.3cm]{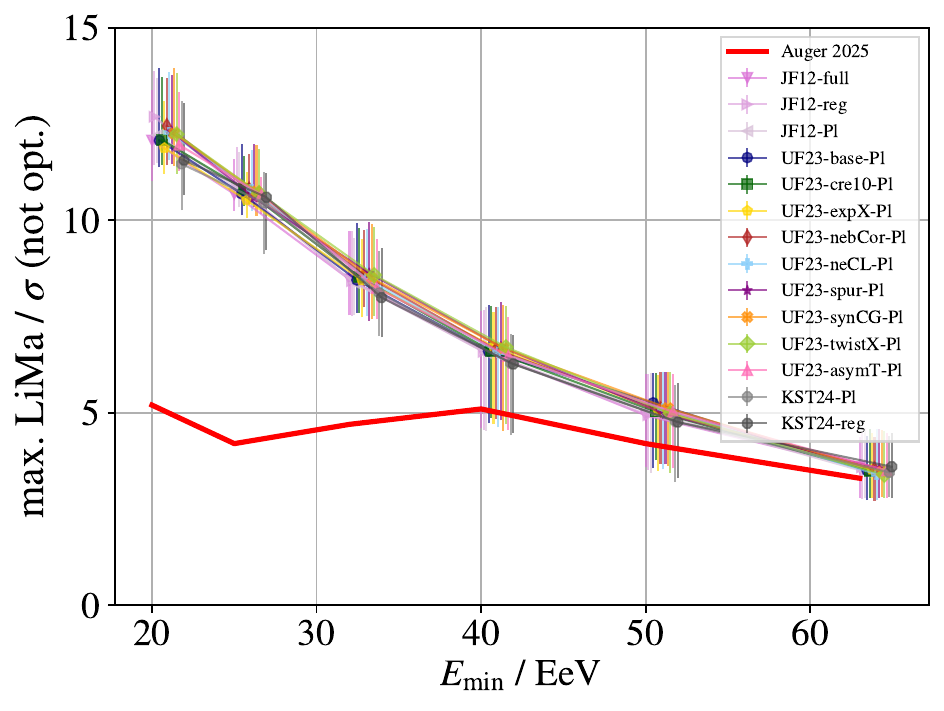}
\includegraphics[height=5.3cm]{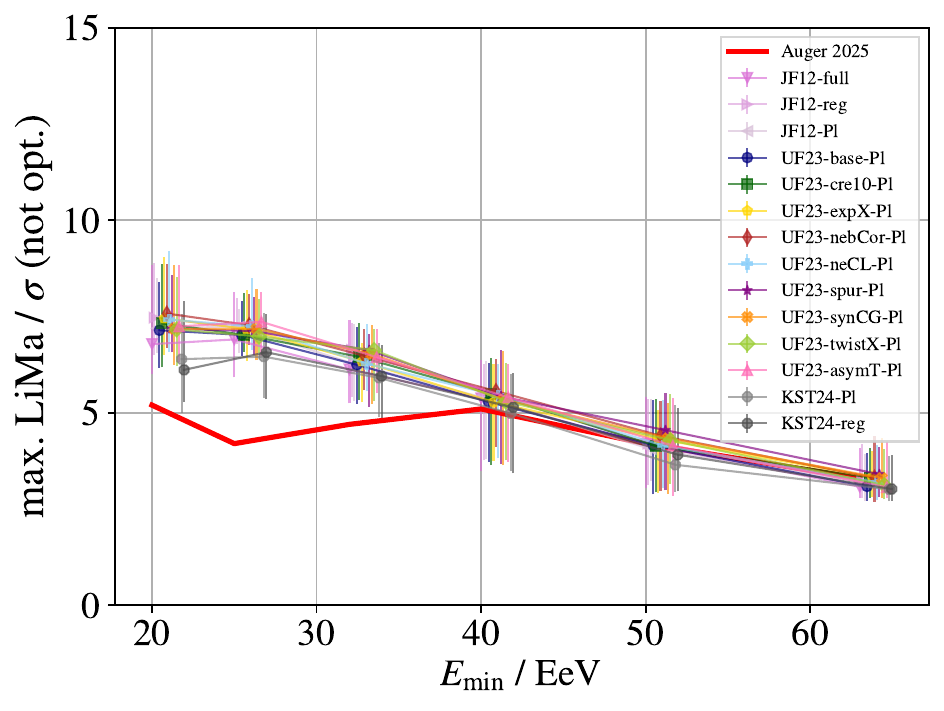}
\caption{Dependency of the maximum LiMa significance on the energy threshold for Cen A as the source with $f=0.05$, $\beta_\mathrm{EGMF}=15$ and $Z=1$ (\textit{left}) or $Z=2$ (\textit{right}). The error bars represent the $90\%$ uncertainty from the 10 random variations per GMF model. The red line denotes the significance evolution observed on data from~\cite{the_pierre_auger_collaboration_flux_2024}. The angular size was kept fixed to $27^\circ$ as in~\cite{the_pierre_auger_collaboration_flux_2024}. The markers are artificially offset on the x-axis for better readability.}
\label{fig:cena_signi}
\end{figure}

The energy dependency of the maximum significance is shown in Fig.~\ref{fig:cena_signi}. For energy thresholds $E_\mathrm{min}\gtrsim40\,\mathrm{EeV}$, the observed energy dependency is well captured by the simulations. For all energy thresholds $E_\mathrm{min}\lesssim40\,\mathrm{EeV}$, however, the maximum significance of the simulations is significantly larger than the observed one~\cite{the_pierre_auger_collaboration_flux_2024} for all GMF models. This indicates that the source energy spectrum should be harder than the Auger spectrum that was used to draw the source energies (see sec.~\ref{sec:simulation}), so that there are less events from the source below $40\,\mathrm{EeV}$. But, because the significance is too large (making the energy evolution of the excess direction more stable, see Fig.~\ref{fig:cena_AD}) instead of too small, this means that the values $\langle\theta\rangle_\mathrm{min}$ shown in Fig.~\ref{fig:cena_constraints} represent an upper limit, so that the constraints on the parameters placed in this work are conservative.

\section{Scenario II - deflected events from a source outside the Centaurus region}  \label{sec:scen2}
In~\cite{he_evidence_2024}, an energy-ordering (\textit{multiplet}) of the events in the excess region was detected, From the event distribution, they calculated an estimate for the coherent and random magnetic field deflections, and conclude that the source should be the \textit{Sombrero galaxy} (M104). Yet, no check was performed to see if the inferred magnetic field displacements are achievable with current GMF models, and what charge the events would need to have. 

\begin{figure}[ht!]
\includegraphics[trim={13cm 0 0 0}, clip, width=0.24\textwidth]{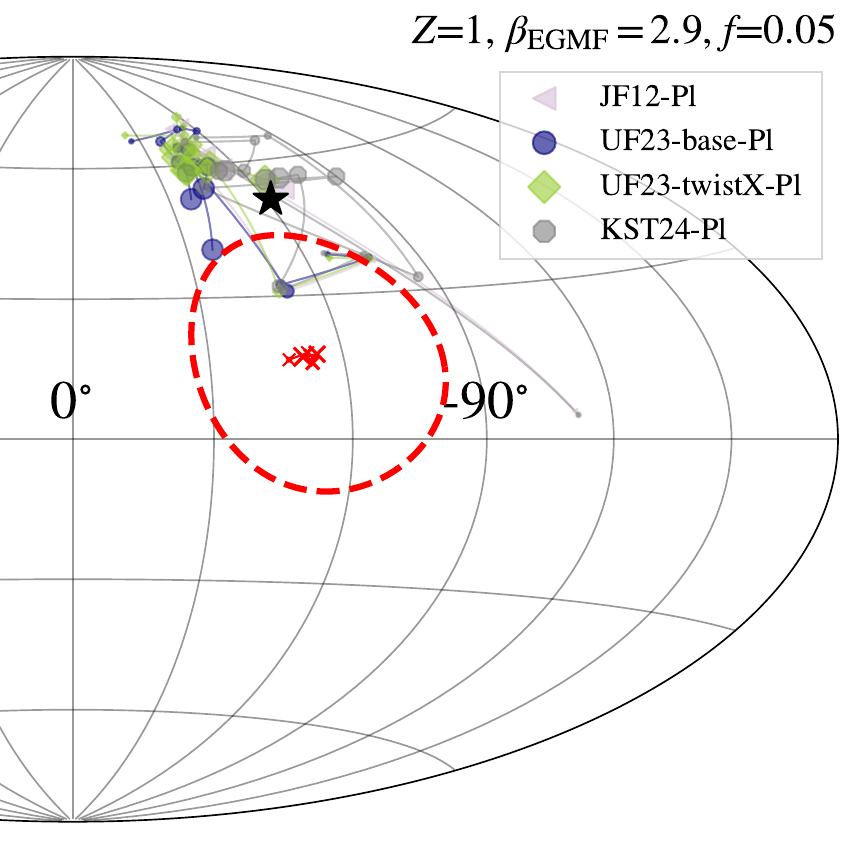}
\includegraphics[trim={13cm 0 0 0}, clip, width=0.24\textwidth]{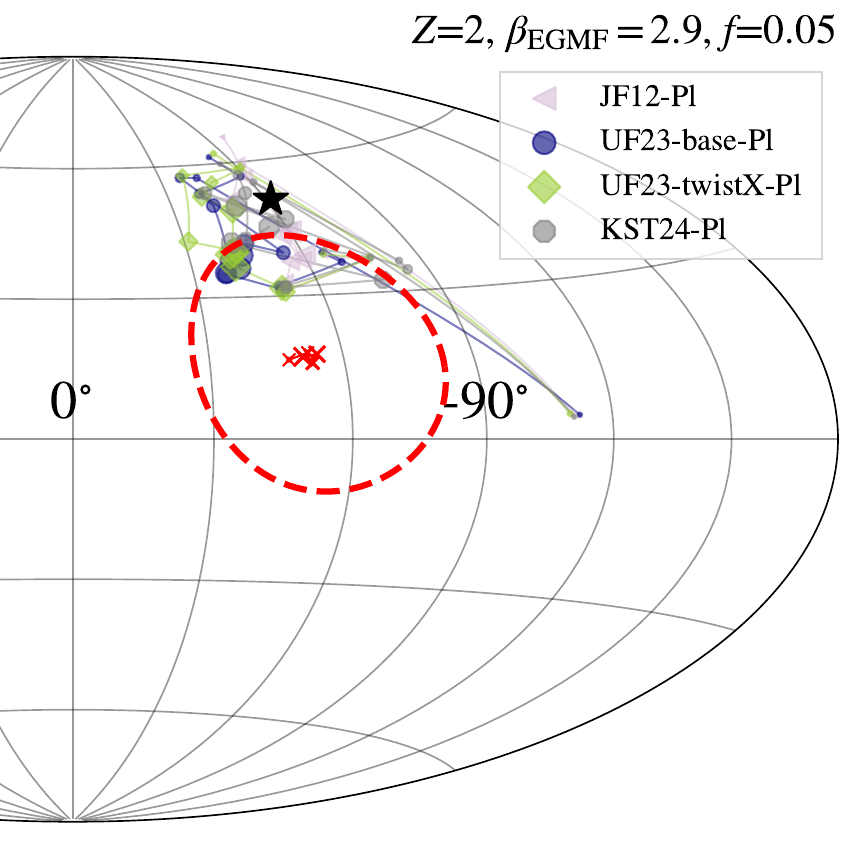}
\includegraphics[trim={13cm 0 0 0}, clip, width=0.24\textwidth]{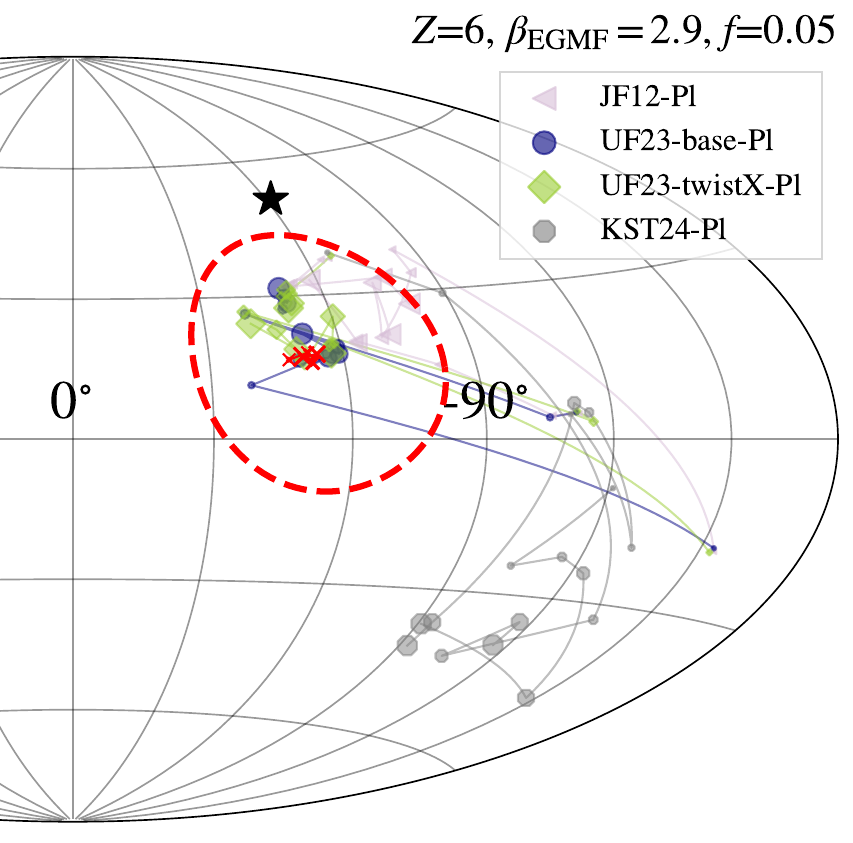}
\includegraphics[trim={13cm 0 0 0}, clip, width=0.24\textwidth]{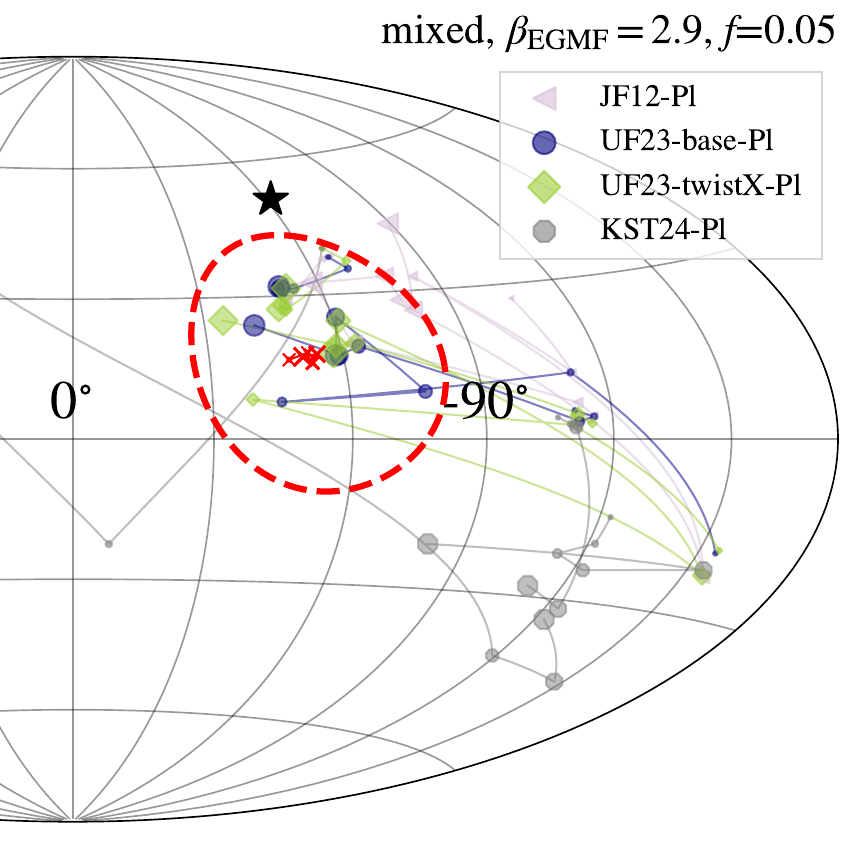}
\caption{Evolution of the excess direction with the energy for random example simulations with the Sombrero galaxy (star marker) as the source, $f=0.05$, $\beta_\mathrm{EGMF}=2.9$, using different charges (see figure title) and four different GMF models. For more details see Fig.~\ref{fig:cena_AD}.}
\label{fig:sombrero_AD}
\end{figure}

Using the same framework as for Cen A in the previous section, the predicted excess directions for the Sombrero galaxy are shown in Fig.~\ref{fig:sombrero_AD} for different charges and GMF models. As expected, too small charge numbers do not lead to large enough coherent displacements. But, for $Z\sim6$ or mixed composition, the simulated excess direction is in very good agreement with the observed one (again for all GMF models apart from \texttt{KST24}). For $Z\simeq12$, small parts of the parameter space are also possible.

\begin{figure}[ht!]
\includegraphics[width=0.49\textwidth]{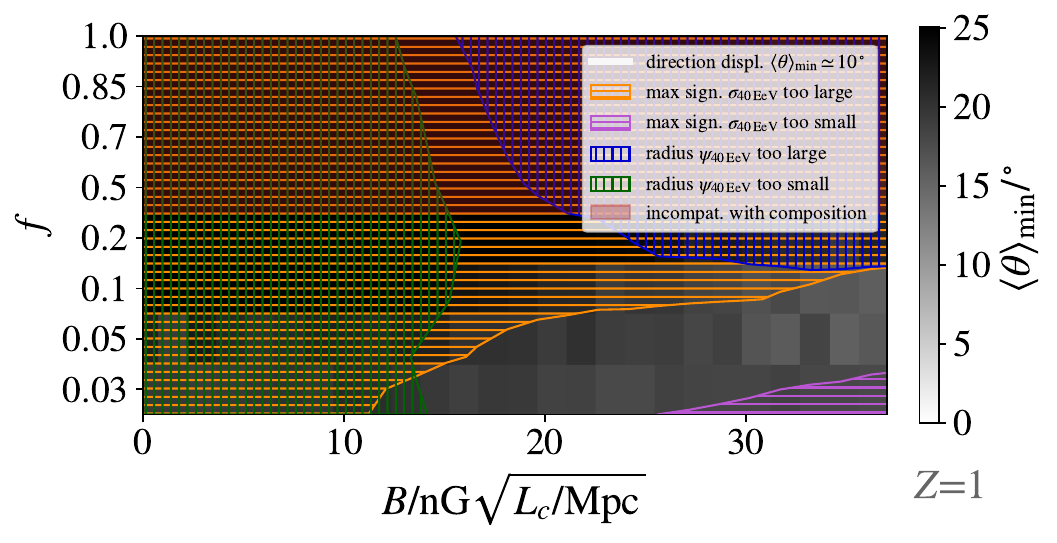}
\includegraphics[width=0.49\textwidth]{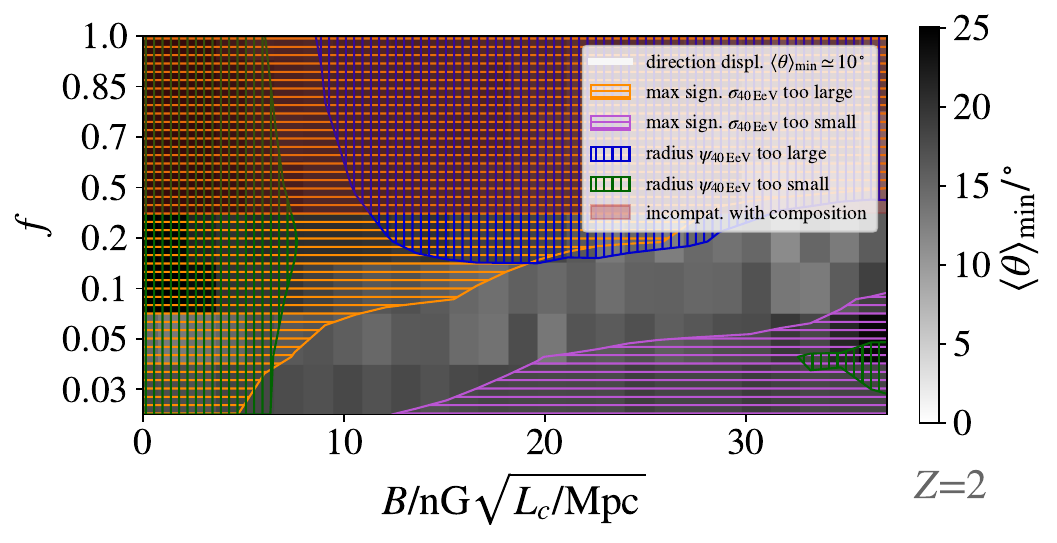}\\
\includegraphics[width=0.49\textwidth]{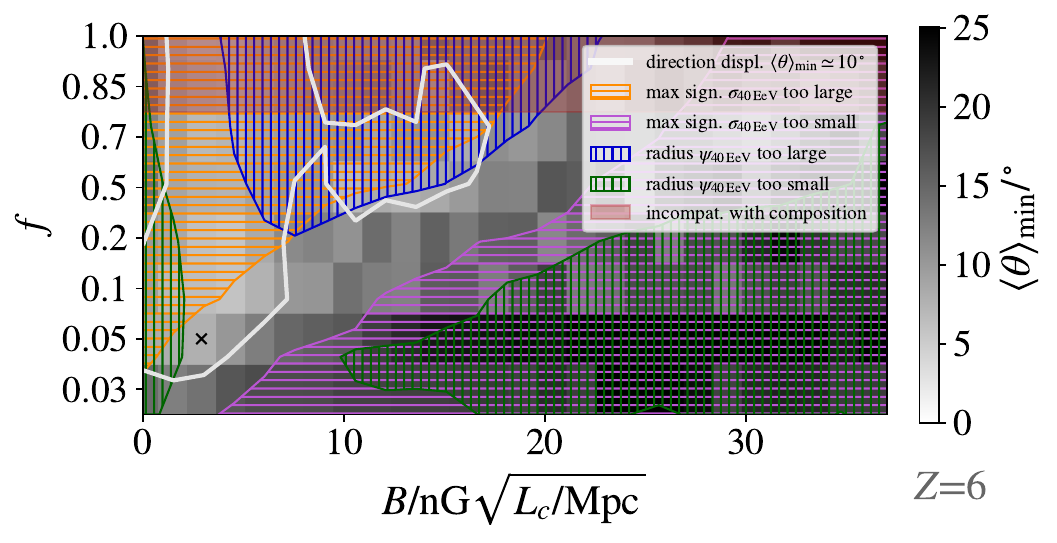}
\includegraphics[width=0.49\textwidth]{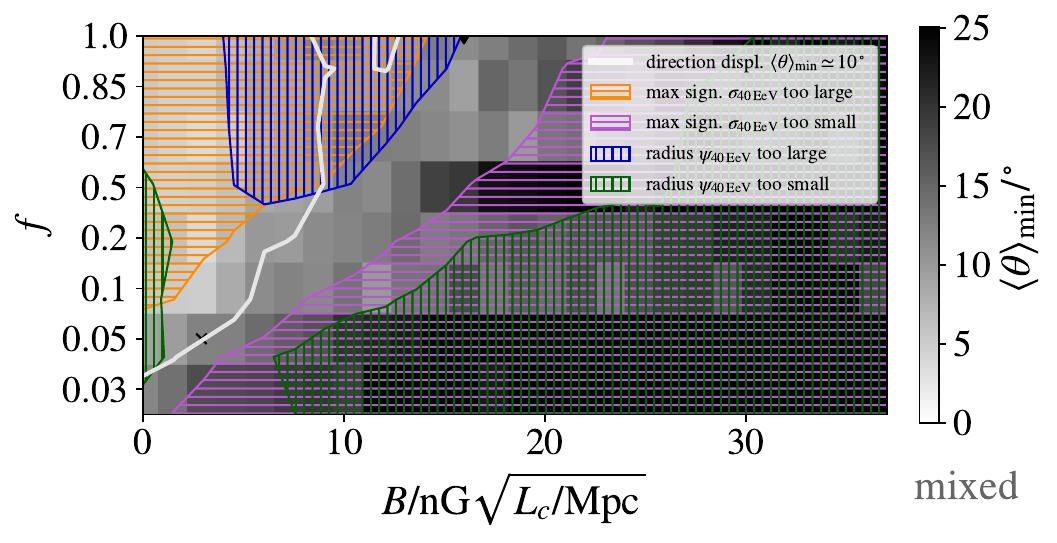}\\
\includegraphics[width=0.49\textwidth]{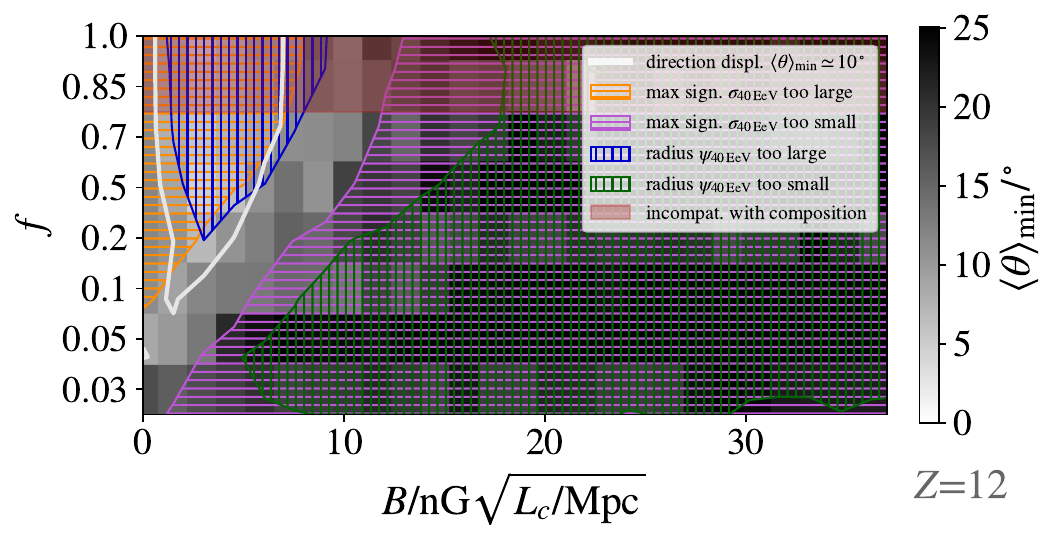}
\includegraphics[width=0.49\textwidth]{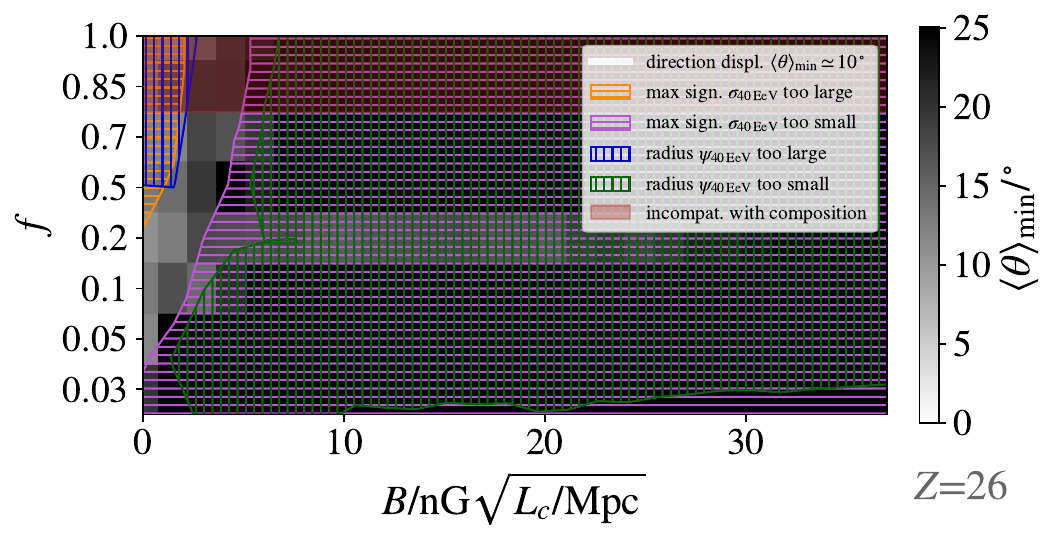}
\caption{Constraints on the parameter space for the Sombrero galaxy as the source of the observed excess. See Fig.~\ref{fig:cena_constraints} for more details.}
\label{fig:sombrero_constraints}
\end{figure}

\begin{figure}[ht!]
\includegraphics[width=0.24\textwidth]{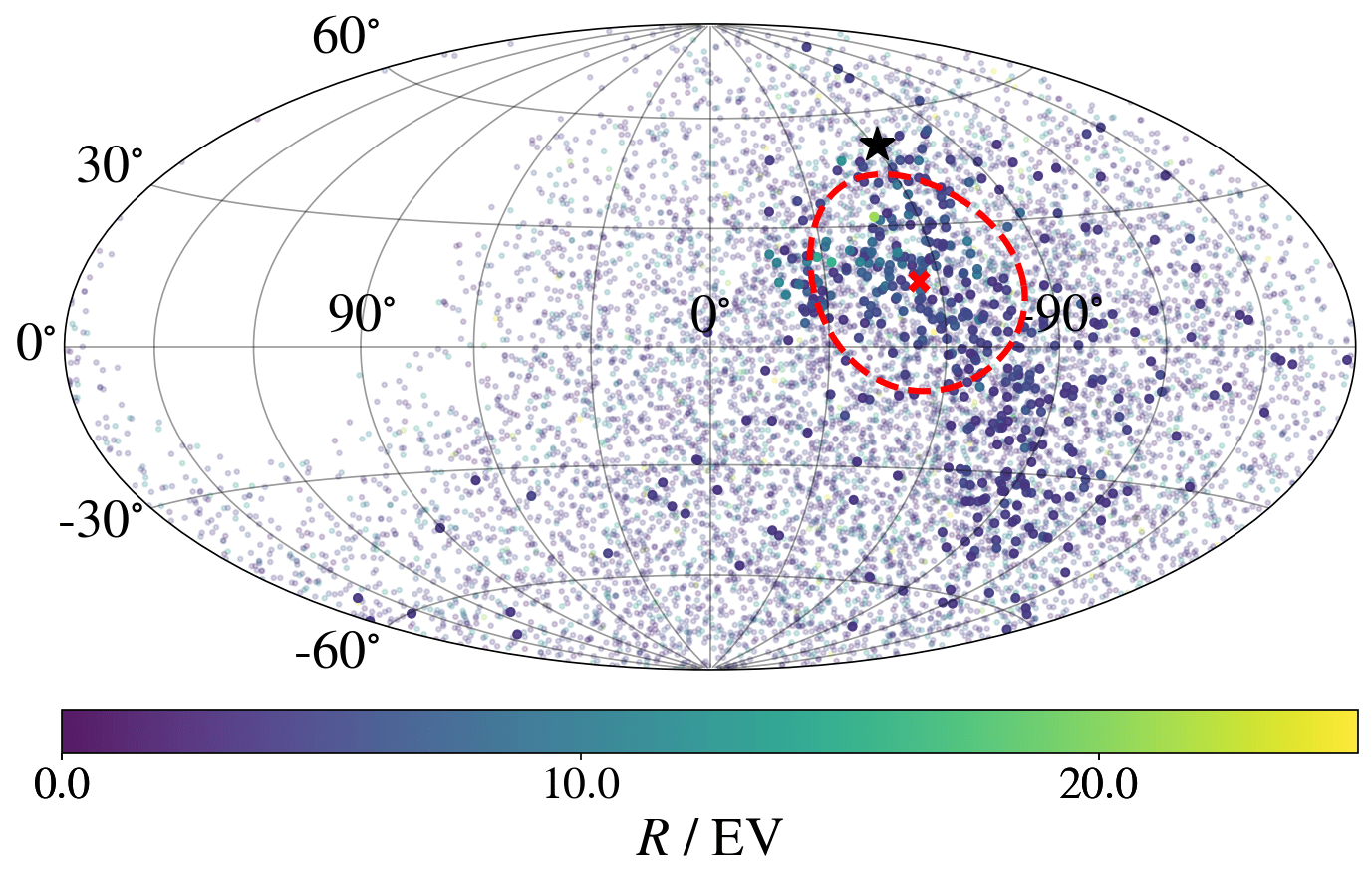}
\includegraphics[width=0.24\textwidth]{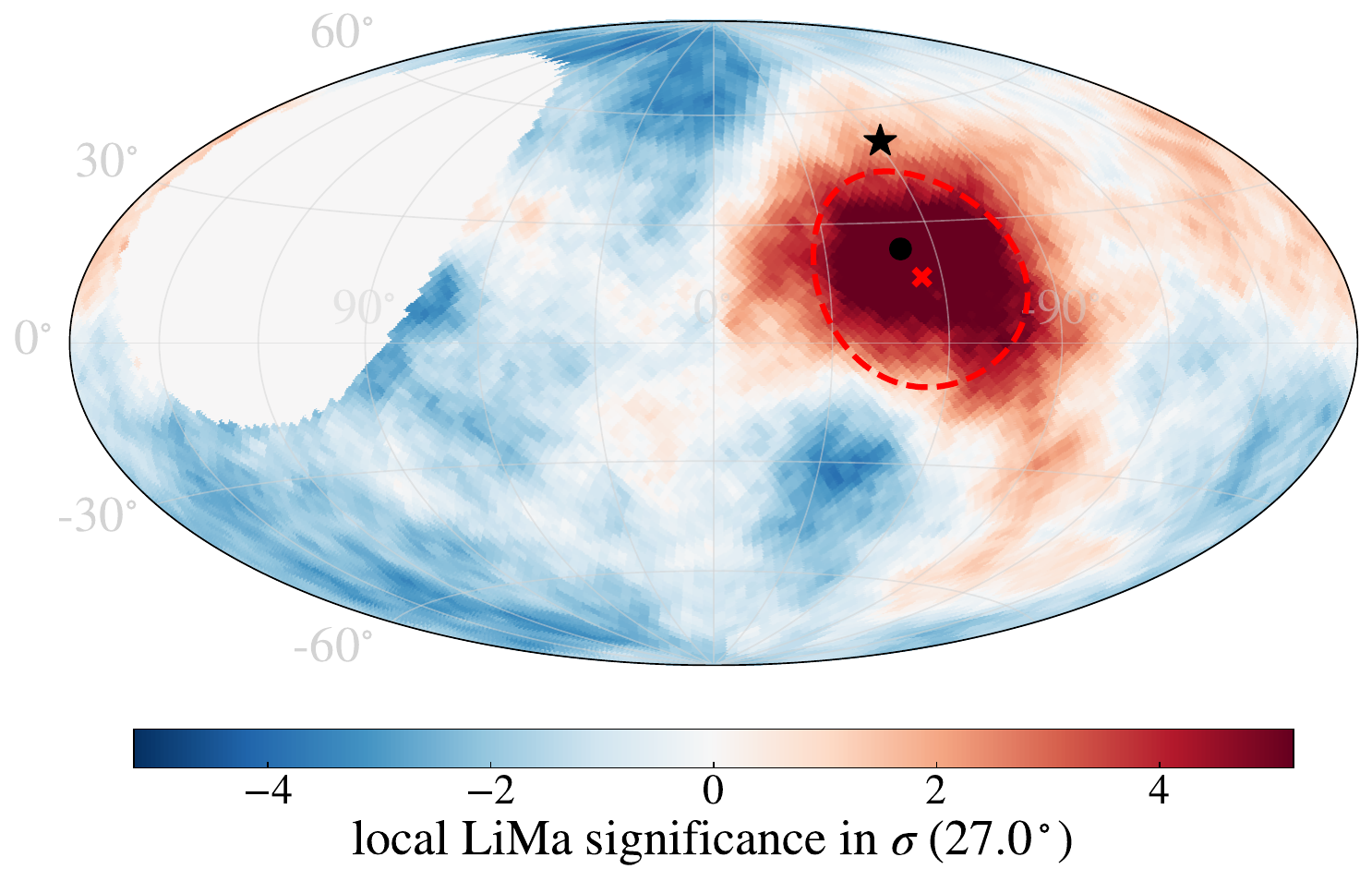}
\includegraphics[width=0.24\textwidth]{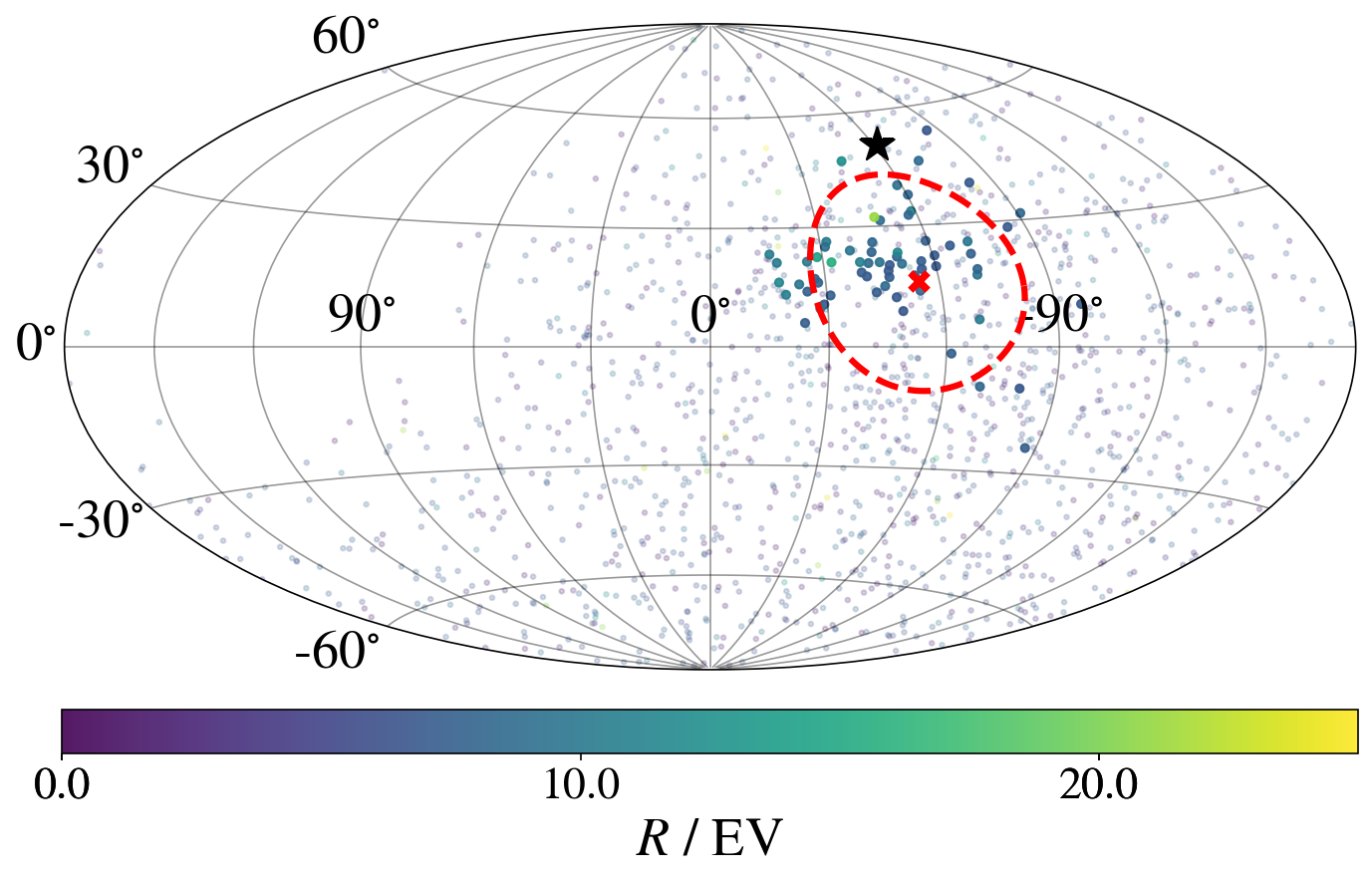}
\includegraphics[width=0.24\textwidth]{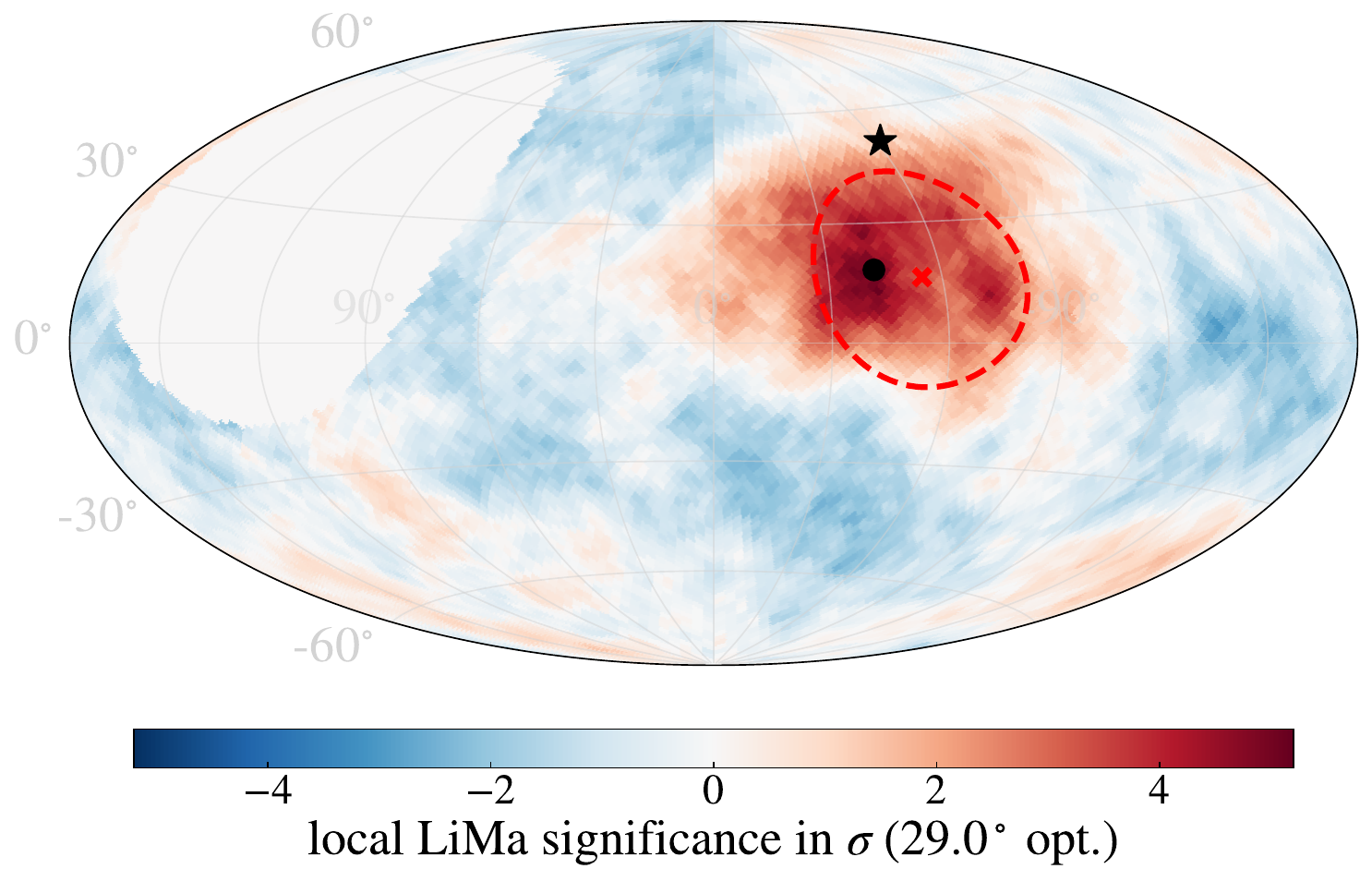}

\includegraphics[width=0.24\textwidth]{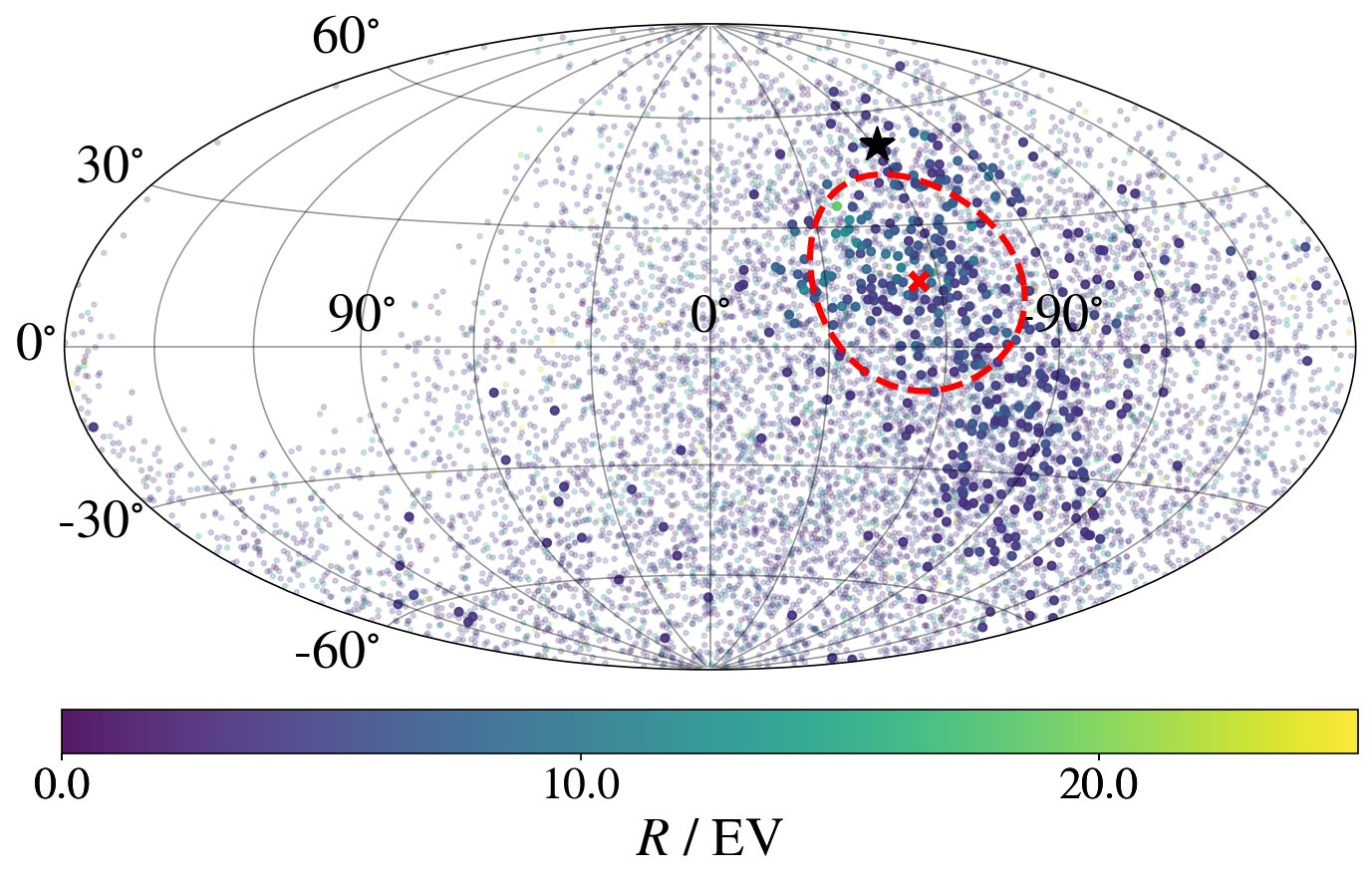}
\includegraphics[width=0.24\textwidth]{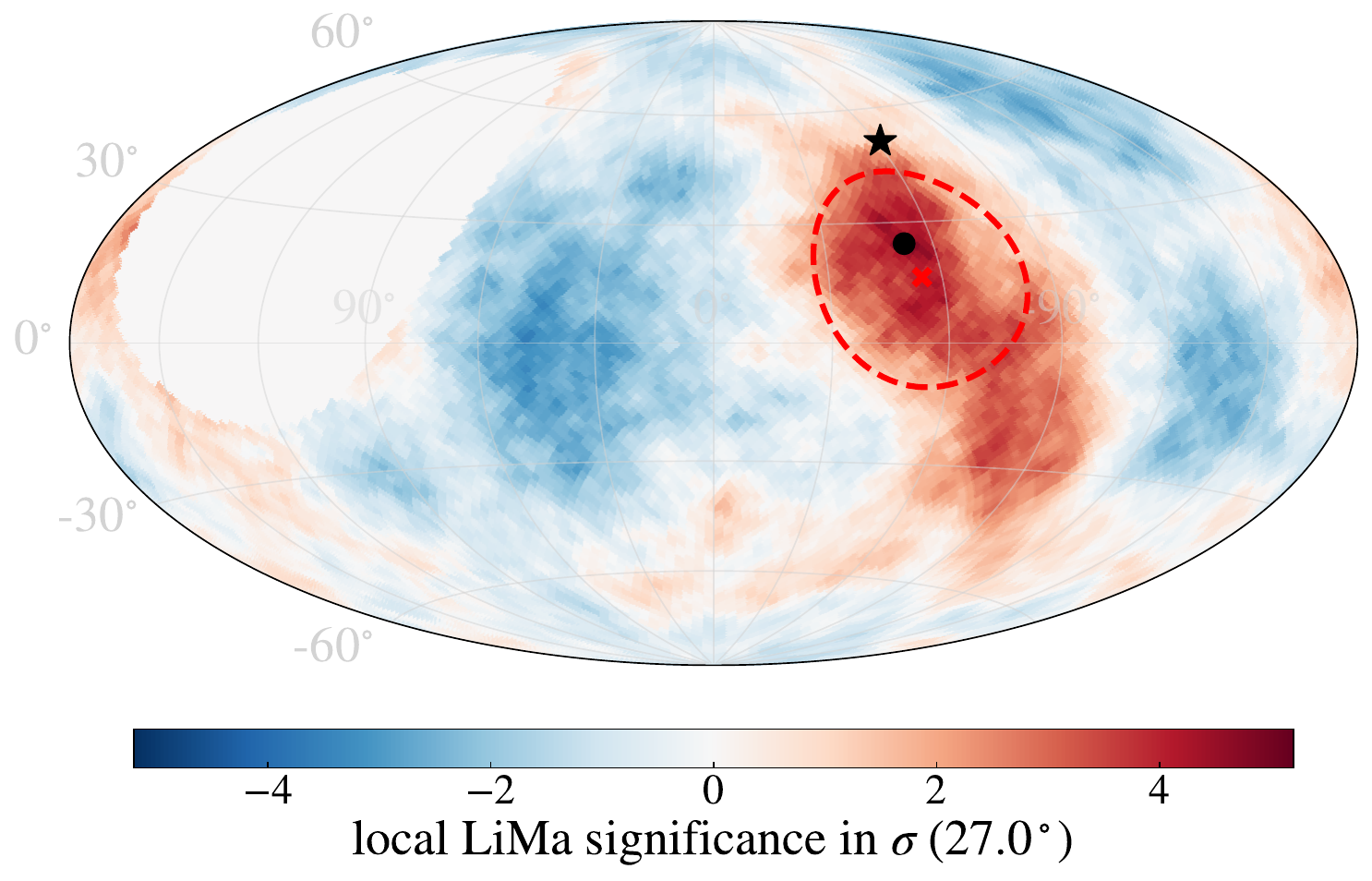}
\includegraphics[width=0.24\textwidth]{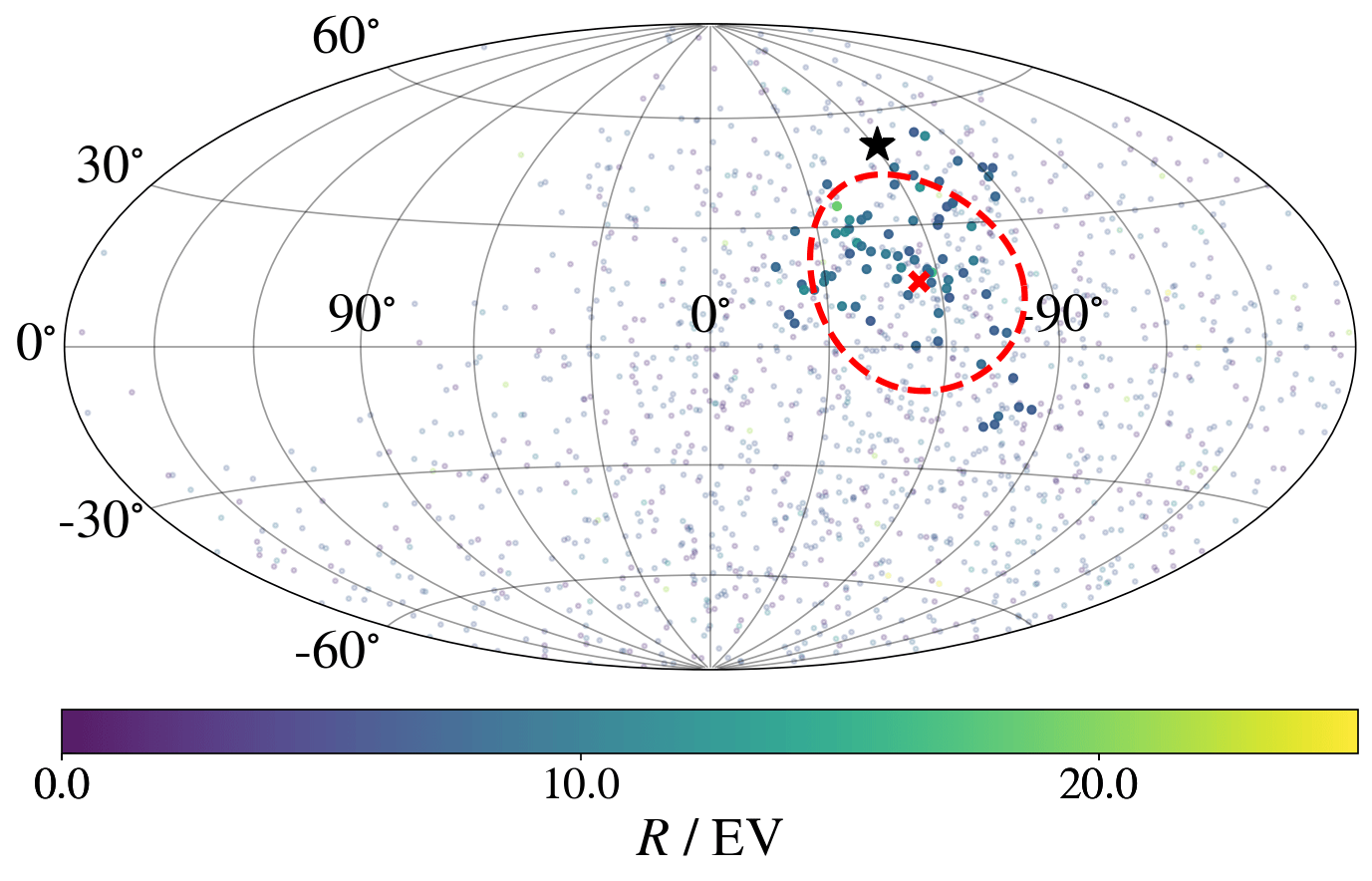}
\includegraphics[width=0.24\textwidth]{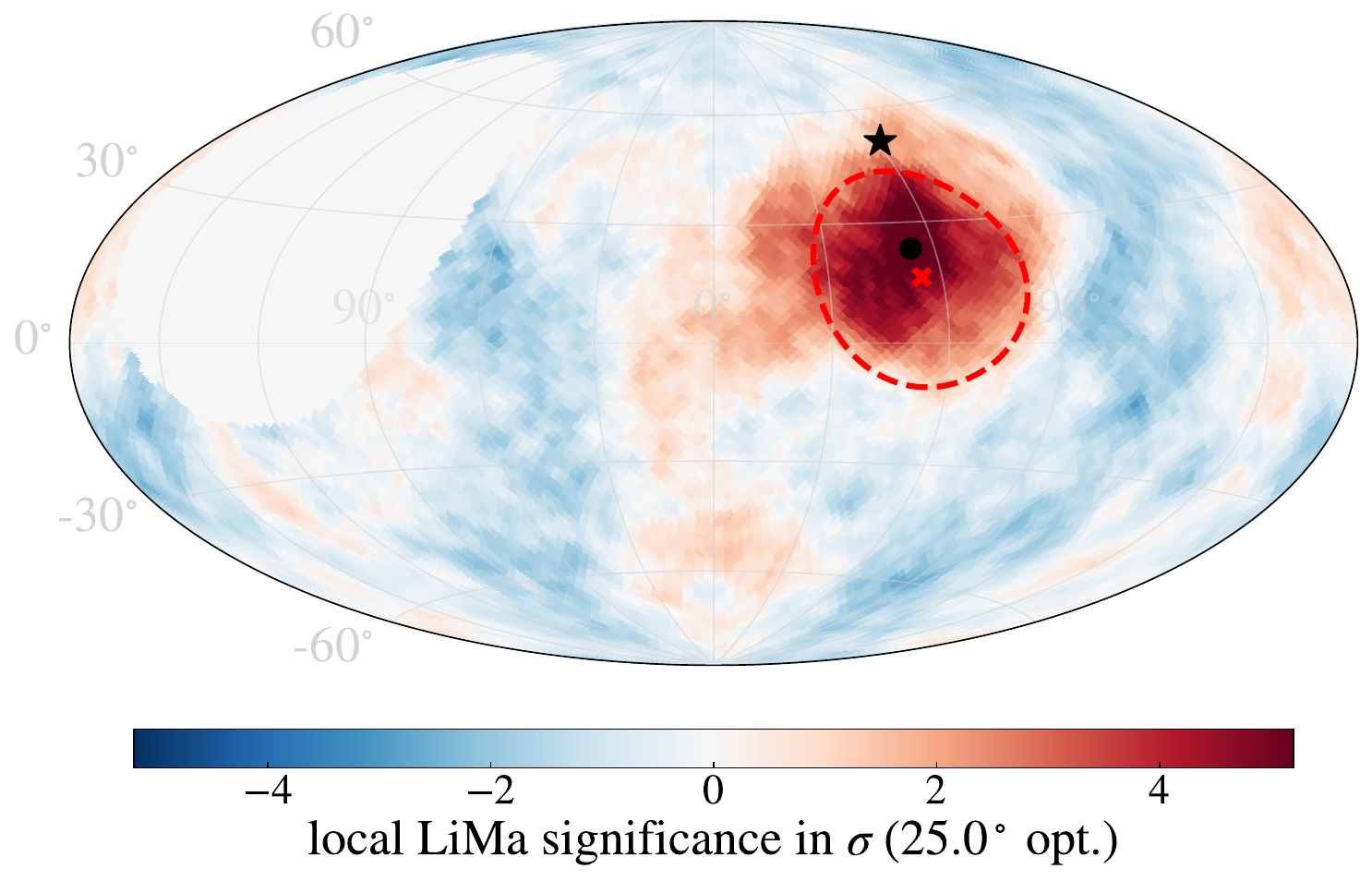}

\includegraphics[width=0.24\textwidth]{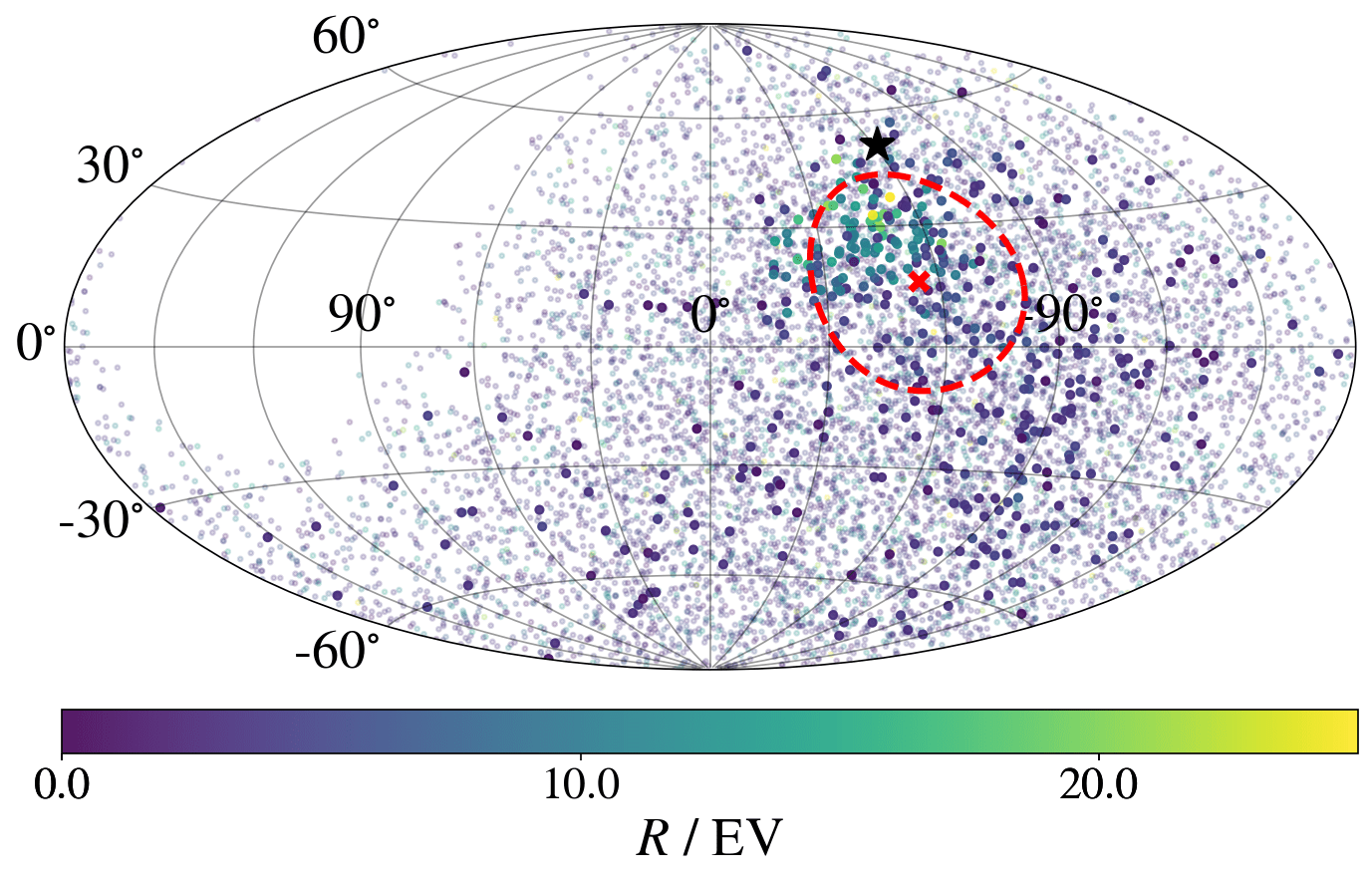}
\includegraphics[width=0.24\textwidth]{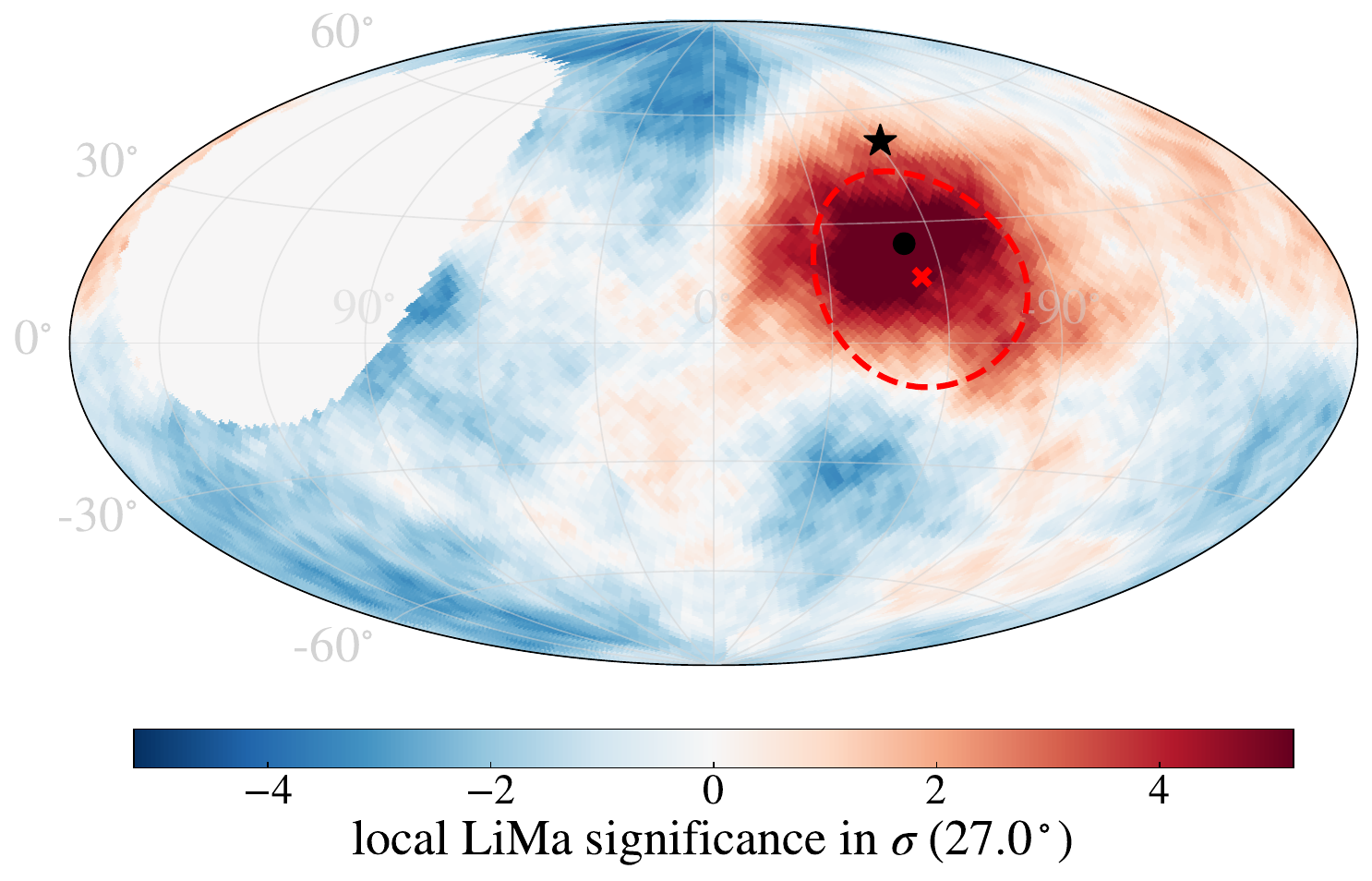}
\includegraphics[width=0.24\textwidth]{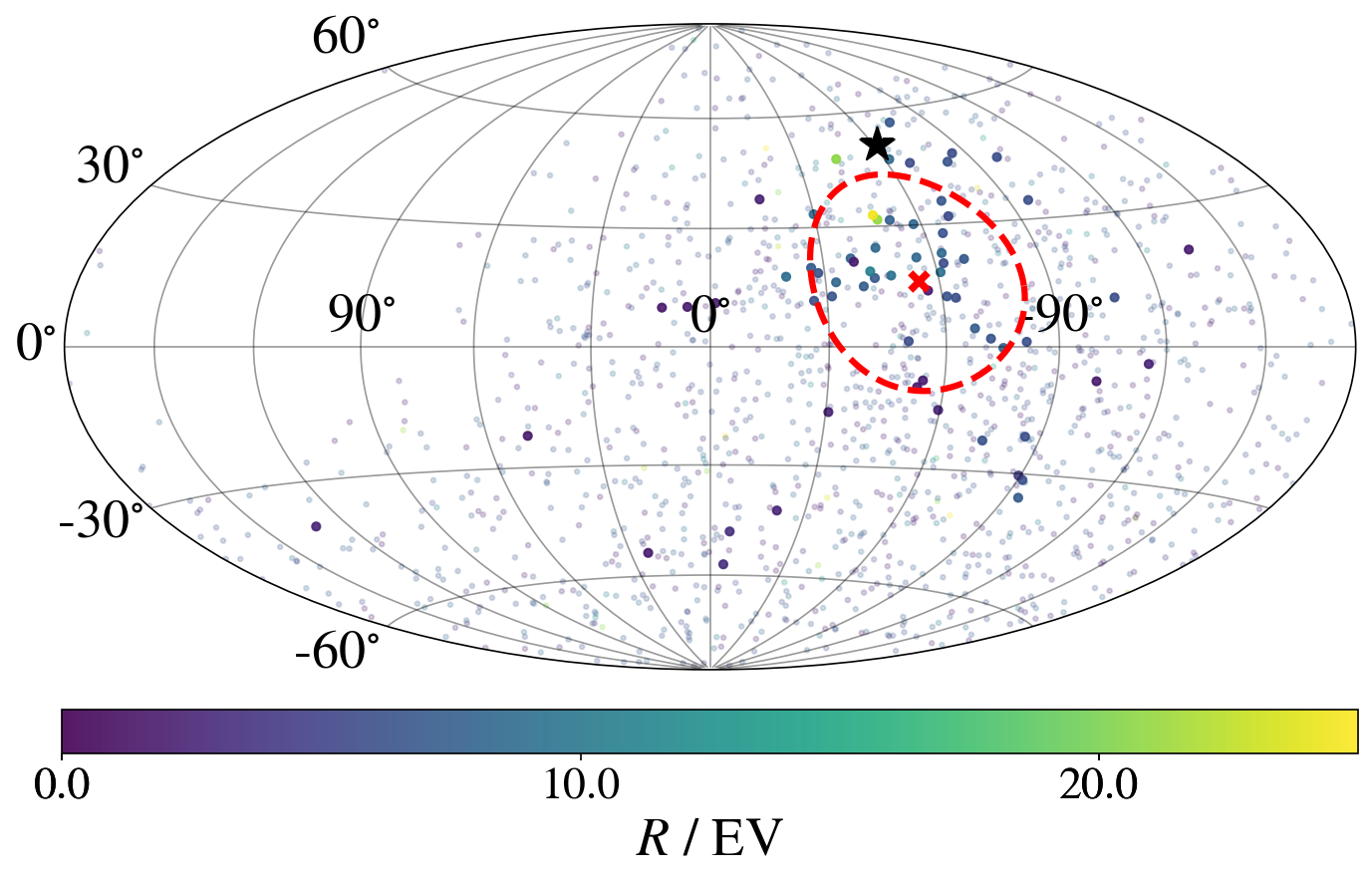}
\includegraphics[width=0.24\textwidth]{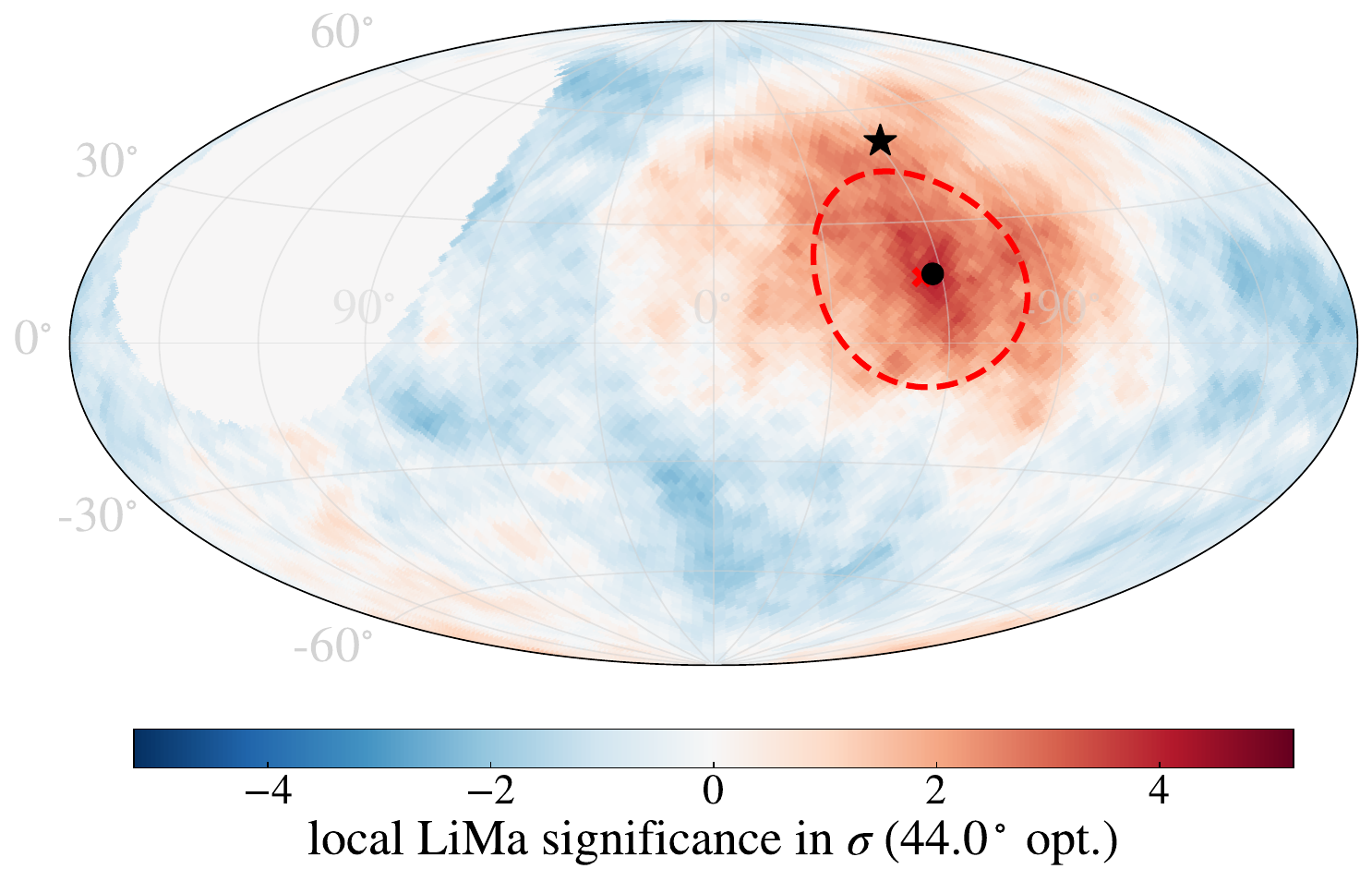}

\includegraphics[width=0.24\textwidth]{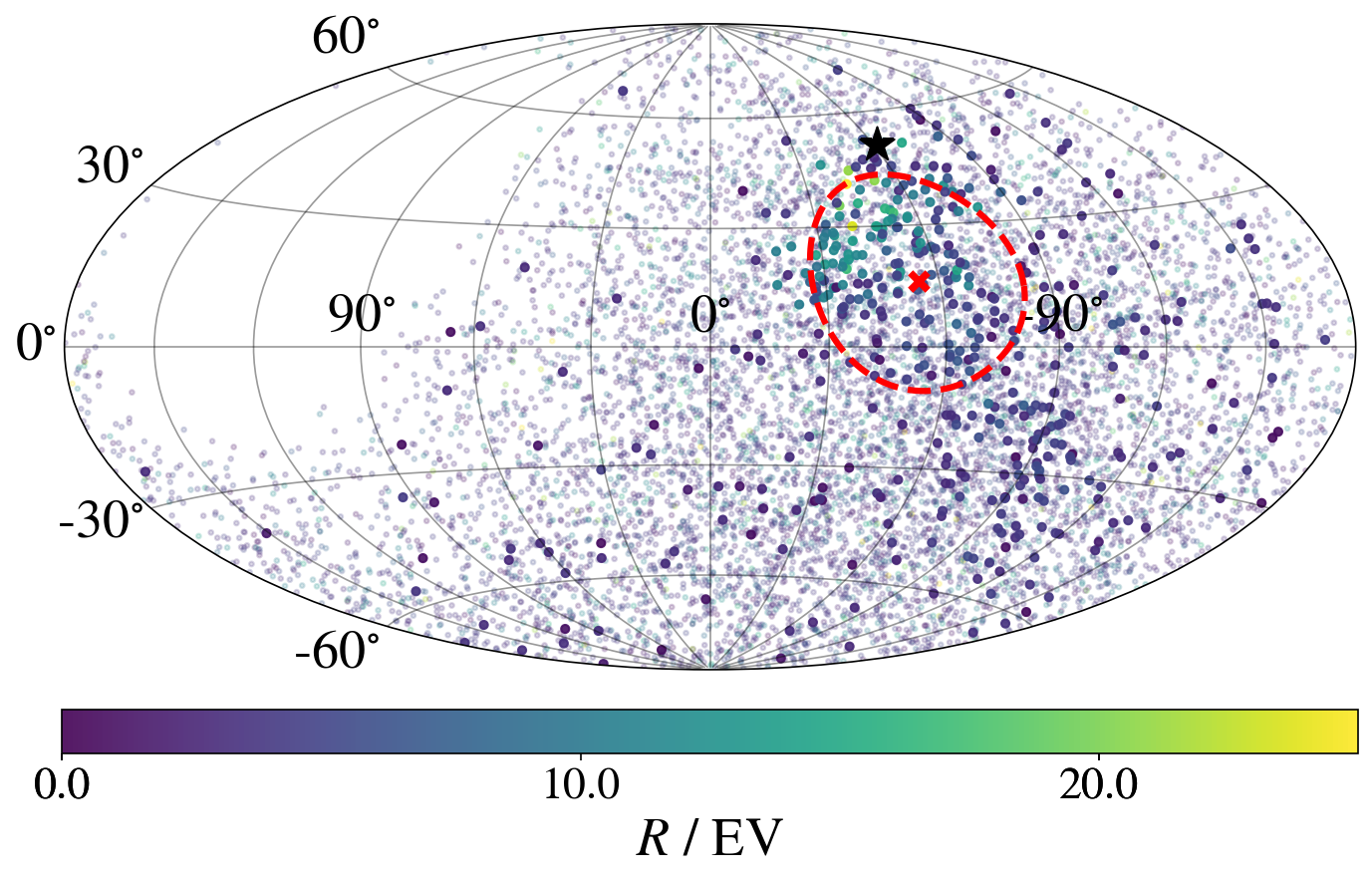}
\includegraphics[width=0.24\textwidth]{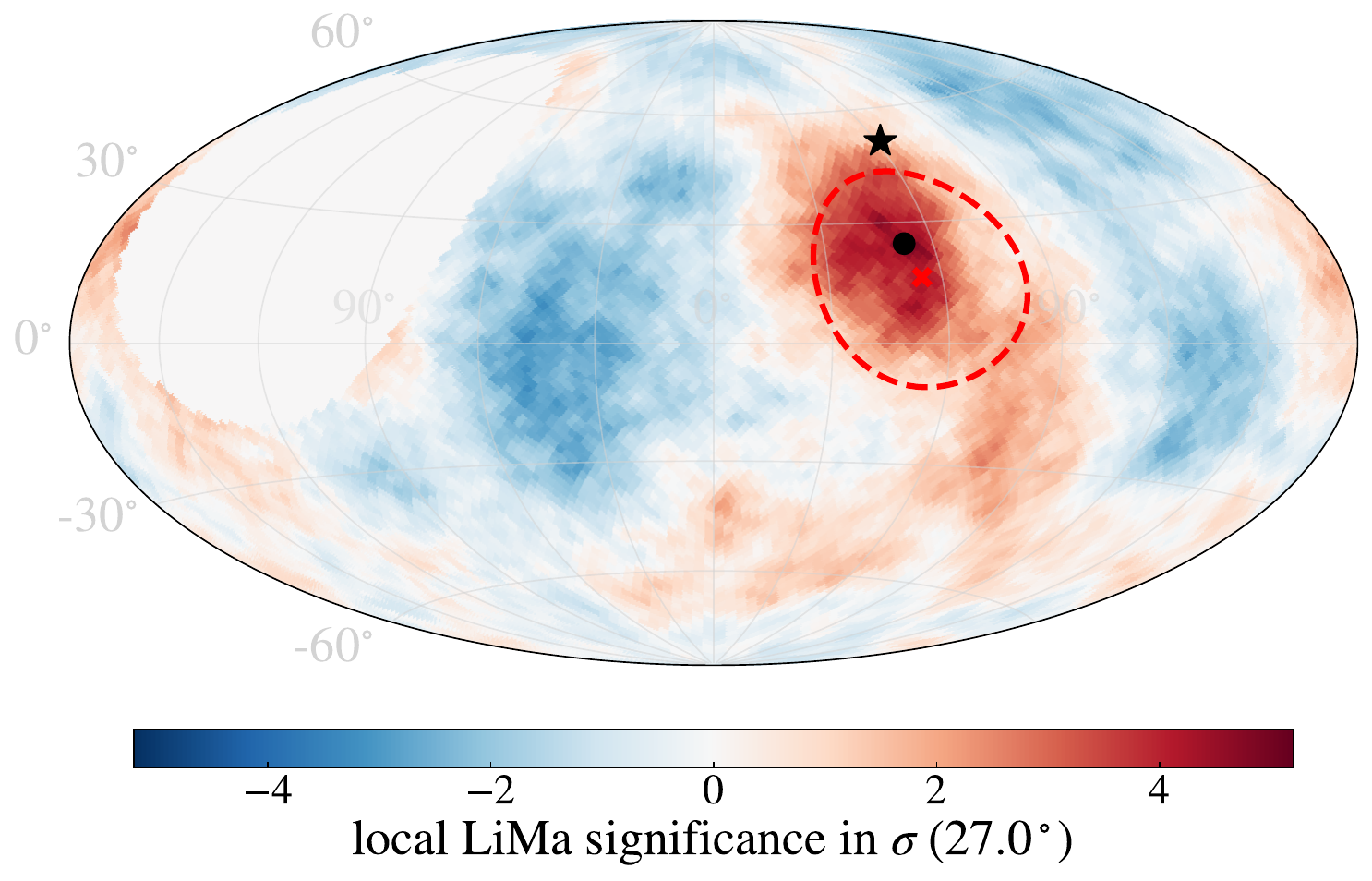}
\includegraphics[width=0.24\textwidth]{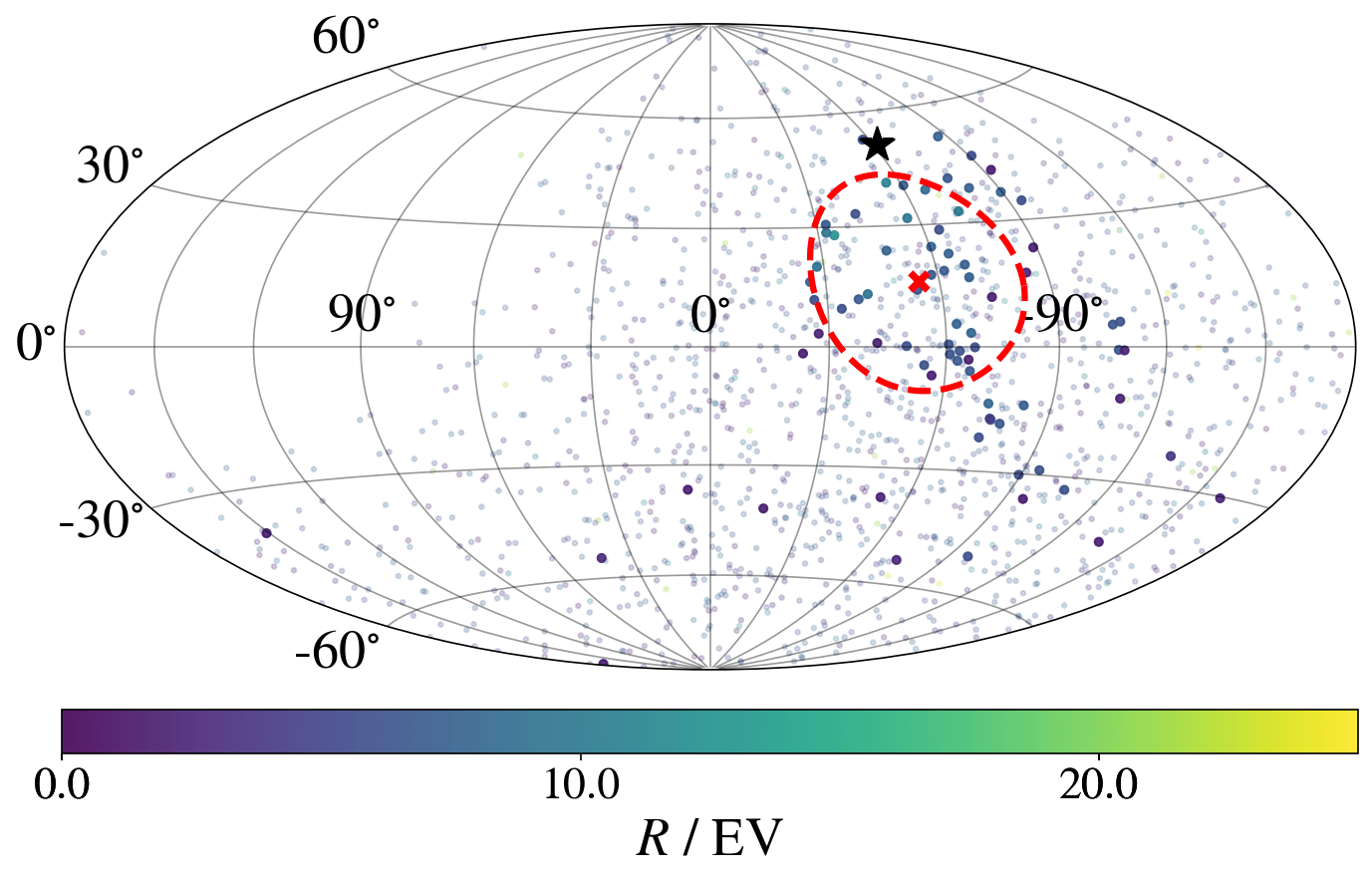}
\includegraphics[width=0.24\textwidth]{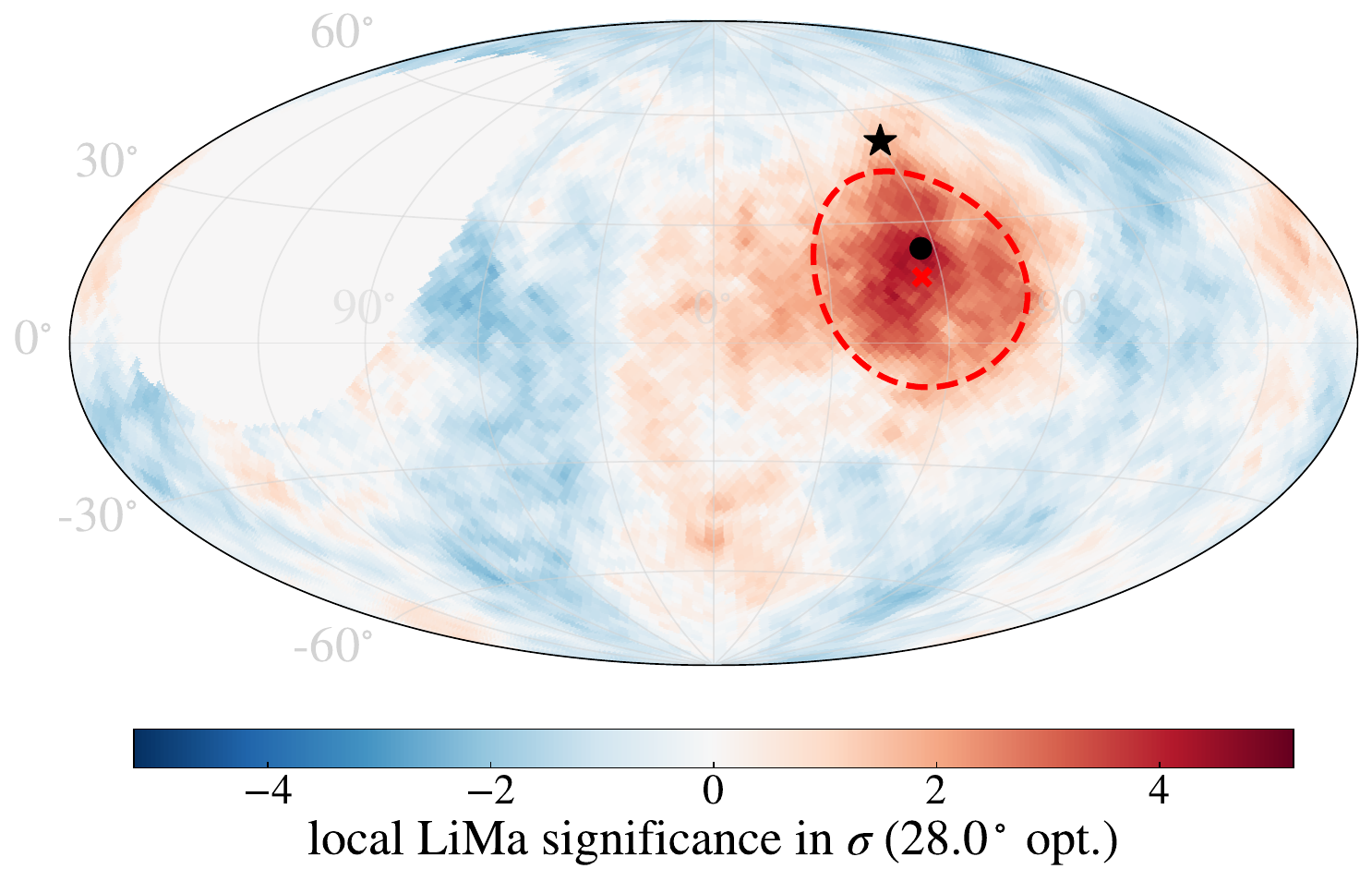}
\caption{Four example simulations with the Sombrero galaxy as the source, with $f=0.05$, $\beta_\mathrm{EGMF}=2.9$ (black cross in Fig.~\ref{fig:sombrero_constraints}) and the \texttt{UF23-base-Pl} GMF model. The \textit{upper two rows} are two examples with $Z=6$, the \textit{lower two rows} show the same two random seeds with mixed composition. The left two figures are for $E_\mathrm{min}=20\,\mathrm{EeV}$, the right two for $E_\mathrm{min}=40\,\mathrm{EeV}$. All skymaps are in Galactic coordinates. The background events are half-transparent. Note that for $E_\mathrm{min}=20\,\mathrm{EeV}$, the angular scale of the LiMa significance scan is fixed to $27^\circ$ as in~\cite{the_pierre_auger_collaboration_flux_2024}.}
\label{fig:sombrero_examples}
\end{figure}

That is also visible in Fig.~\ref{fig:sombrero_constraints} summarizing the constraints on the parameter values. Due to the Sombrero galaxy being further away than Cen A in combination with larger scattering by the GMF for the larger charge number, the EGMF has to be weaker to reproduce the observations. A good fit is reached for $\beta_\mathrm{EGMF}\simeq3$, $f=0.05$, and $Z=6$ or mixed composition (black cross in Fig.~\ref{fig:sombrero_AD}). Example simulations for that case are depicted in Fig.~\ref{fig:sombrero_examples}. Both direction and angular scale are well reproduced, and for this scenario, also the energy dependency of the significance is well reproduced.

Thus, the claim from~\cite{he_evidence_2024} that the Cen A overdensity could be generated by the Sombrero galaxy is supported by our simulations, using either the \texttt{UF23} or \texttt{JF12} GMF models - at least if the charge emitted by the source is close to $Z=6$. The inferred very small EGMF strength of $\beta_\mathrm{EGMF}\approx4\times10^{-3}$ from~\cite{he_evidence_2024} (converting their quantity $A_\mathrm{ran}$ to the formalism used in this work) is however not optimal according to our simulations. Larger EGMF strengths $\beta_\mathrm{EGMF}\approx1-10$ allow for a better agreement with both the excess direction, as well as the the angular scale. This implies that the method used in~\cite{he_evidence_2024} may not be able to identify all events coming from the Sombrero galaxy, especially not the ones which are too far from the major axis of the multiplet through turbulent blurring.

In addition to the Sombrero galaxy, we also tested if the overdensity could be produced by galaxies at even larger angular distances to the observed excess, specifically Virgo A. We find that large parts of the parameter space are forbidden, and that it is extremely difficult to reproduce the excess direction. Only with the \texttt{UF23-twistX} GMF model and $Z=12$ is it even possible to get the overall excess direction right, but still it varies so much that only by coincidence would it be possible to reproduce the observed stability with the energy threshold. The corresponding figures are provided in~\ref{app:virgoA}.

\section{Scenario III - one dominant source above the ankle} \label{sec:scen3}
In Fig.~\ref{fig:cena_constraints} and Fig.~\ref{fig:sombrero_constraints}, it is visible that also simulations with extremely large values for the signal fraction up to $f=1$ can reproduce the observations. 
If the whole flux is dominated by one source, the mass composition emitted by that source has to be in agreement with Auger measurements. For that reason, in the following only the mixed composition model is discussed.

In order for a single source at a distance of around 4\,Mpc to supply the whole UHECR flux $\gtrsim10\,\mathrm{EeV}$ (in the absence of magnetic field deflections), a source luminosity of $\mathcal{L}_\mathrm{UHECR>10\,EeV}\simeq5\times10^{41}\,\mathrm{erg}\,\mathrm{s}^{-1}$ is required~\cite{mollerach_case_2024}. The power of the Cen A's jets is estimated to be $\sim10^{43}\,\mathrm{erg} \,\mathrm{s}^{-1}$~\cite{wykes_mass_2013, eichmann_explaining_2022}. Thus, Cen A is able to supply the UHECR flux above the ankle for reasonable conversion efficiency of jet power to UHECR power, especially considering that there are indications of higher source activity in the past~\cite{Matthews_2018, taylor_uhecr_2023}. In addition, Cen A fulfills the Hillas-Lovelace condition and can hence produce UHECRs of the observed highest energies~\cite{eichmann_explaining_2022}. Note that the expected neutrino flux from the jets of Cen A is significantly below the reach of current observatories and can at this point thus not be used to challenge this scenario~\cite{Mbarek_2025}.
For the Sombrero galaxy, the jet power is around $\sim2\times 10^{42}\,\mathrm{erg} \,\mathrm{s}^{-1}$, and due to its larger distance, it would need to produce $\mathcal{L}_\mathrm{UHECR>10\,EeV}\simeq5\times10^{41}\,\mathrm{erg}\,\mathrm{s}^{-1} \cdot (9 \mathrm{Mpc} / 4 \mathrm{Mpc})^2 \simeq 2\times10^{42}\,\mathrm{erg}\,\mathrm{s}^{-1}$. Thus, 100\% efficiency would be required, making the Sombrero galaxy an unreasonable choice for the only UHECR flux above the ankle. The same is true for Virgo A / M87. Additionally, it is not plausible, why in such a case the more powerful and more nearby AGN Cen A would not contribute significantly more to the UHECR flux. 
For the starburst galaxies NGC4945 and M83, and SBGs in general, the acceleration mechanism is still quite unclear, and their expected UHECR power and maximum energy are under debate, but likely too small to produce very large UHECR fluxes above the ankle~\cite{Romero_2018, Matthews_2018}.

Thus, while only Cen A is believed to be able to supply the whole UHECR flux above the ankle, in this section we will still also show the Sombrero galaxy as an alternative - even though the parameter space close to $f=1$ is forbidden for the Sombrero galaxy because it does not reproduce the excess direction well enough for all energy thresholds - simply to demonstrate how in the case of very large EGMF blurring it is not possible to distinguish anymore between the candidates in the vicinity of the excess direction. Only for Virgo A to dominate the sky above the ankle, deflections are not large enough for the charge number $Z=6$ around $E=40\,\mathrm{EeV}$ dominating the mixed-composition model (see Fig.~\ref{fig:ADs}, and Fig.~\ref{fig:virgoA_ADs} in the appendix). Even when allowing for a possible shift of this dominant mass towards heavier elements (due to systematic uncertainties or uncertainties of the hadronic interaction models) as in~\cite{vicha_heavy-metal_2025}, the excess direction is not well reproduced with $f=1$ as visible in Fig.~\ref{fig:virgoA_ADs}. Also for NGC 4945, it is very difficult to reproduce the excess direction well for $f=1$. For M83, however, the excess direction fits very well for $f=1$, see~\ref{app:m83}.

\begin{figure}[ht!]
\includegraphics[trim={13.5cm 0 0 0}, clip, height=4.9cm]{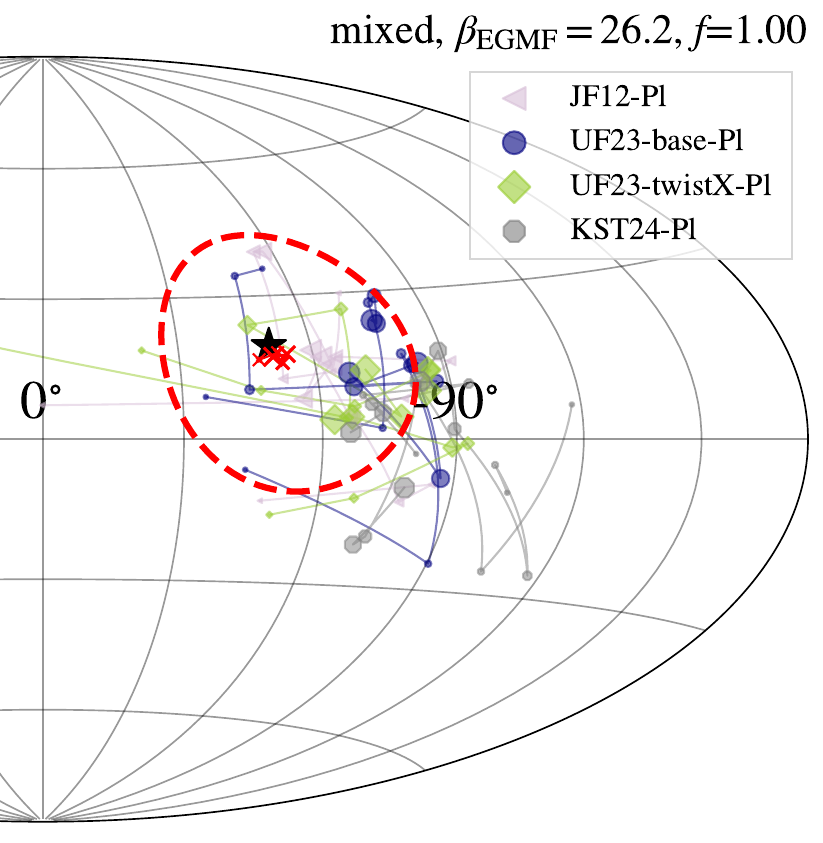}
\includegraphics[trim={13.5cm 0 0 0}, clip, height=4.9cm]{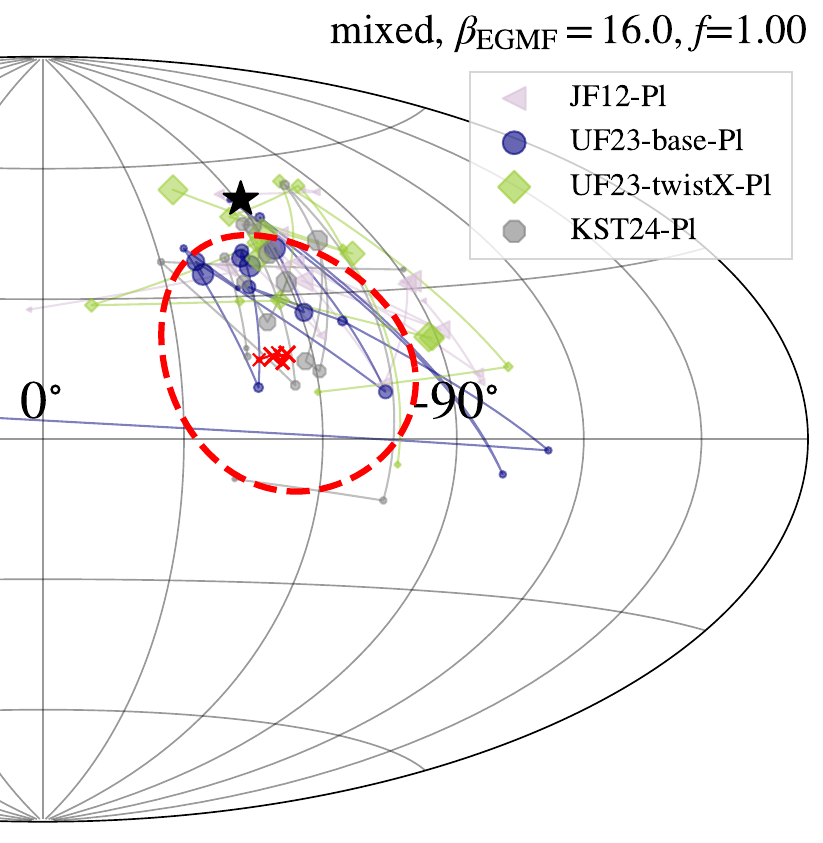}
\includegraphics[height=5.3cm]{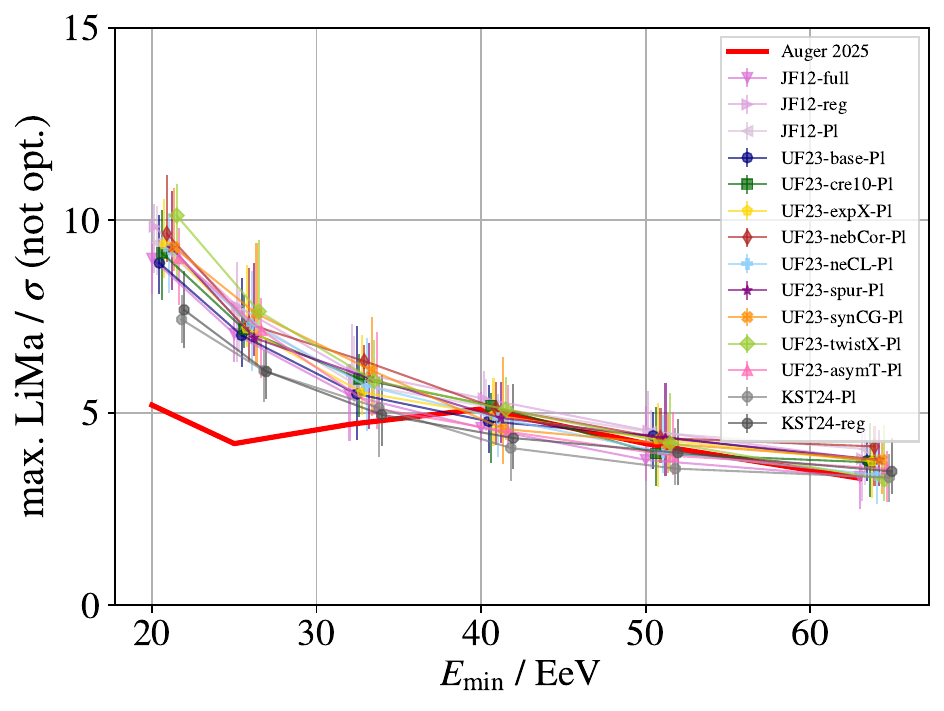}
\caption{Evolution of the excess direction with the energy for example simulations with Cen A (\textit{left}, $\beta_\mathrm{EGMF}=26.2$) and the Sombrero galaxy (\textit{middle}, $\beta_\mathrm{EGMF}=16$) as the only source with $f=1$ and mixed composition, using four different GMF models. For more details see Fig.~\ref{fig:cena_AD}. The \textit{right} figure shows the evolution of the maximum LiMa significance with the energy threshold for the case of Cen A with $\beta_\mathrm{EGMF}=26.2$. The error bars represent the $90\%$ uncertainty from the 10 random variations per GMF model. The red line denotes the significance evolution observed on data from~\cite{the_pierre_auger_collaboration_flux_2024}. The angular size was kept fixed to $27^\circ$ as in~\cite{the_pierre_auger_collaboration_flux_2024}.}
\label{fig:f1}
\end{figure}

The EGMF has to be $\beta_\mathrm{EGMF}\approx20-30$ for Cen A or M83, and $\beta_\mathrm{EGMF}\approx15-20$ for the Sombrero galaxy if the signal fraction is close to $f=1$ (due to their difference in distance) as visible in Fig.~\ref{fig:cena_constraints} and Fig.~\ref{fig:sombrero_constraints}. Note that this is slightly larger than the value $\beta_\mathrm{EGMF}=5-15$ inferred in~\cite{mollerach_case_2024} where Cen A as the only source above the ankle was fit to energy spectrum and mass composition data.
The direction of the excess is not perfectly reproduced for either source candidate as visible in Fig.~\ref{fig:f1} (\textit{left} and \textit{middle}), but for both candidates it is possible to find simulations roughly matching the observations. Example simulations are provided in Fig.~\ref{fig:cena_f1} and Fig.~\ref{fig:sombrero_f1} in the appendix. 

The evolution of the maximum LiMa significance with the energy threshold is shown in Fig.~\ref{fig:f1} (\textit{right}) for Cen A (for Sombrero it looks almost the same). For energy thresholds below $E_\mathrm{min}=30\,\mathrm{EeV}$, the significance is too large and not in agreement with the data at more than 90\%\,C.L. for any GMF model. Note that decreasing the signal fraction only shifts the significance up and down for all energy thresholds without changing the shape of the curve. This implies that another source contribution is necessary in the energy range below $30\,\mathrm{EeV}$ in order to decrease the maximum significance in the Centaurus region enough to be in agreement with the data. One possibility is that part of the cosmic rays from the source are fully isotropized due to the strong EGMF. Such a second contribution would depend on the emission history of the source and details of the EGMF and is not covered by the simplified Fisher distribution (eq.~\ref{eq:egmf}) as discussed above.

\begin{figure}[ht]
\subfloat[Cen A, \texttt{UF23-base-Pl},\\ $\beta_\mathrm{EGMF}=26.2$]{\includegraphics[width=0.32\textwidth]{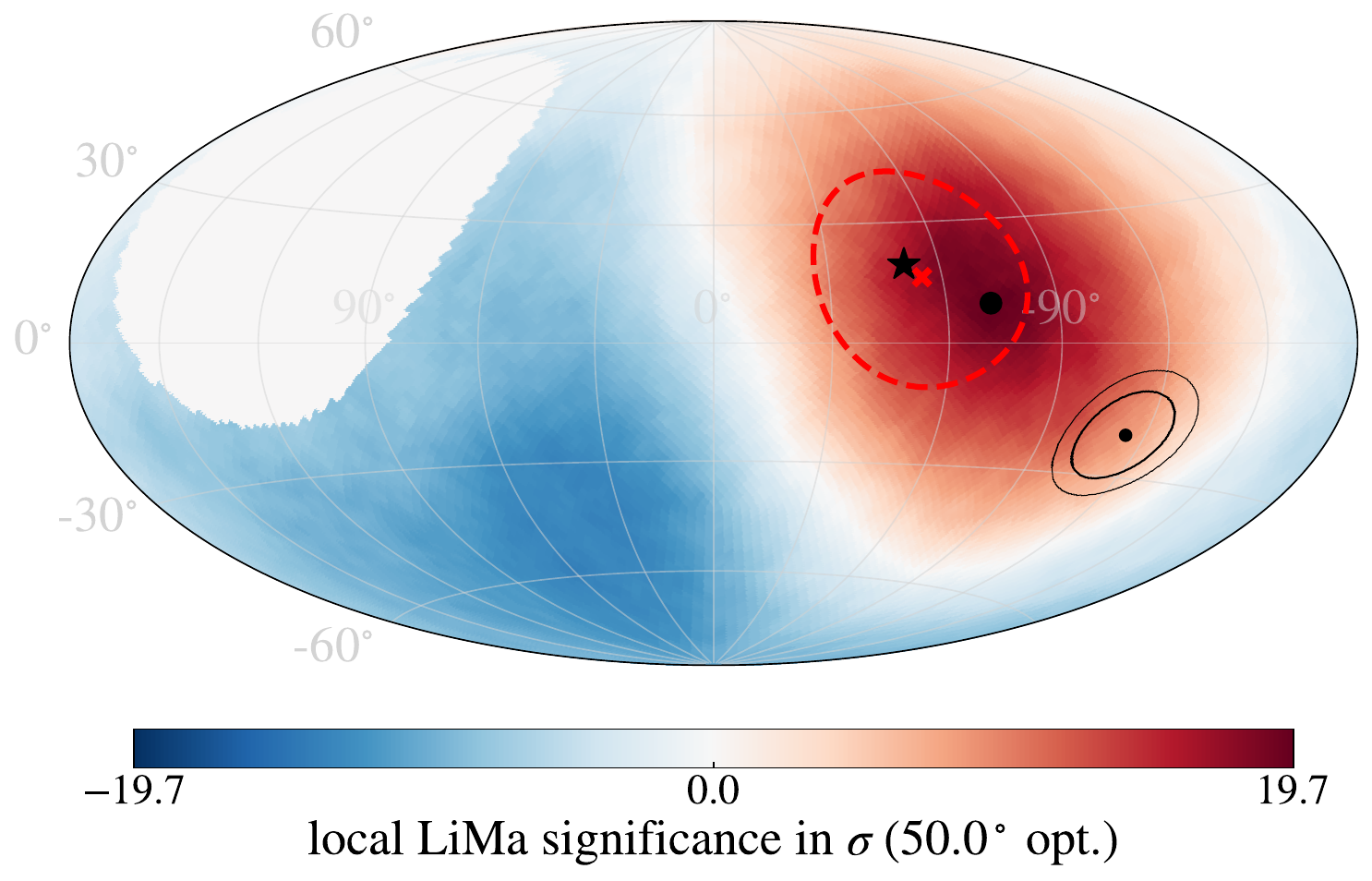}}
\subfloat[Cen A, \texttt{UF23-twistX-Pl},\\ $\beta_\mathrm{EGMF}=26.2$]{\includegraphics[width=0.32\textwidth]{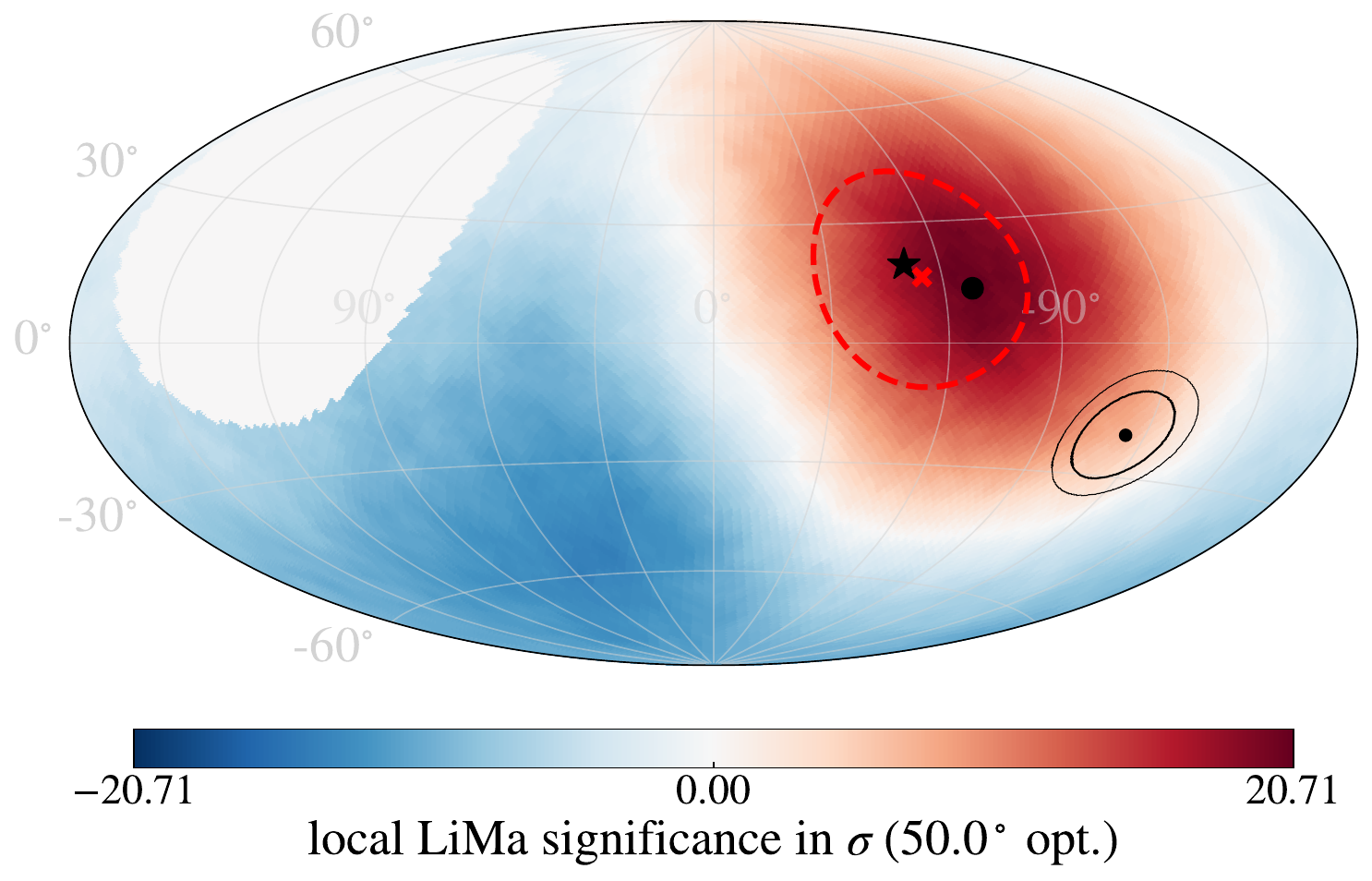}}
\subfloat[Sombrero, \texttt{UF23-base-Pl},\\ $\beta_\mathrm{EGMF}=16$]{\includegraphics[width=0.32\textwidth]{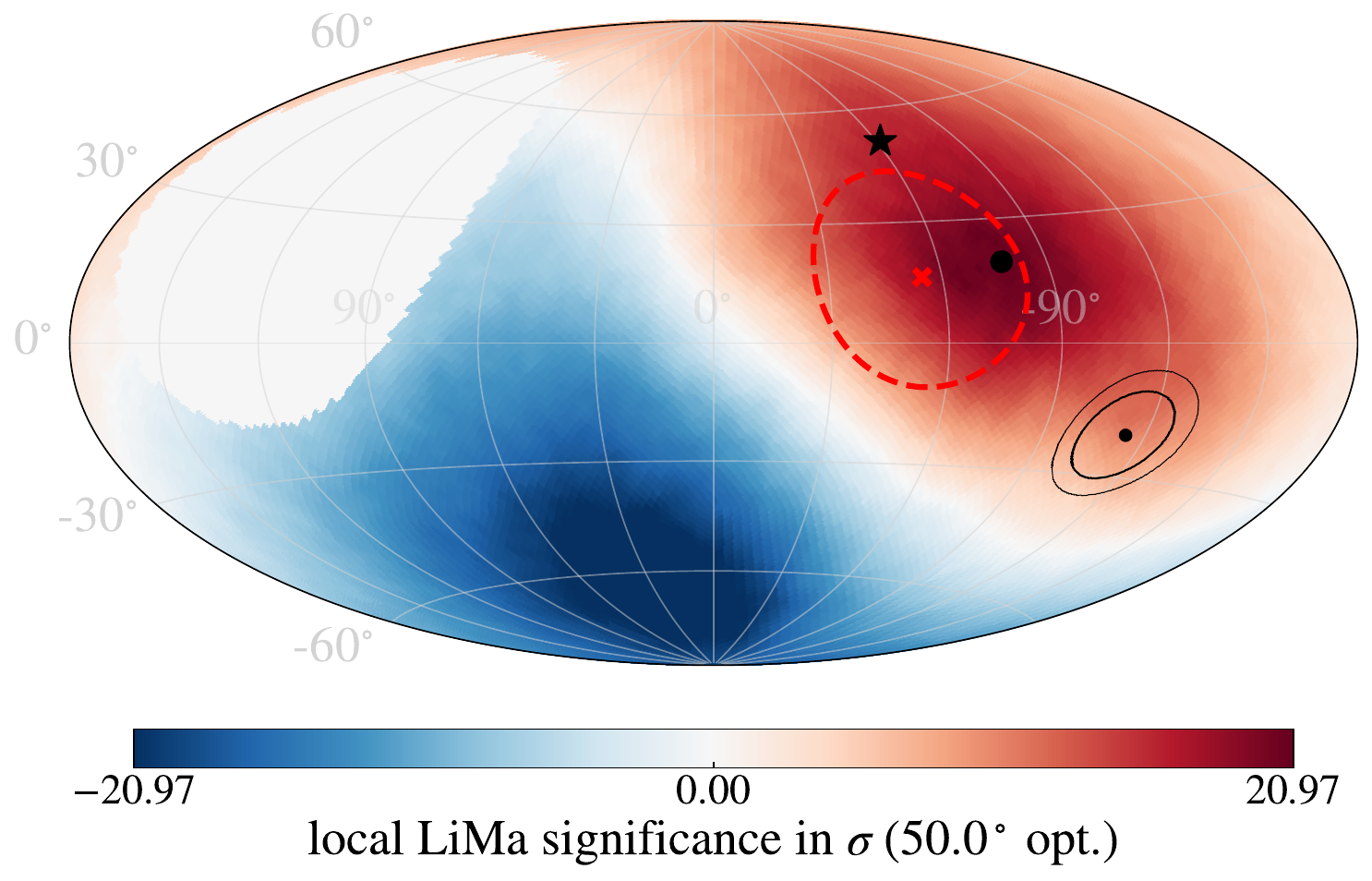}}
\caption{Examples for the LiMa map $>8\,\mathrm{EeV}$ for scenario III. The source (star marker) and GMF model is indicated below each figure. The direction of the maximum significance (black round marker) is not close to the direction of the observed dipole~\cite{abdul_halim_large-scale_2024} (black 1 and $2\sigma$ contours). Note that $50^\circ$ corresponds to the largest angular size part of the scan. Due to the large statistics of $\mathcal{O}(50,000)$ events~\cite{abdul_halim_large-scale_2024}, the LiMa maps look very similar for each simulation with the same source.}
\label{fig:ADs8}
\end{figure}

Another possibility is that other sources start contributing at lower energies. To look into this further, we extended the energy range of the simulations down to 8\,EeV to see if the direction of the maximum LiMa significance is close to the direction of the dipole that dominates the flux at this energy~\cite{the_pierre_auger_collaboration_a_aab_et_al_observation_2017, abdul_halim_large-scale_2024}. Three examples are shown in Fig.~\ref{fig:ADs8}. For both Cen A as well as the Sombrero galaxy as the only source and a mixed composition, the direction of the dipole is not well reproduced. While any finetuning is outside of the scope of this work (focused on the overdensity in the Centaurus region at higher energies), the disagreement of the flux maximum $>8\,\mathrm{EeV}$ of the simulations and the measured dipole direction may imply that the other source contribution between 8 and 30\,EeV has to have some directionality that moves the dipole more towards the Galactic south. That implies that the dipole is not generated by one source alone - but that instead at least part of the flux between 8 and 30\,EeV has to come from other sources. To get the maximum flux direction into better agreement with the measured dipole direction, those sources could either be numerous and follow the large-scale structure as e.g. explored in~\cite{bister_large-scale_2024, bister_constraints_2024} or be dominated by few other source candidates in the Galactic south (like Fornax A), see e.g.~\cite{eichmann_explaining_2022}.

\begin{figure}[ht]
\subfloat[data, from~\cite{abdul_halim_catalog_2023}]{\includegraphics[width=0.32\textwidth]{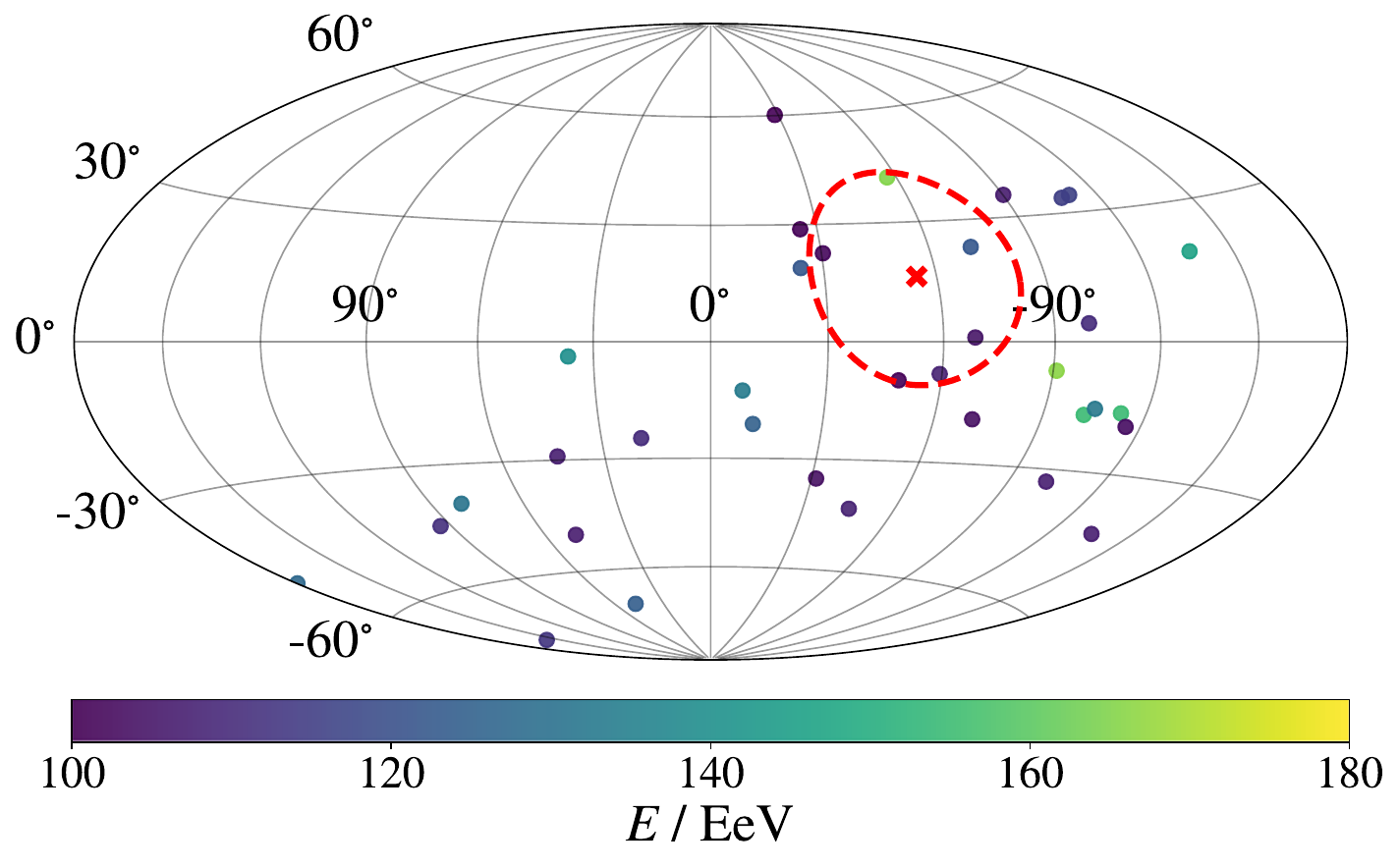}}
\subfloat[Cen A, \texttt{UF23-base-Pl},\\ $\beta_\mathrm{EGMF}=26.2$]{\includegraphics[width=0.32\textwidth]{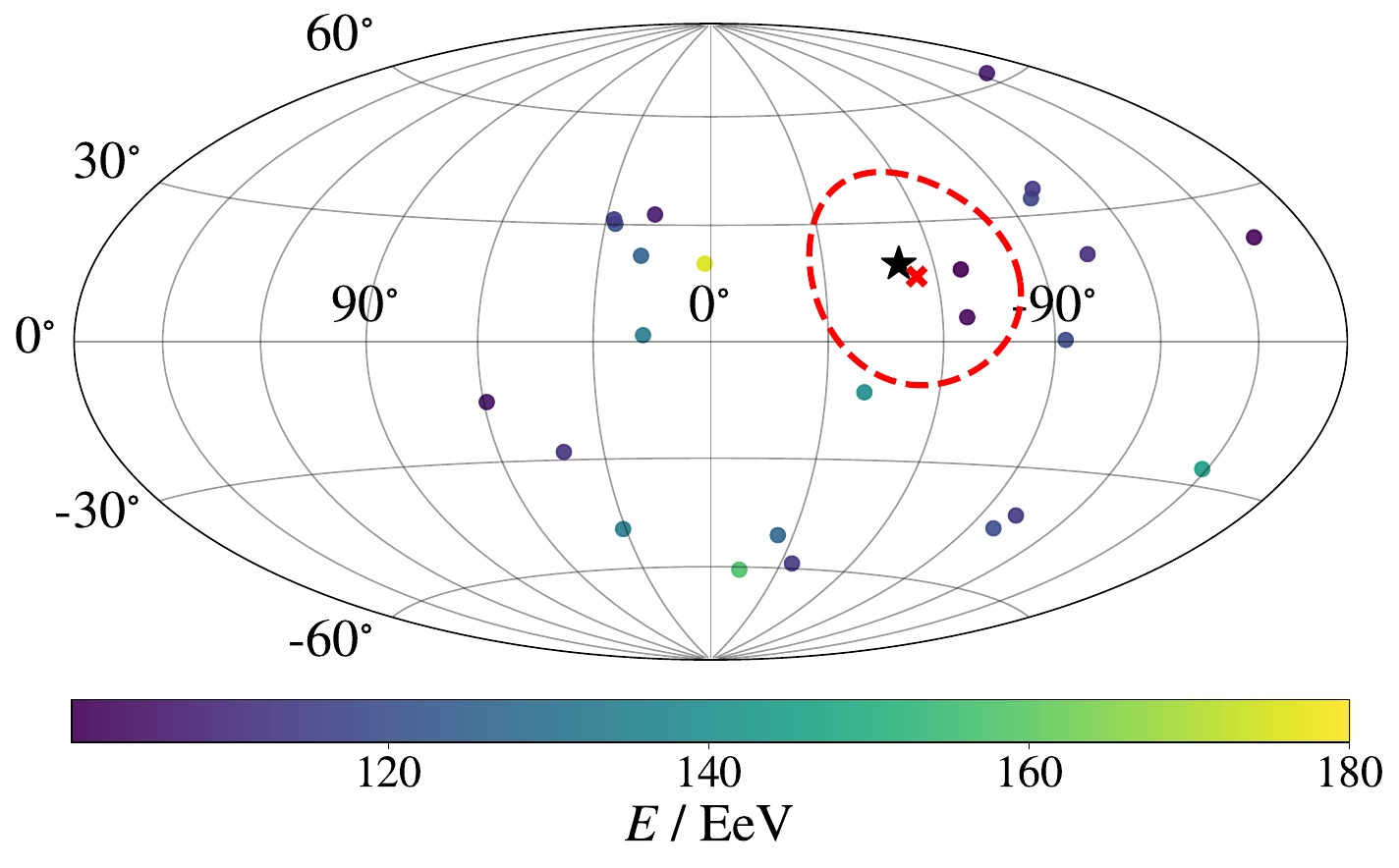}}
\subfloat[Sombrero \texttt{UF23-base-Pl},\\ $\beta_\mathrm{EGMF}=17$]{\includegraphics[width=0.32\textwidth]{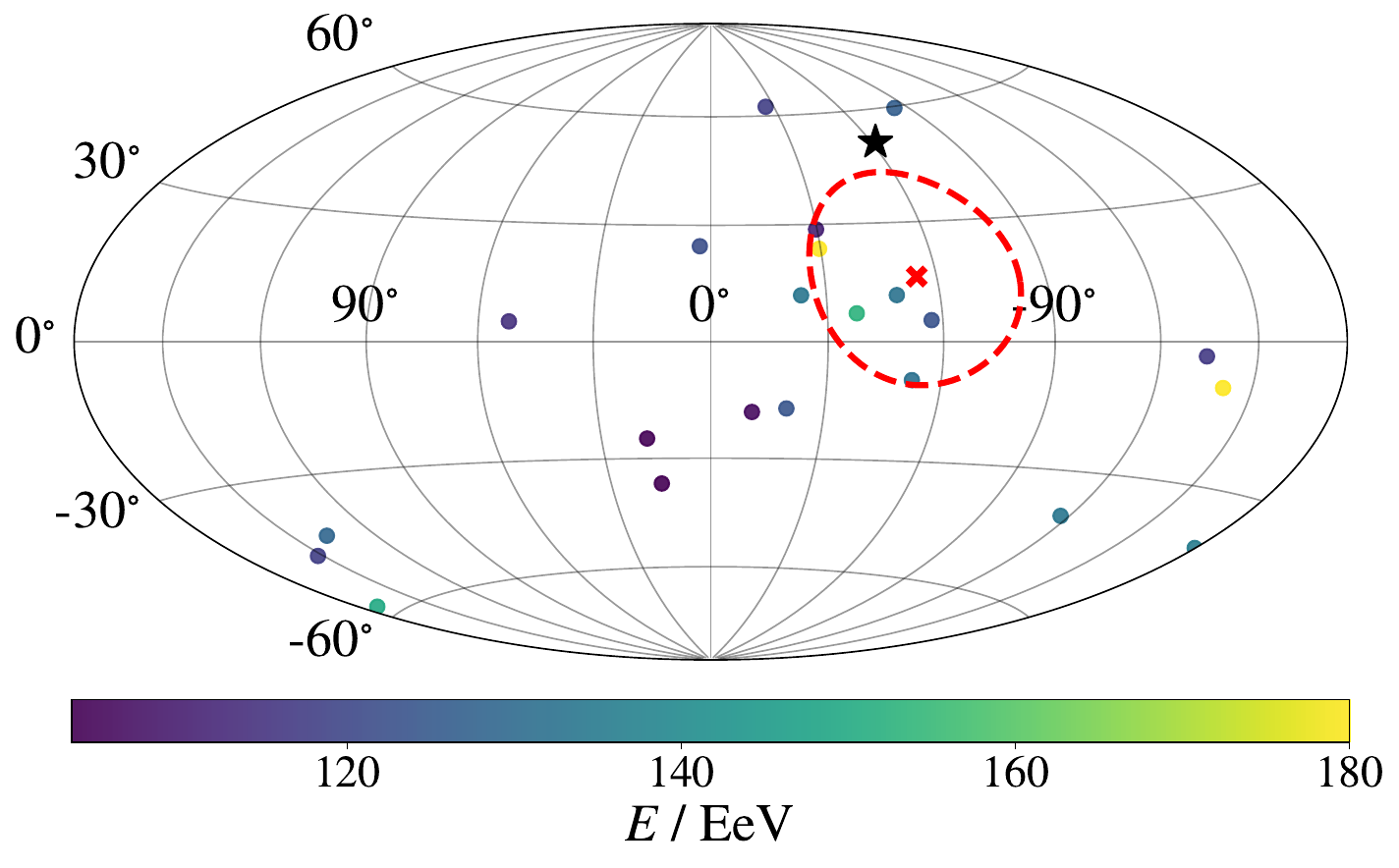}}
\caption{Examples for the distribution of events with $E>100\,\mathrm{EeV}$. The source and GMF model (scenario III with $f=1$) is indicated below each figure. The \textit{left} figure displays the data from~\cite{abdul_halim_catalog_2023}.}
\label{fig:scatter100}
\end{figure}

While the dipole at lower energies is not well described by the single-source scenario III, the highest energy events could instead well be generated by one source. Even though the distribution of the measured events is surprisingly isotropic~\cite{abdul_halim_catalog_2023, Bianciotto_ICRC_2025} as shown Fig.~\ref{fig:scatter100}, (\textit{left}), it is straightforward to reproduce such a distribution with the same fit parameters that describe the overdensity well at $\sim40\,\mathrm{EeV}$ as visible in Fig.~\ref{fig:scatter100} (\textit{middle} and \textit{right}). Thus, no ultra-heavy elements~\cite{zhang_ultraheavy_2024}, transient events in regular galaxies~\cite{farrar_binary_2025} or beyond-standard-model physics is required to explain the distribution of the highest energy events in a scenario where a nearby galaxy dominates the flux in combination with a strong EGMF between that source and the Milky Way.

\section{Summary and Outlook} \label{sec:conclusion}
\begin{figure}[ht]
\centering
\includegraphics[width=0.7\textwidth]{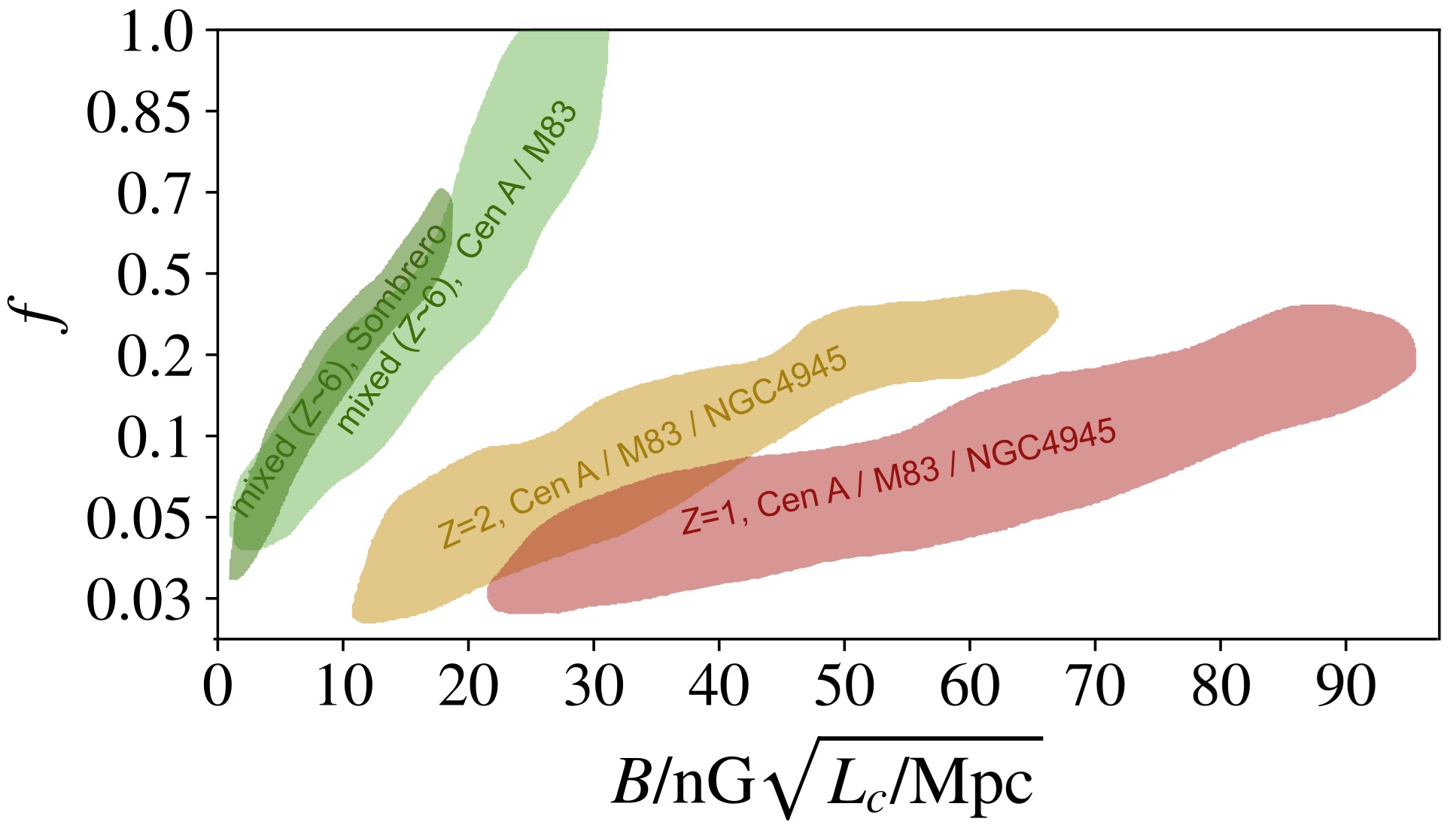}
\caption{Summarized constraints on the signal fraction $f$ and EGMF to be compatible with the UHECR data $\gtrsim30\,\mathrm{EeV}$.}
\label{fig:summary}
\end{figure}

The most significant intermediate-scale overdensity in the arrival flux of UHECRs~\cite{ g_golup_on_behalf_of_the_pierre_auger_collaboration_update_2024, the_pierre_auger_collaboration_flux_2024, the_pierre_auger_collaboration_p_abreu_arrival_2022} is in a region of the sky where multiple interesting source candidates reside - like the radio galaxy Cen A as well as the starburst galaxies NGC 4945 and M83. Recent proposals also suggested that the excess is instead generated by deflected events from the Virgo cluster or the Sombrero galaxy.
The strength, direction, angular scale and energy evolution of the overdensity can be used to verify if these source candidates can indeed generate the overdensity, and to place constraints on the emitted charge, the signal fraction, and the EGMF between the respective source and Earth. We marginalize over multiple models of the GMF, including also different random field models, to get conservative constraints that are summarized in the following, divided into the three studied scenarios:
\begin{itemize}
    \item \textit{Scenario I}: A subdominant source close to the excess direction, namely Cen A, NGC 4945 or M83 could generate the overdensity for charges $Z\lesssim6$ or a mixed composition dominated by $Z=6$. The EGMF between the respective source and Earth has to be very strong for lighter elements, $20\lesssim B/\mathrm{nG} \sqrt{L_c/\mathrm{Mpc}}\lesssim100$ for $Z=1$, $10\lesssim B/\mathrm{nG} \sqrt{L_c/\mathrm{Mpc}}\lesssim70$ for $Z=2$ and $1\lesssim B/\mathrm{nG} \sqrt{L_c/\mathrm{Mpc}}\lesssim20$ for $Z\sim6$. It is not possible to differentiate between the source candidates because of the large EGMF blurring in combination with uncertainties from the GMF and mass composition. The \texttt{KST24} GMF model shows larger coherent deflections and is only compatible with the excess direction for $Z=1$.
    \item \textit{Scenario II}: A subdominant source at larger angular distance to the Centaurus region, like the Sombrero galaxy as proposed in~\cite{he_evidence_2024}, requires larger coherent deflections and therefore a charge of $Z\sim6$. For the Sombrero galaxy to be the (subdominant) source of the excess, the EGMF has to be $1\lesssim B/\mathrm{nG} \sqrt{L_c/\mathrm{Mpc}}\lesssim20$. Again, the \texttt{KST24} GMF model leads to too large deflections and is not compatible with observations in this case. For the other GMF models, the predicted overdensity direction is however in good agreement with the measured one. Only the inferred EGMF blurring is stronger than the one found in~\cite{he_evidence_2024} which indicates that the method to search for multiplets used in~\cite{he_evidence_2024} may not able to capture all source events when the blurring is of the same size as the coherent deflections.
    Sources even further from the excess direction like Virgo A require even larger charges $Z\sim12$, which in turn leads to large variations of the excess direction with the energy that are not observed for the data~\cite{the_pierre_auger_collaboration_flux_2024}.
    \item \textit{Scenario III}: The data $\gtrsim30\,\mathrm{EeV}$ (both the Centaurus region overdensity and highest-energy events) are also in agreement with a single source dominating the whole UHECR flux. To describe the Centaurus excess properly, that source could be Cen A or M83 with an EGMF of $20\lesssim B/\mathrm{nG} \sqrt{L_c/\mathrm{Mpc}}\lesssim35$. Note however that only Cen A is expected to be powerful enough to supply the whole UHECR flux above the ankle~\cite{mollerach_case_2024, Matthews_2018, taylor_uhecr_2023}. The composition in this case has to be mixed (Nitrogen-dominated) to be in agreement with Auger mass composition measurements~\cite{Mayotte_ICRC_2025}. Below $\sim30\,\mathrm{EeV}$, another component has to arise because the single-source scenario leads to a too strong excess and a dipole direction not in agreement with measurements~\cite{abdul_halim_large-scale_2024}. That contribution is likely from other source candidates - that e.g. follow the large-scale structure - but could also stem from fully-isotropized events from the same source.
\end{itemize}
The constraints on the signal fraction and EGMF placed in this work are summarized in Fig.~\ref{fig:summary}. The EGMF as a whole, as well as in our local neighborhood, is not well constrained from observations and theory~\cite{durrer_cosmological_2013, vazza_magnetogenesis_2021}.
An upper bound on the field strength in voids $\mathcal{O}(1\,\mathrm{nG})$ can be placed using rotational measures~\cite{pshirkov_new_2016}, but stronger fields are expected for the denser region between the Milky Way and nearby source candidates like Cen A. The estimates for the EGMF derived in this work are thus in accordance with expectations and represent the currently best estimate of the EGMF in our cosmic vicinity. Note that for none of the source candidates the observations are in agreement with an EGMF of zero, mostly because then the search radius is significantly smaller than the measured value of $27^\circ$.

While it is at this point not possible to differentiate which source is responsible for the overdensity in the Centaurus region, future measurements of the mass composition in that region could help to differentiate between the different scenarios. One possibility would be the analysis proposed in~\cite{Apollonio_ICRC_2025}. Evaluating the significance for subsets of the data based on rigidity-dependent cuts leads to very different significances for the different mass composition models tested in this work, see~\cite{Apollonio_ICRC_2025}. Thus, once it is known if the overdensity contains predominently light particles or not, Fig.~\ref{fig:summary} can be used to draw conclusions about the EGMF. Another option would be to refine the search for multiplets~\cite{he_evidence_2024, the_pierre_auger_collaboration_a_aab_et_al_search_2020} to be able to consider also larger turbulent blurrings. Additionally, a search for rigidity-ordering instead of just energy-ordering could help differentiate between a mixed-composition scenario and one where the source emits predominantly one element.

\paragraph{\textbf{Acknowledgements:}}
I thank my colleagues from the Pierre Auger Collaboration for valuable discussions; Esteban Roulet, Glennys Farrar, Armando di Matteo, Charles Timmermans, and the anonymous referee for comments on this manuscript; and Alexander Korochkin and Michael Unger for providing their GMF models.


\appendix

\section{Simulated arrival directions} 
\noindent
\begin{minipage}{\textwidth}
\centering
\includegraphics[trim={12.2cm 0 0 0}, clip, width=0.19\textwidth]{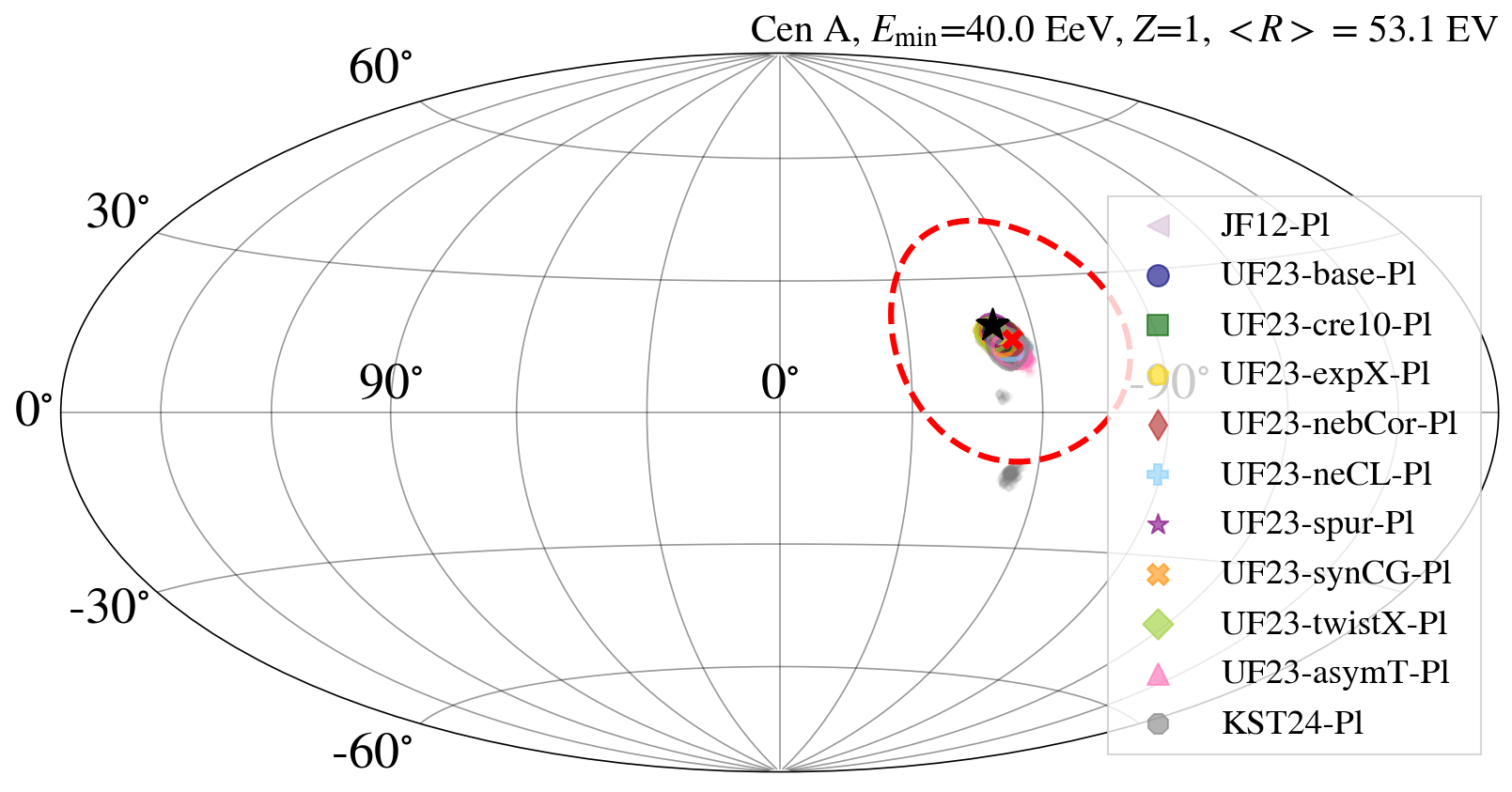}
\includegraphics[trim={12.2cm 0 0 0}, clip, width=0.19\textwidth]{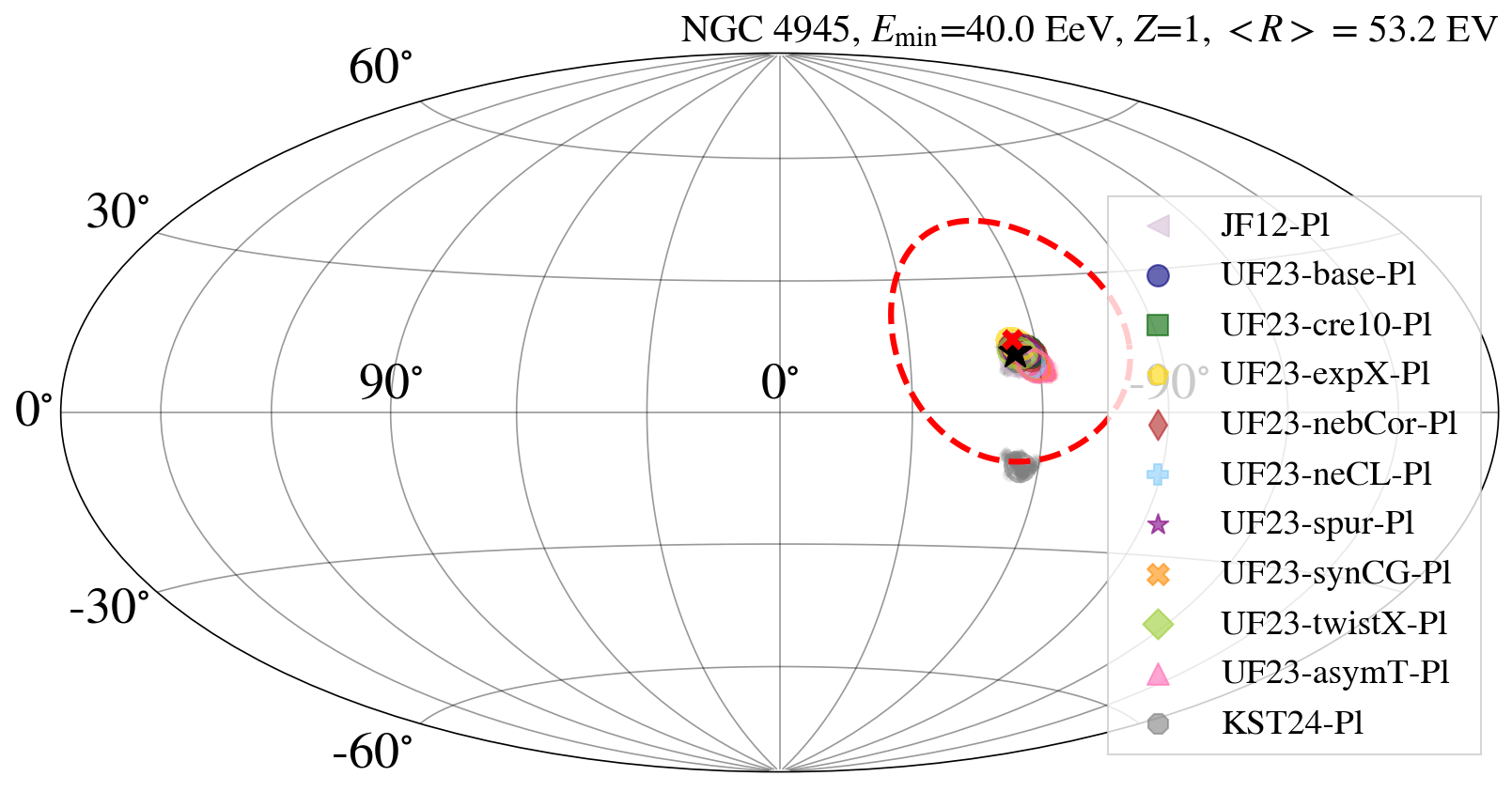}
\includegraphics[trim={12.2cm 0 0 0}, clip, width=0.19\textwidth]{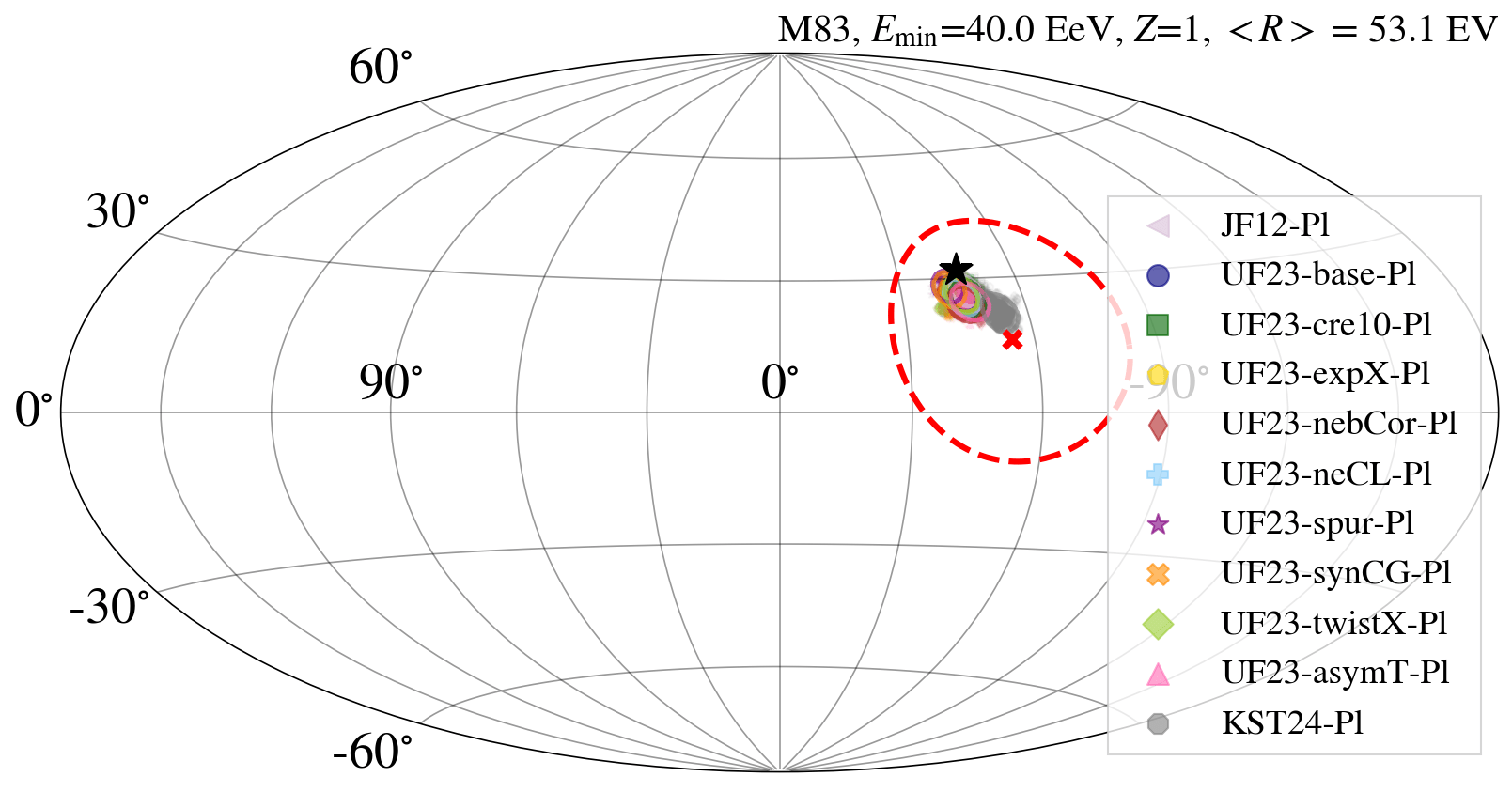}
\includegraphics[trim={12.2cm 0 0 0}, clip, width=0.19\textwidth]{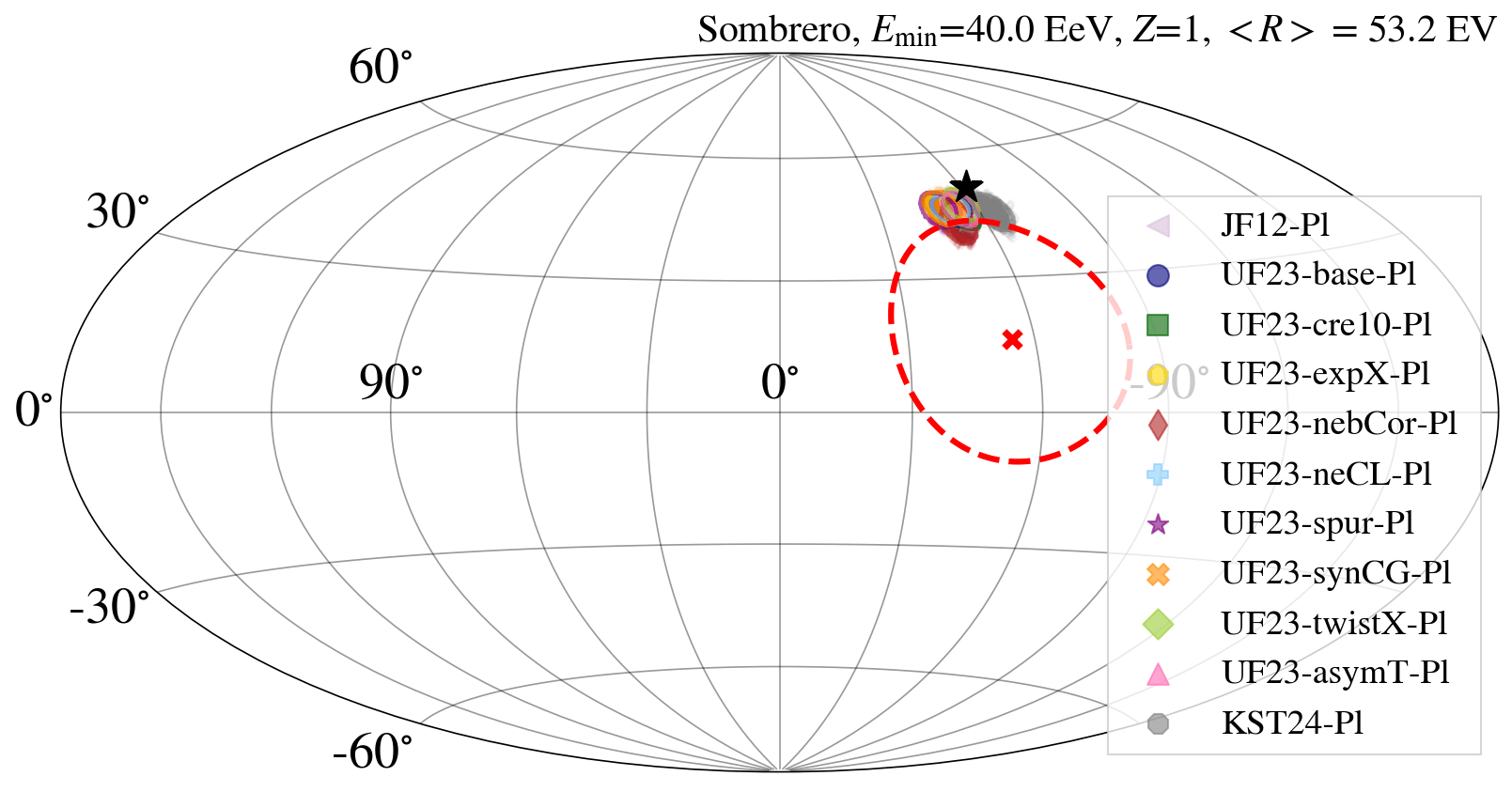}
\includegraphics[trim={12.2cm 0 0 0}, clip, width=0.19\textwidth]{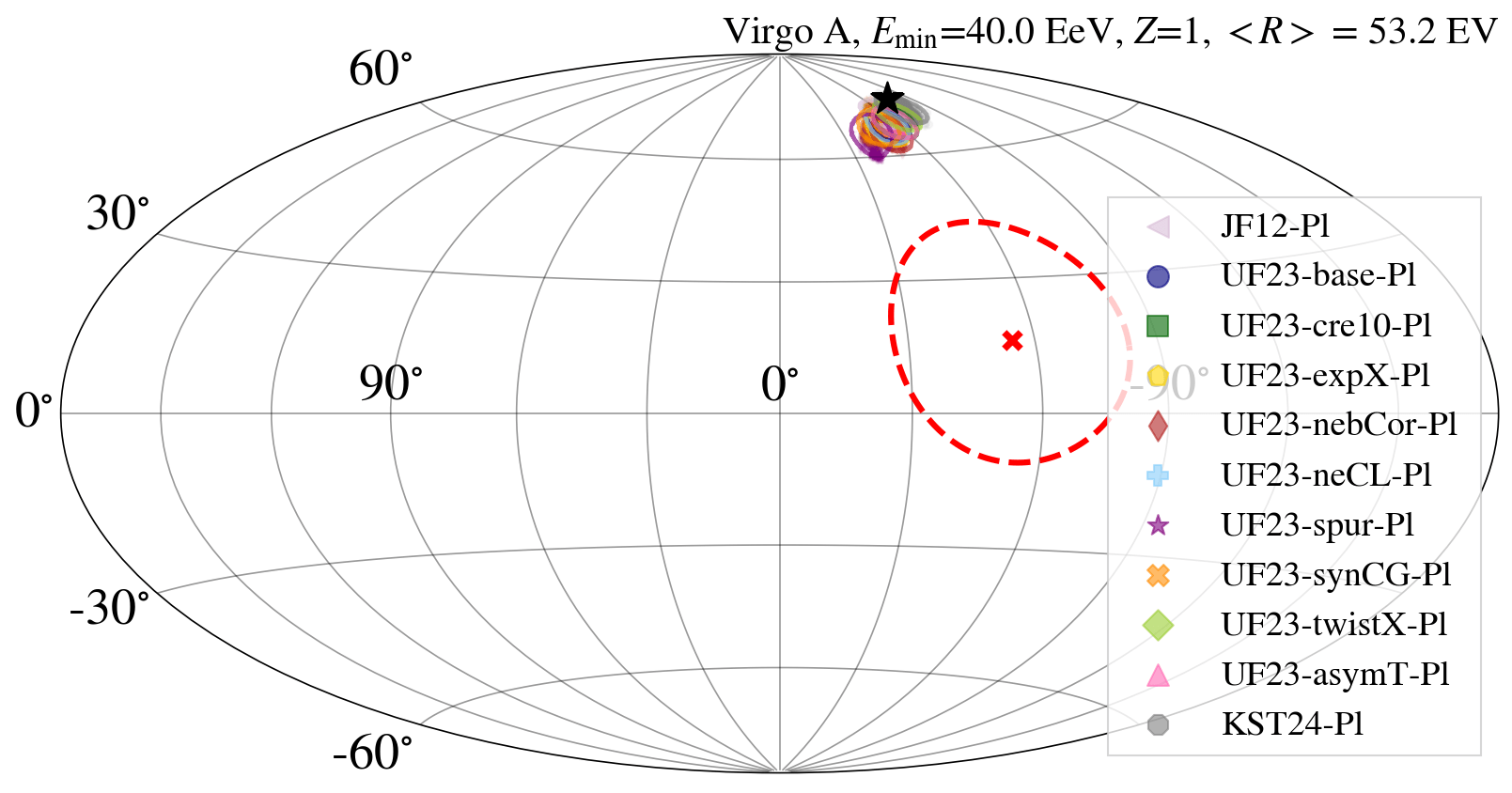}\\ 

\includegraphics[trim={12.2cm 0 0 0}, clip, width=0.19\textwidth]{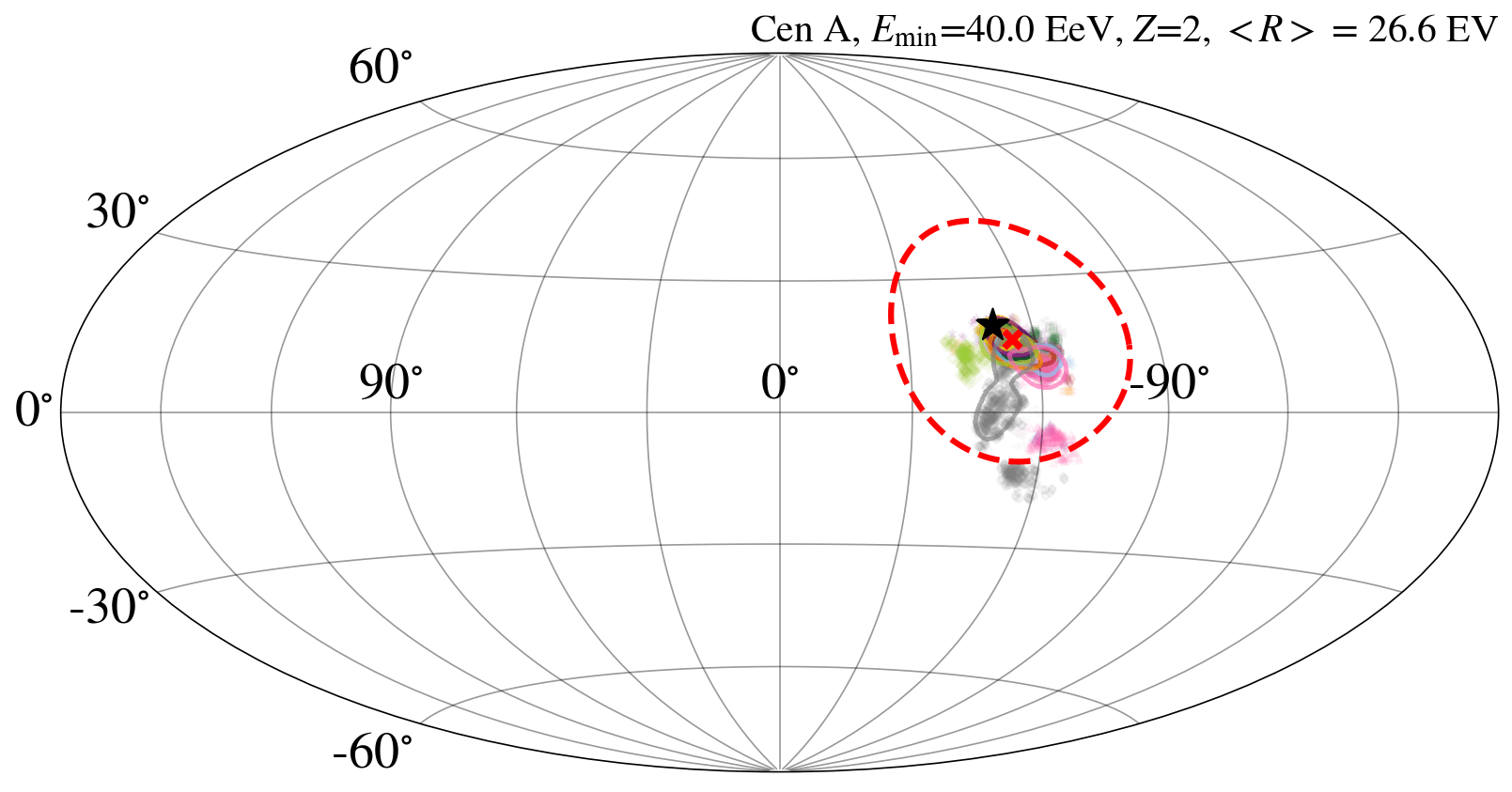}
\includegraphics[trim={12.2cm 0 0 0}, clip, width=0.19\textwidth]{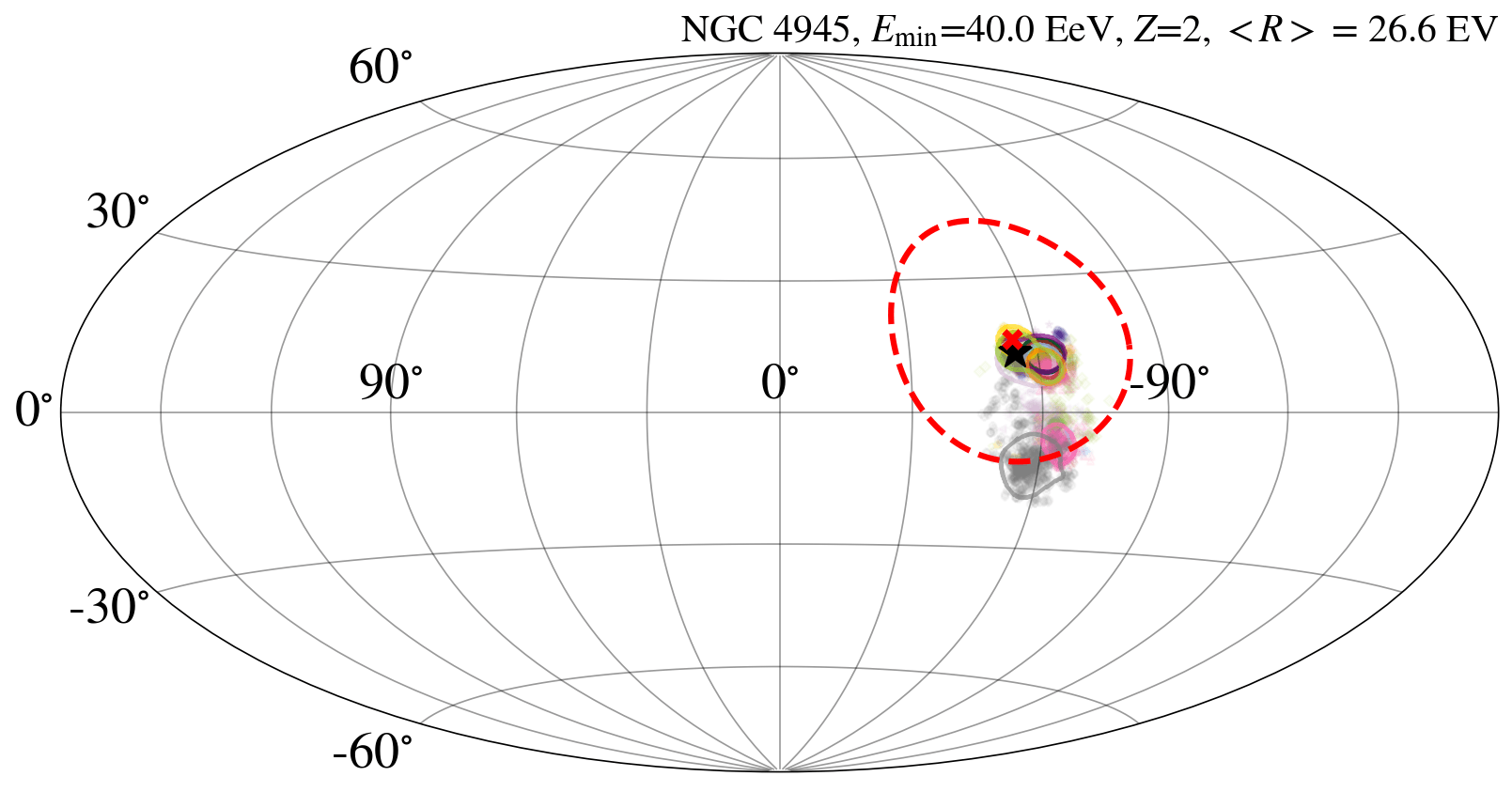}
\includegraphics[trim={12.2cm 0 0 0}, clip, width=0.19\textwidth]{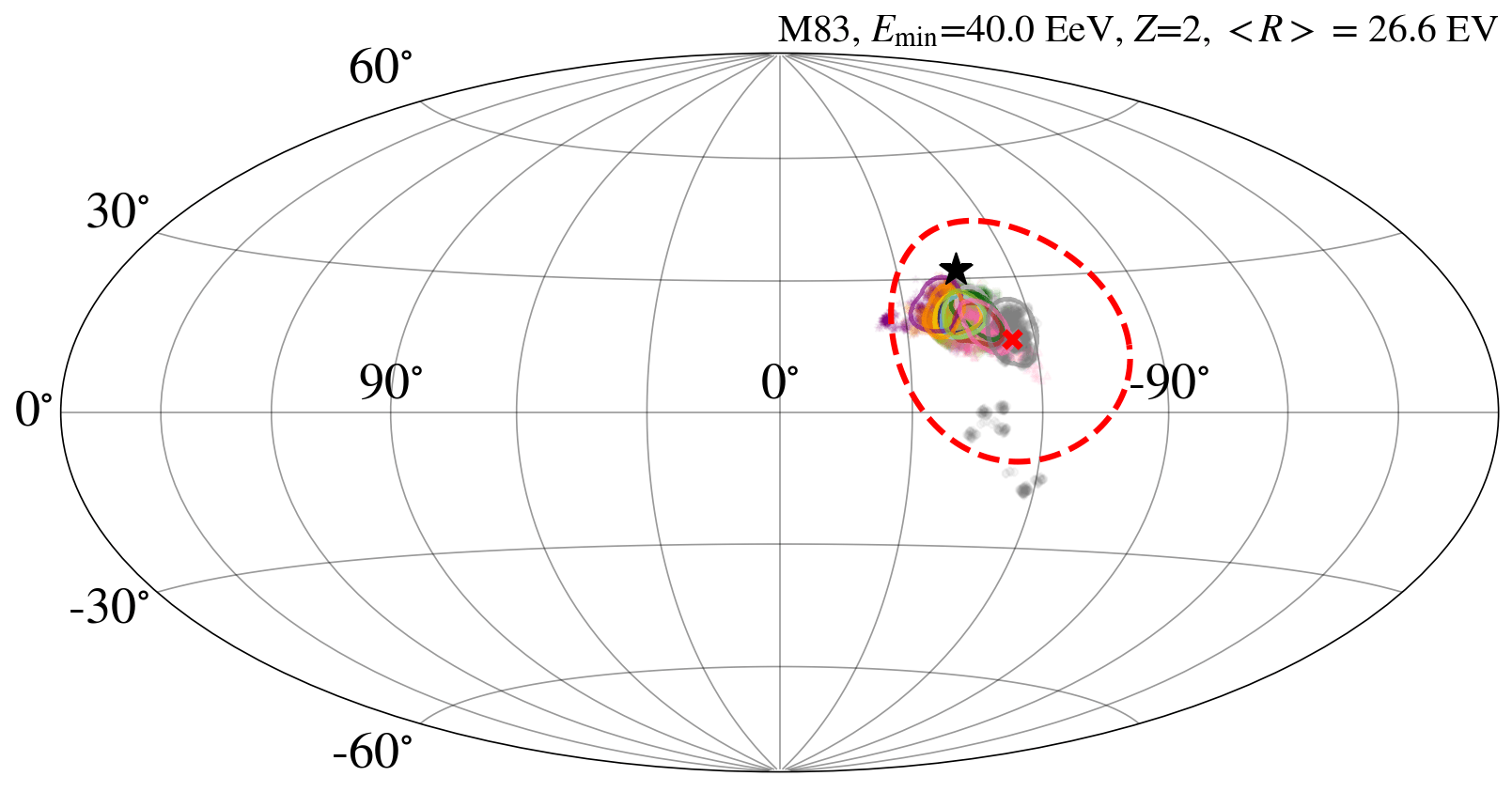}
\includegraphics[trim={12.2cm 0 0 0}, clip, width=0.19\textwidth]{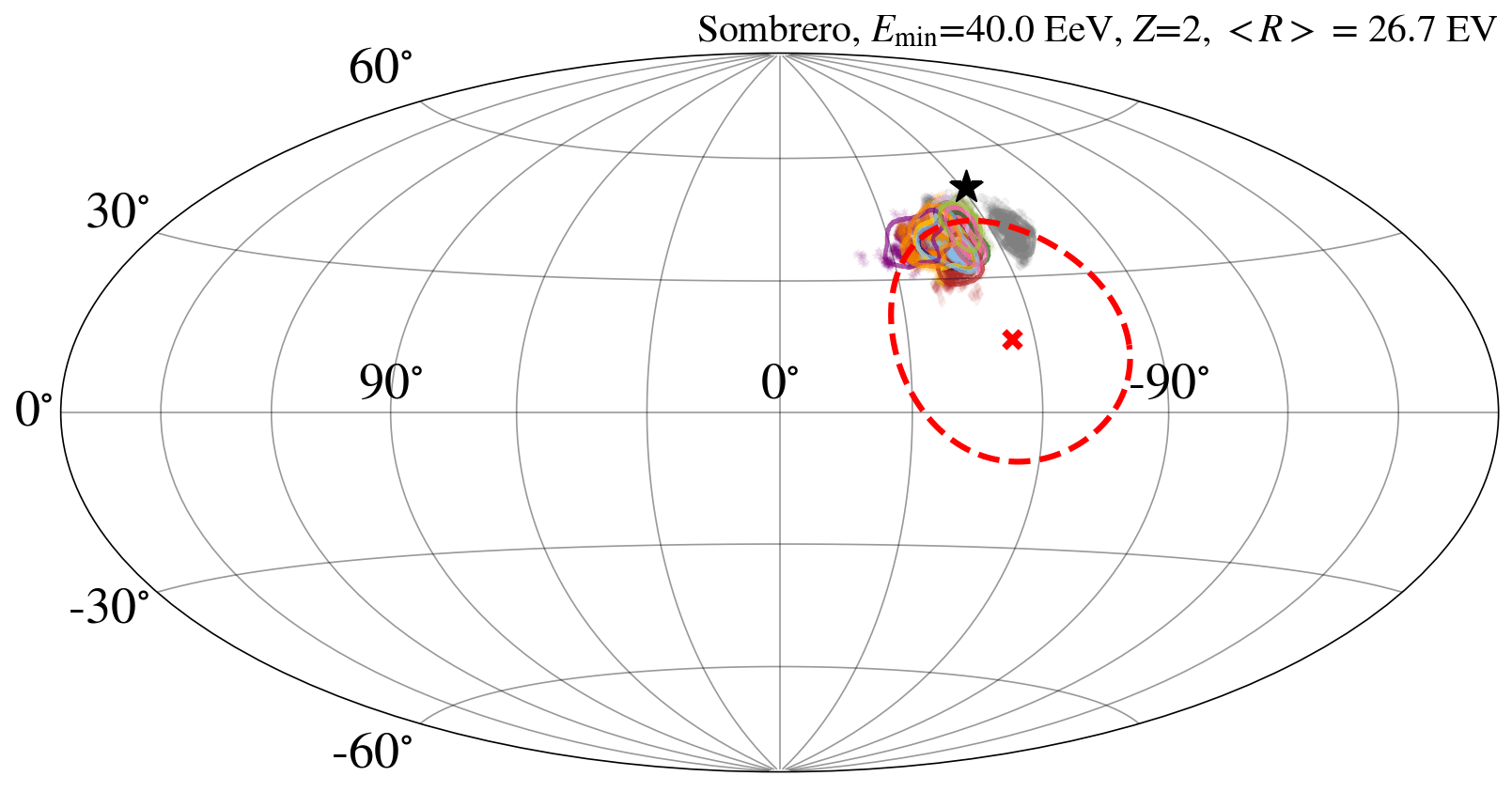}
\includegraphics[trim={12.2cm 0 0 0}, clip, width=0.19\textwidth]{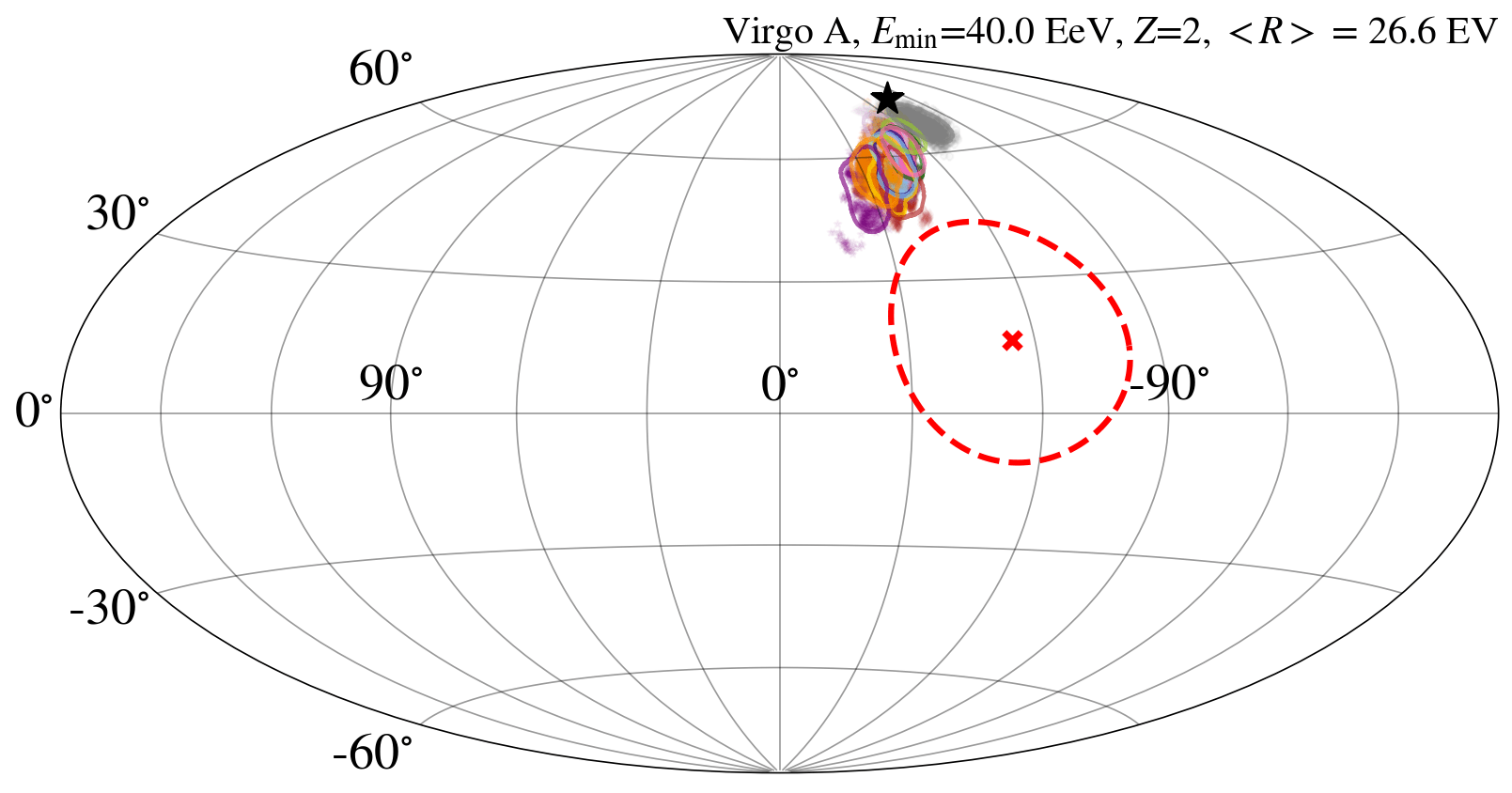}\\

\includegraphics[trim={12.2cm 0 0 0}, clip, width=0.19\textwidth]{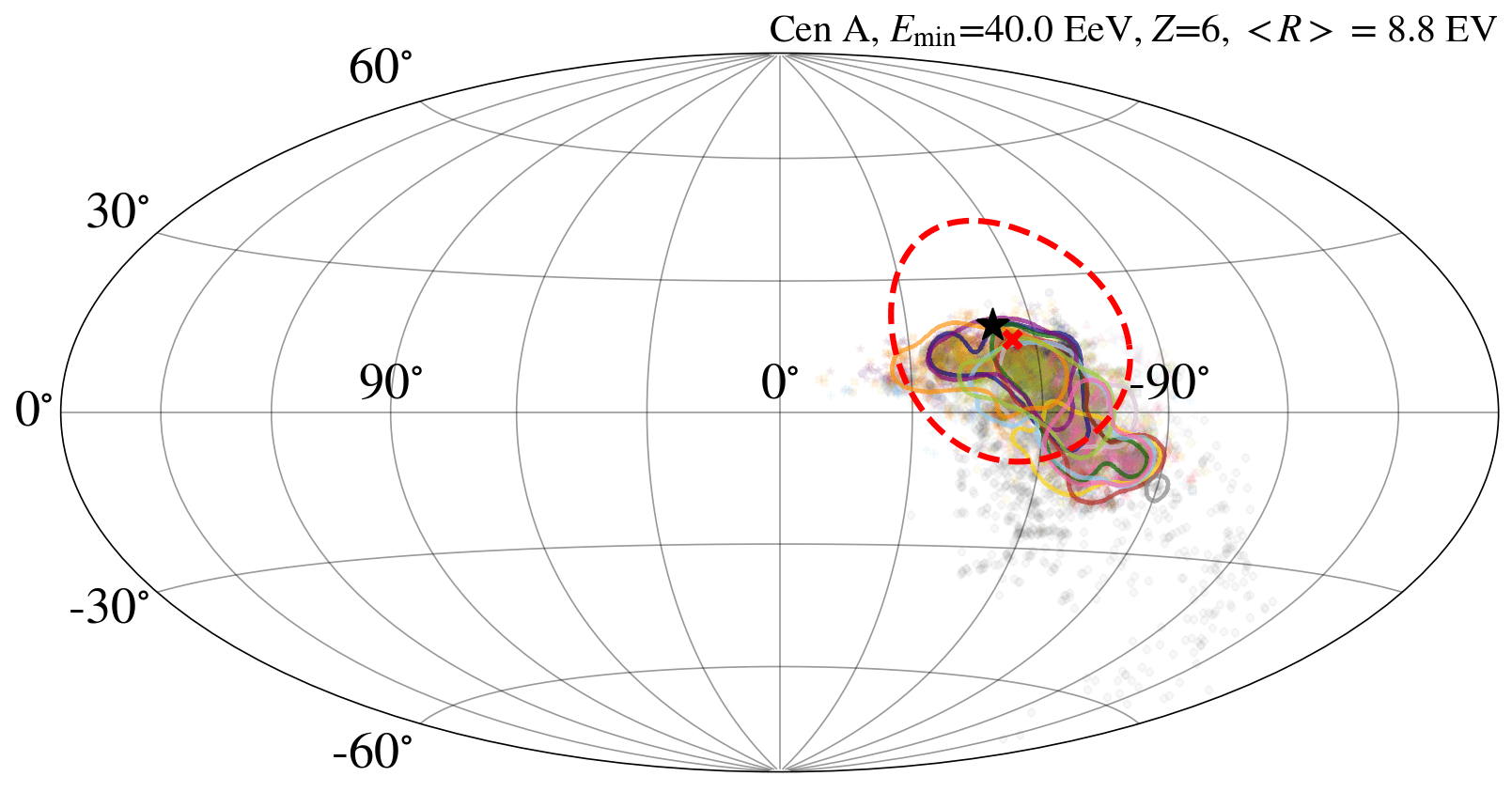}
\includegraphics[trim={12.2cm 0 0 0}, clip, width=0.19\textwidth]{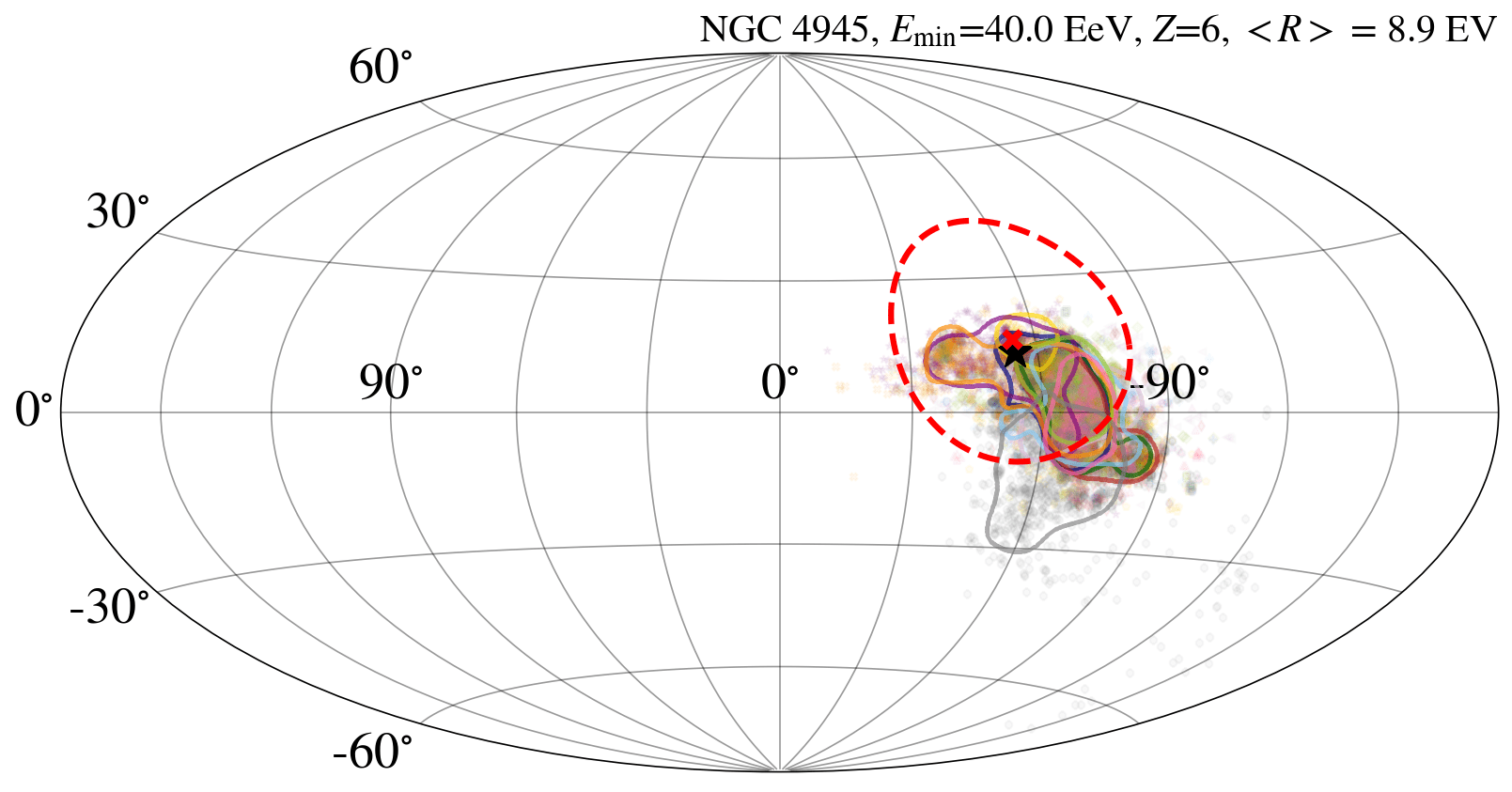}
\includegraphics[trim={12.2cm 0 0 0}, clip, width=0.19\textwidth]{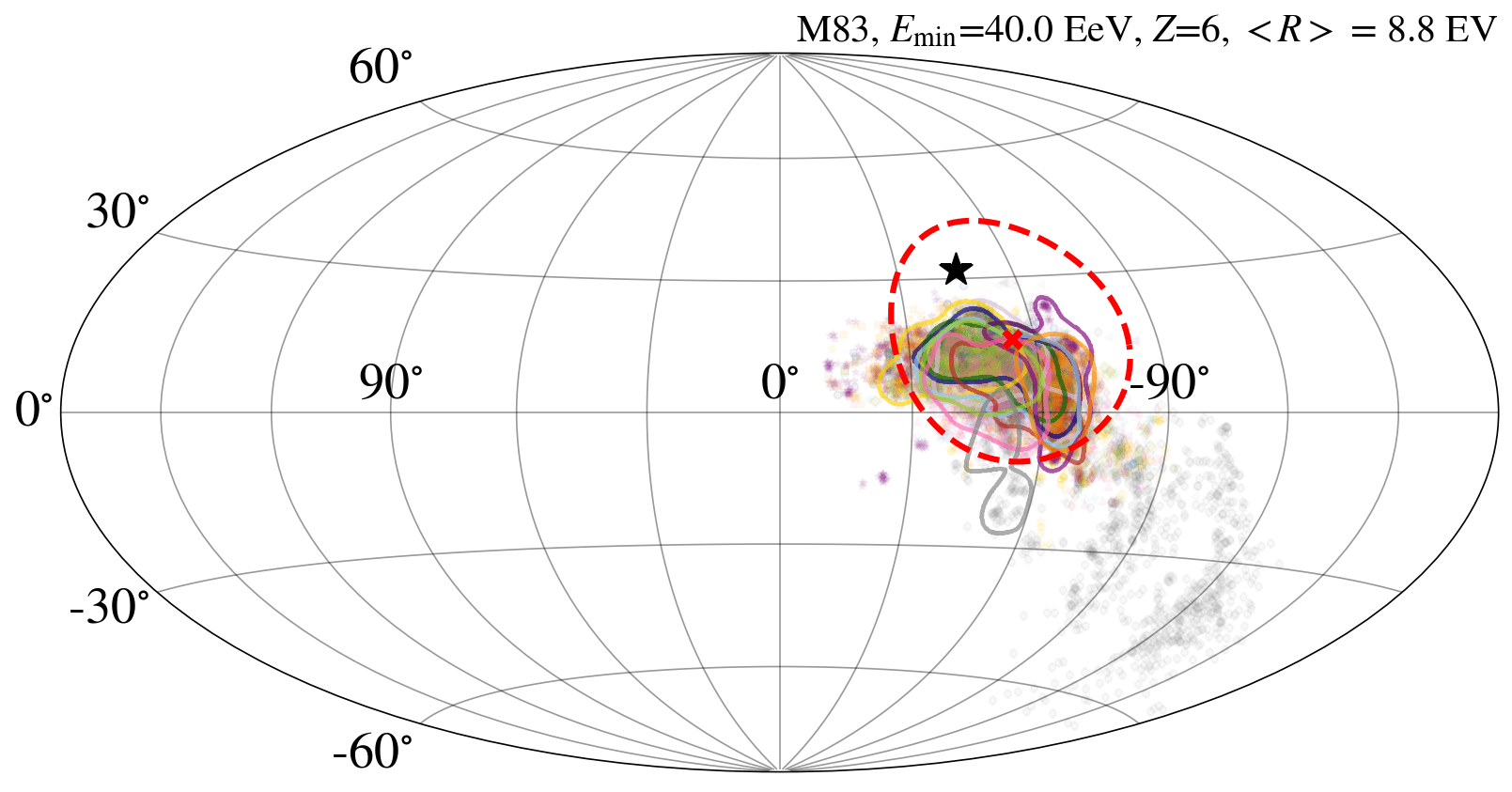}
\includegraphics[trim={12.2cm 0 0 0}, clip, width=0.19\textwidth]{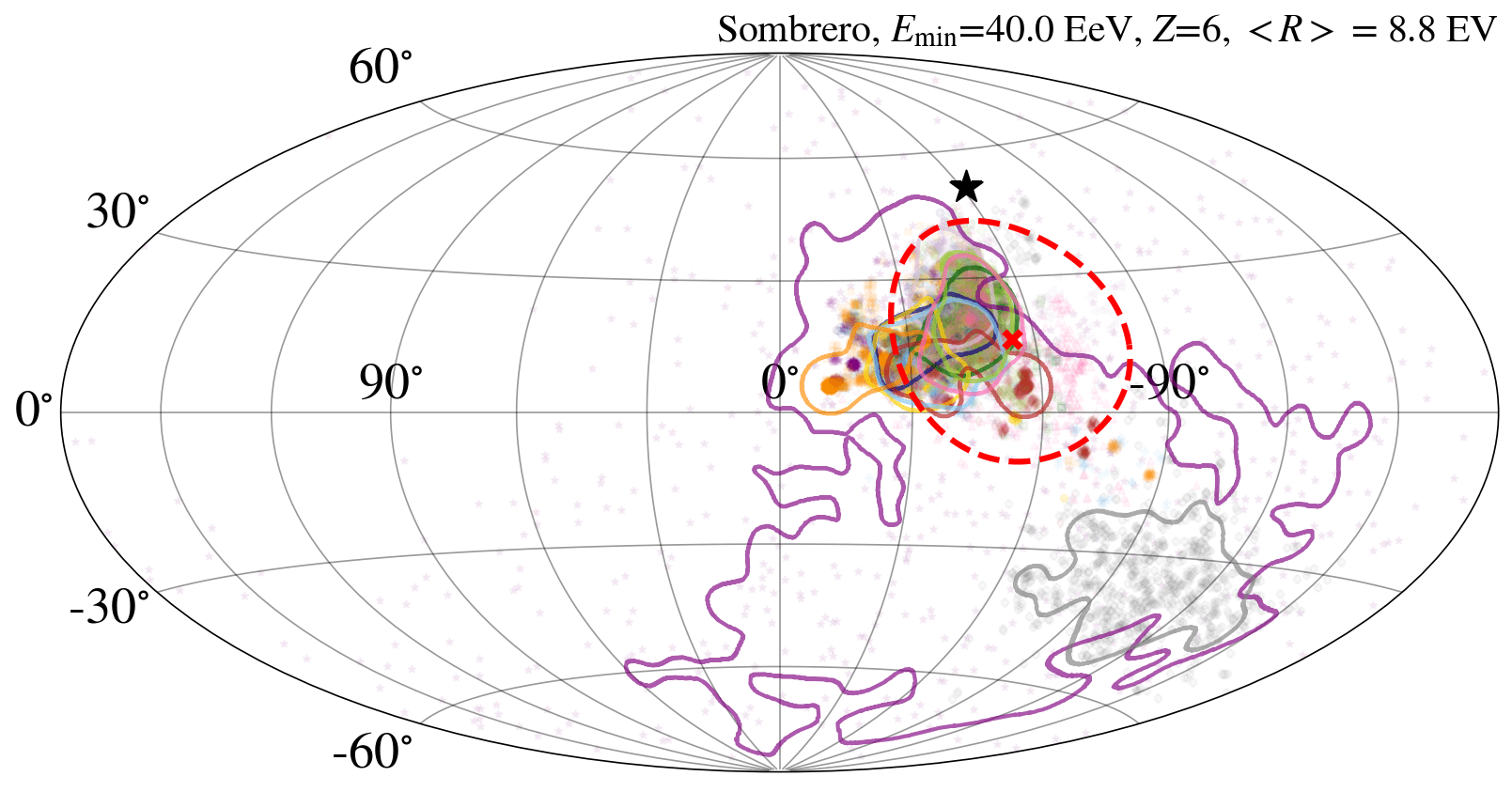}
\includegraphics[trim={12.2cm 0 0 0}, clip, width=0.19\textwidth]{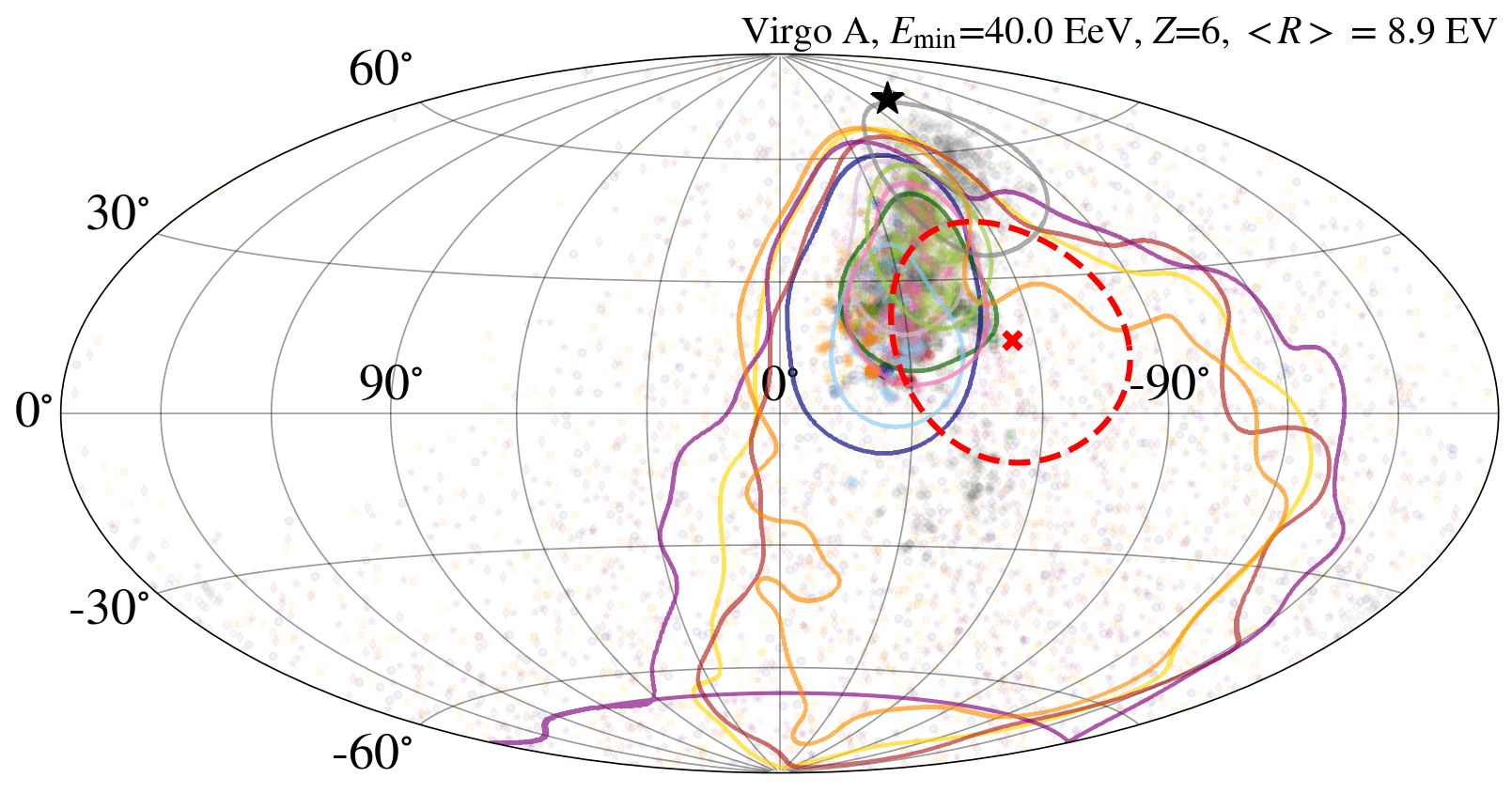}\\

\includegraphics[trim={12.2cm 0 0 0}, clip, width=0.19\textwidth]{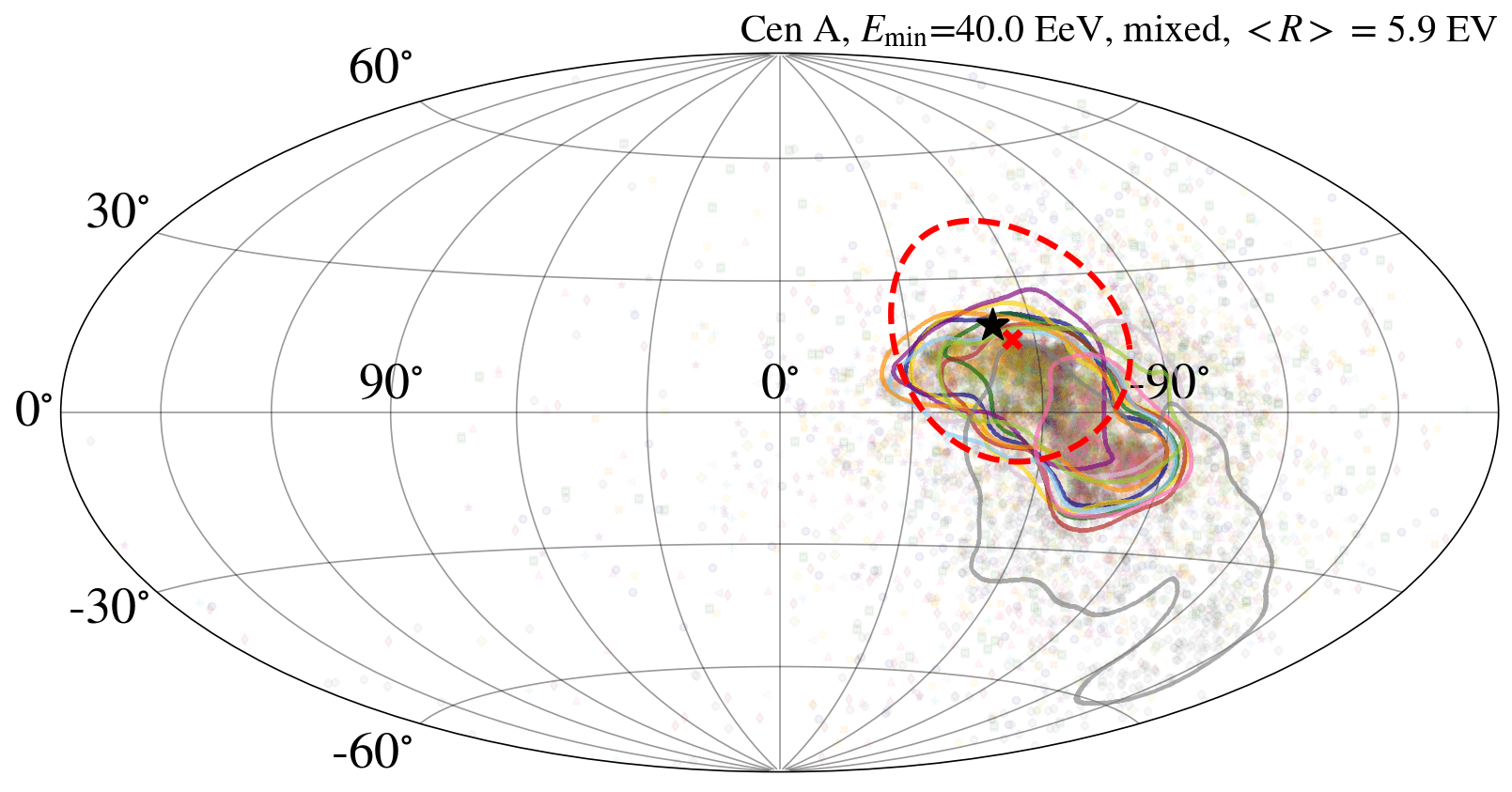}
\includegraphics[trim={12.2cm 0 0 0}, clip, width=0.19\textwidth]{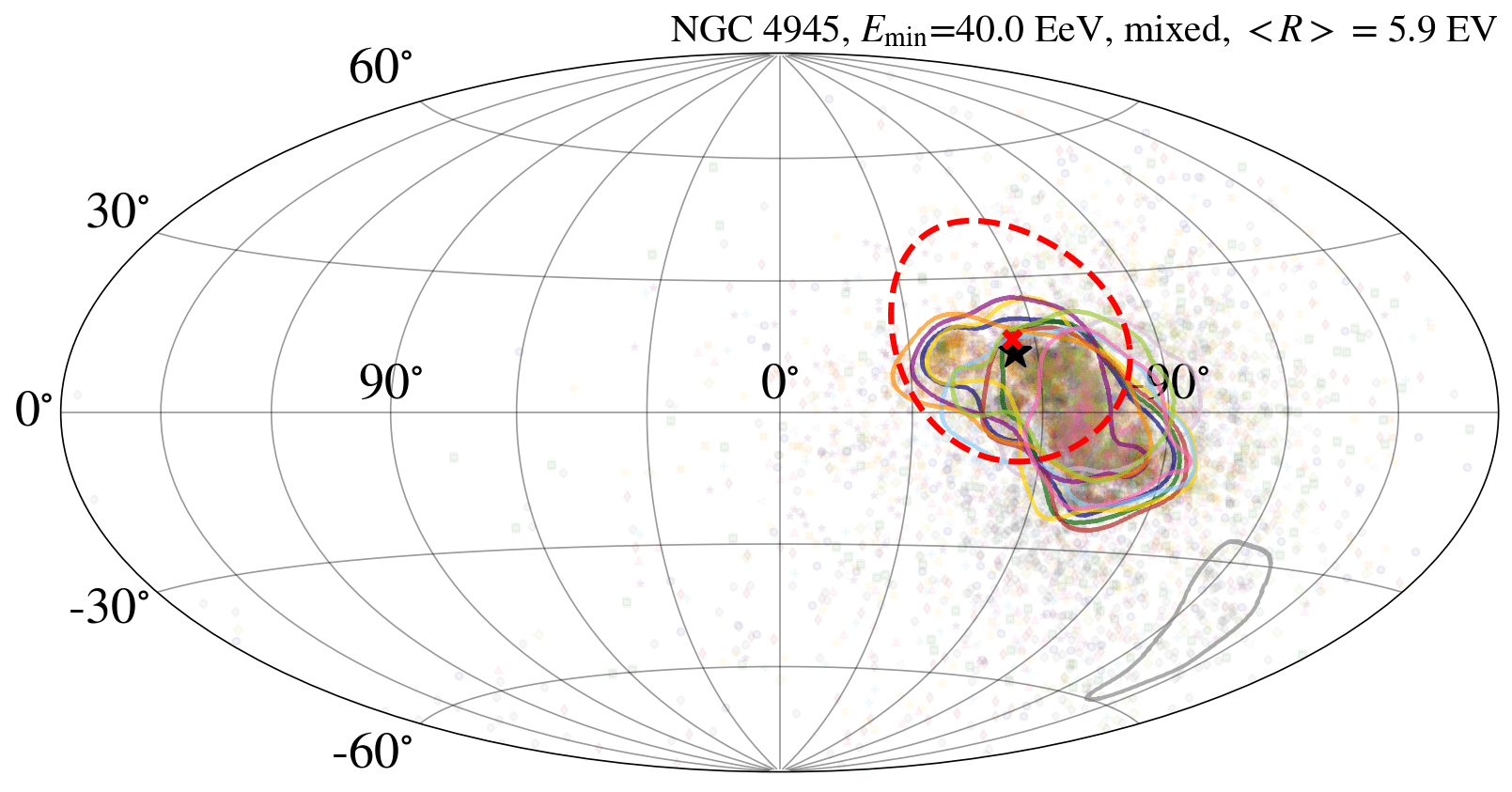}
\includegraphics[trim={12.2cm 0 0 0}, clip, width=0.19\textwidth]{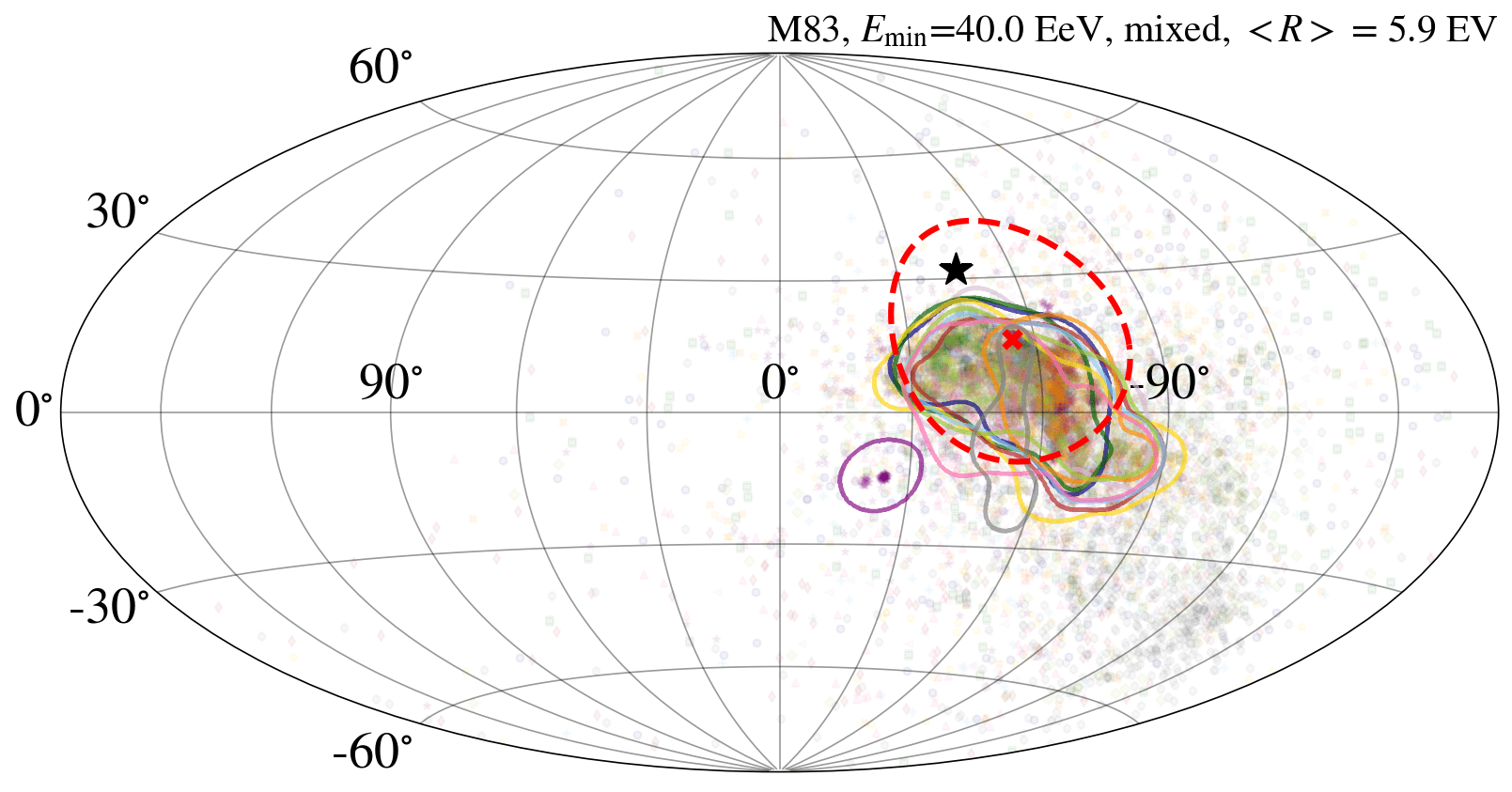}
\includegraphics[trim={12.2cm 0 0 0}, clip, width=0.19\textwidth]{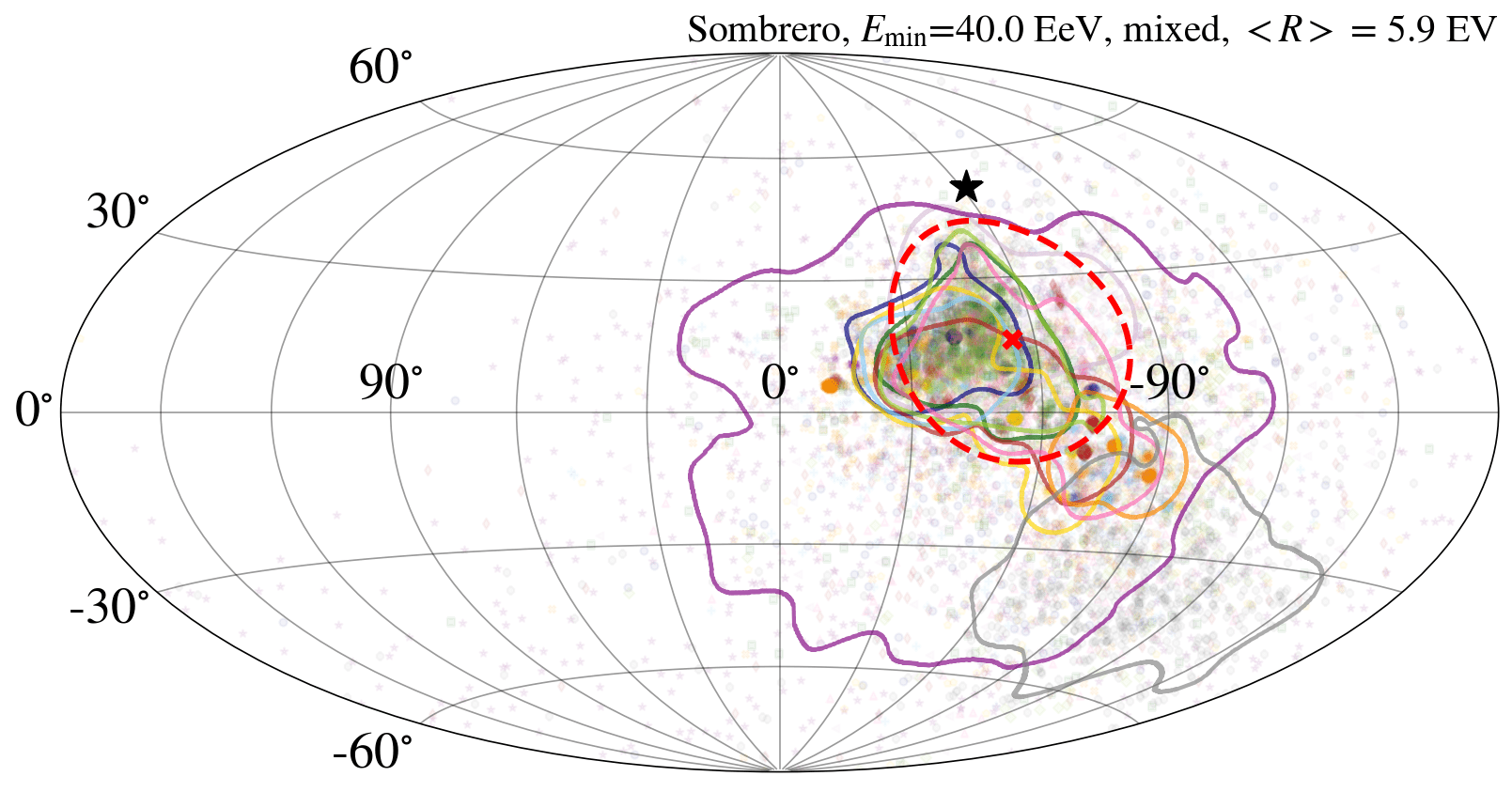}
\includegraphics[trim={12.2cm 0 0 0}, clip, width=0.19\textwidth]{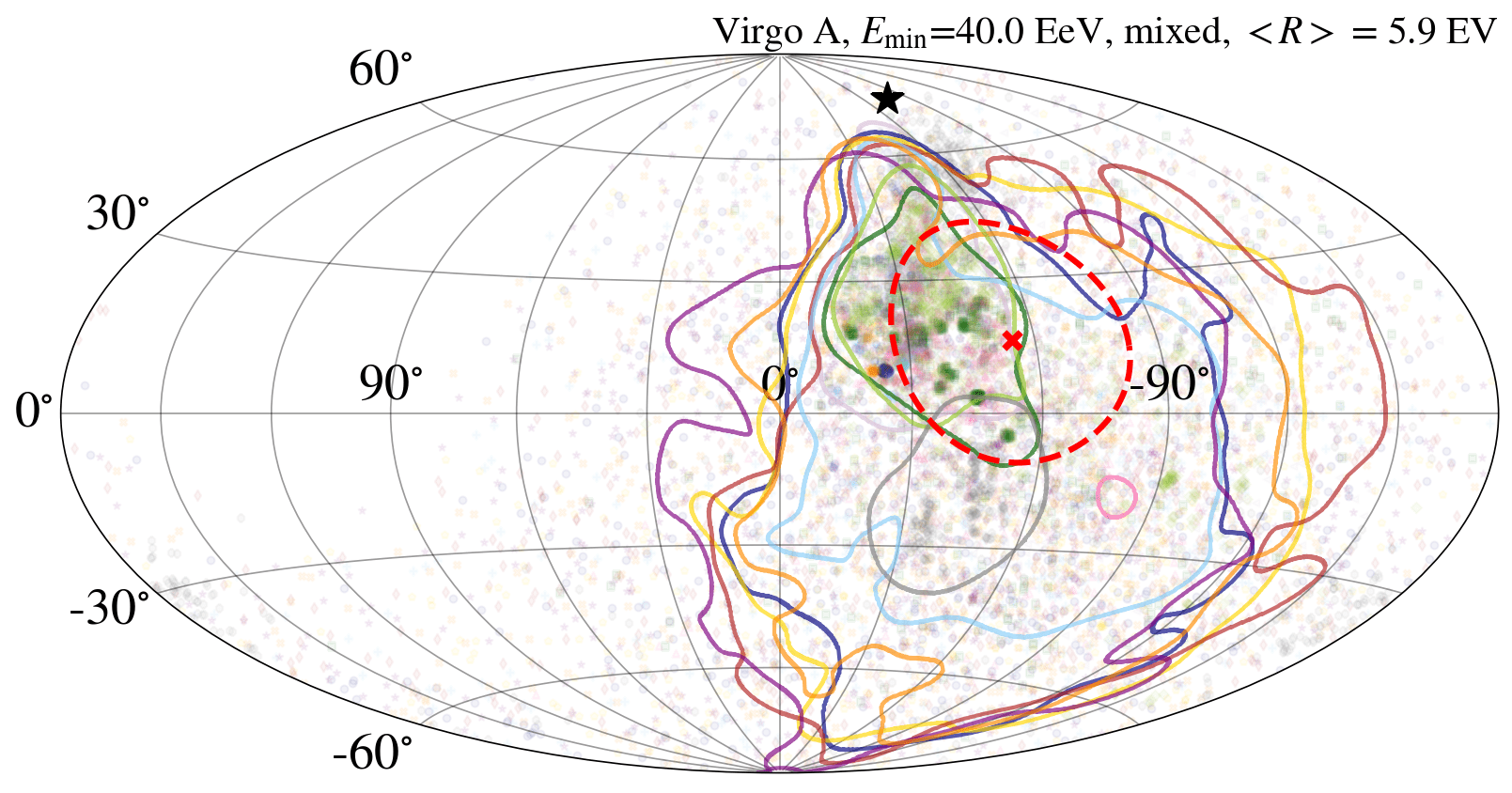}\\

\includegraphics[trim={12.2cm 0 0 0}, clip, width=0.19\textwidth]{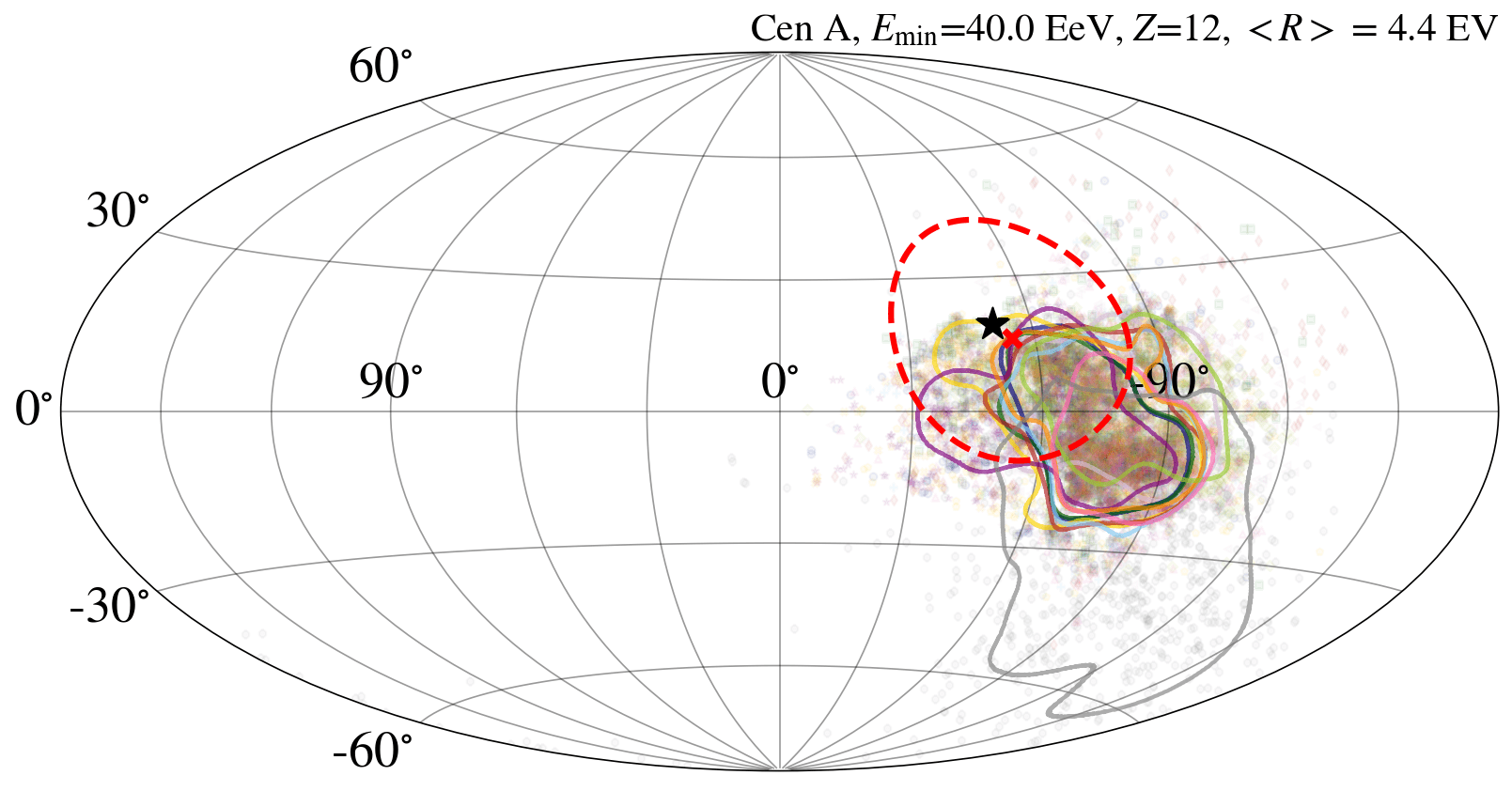}
\includegraphics[trim={12.2cm 0 0 0}, clip, width=0.19\textwidth]{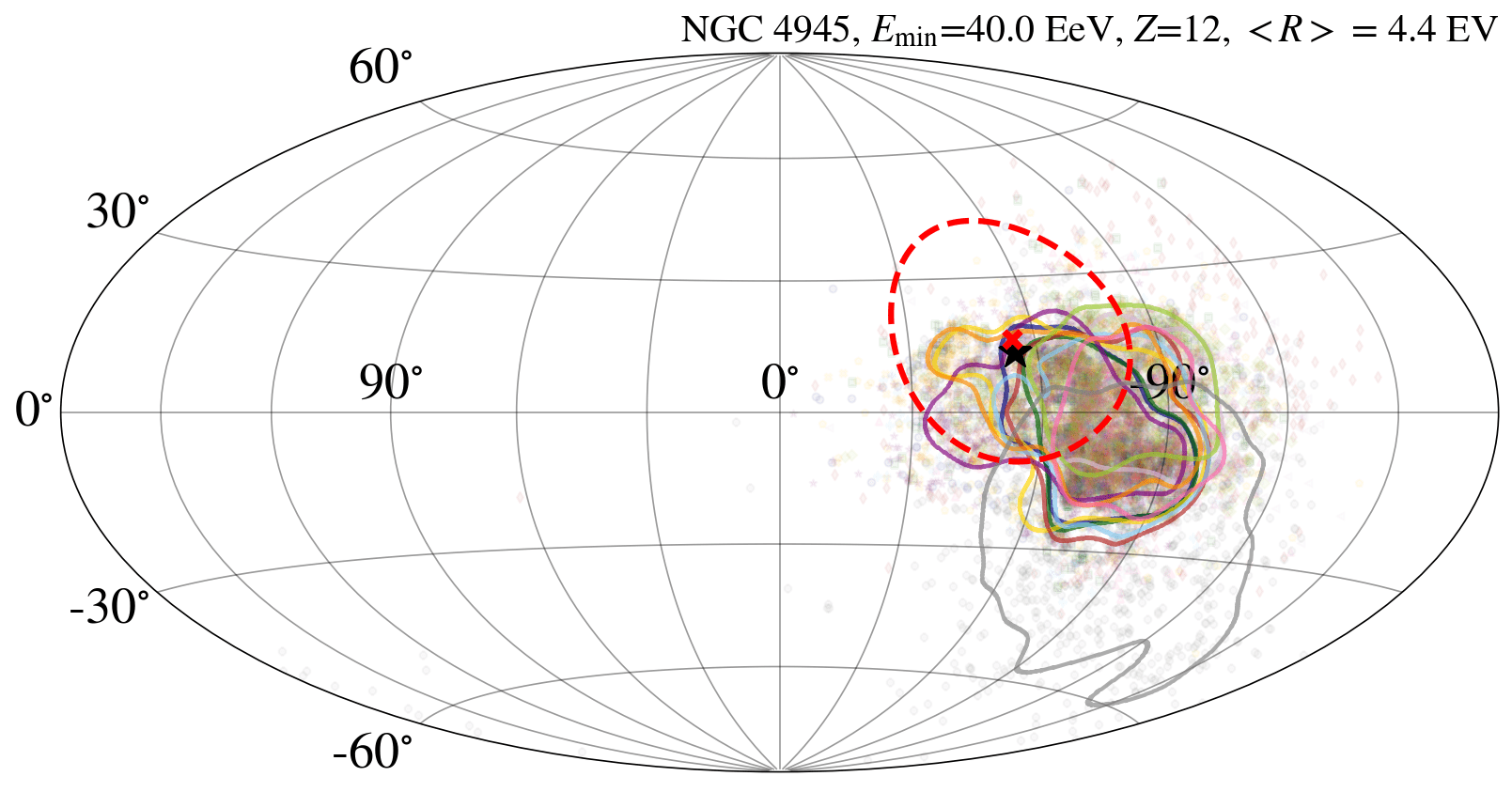}
\includegraphics[trim={12.2cm 0 0 0}, clip, width=0.19\textwidth]{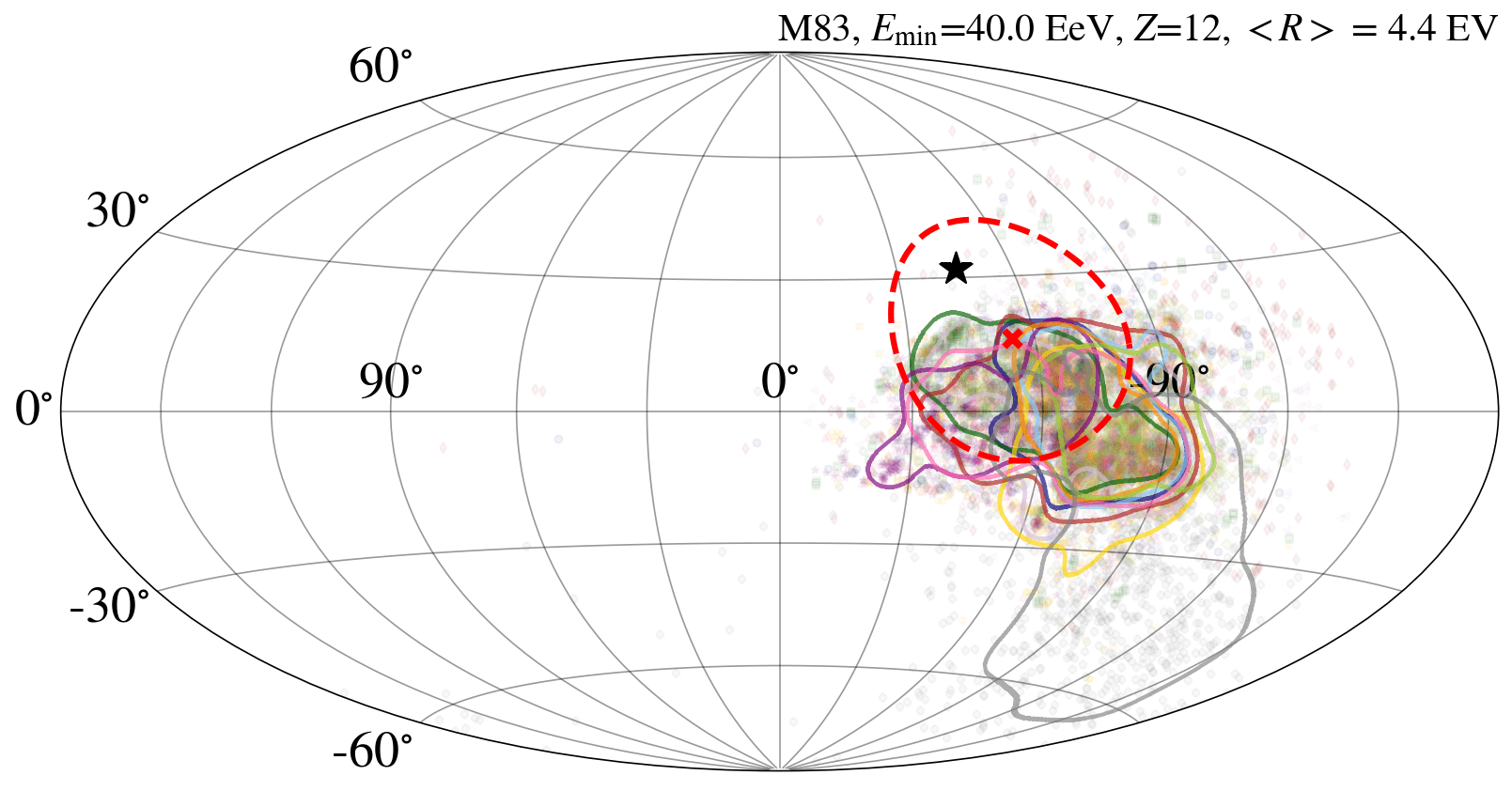}
\includegraphics[trim={12.2cm 0 0 0}, clip, width=0.19\textwidth]{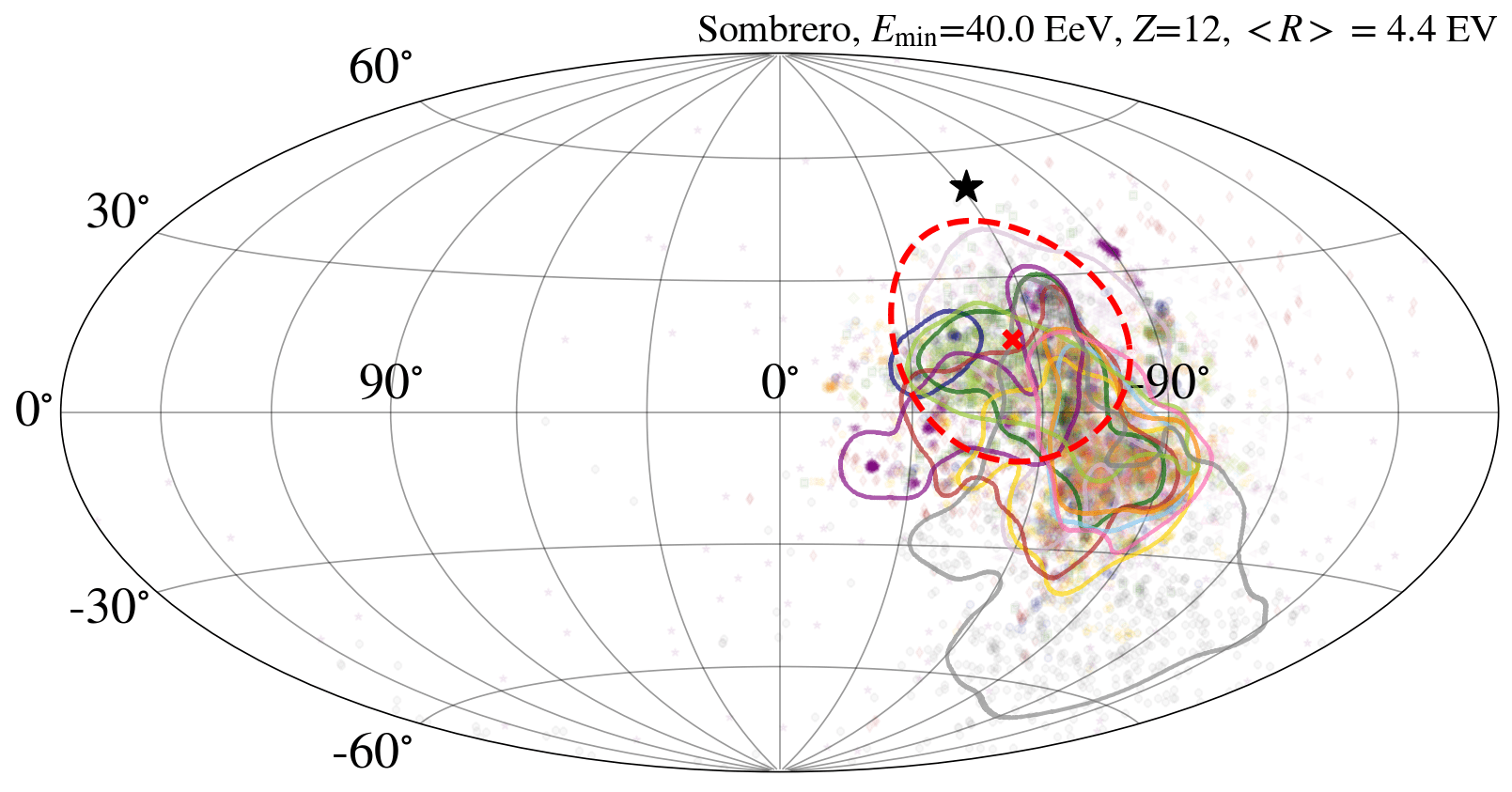}
\includegraphics[trim={12.2cm 0 0 0}, clip, width=0.19\textwidth]{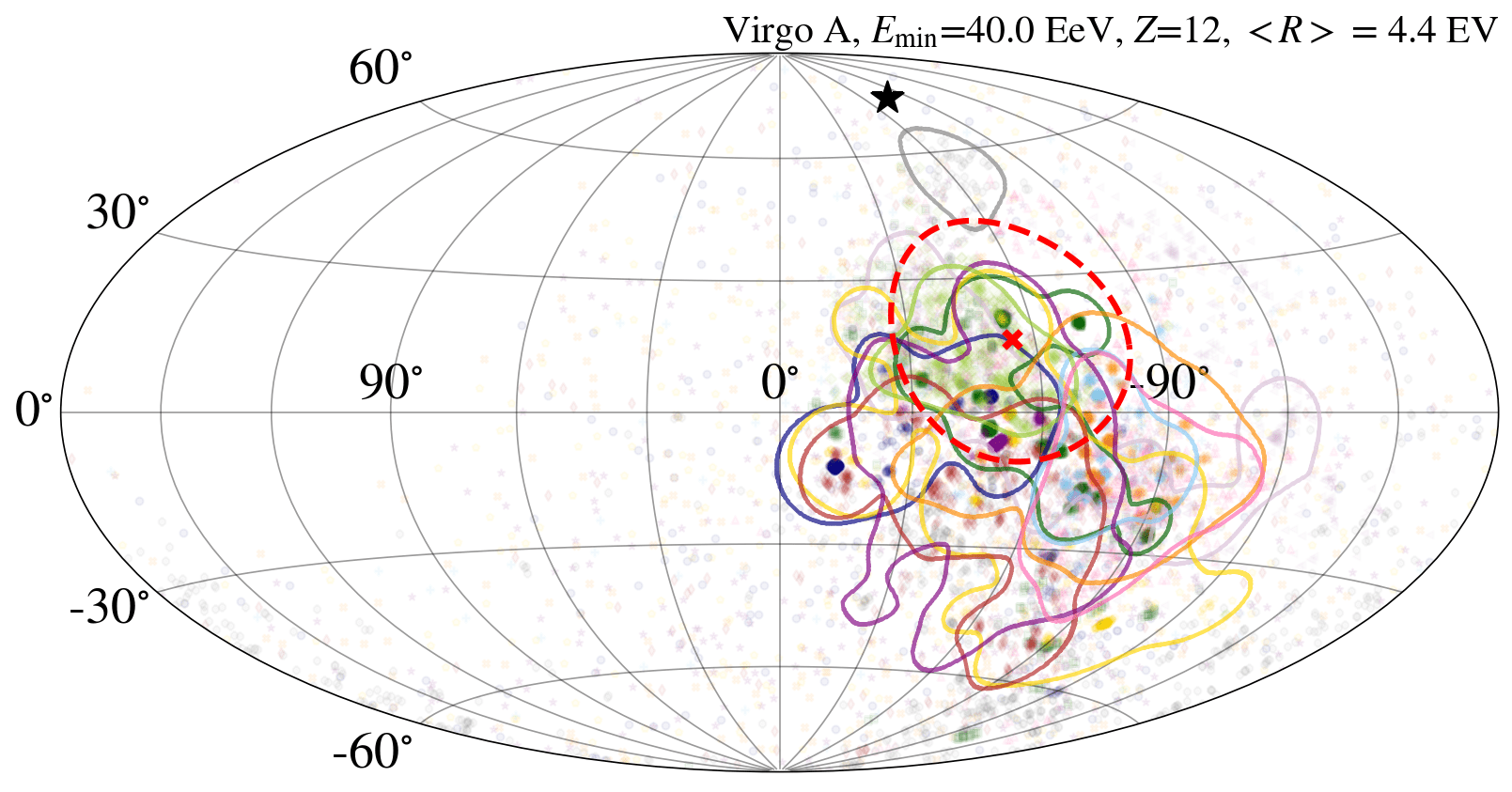}\\

\includegraphics[trim={12.2cm 0 0 0}, clip, width=0.19\textwidth]{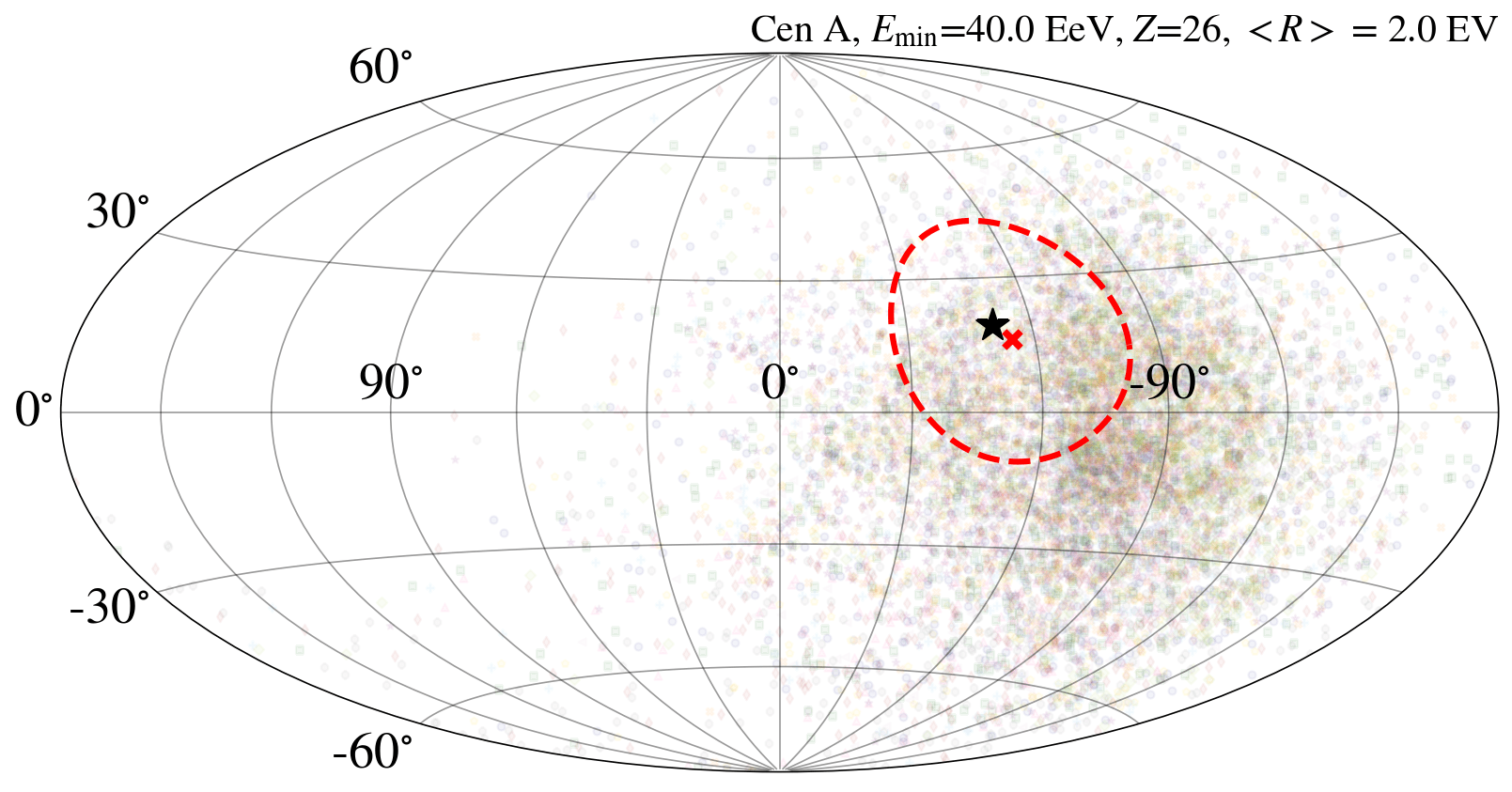}
\includegraphics[trim={12.2cm 0 0 0}, clip, width=0.19\textwidth]{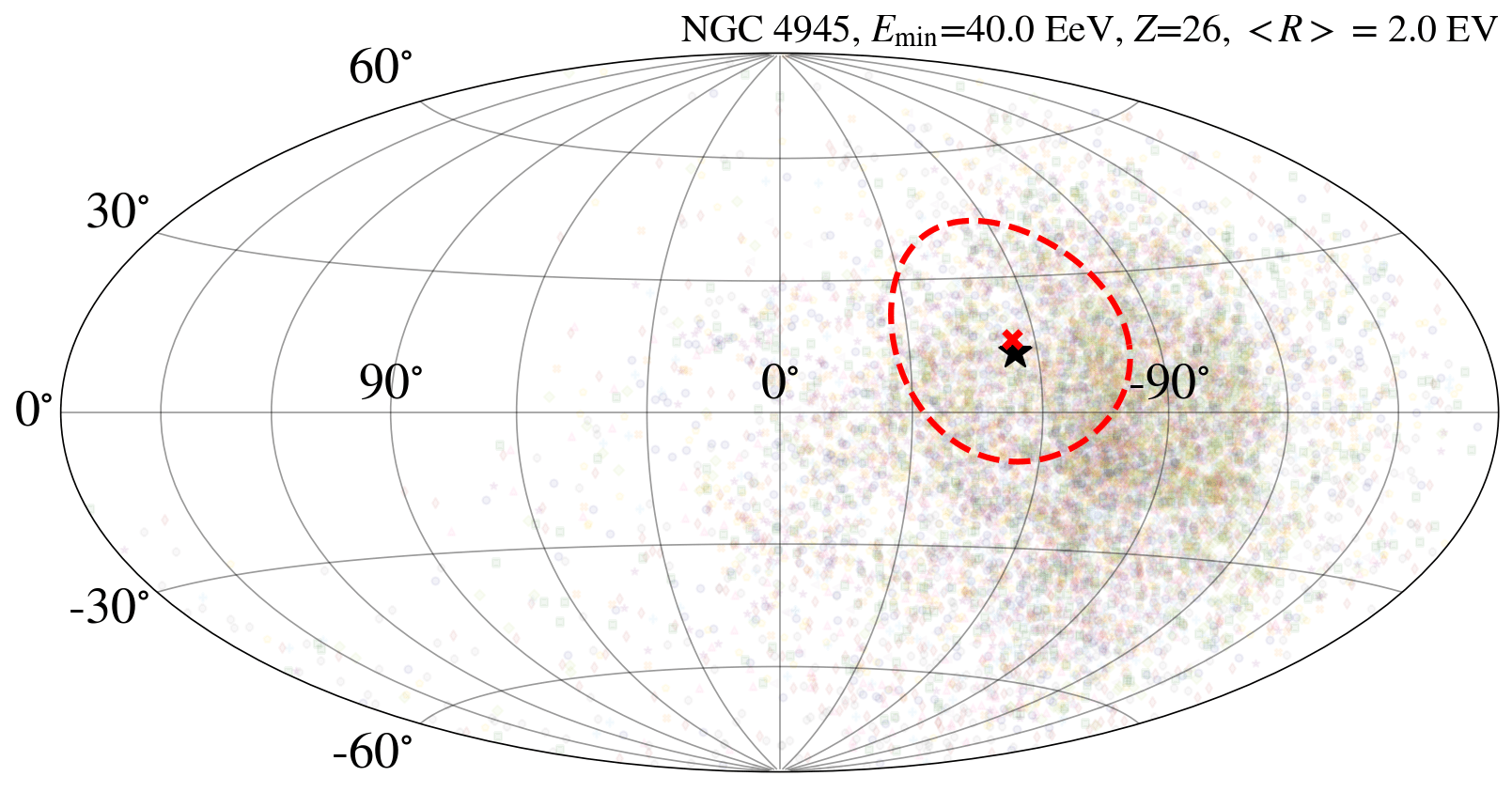}
\includegraphics[trim={12.2cm 0 0 0}, clip, width=0.19\textwidth]{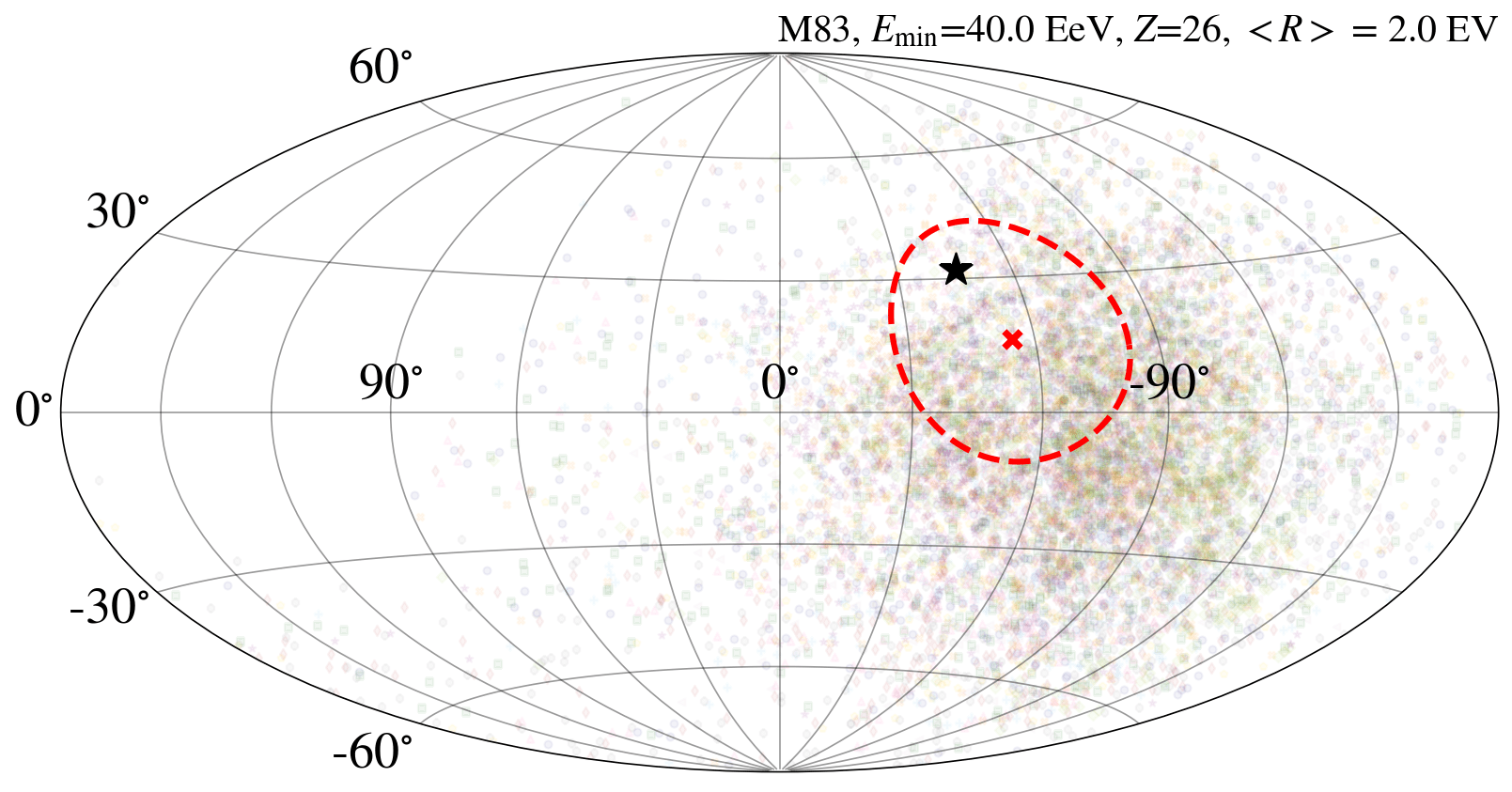}
\includegraphics[trim={12.2cm 0 0 0}, clip, width=0.19\textwidth]{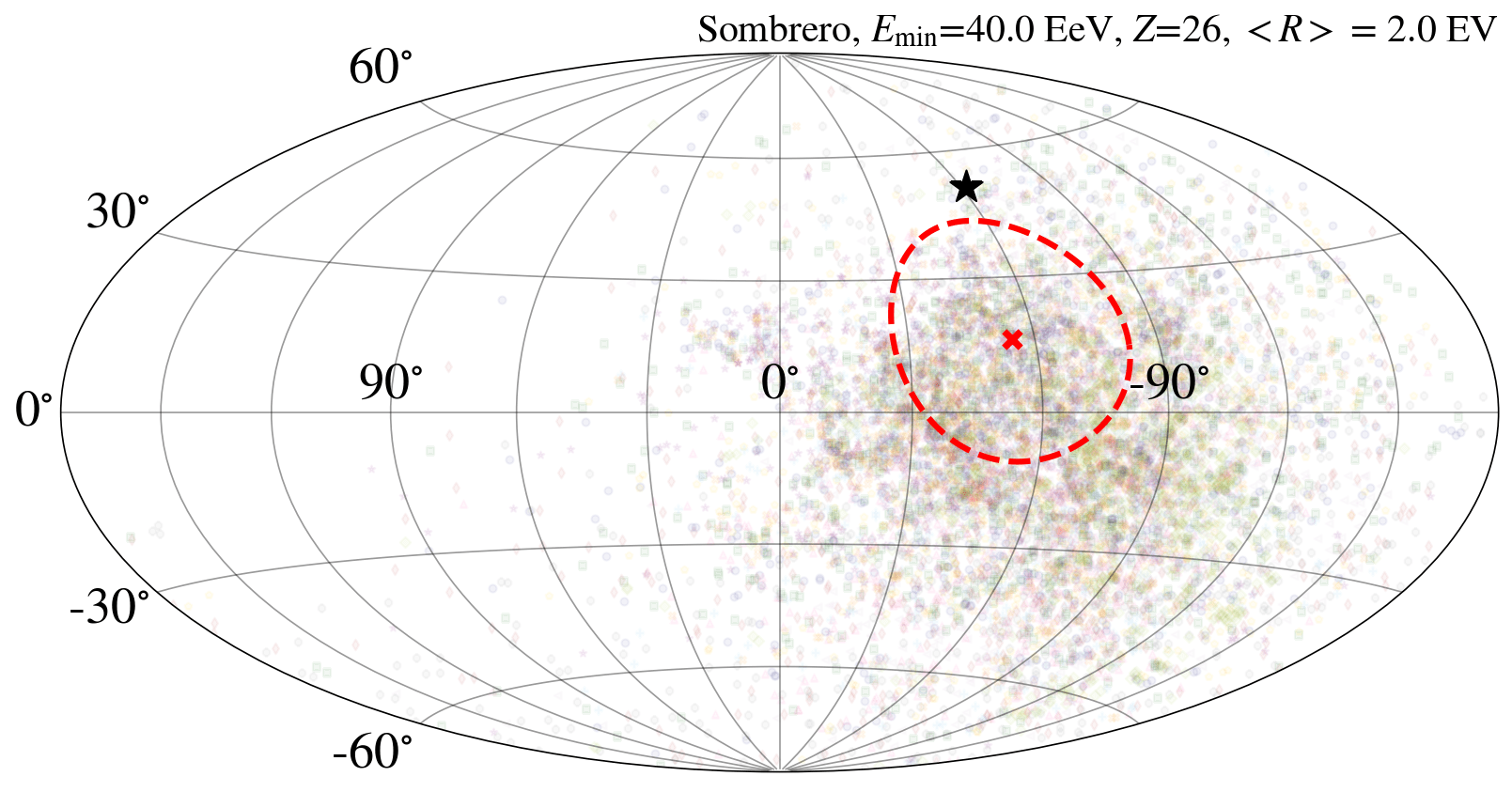}
\includegraphics[trim={12.2cm 0 0 0}, clip, width=0.19\textwidth]{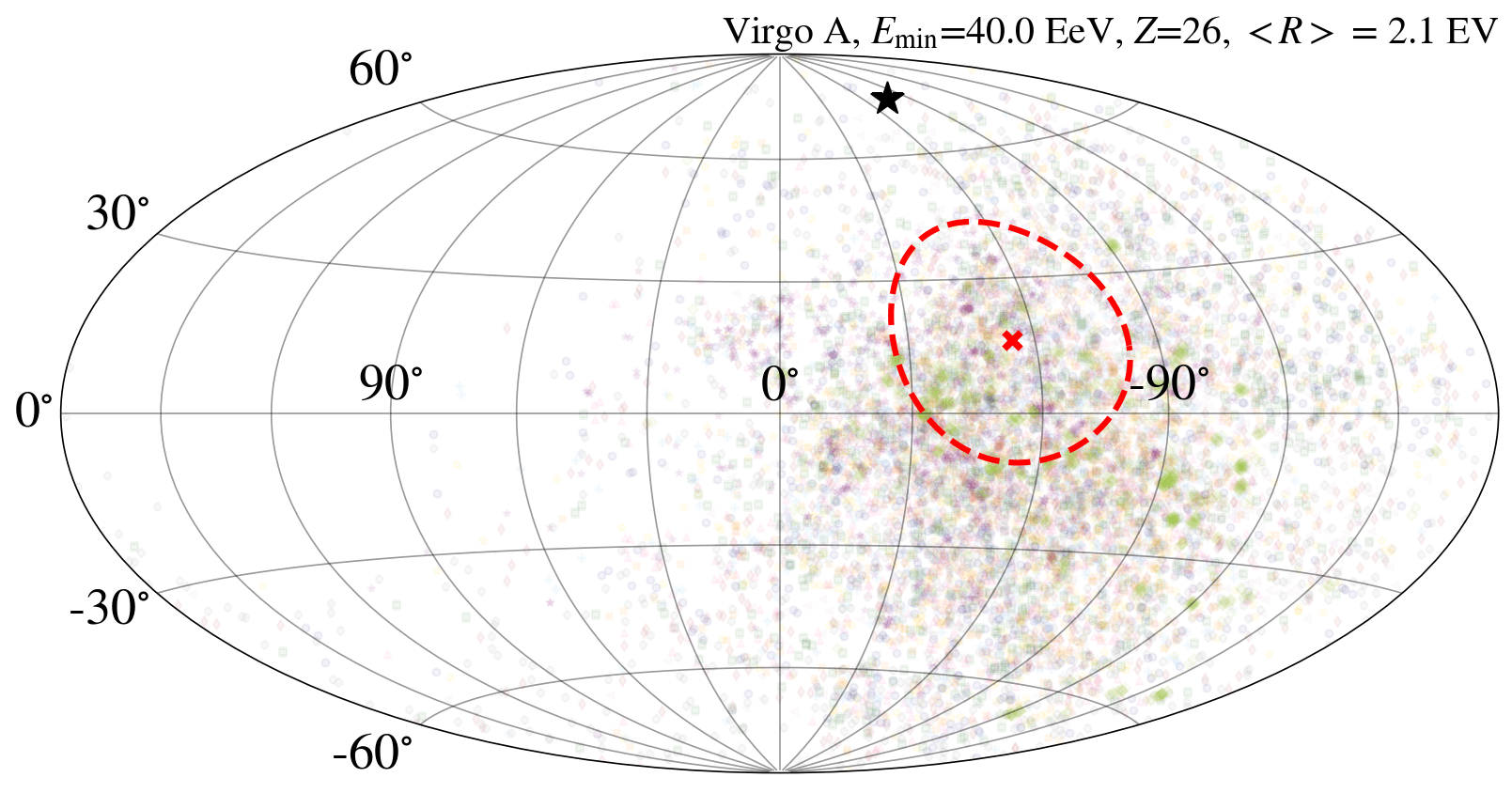}\\
\captionof{figure}{Simulated arrival directions for the five different source candidates tested in this work (columns), and different charge numbers (rows). All figures are for $E_\mathrm{min}=40\,\mathrm{EeV}$ and no EGMF blurring. The contours contain roughly 90\% of the distribution and are only meant to guide the eye. For some cases, e.g. Virgo A with $Z=12$, \textit{multiple images} are visible, an effect of the GMF.}
\label{fig:ADs}
\end{minipage}

\section{Alternative sources in the Centaurus region: NGC 4945} \label{app:ngc4945}
\noindent
\begin{minipage}{\textwidth}
\centering
\includegraphics[width=0.49\textwidth]{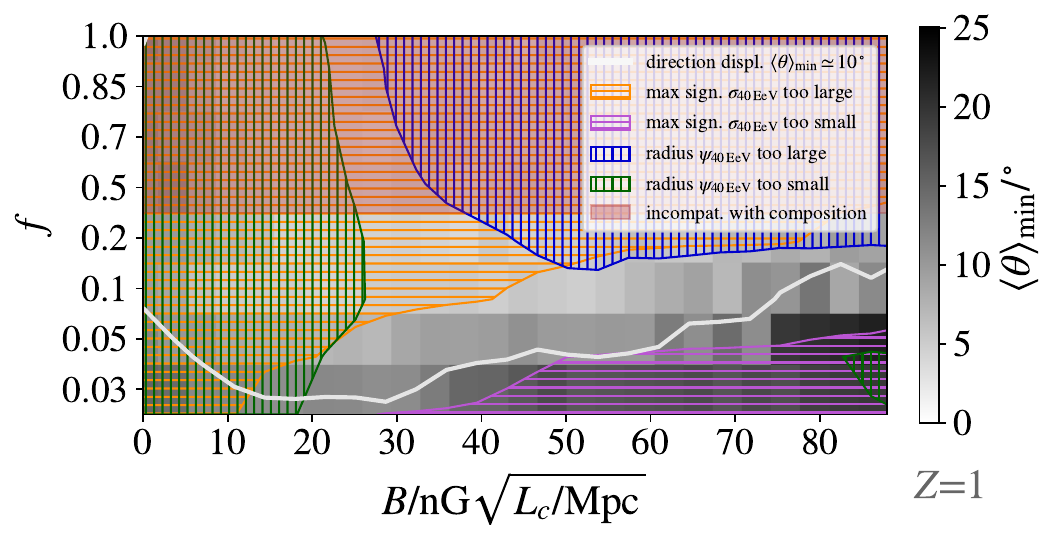}
\includegraphics[width=0.49\textwidth]{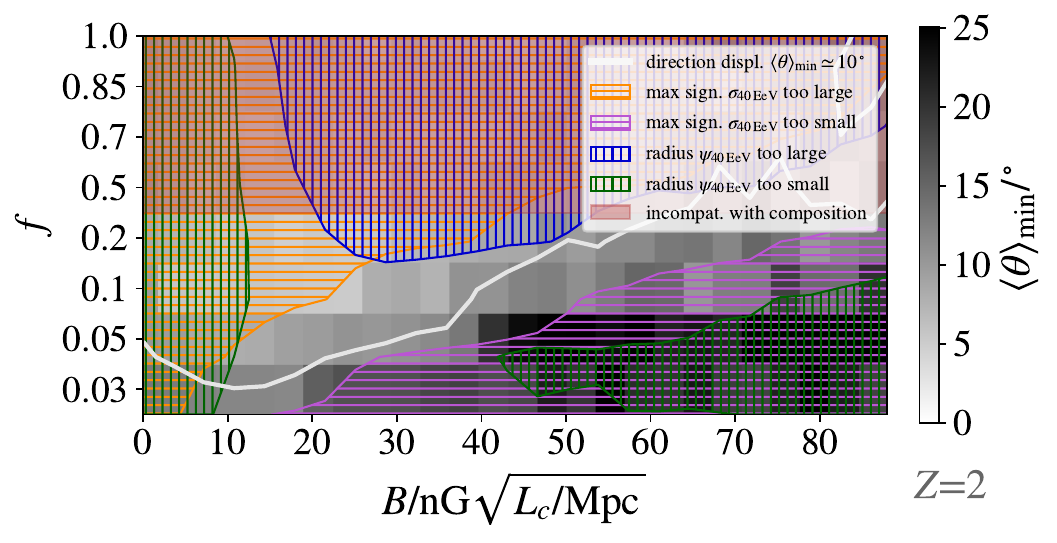}

\includegraphics[width=0.49\textwidth]{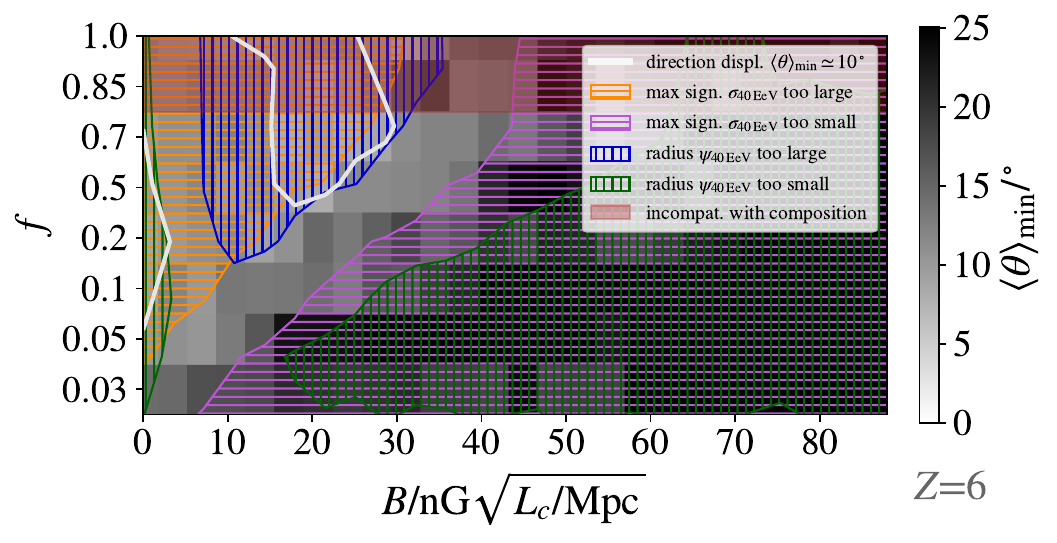}
\includegraphics[width=0.49\textwidth]{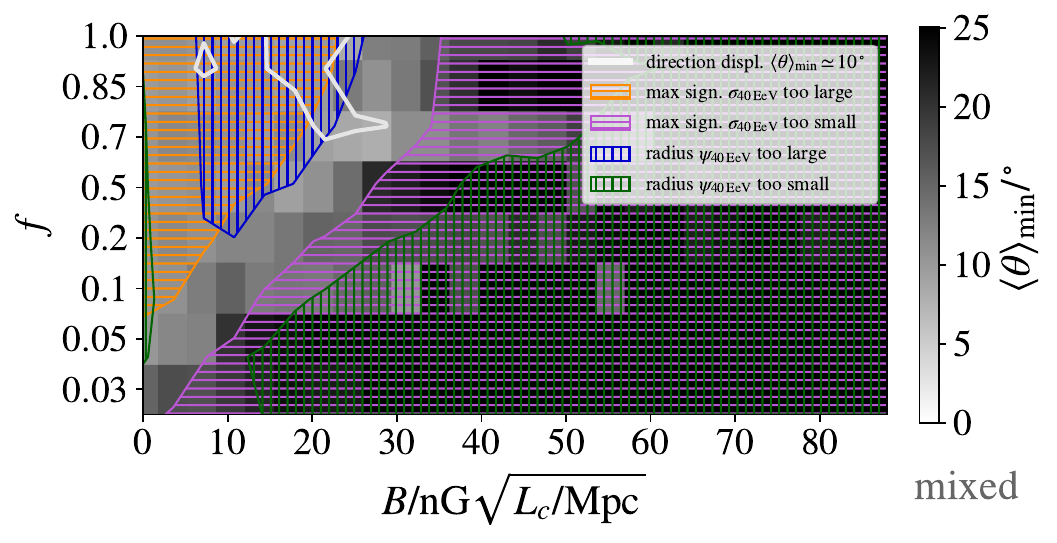}

\includegraphics[width=0.49\textwidth]{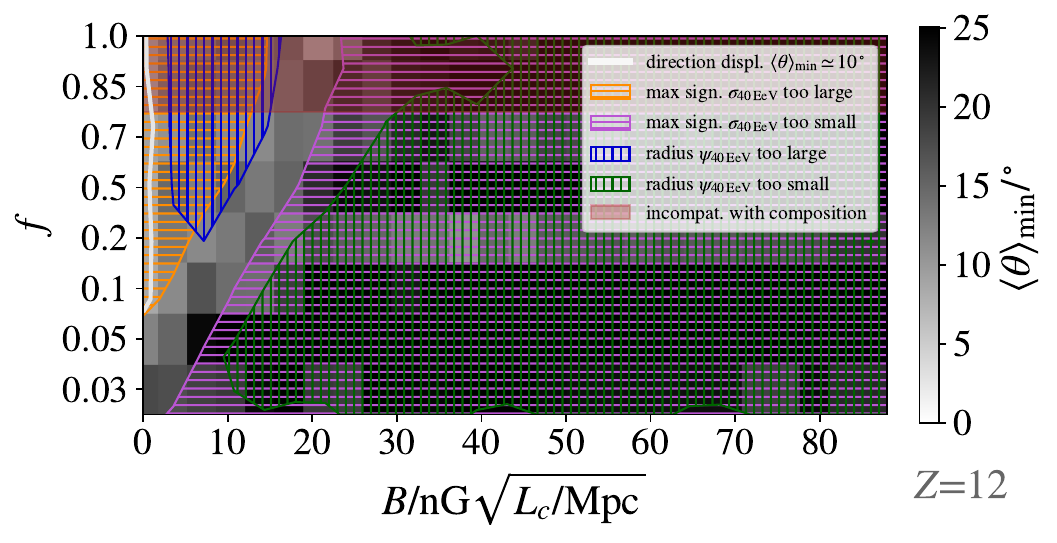}
\includegraphics[width=0.49\textwidth]{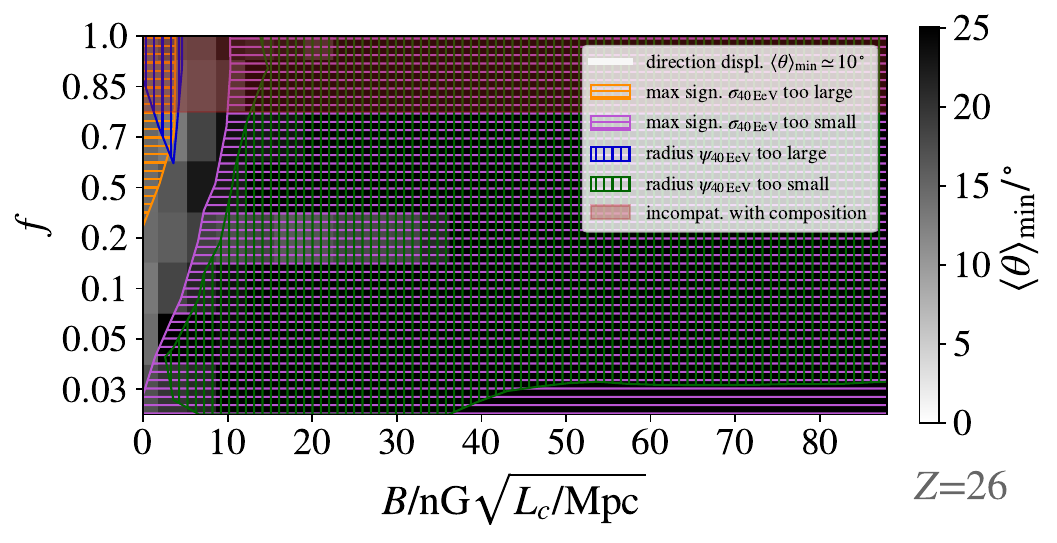}
\captionof{figure}{Constraints on the parameter space for NGC 4945 as the source of the observed excess. See Fig.~\ref{fig:cena_constraints} for more details.}
\label{fig:}
\end{minipage}

\vspace{1cm}
\noindent
\begin{minipage}{\textwidth}
\centering
\includegraphics[trim={13cm 0 0 0}, clip, width=0.24\textwidth]{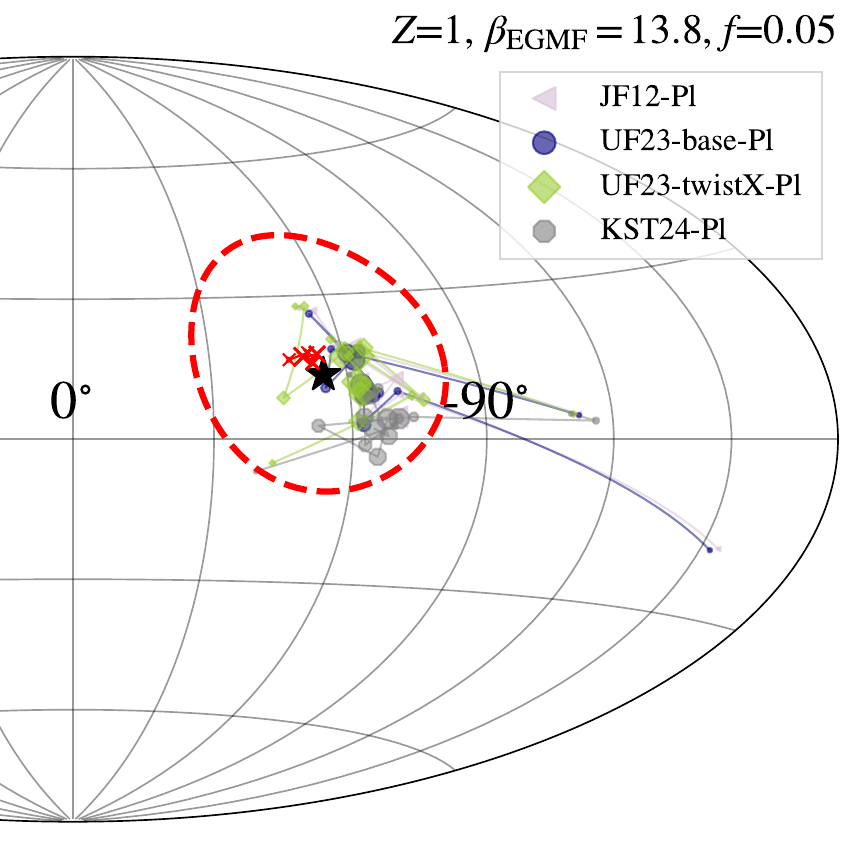}
\includegraphics[trim={13cm 0 0 0}, clip, width=0.24\textwidth]{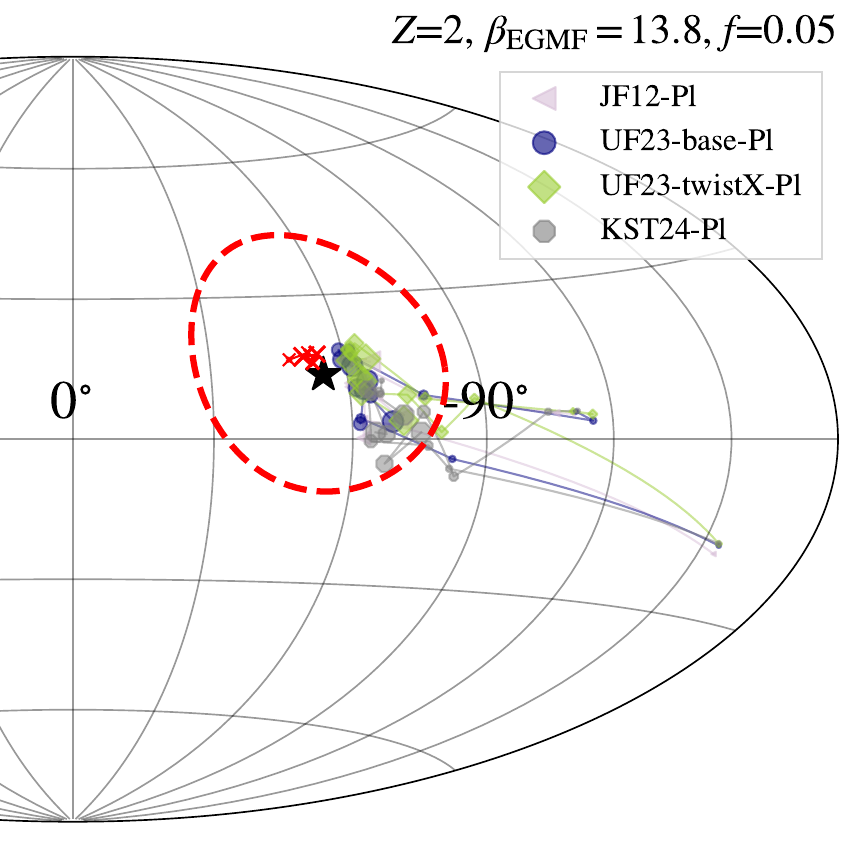}
\includegraphics[trim={13cm 0 0 0}, clip, width=0.24\textwidth]{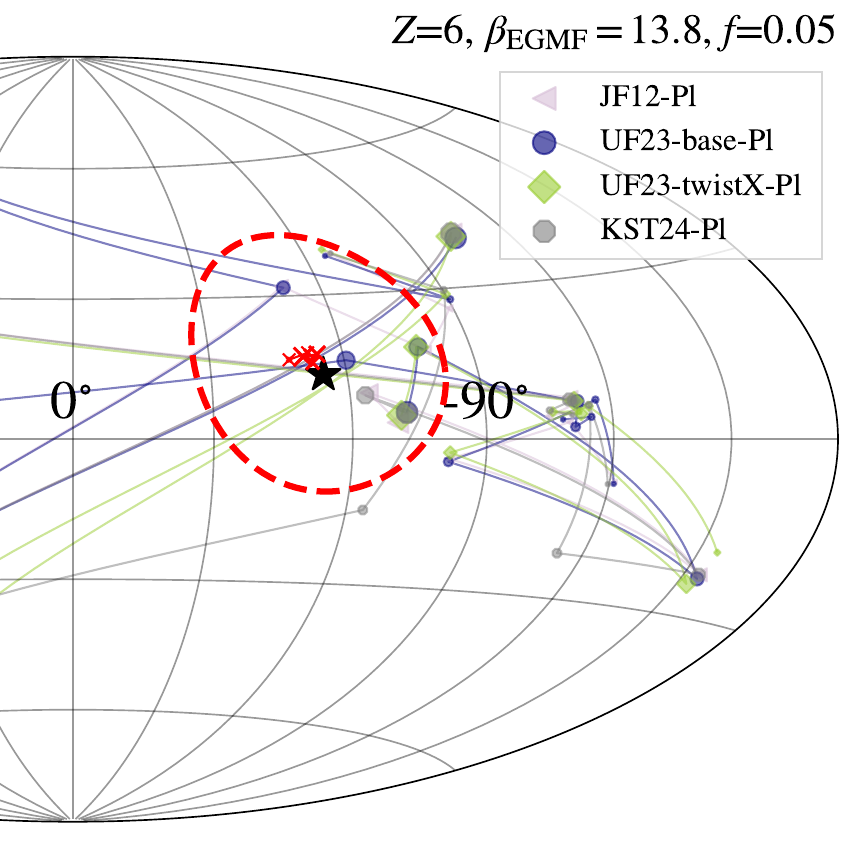}
\includegraphics[trim={13cm 0 0 0}, clip, width=0.24\textwidth]{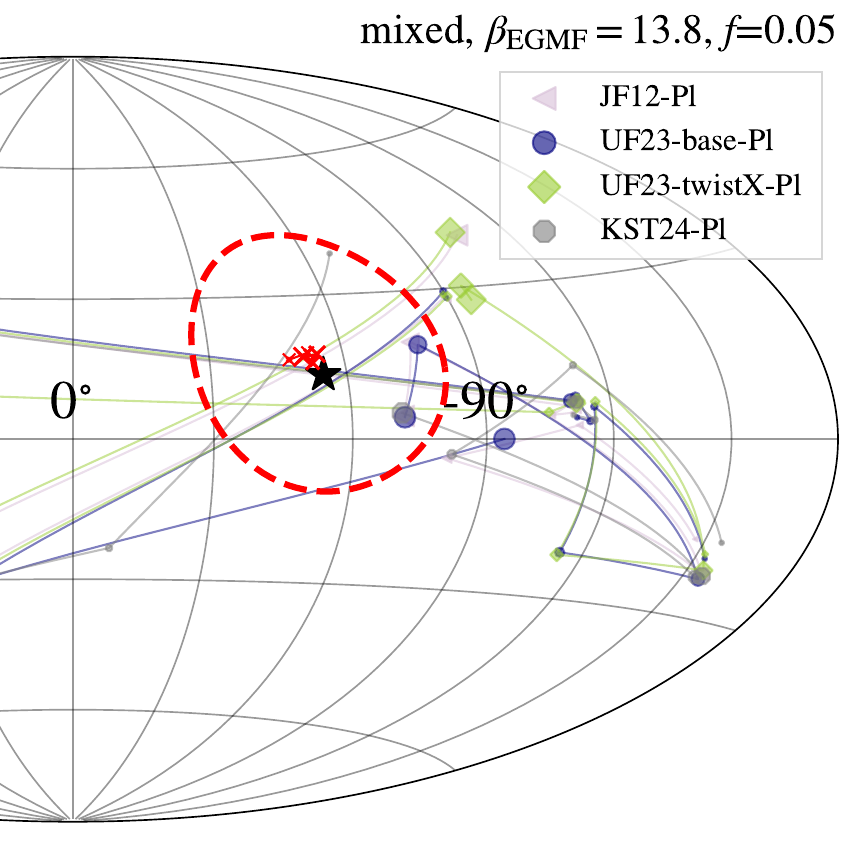}
\captionof{figure}{Energy evolution of the excess direction for NGC 4945 as the source of the observed excess. See Fig.~\ref{fig:cena_AD} for more details.}
\label{fig:}
\end{minipage}

\section{Alternative sources in the Centaurus region: M83} \label{app:m83}

\noindent
\begin{minipage}{\textwidth}
\centering
\includegraphics[width=0.49\textwidth]{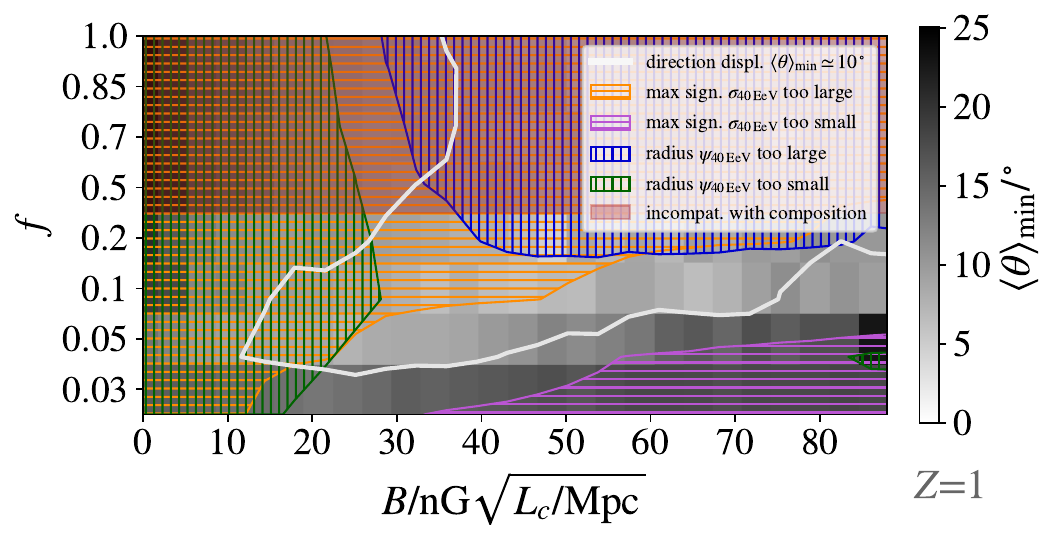}
\includegraphics[width=0.49\textwidth]{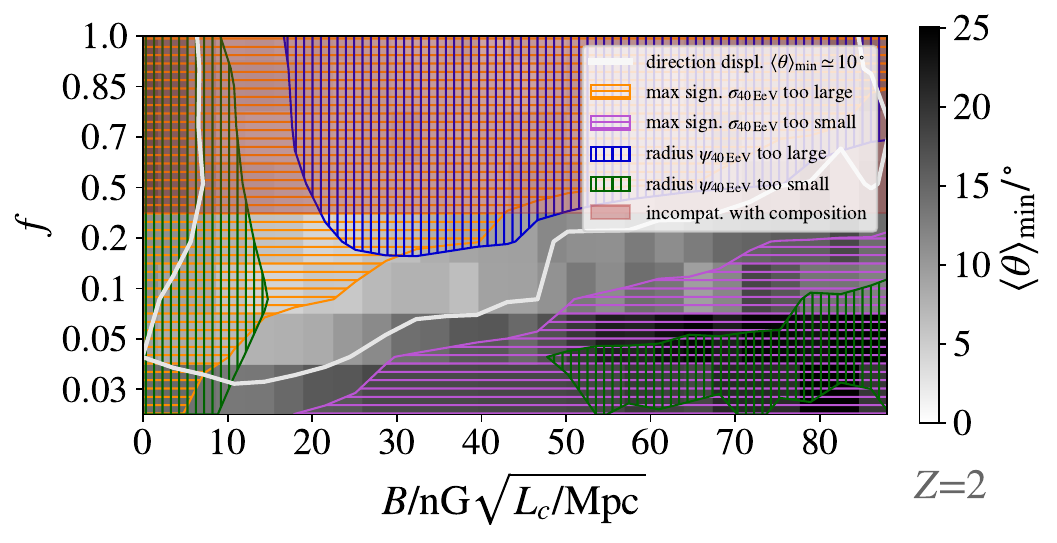}
\includegraphics[width=0.49\textwidth]{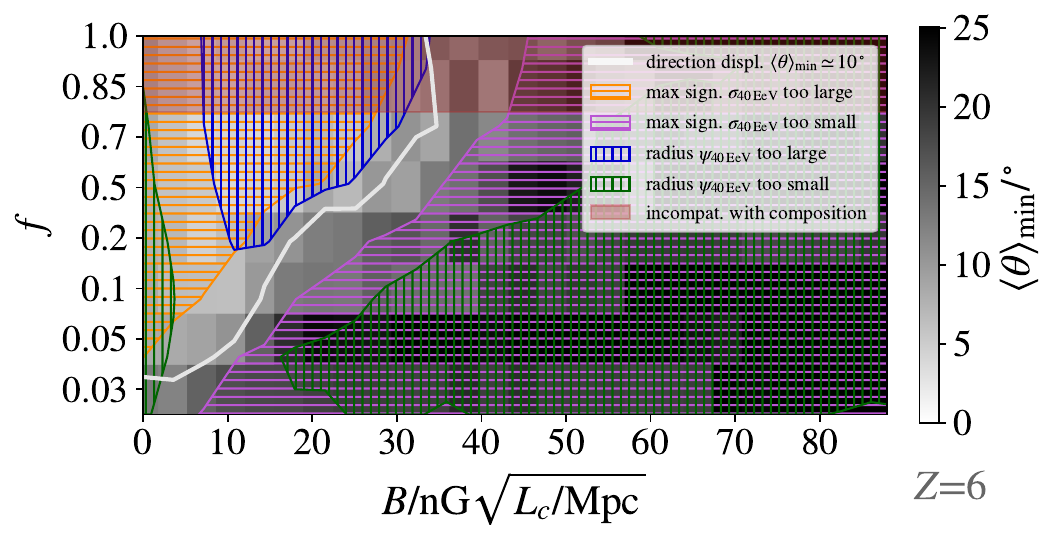}
\includegraphics[width=0.49\textwidth]{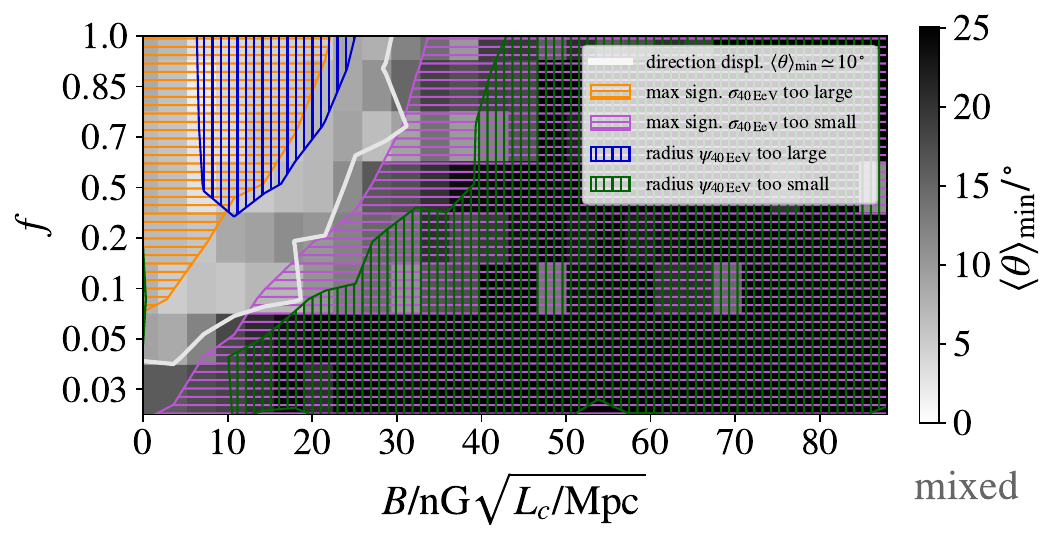}
\includegraphics[width=0.49\textwidth]{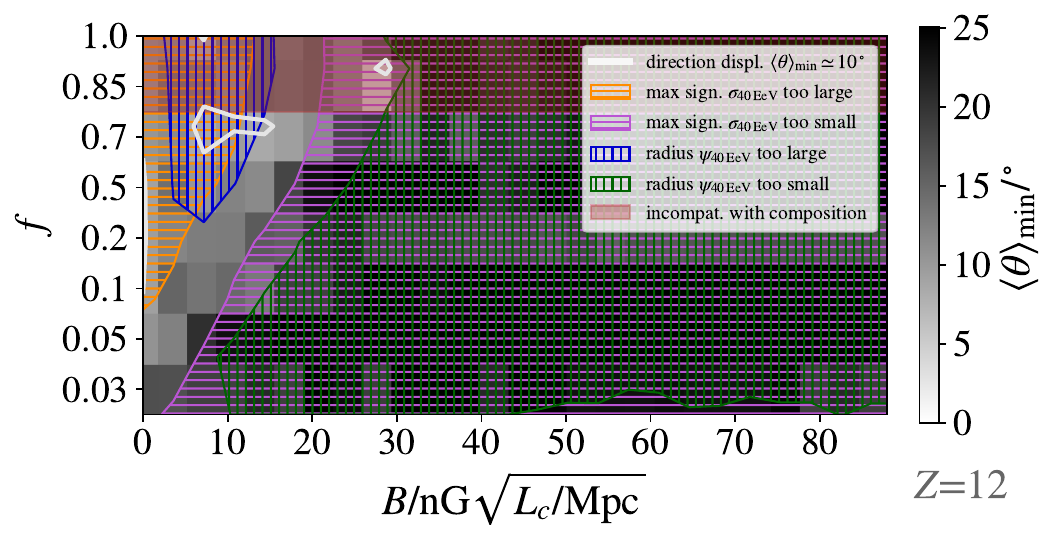}
\includegraphics[width=0.49\textwidth]{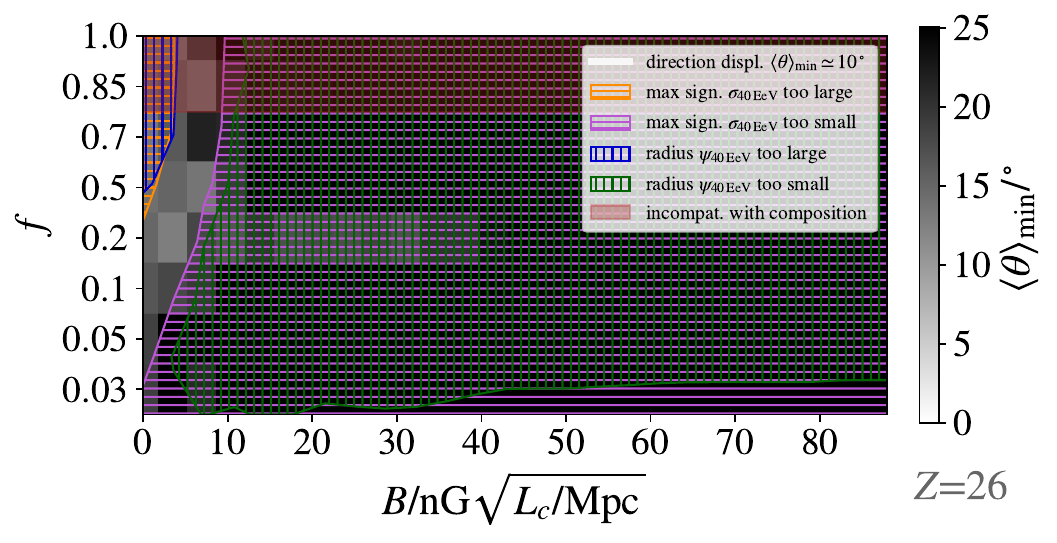}
\captionof{figure}{Constraints on the parameter space for M83 as the source of the observed excess. See Fig.~\ref{fig:cena_constraints} for more details.}
\label{fig:}
\end{minipage}

\vspace{1cm}
\noindent
\begin{minipage}{\textwidth}
\centering
\includegraphics[trim={13cm 0 0 0}, clip, width=0.24\textwidth]{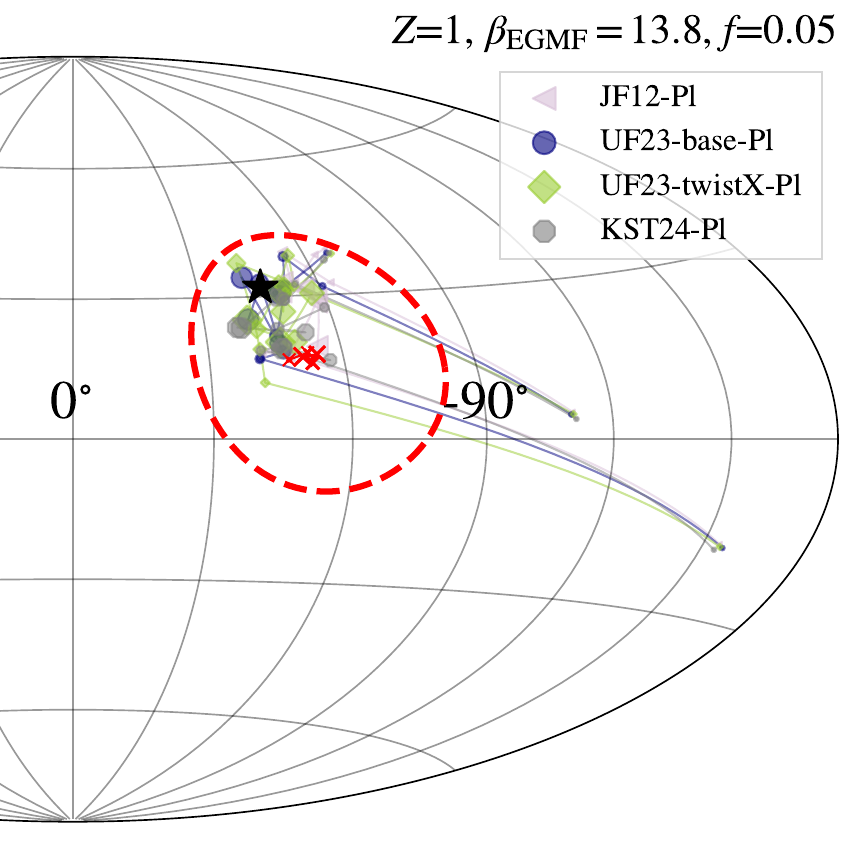}
\includegraphics[trim={13cm 0 0 0}, clip, width=0.24\textwidth]{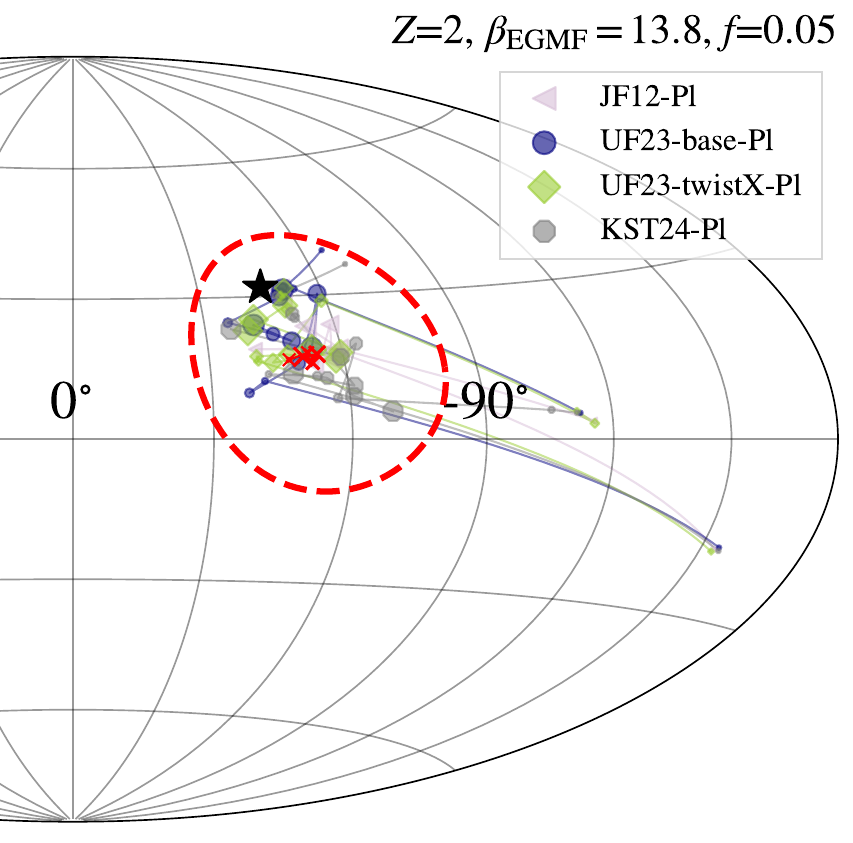}
\includegraphics[trim={13cm 0 0 0}, clip, width=0.24\textwidth]{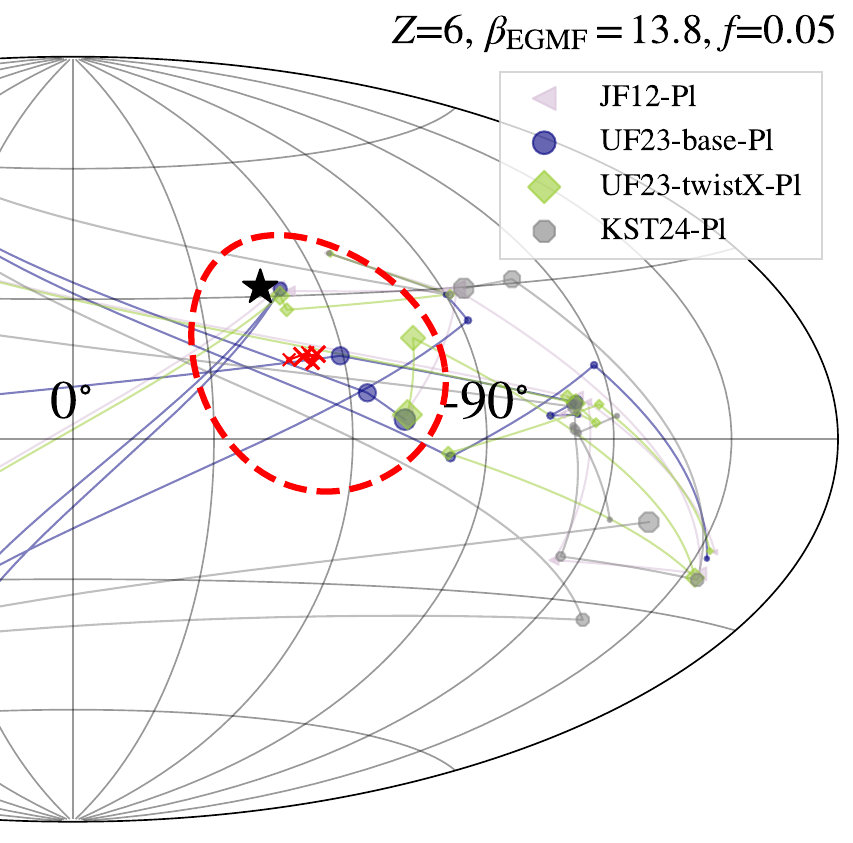}
\includegraphics[trim={13cm 0 0 0}, clip, width=0.24\textwidth]{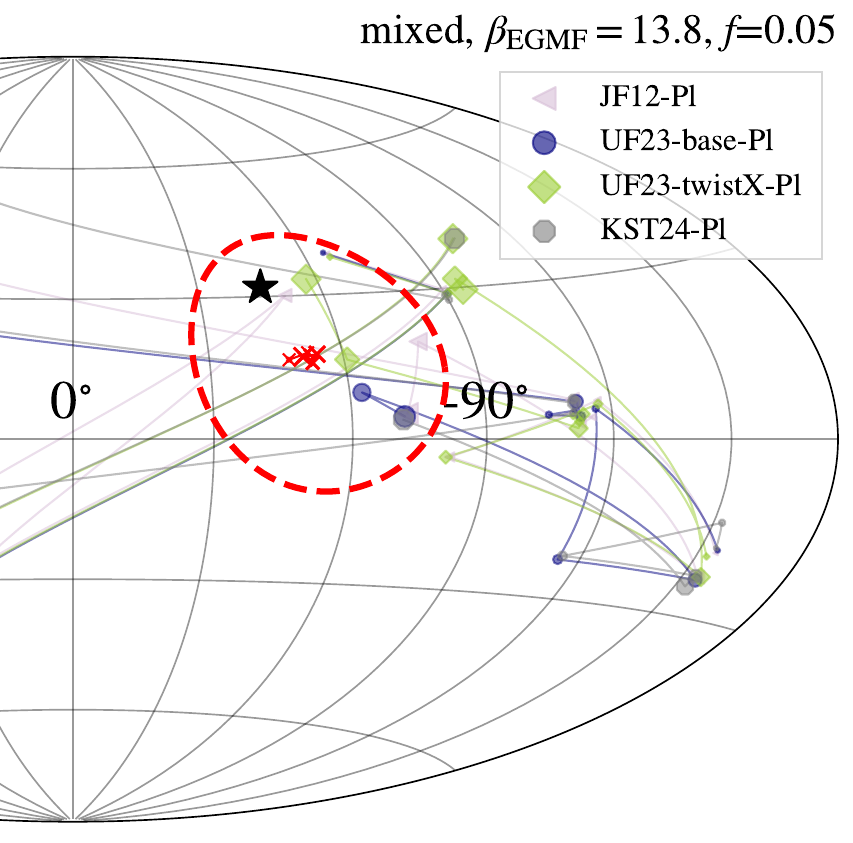}
\captionof{figure}{Energy evolution of the excess direction for M83 as the source of the observed excess. See Fig.~\ref{fig:cena_AD} for more details.}
\label{fig:}
\end{minipage}

\section{Alternative sources outside the Centaurus region: Virgo A / M87} \label{app:virgoA}
\noindent
\begin{minipage}{\textwidth}
\centering
\includegraphics[width=0.49\textwidth]{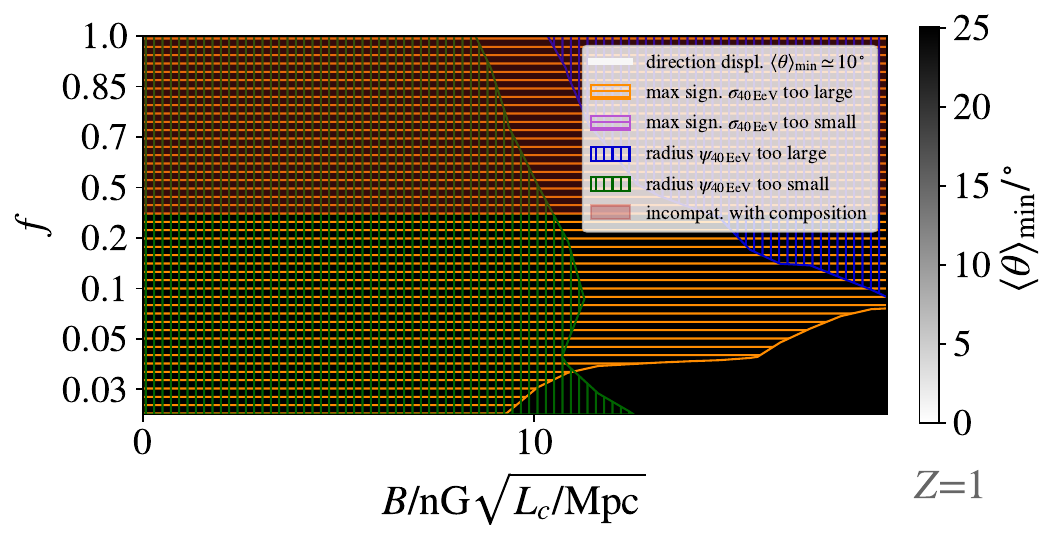}
\includegraphics[width=0.49\textwidth]{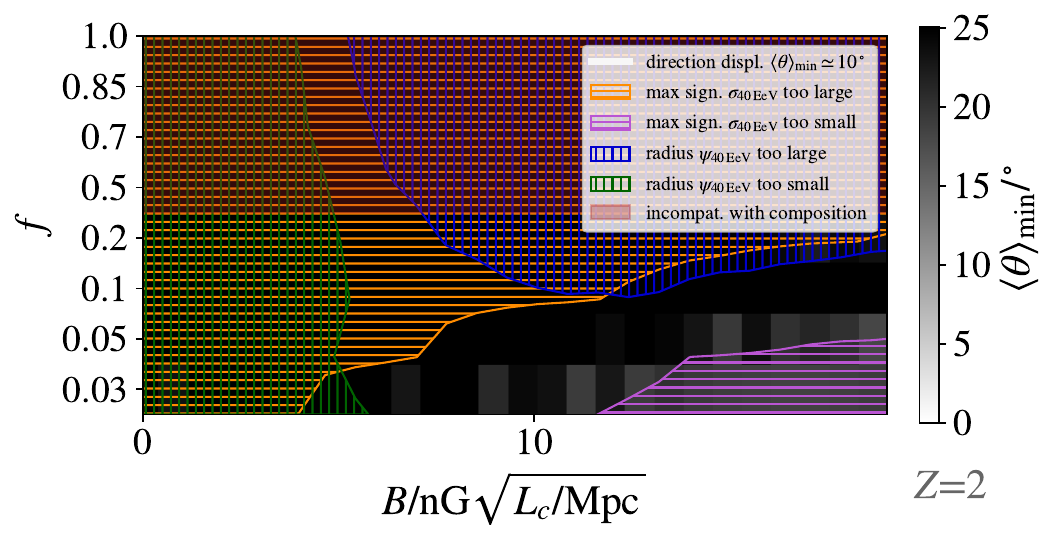}

\includegraphics[width=0.49\textwidth]{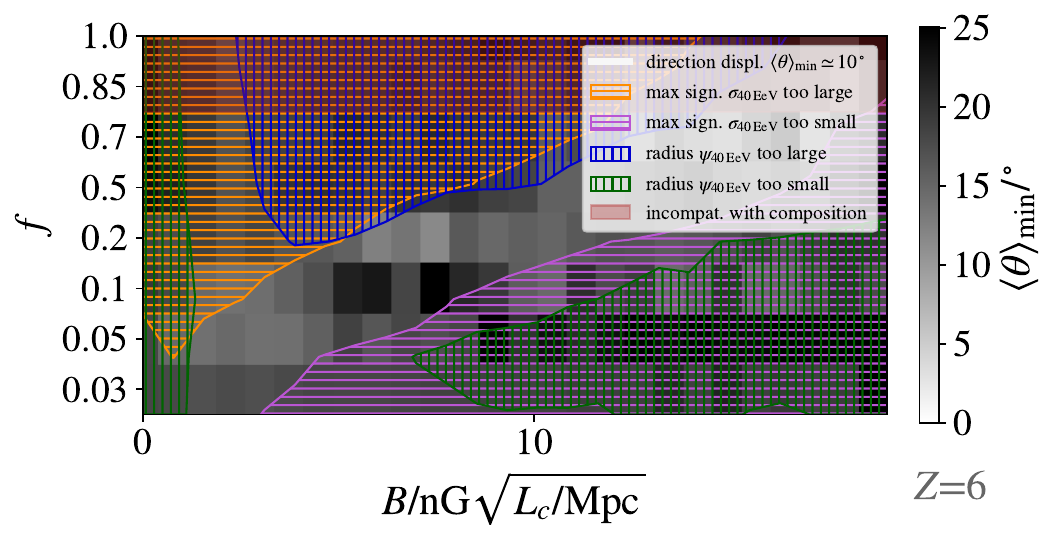}
\includegraphics[width=0.49\textwidth]{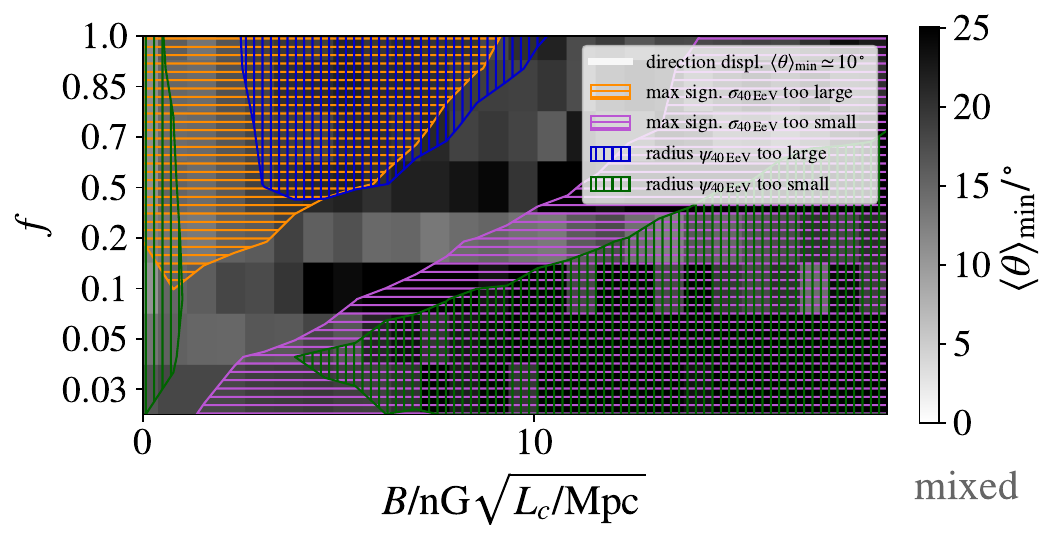}

\includegraphics[width=0.49\textwidth]{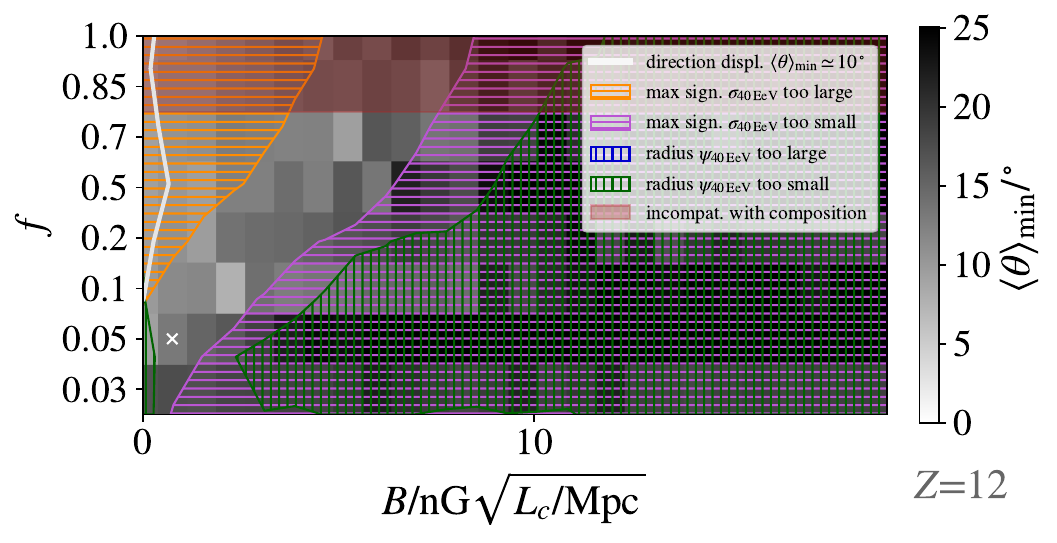}
\includegraphics[width=0.49\textwidth]{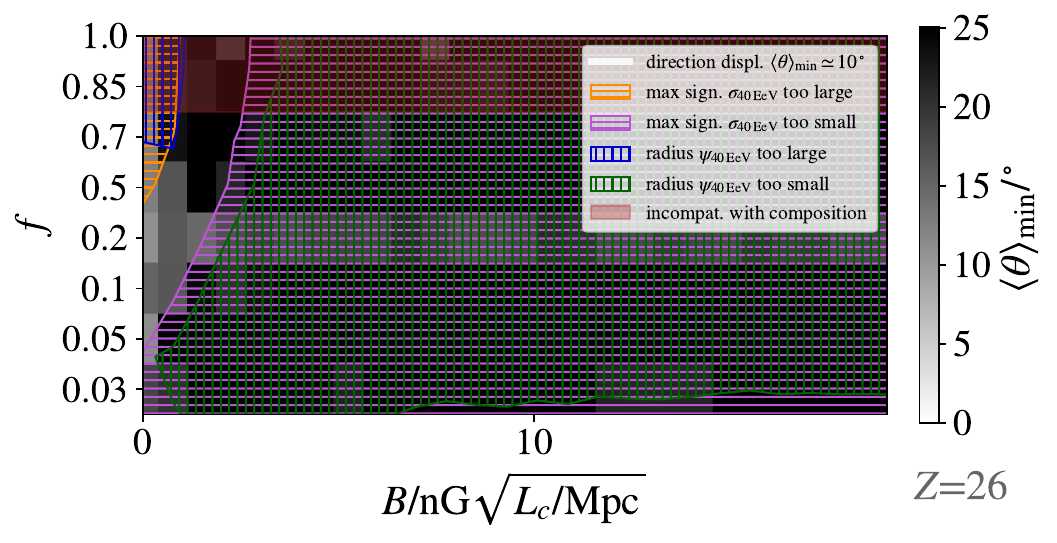}
\captionof{figure}{Constraints on the parameter space for Virgo A as the source of the observed excess. See Fig.~\ref{fig:cena_constraints} for more details.}
\label{fig:virgoA_constraints}
\end{minipage}

\noindent
\begin{minipage}{\textwidth}
\centering
\includegraphics[trim={11.5cm 0 0 0}, clip, width=0.33\textwidth]{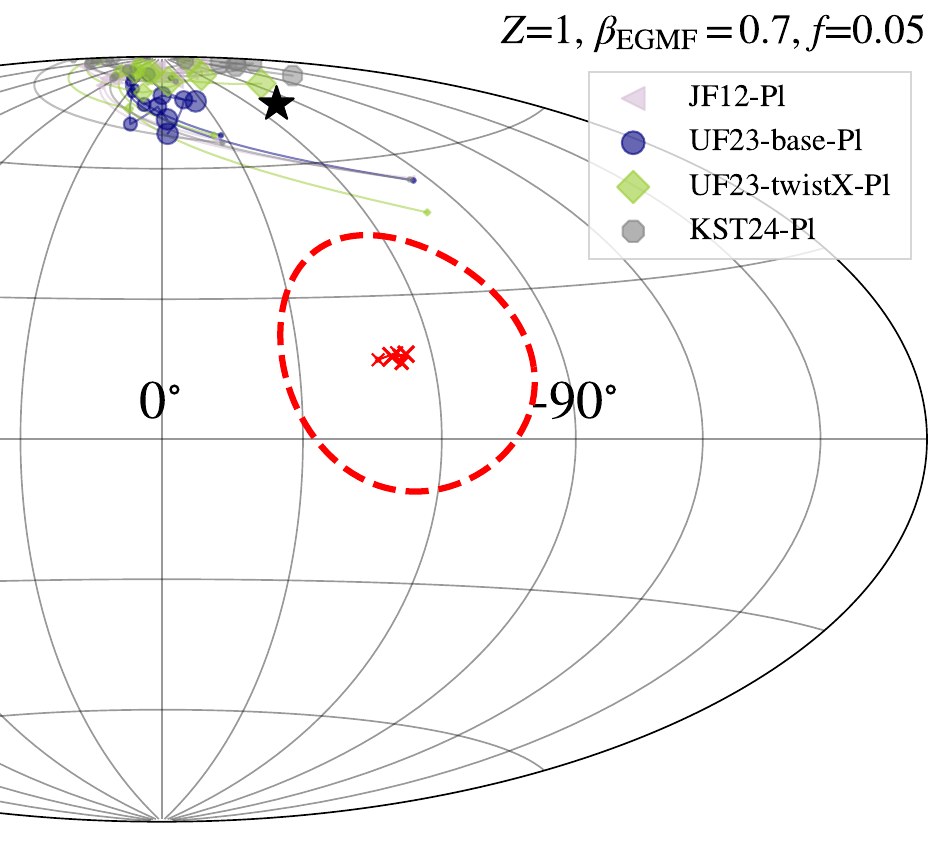}
\includegraphics[trim={11.5cm 0 0 0}, clip, width=0.33\textwidth]{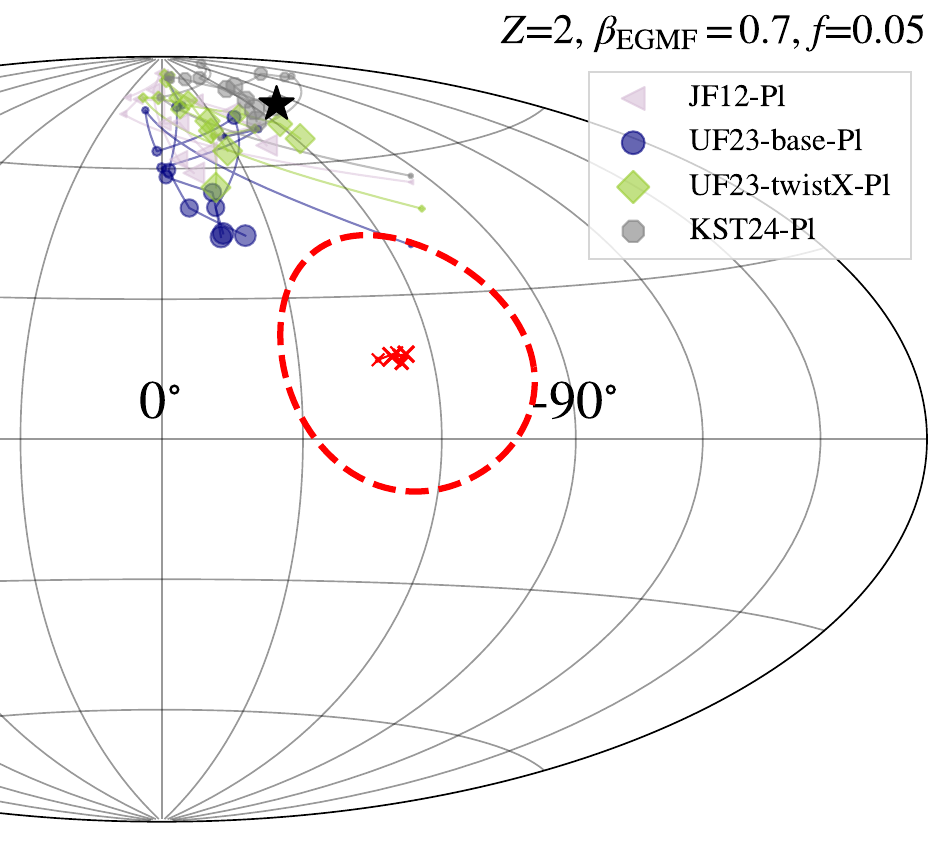}
\includegraphics[trim={11.5cm 0 0 0}, clip, width=0.33\textwidth]{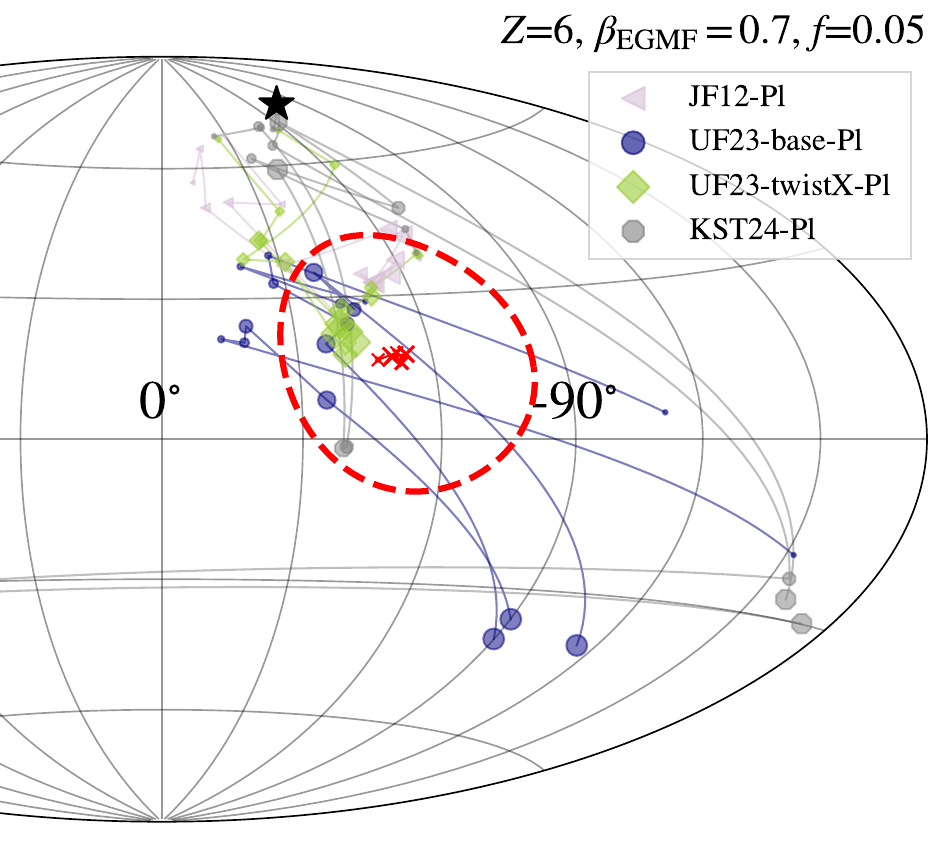}

\includegraphics[trim={11.5cm 0 0 0}, clip, width=0.33\textwidth]{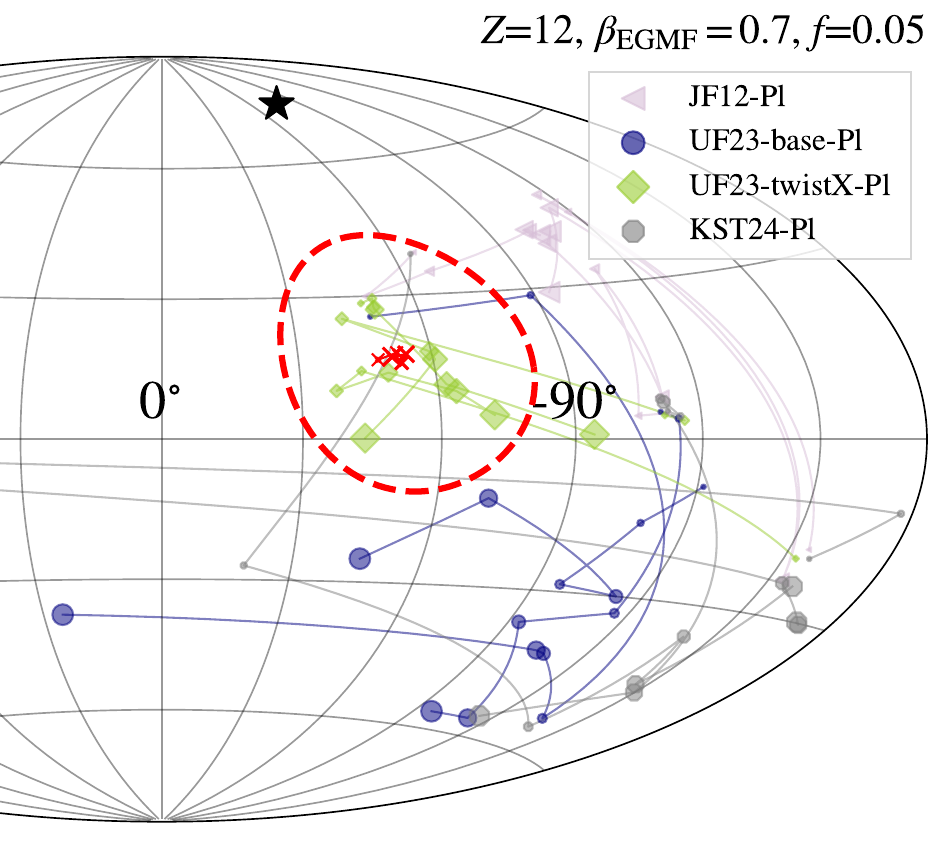}
\includegraphics[trim={11.5cm 0 0 0}, clip, width=0.33\textwidth]{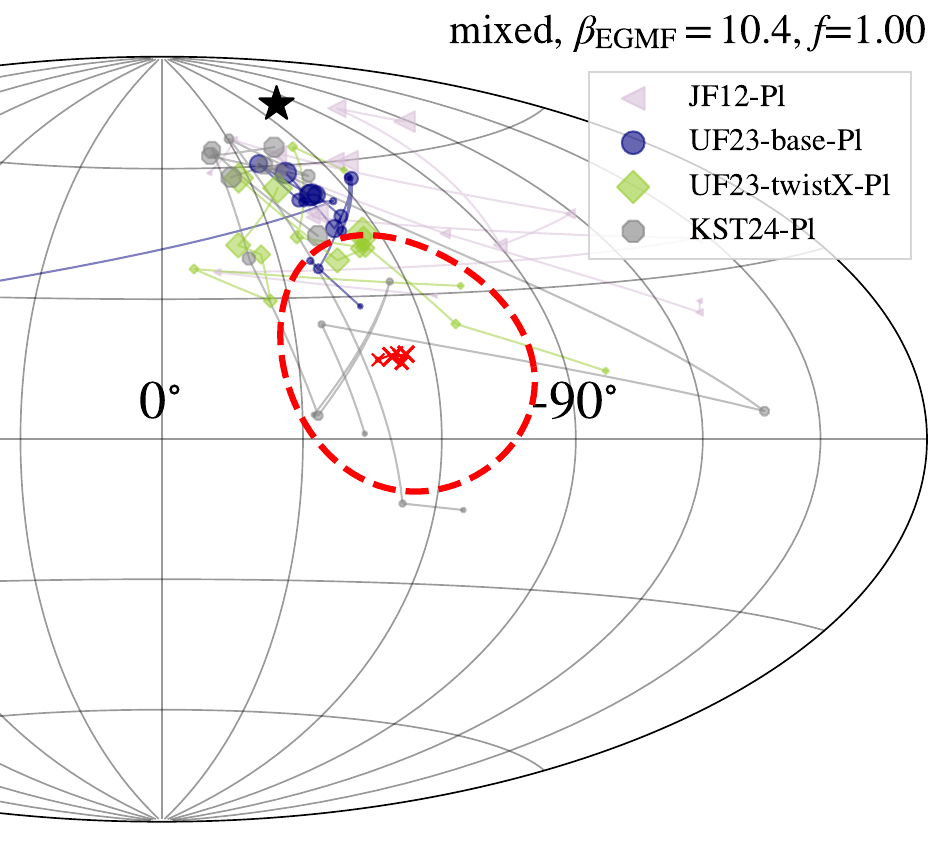}
\includegraphics[trim={11.5cm 0 0 0}, clip, width=0.33\textwidth]{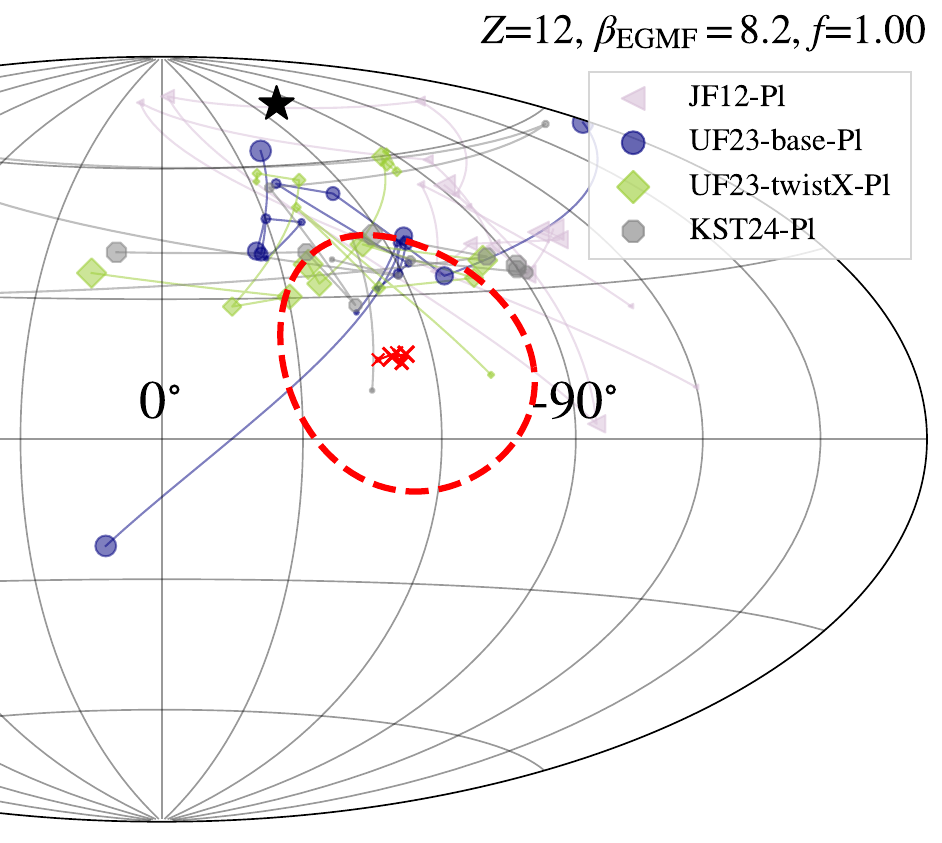}
\captionof{figure}{Energy evolution of the excess direction for Virgo A as the source of the observed excess. See Fig.~\ref{fig:cena_AD} for more details. Note that the figures show two different scenarios: the first four figures are for \textit{Scenario II} with a subdominant source contribution, and the last two for \textit{Scenario III} with $f=1$, see figure titles.}
\label{fig:virgoA_ADs}
\end{minipage}

\begin{figure}[h!]
\includegraphics[width=0.24\textwidth]{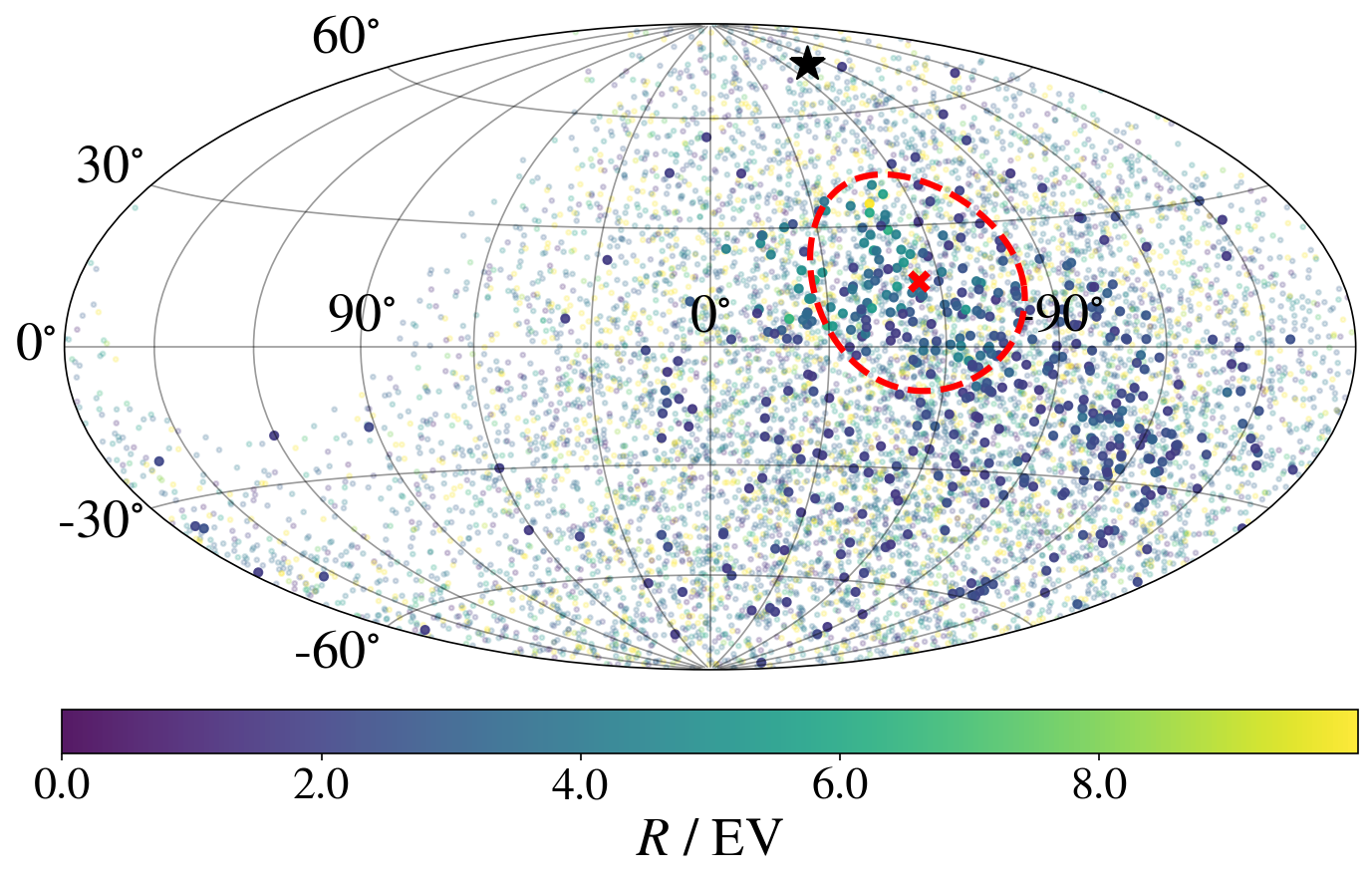}
\includegraphics[width=0.24\textwidth]{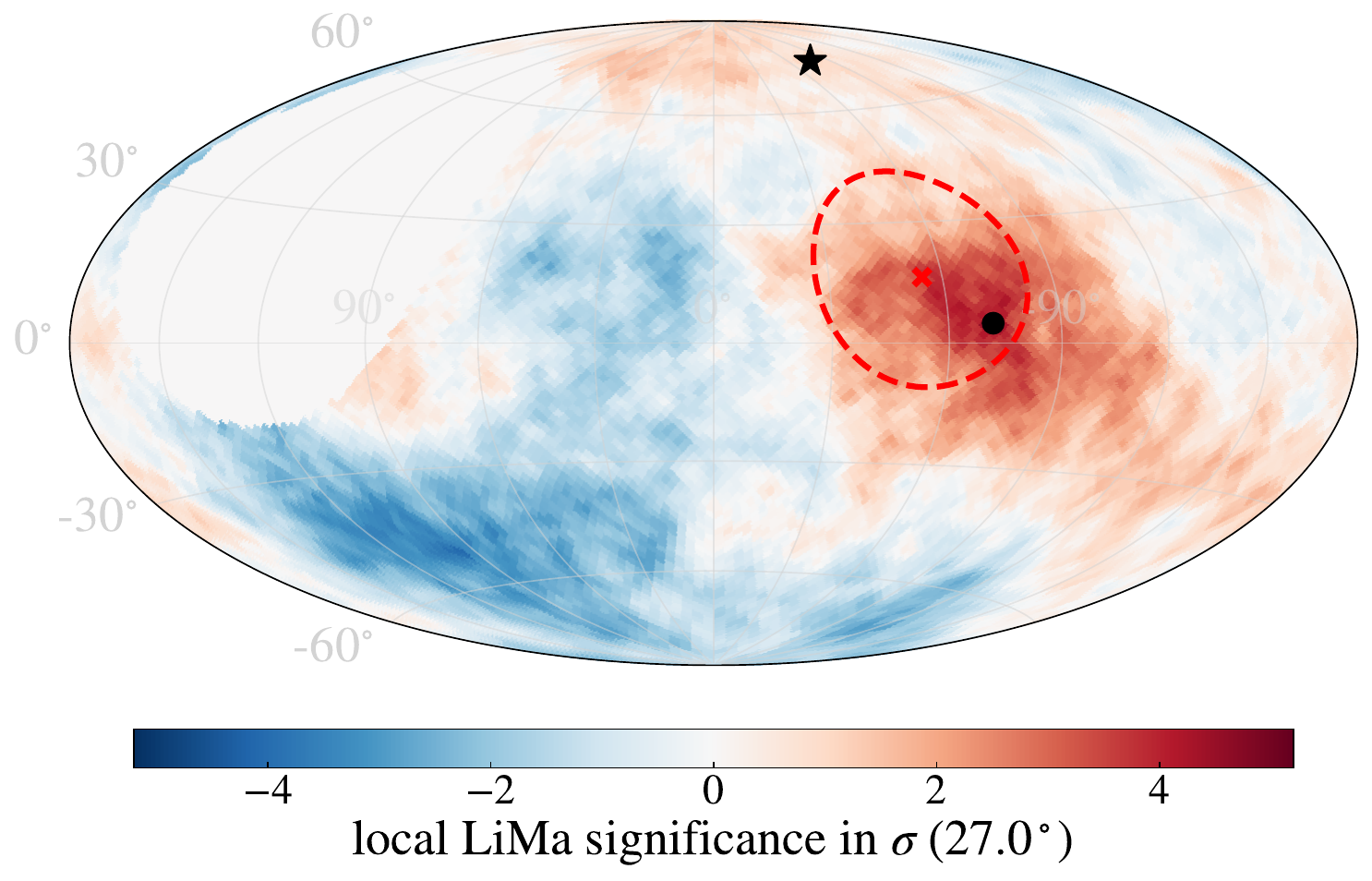}
\includegraphics[width=0.24\textwidth]{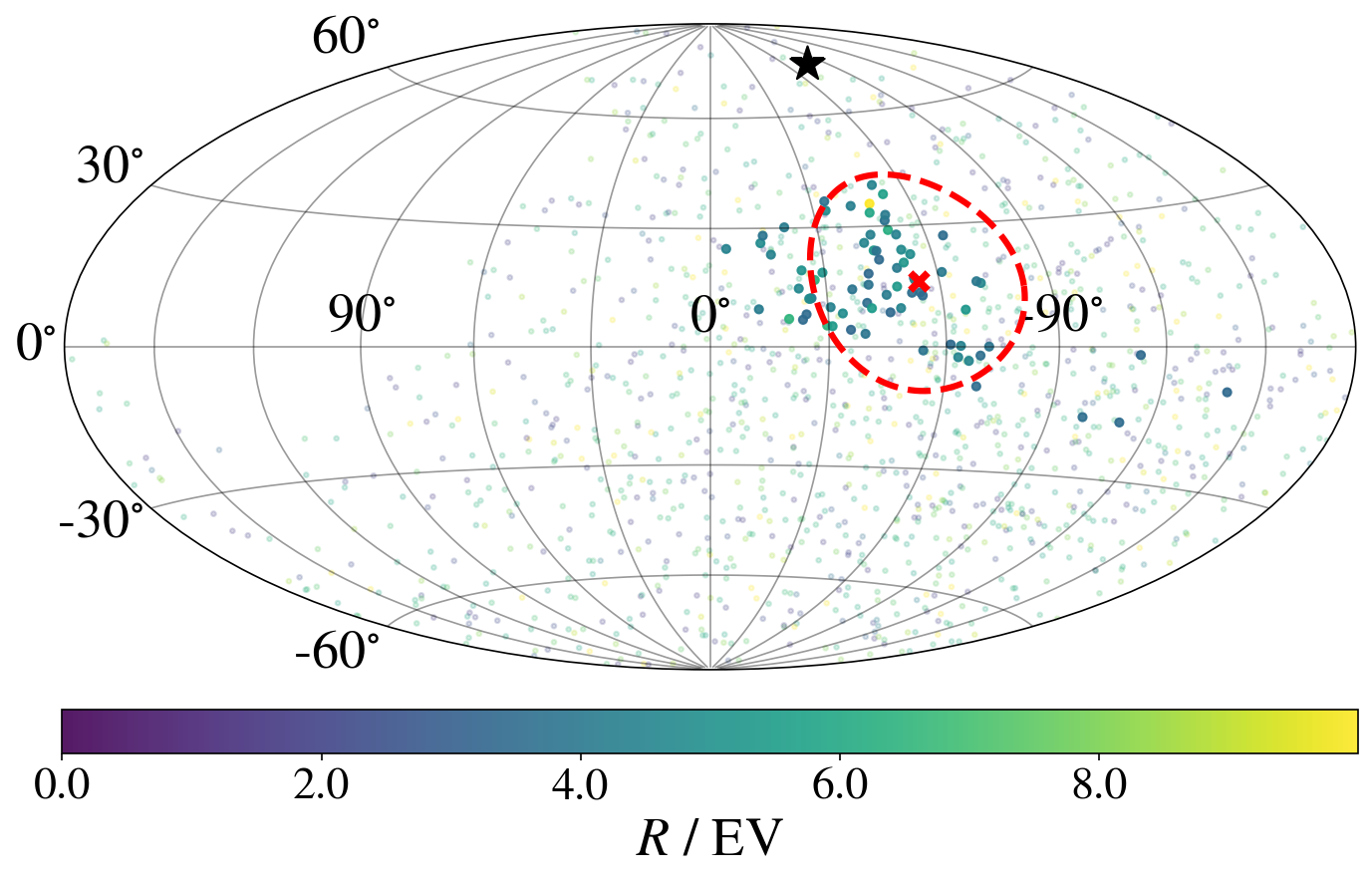}
\includegraphics[width=0.24\textwidth]{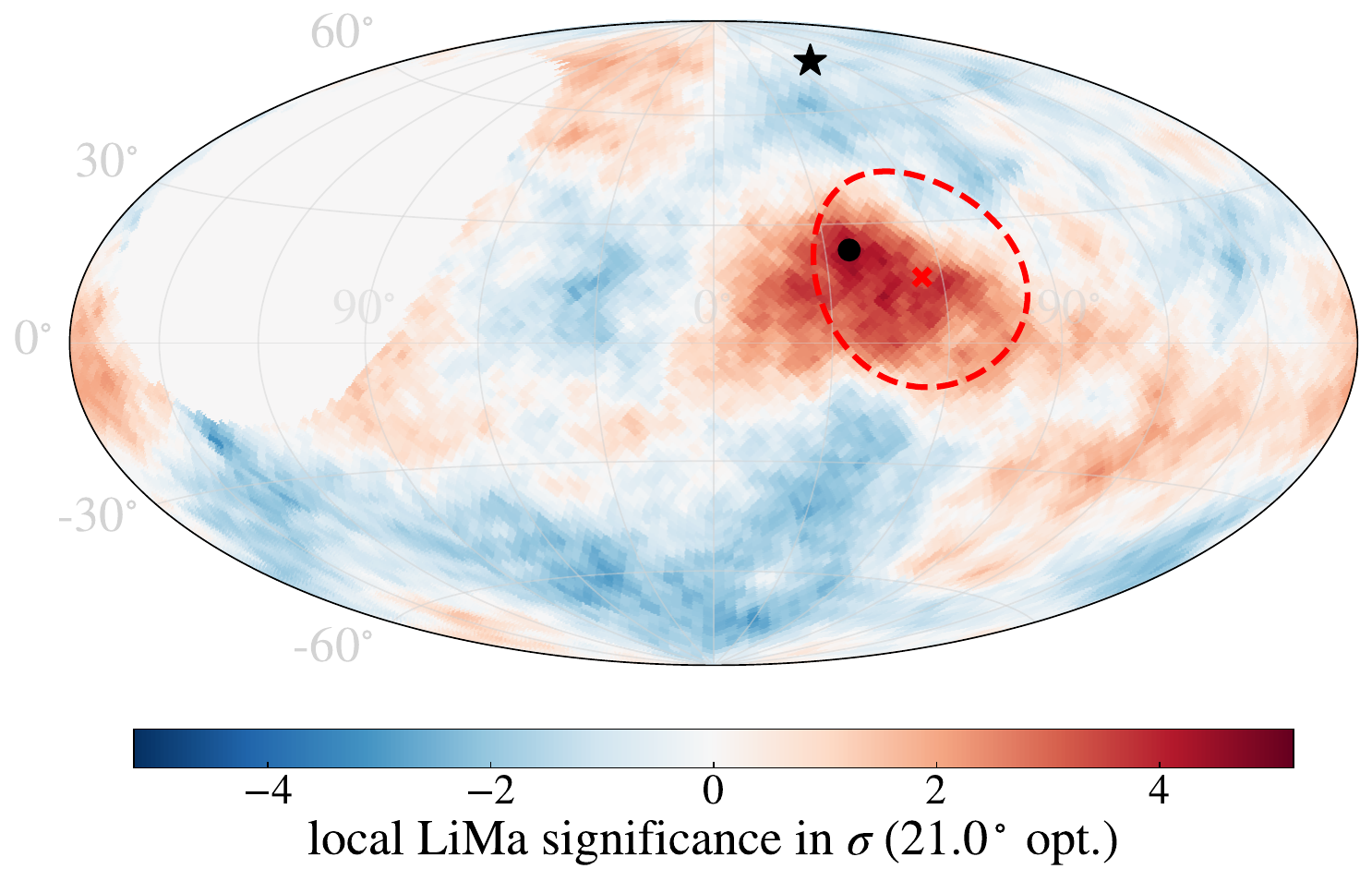}

\includegraphics[width=0.24\textwidth]{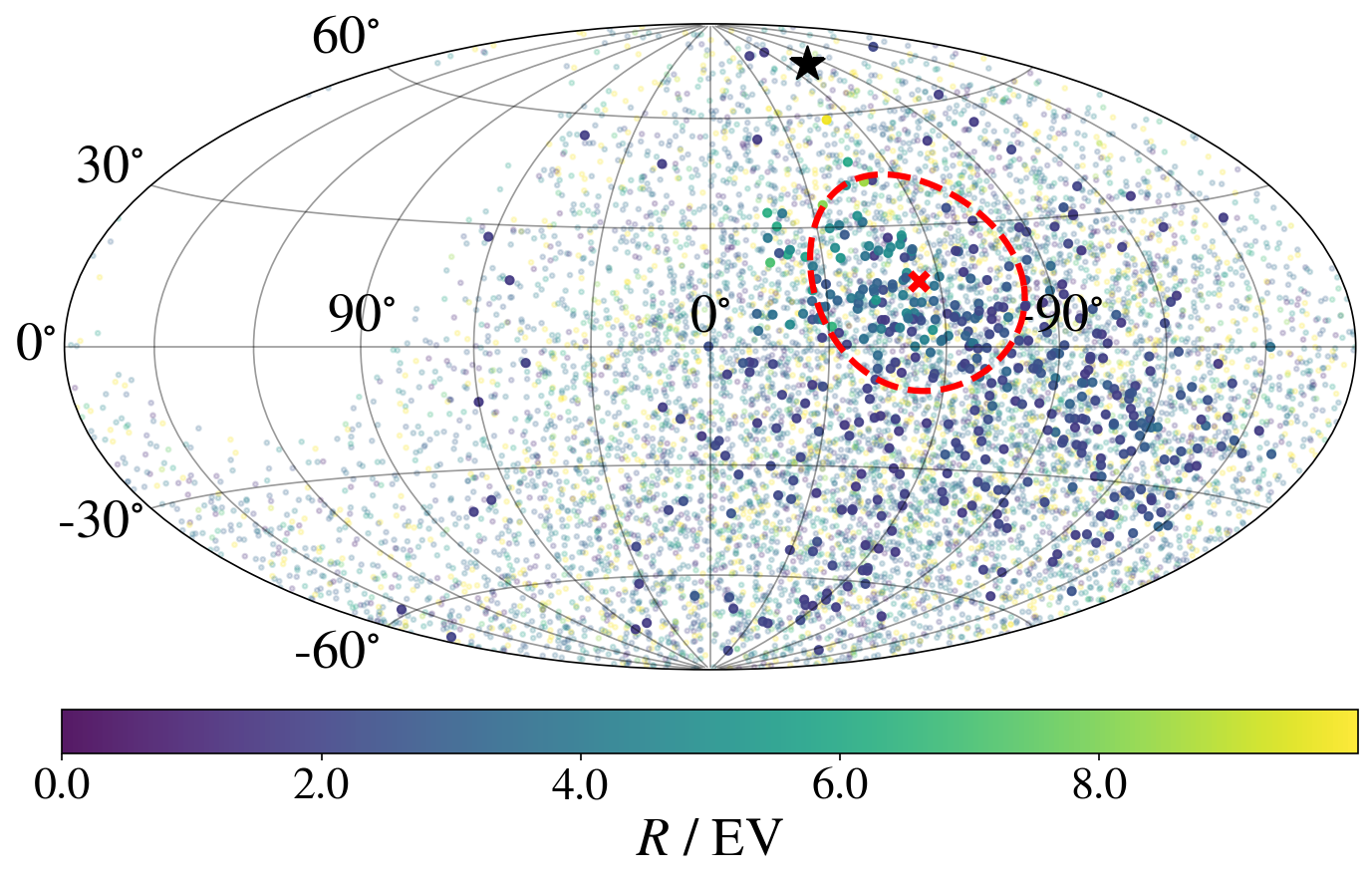}
\includegraphics[width=0.24\textwidth]{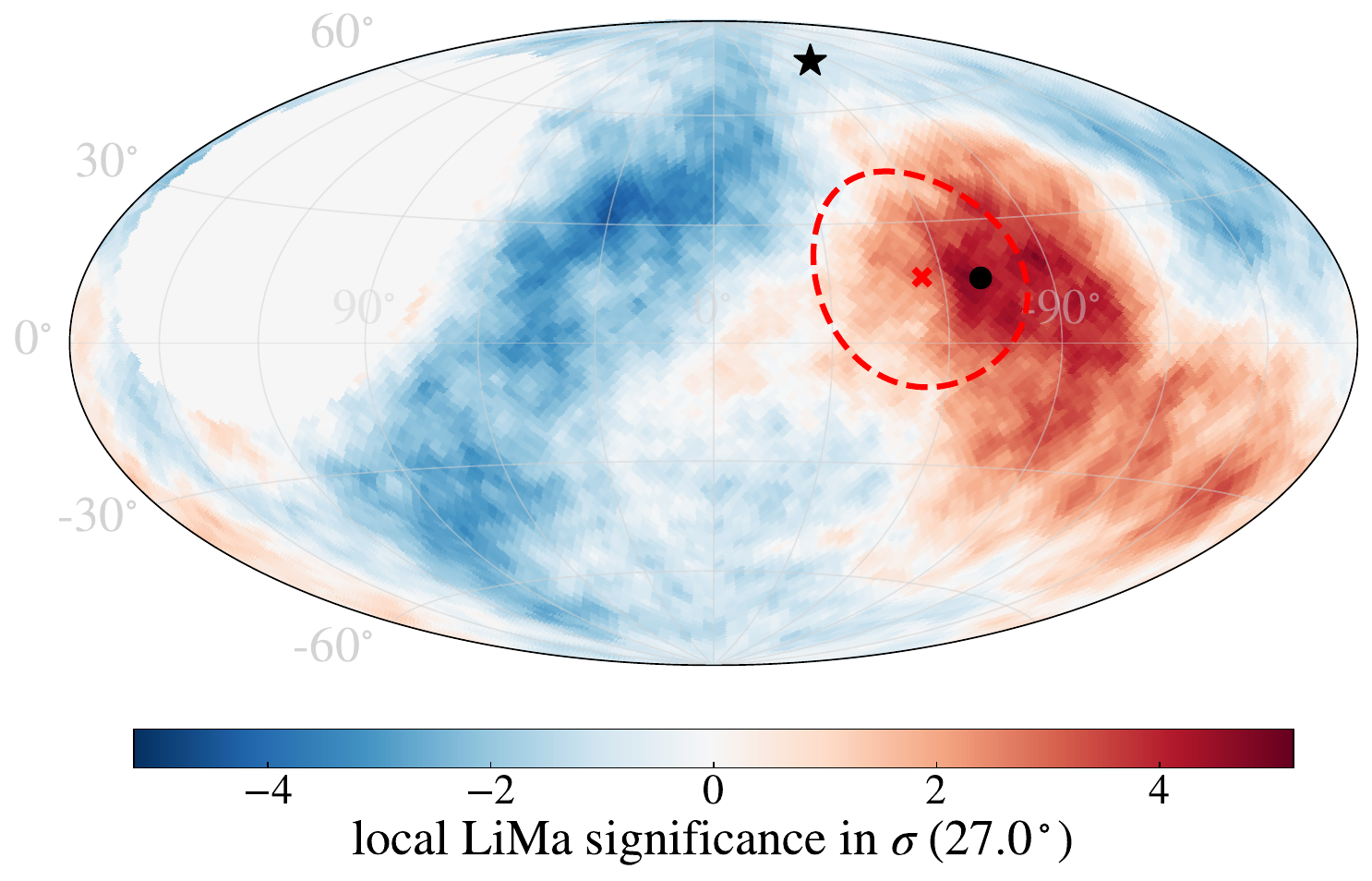}
\includegraphics[width=0.24\textwidth]{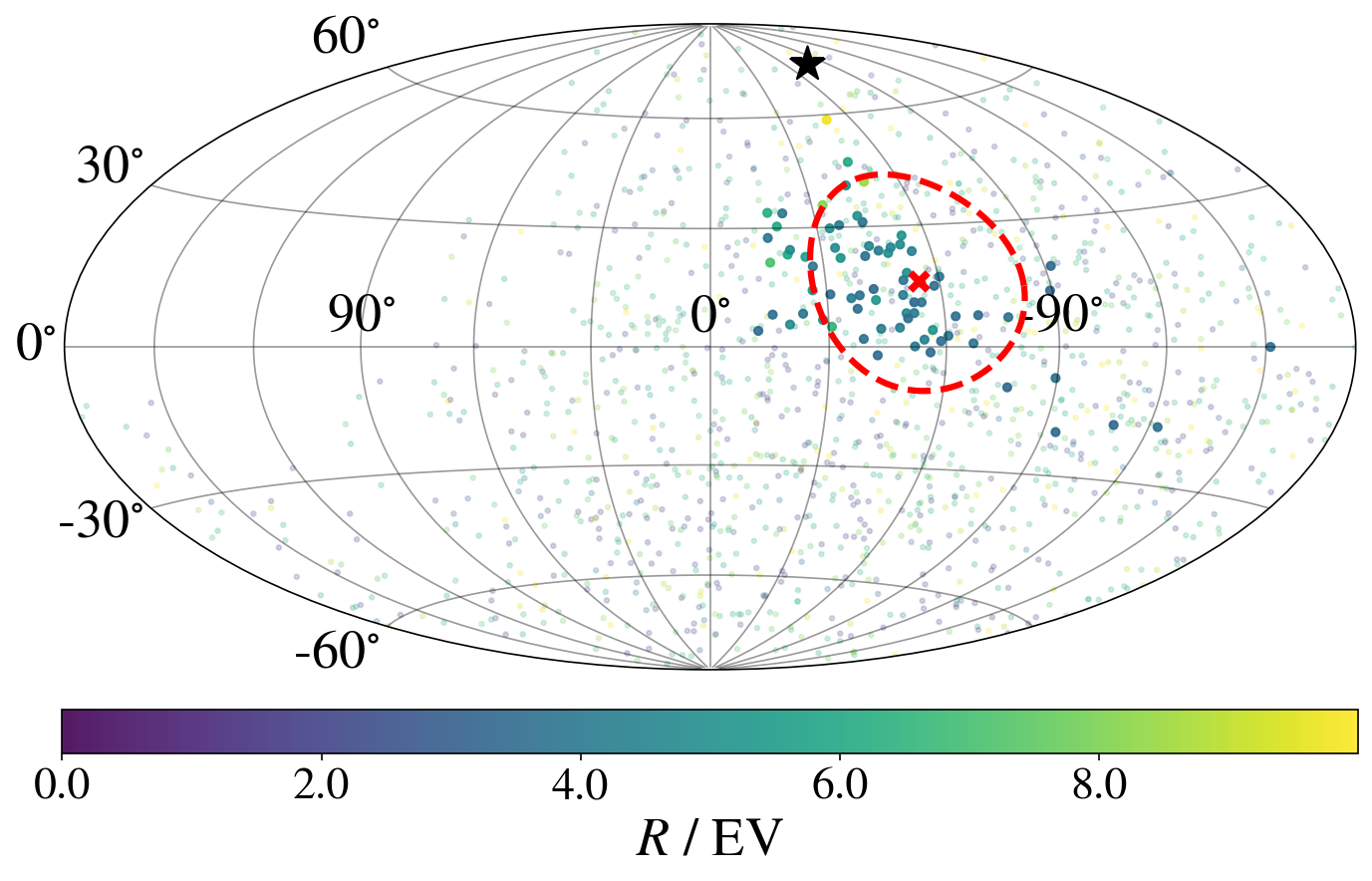}
\includegraphics[width=0.24\textwidth]{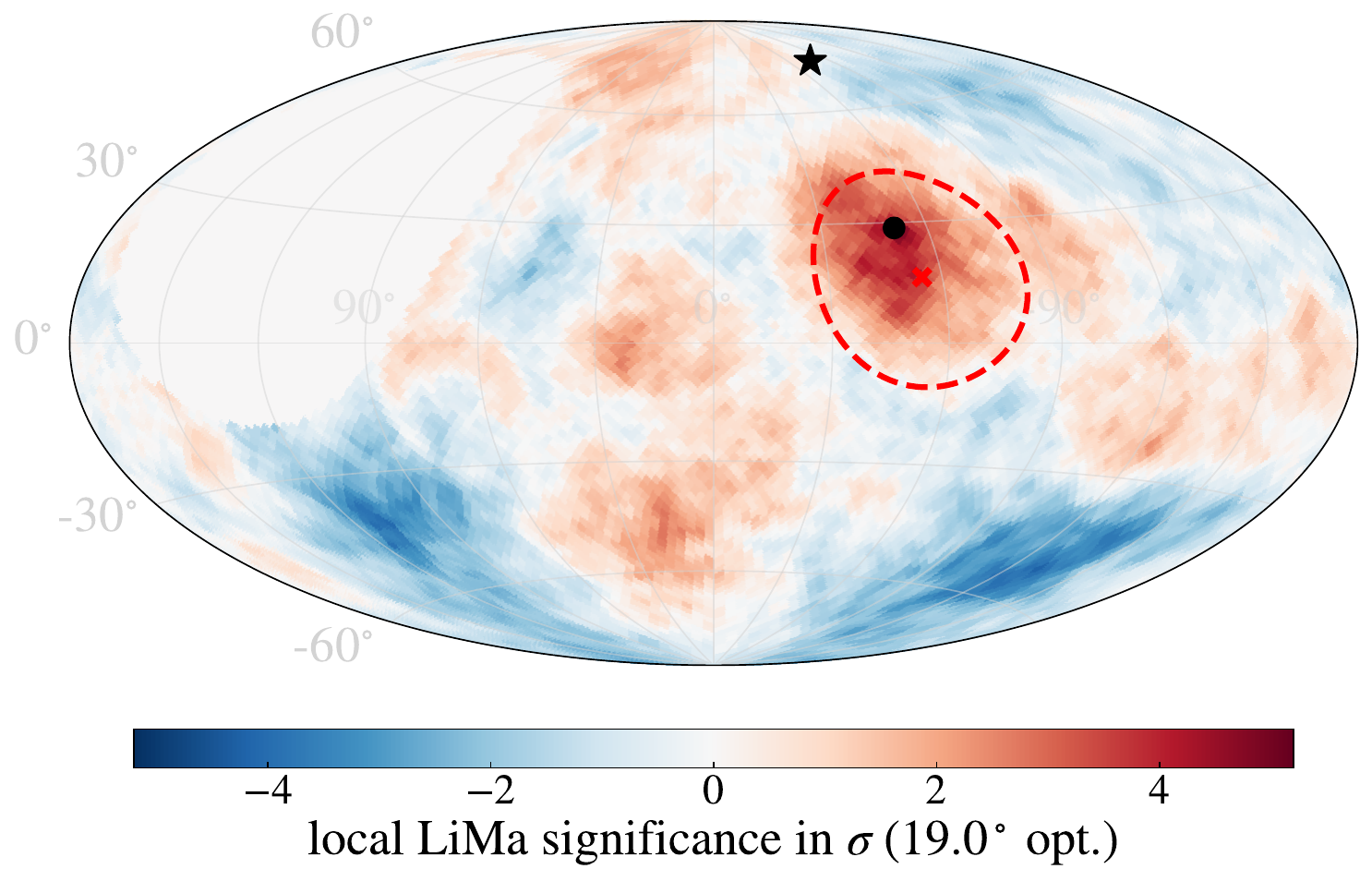}
\caption{Two example simulations with Virgo A as the source (indicated by the black star), with $Z=12$, $f=0.05$, $\beta_\mathrm{EGMF}=3$ (white cross in Fig.~\ref{fig:virgoA_constraints}) and using the \texttt{UF23-twistX-Pl} GMF model. Each row shows one simulations, where the left two figures are for $E_\mathrm{min}=20\,\mathrm{EeV}$ and the right two for $E_\mathrm{min}=40\,\mathrm{EeV}$. All skymaps are in Galactic coordinates. See Fig.~\ref{fig:cena_AD} for more details.}
\label{fig:virgoA_examples}
\end{figure}

\section{Example simulations for Scenario III with $f=1$}
\noindent
\begin{minipage}{\textwidth}
\centering
\includegraphics[width=0.24\textwidth]{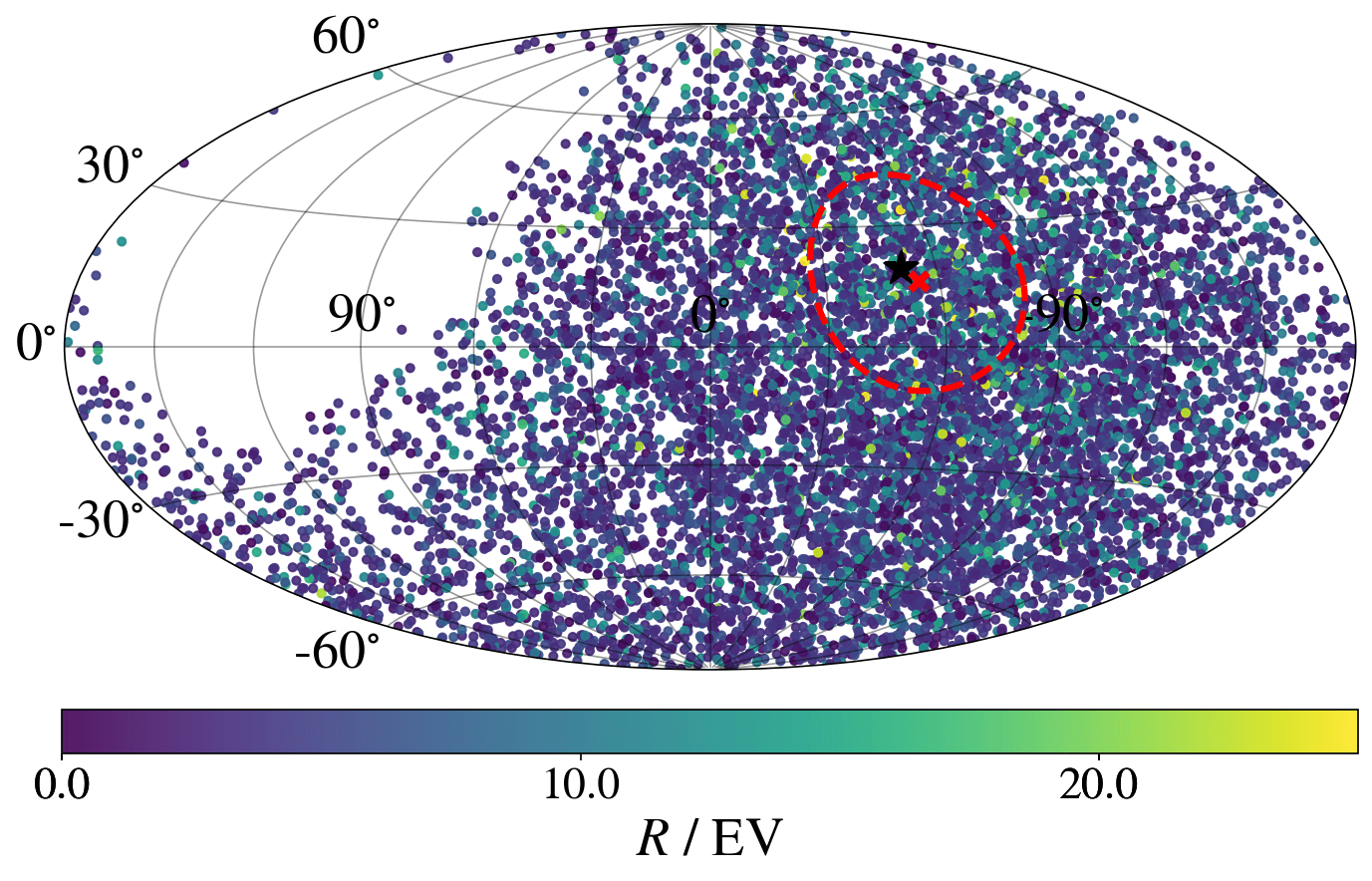}
\includegraphics[width=0.24\textwidth]{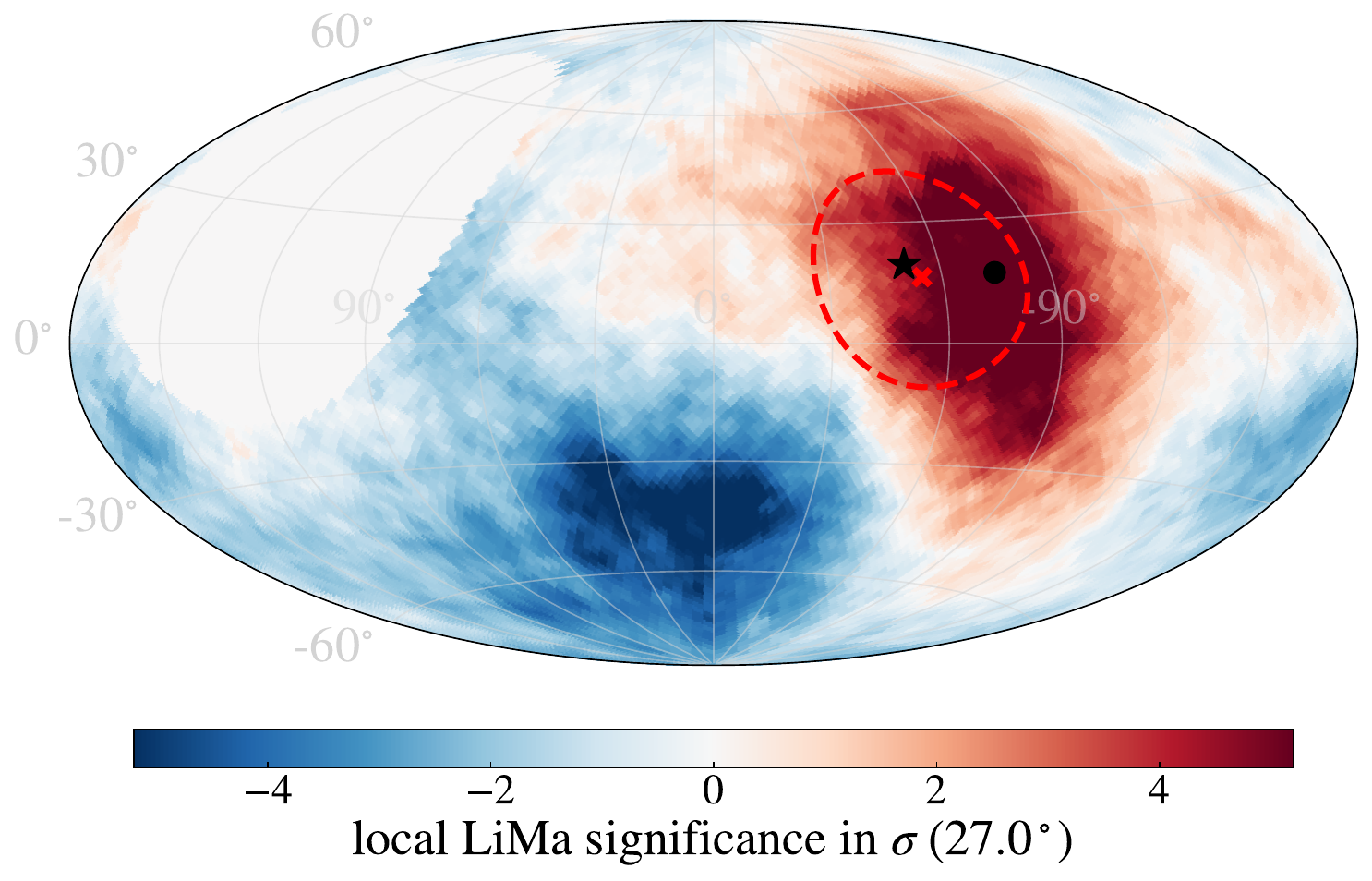}
\includegraphics[width=0.24\textwidth]{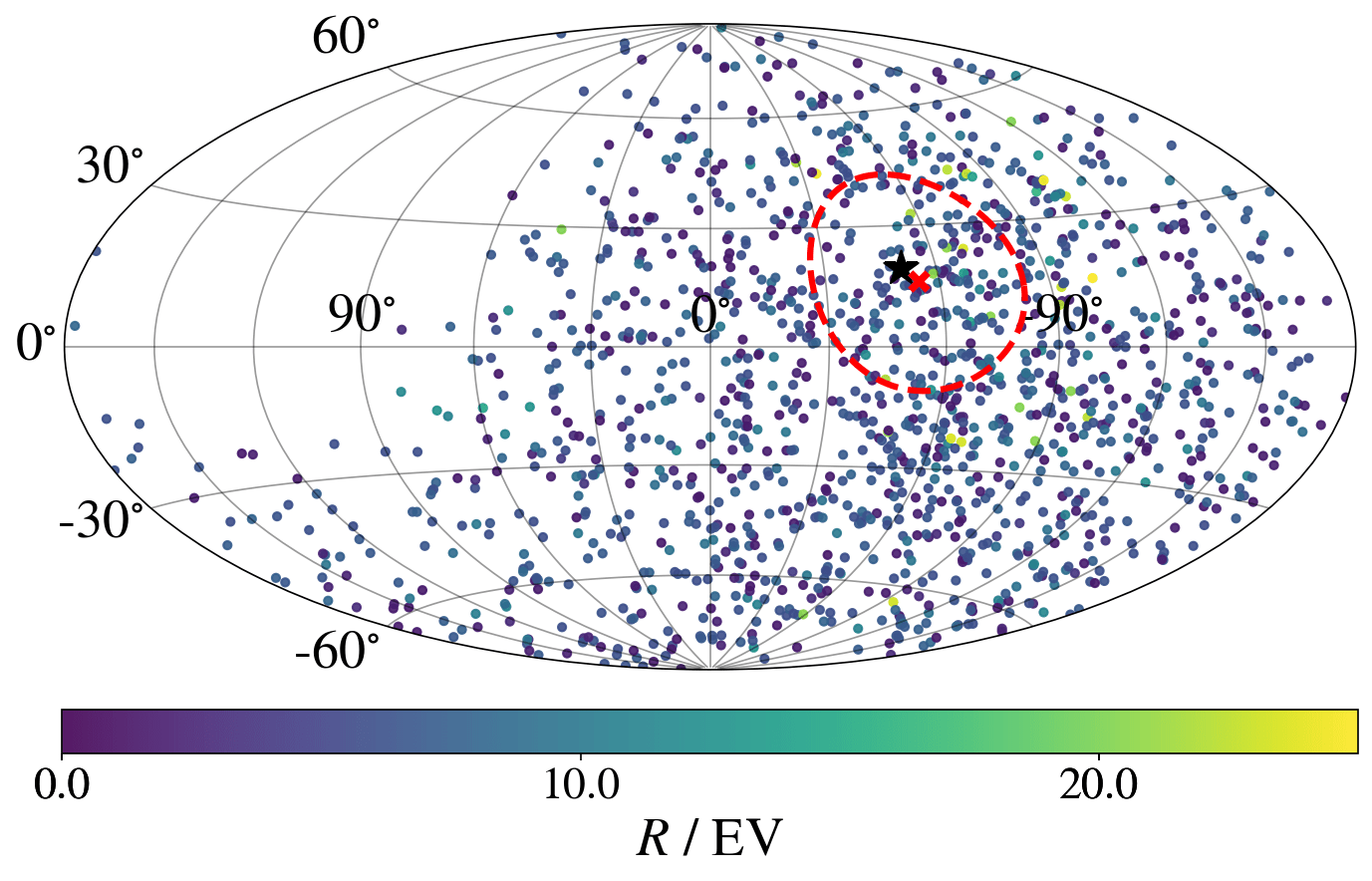}
\includegraphics[width=0.24\textwidth]{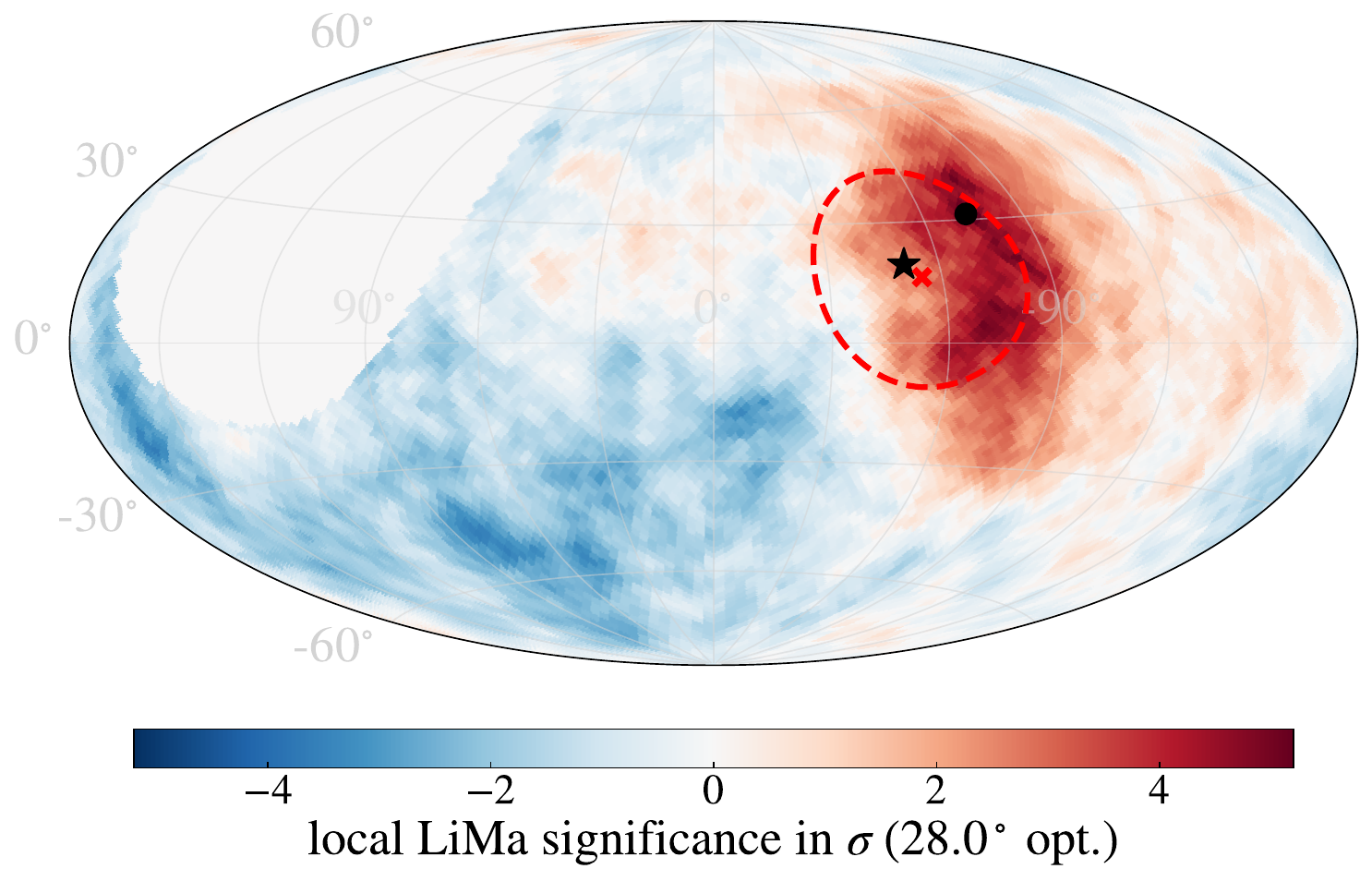}

\includegraphics[width=0.24\textwidth]{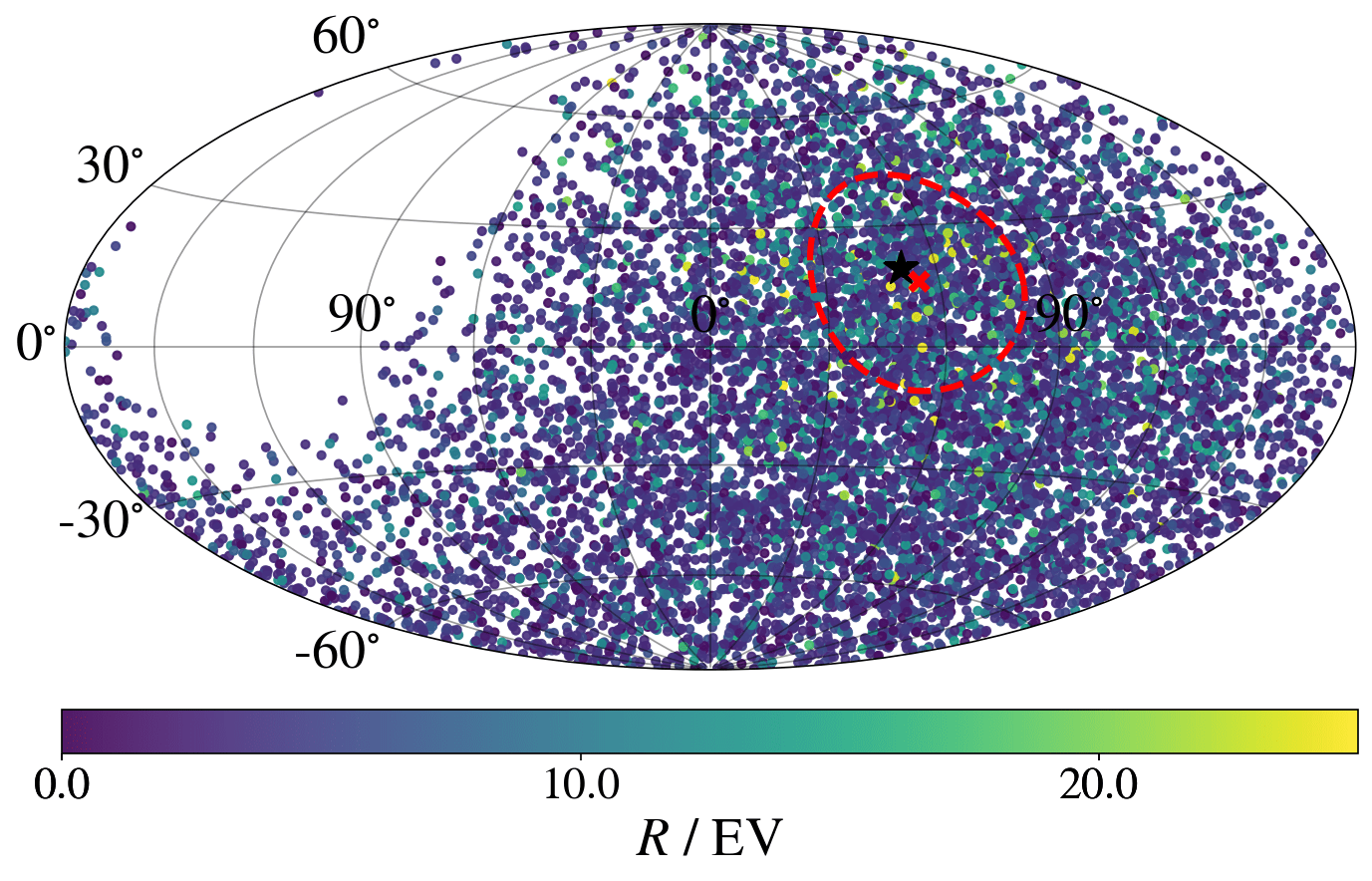}
\includegraphics[width=0.24\textwidth]{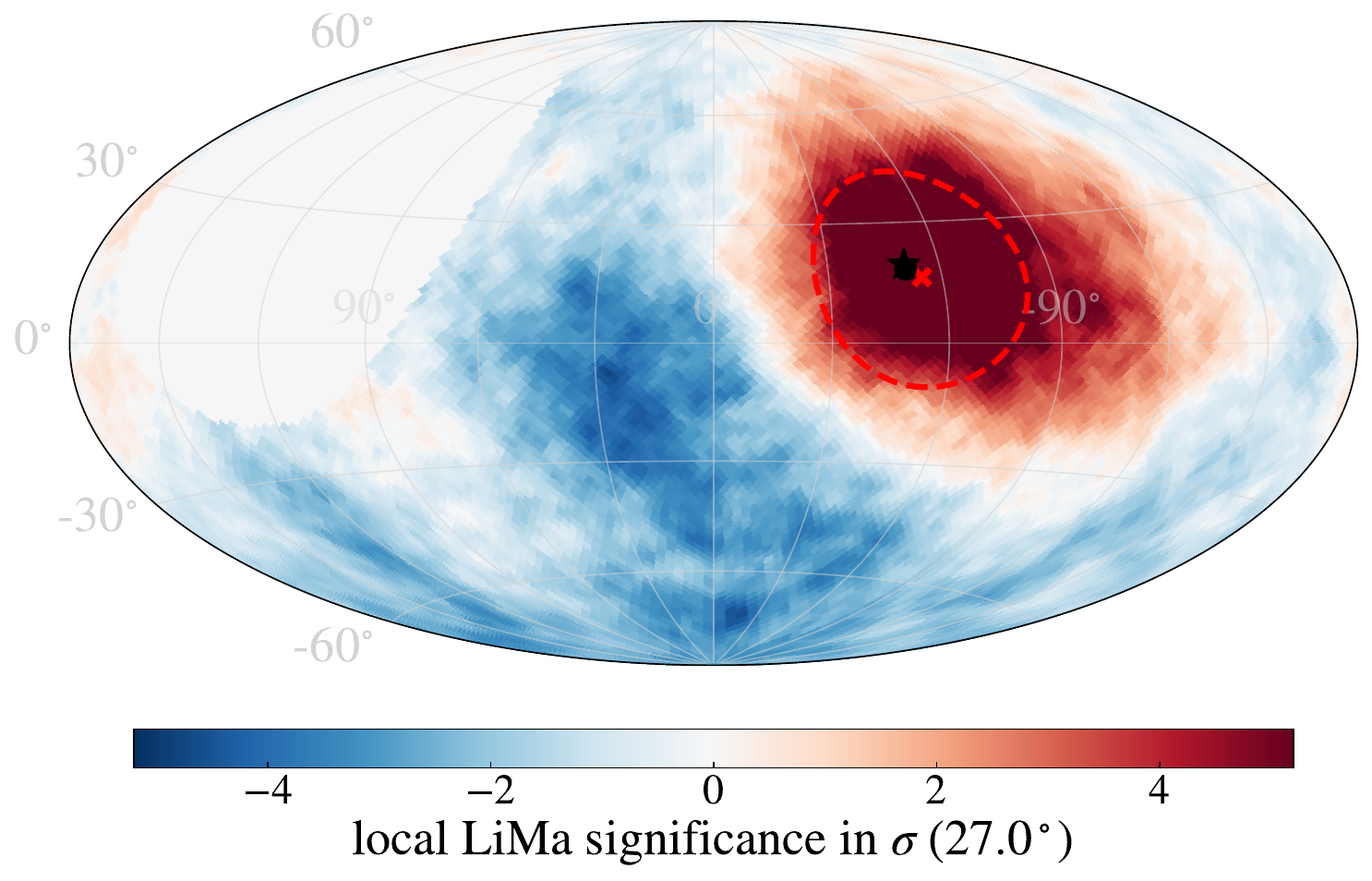}
\includegraphics[width=0.24\textwidth]{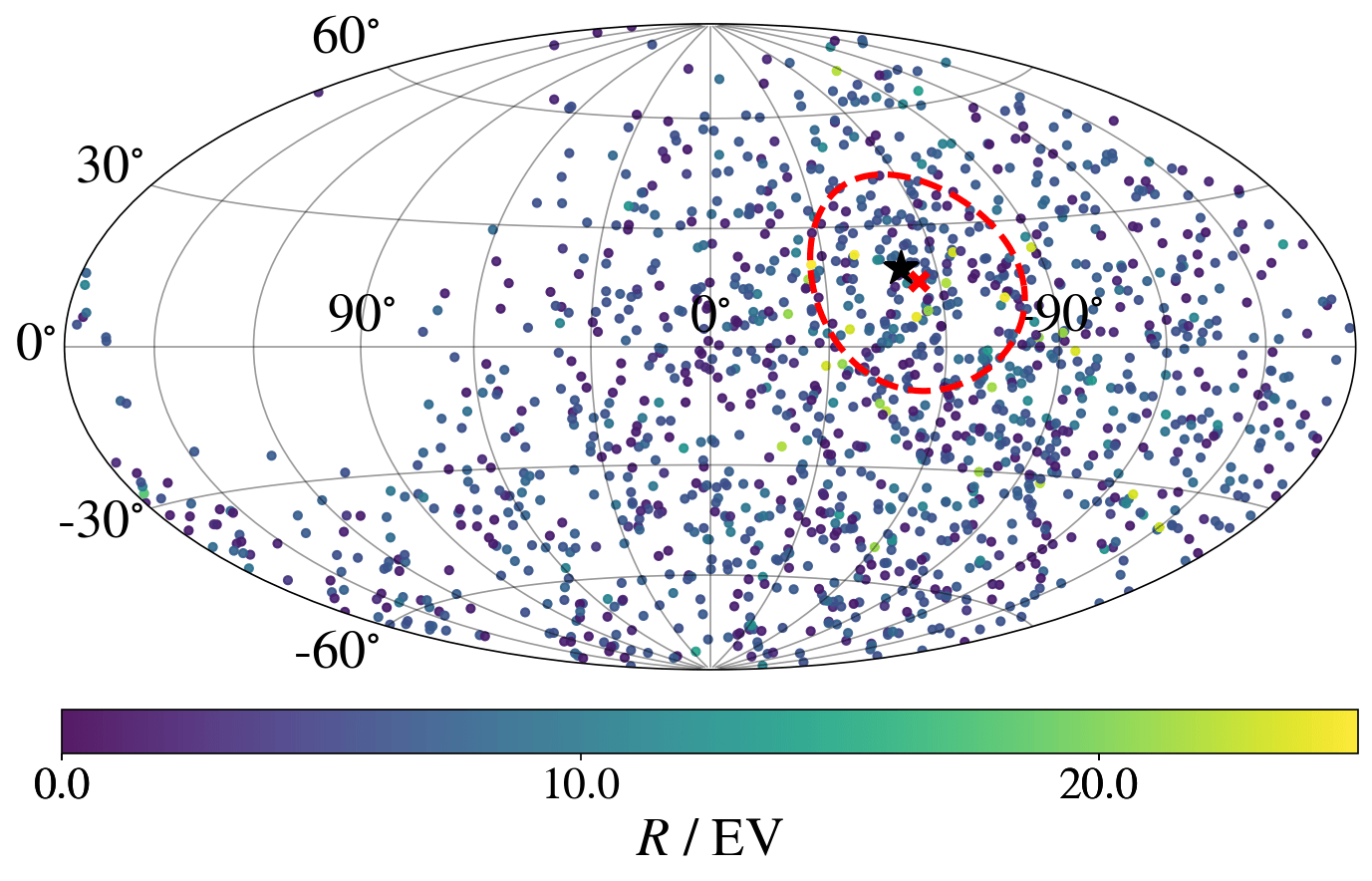}
\includegraphics[width=0.24\textwidth]{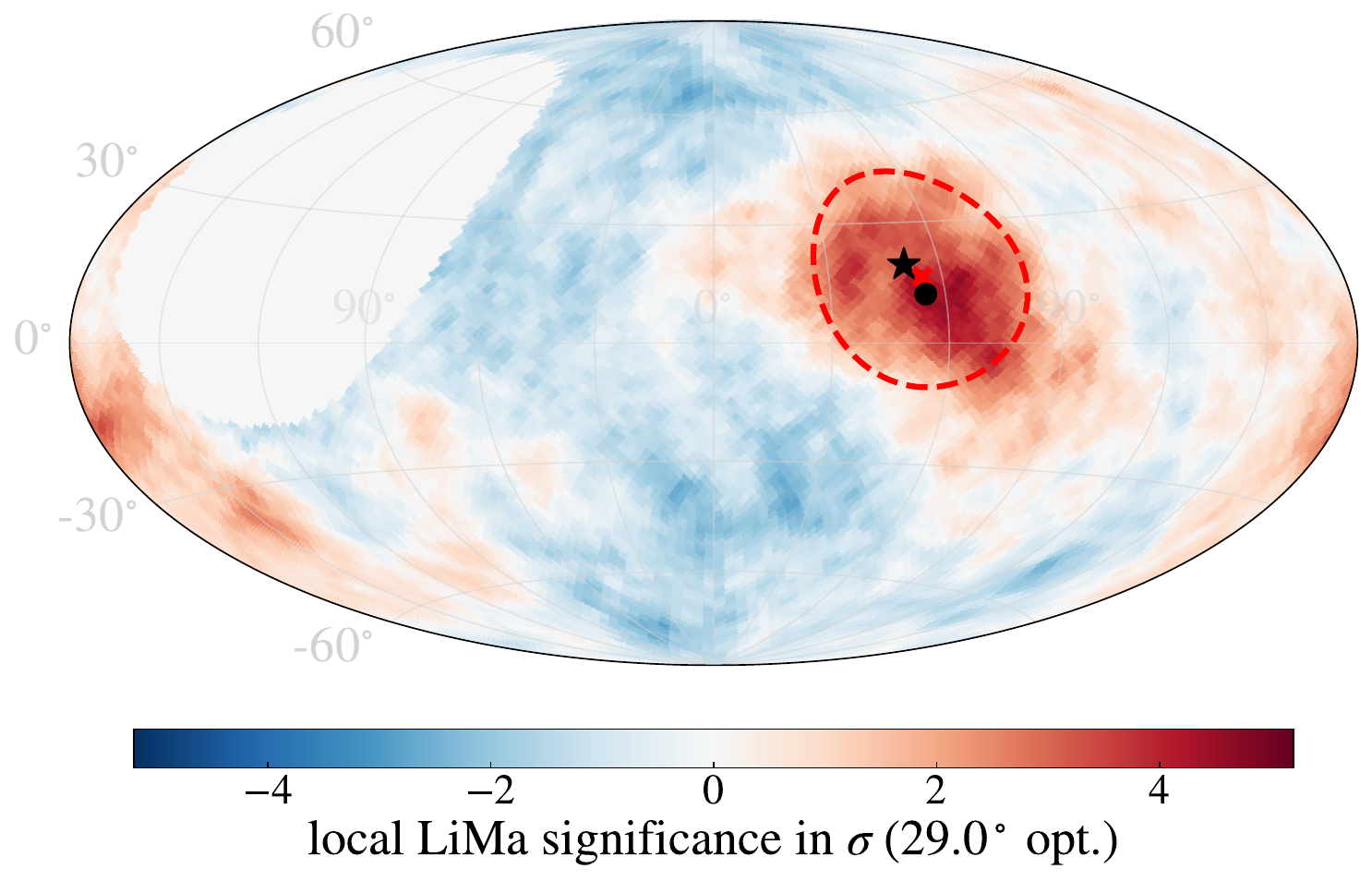}
\captionof{figure}{Two example simulations with Cen A as the only source (indicated by the black star), with $f=1$, mixed composition, $\beta_\mathrm{EGMF}=26.2$ and using the \texttt{UF23-base-Pl} (\textit{upper row}) and the \texttt{UF23-twistX-Pl} (\textit{lower row}) GMF models. The left two figures are for $E_\mathrm{min}=20\,\mathrm{EeV}$ and the right two for $E_\mathrm{min}=40\,\mathrm{EeV}$. All skymaps are in Galactic coordinates. See Fig.~\ref{fig:cena_AD} for more details.}
\label{fig:cena_f1}
\end{minipage}

\begin{figure}[ht]
\includegraphics[width=0.24\textwidth]{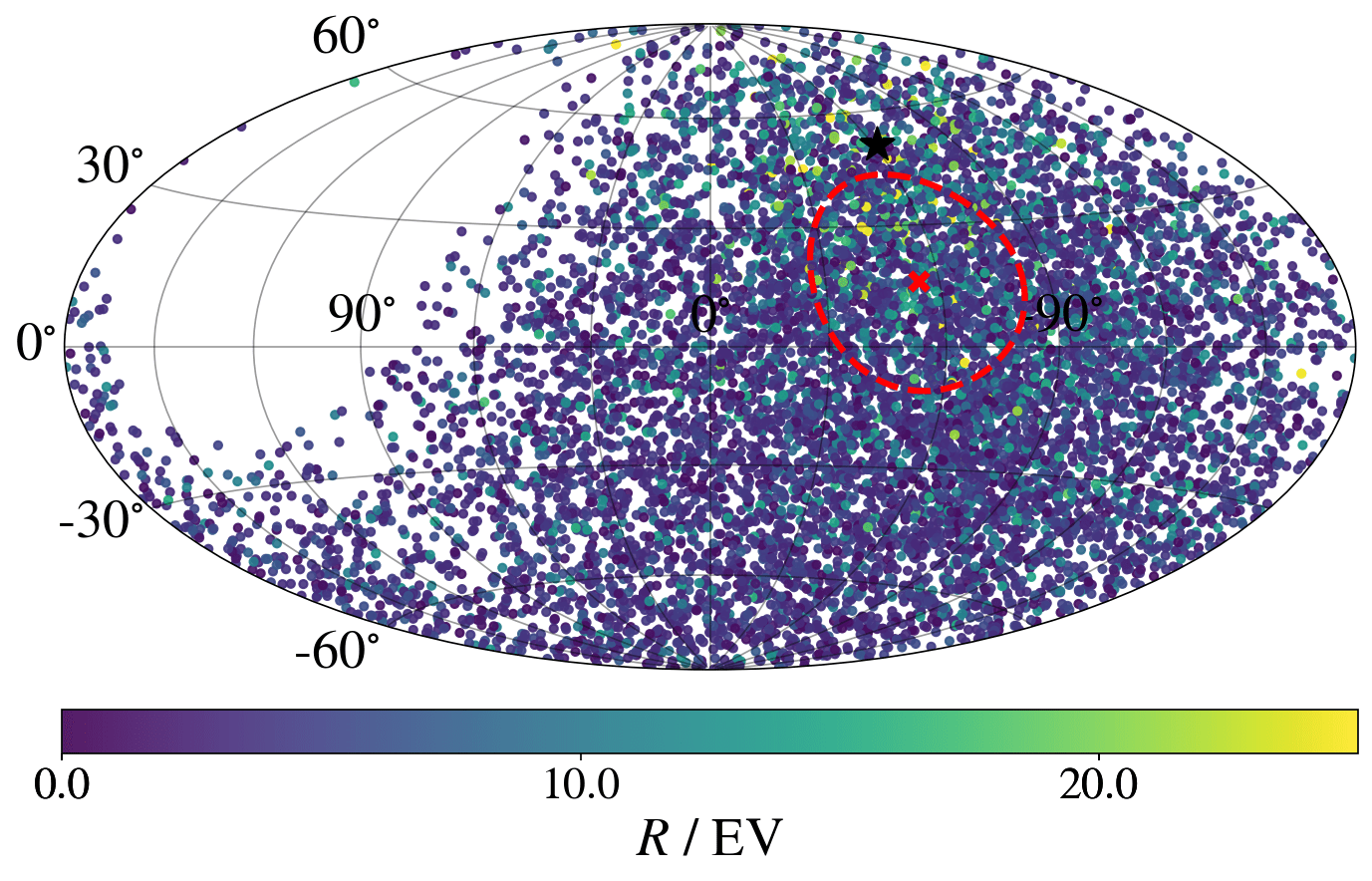}
\includegraphics[width=0.24\textwidth]{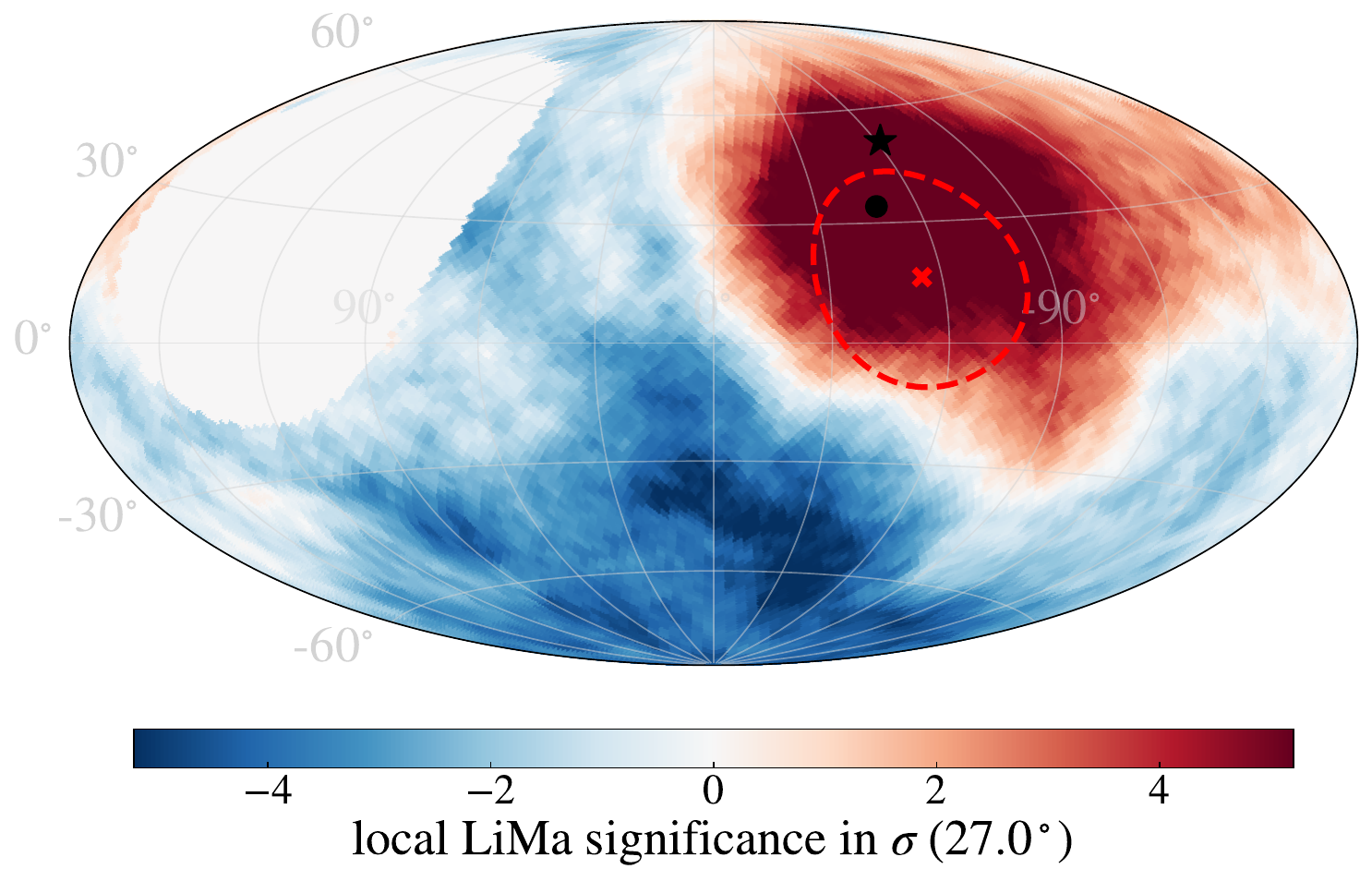}
\includegraphics[width=0.24\textwidth]{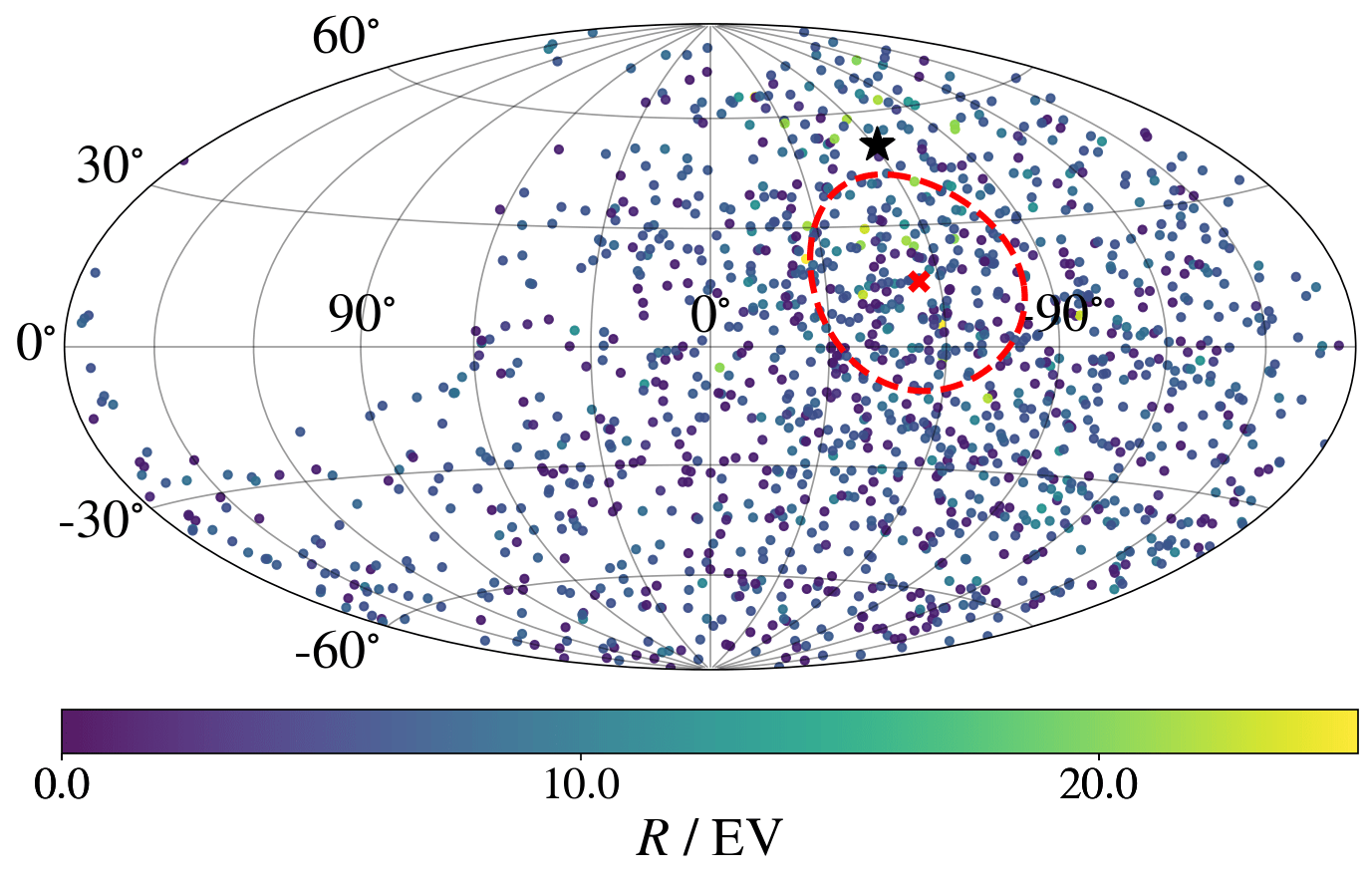}
\includegraphics[width=0.24\textwidth]{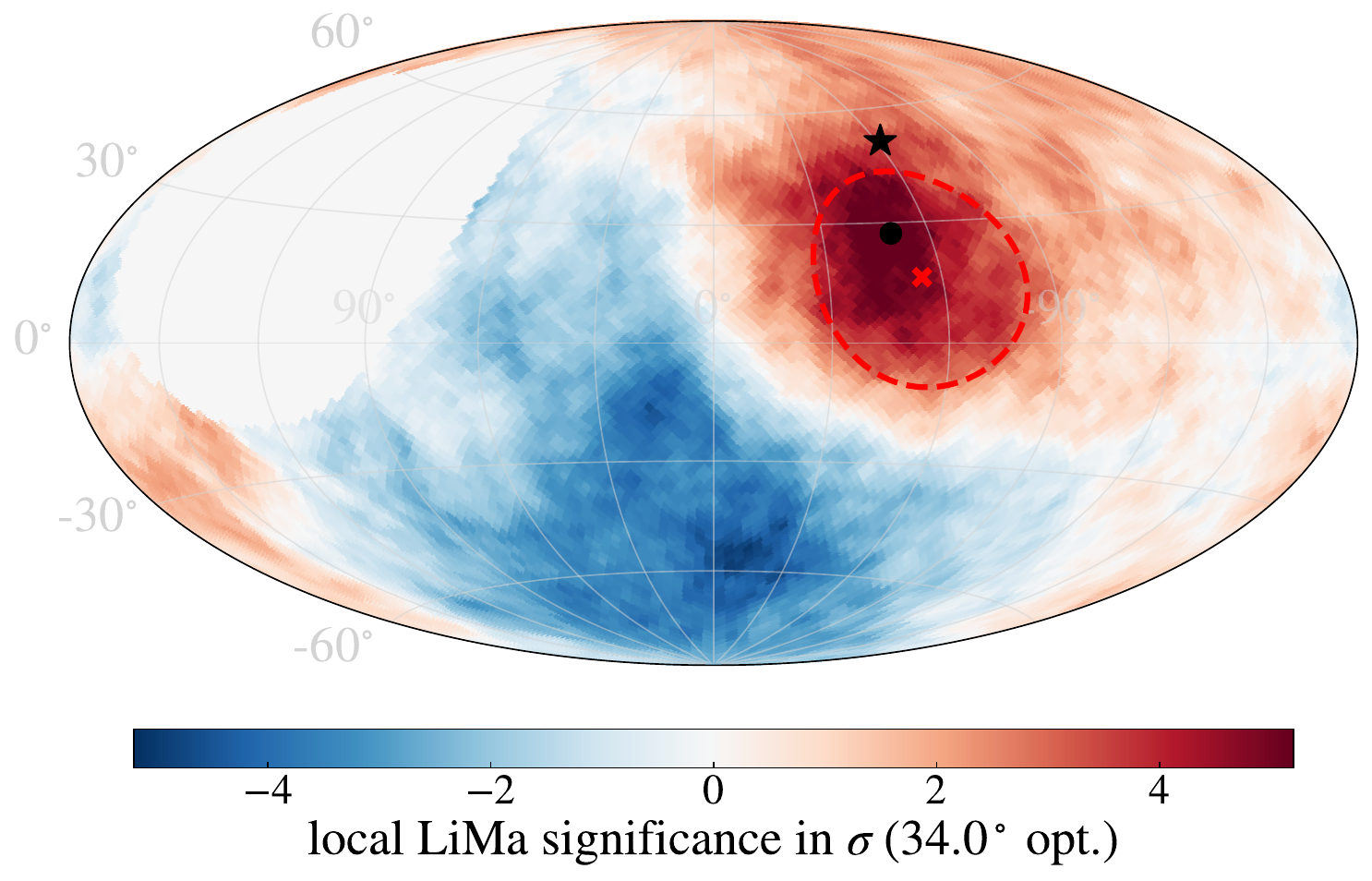}

\includegraphics[width=0.24\textwidth]{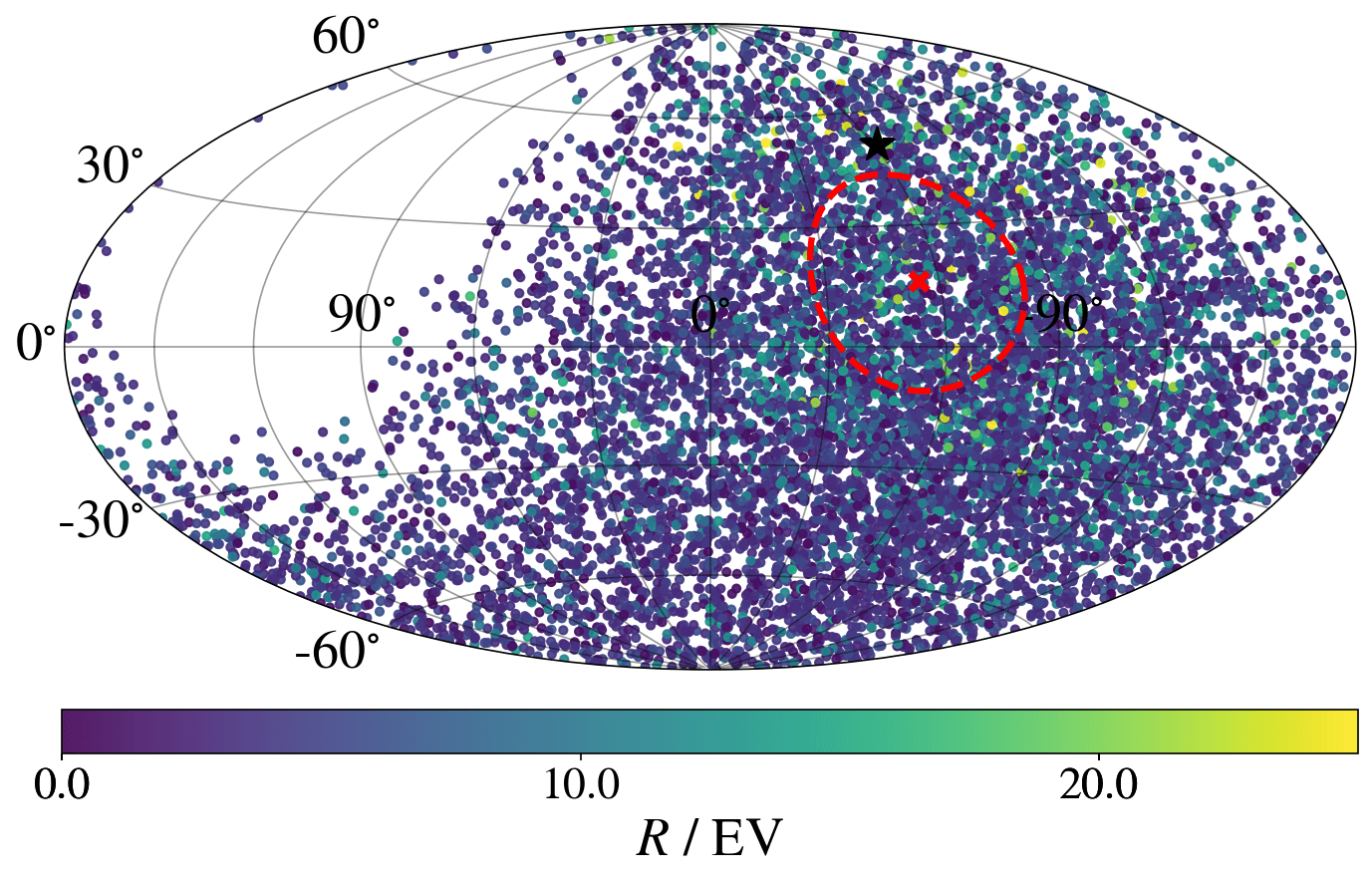}
\includegraphics[width=0.24\textwidth]{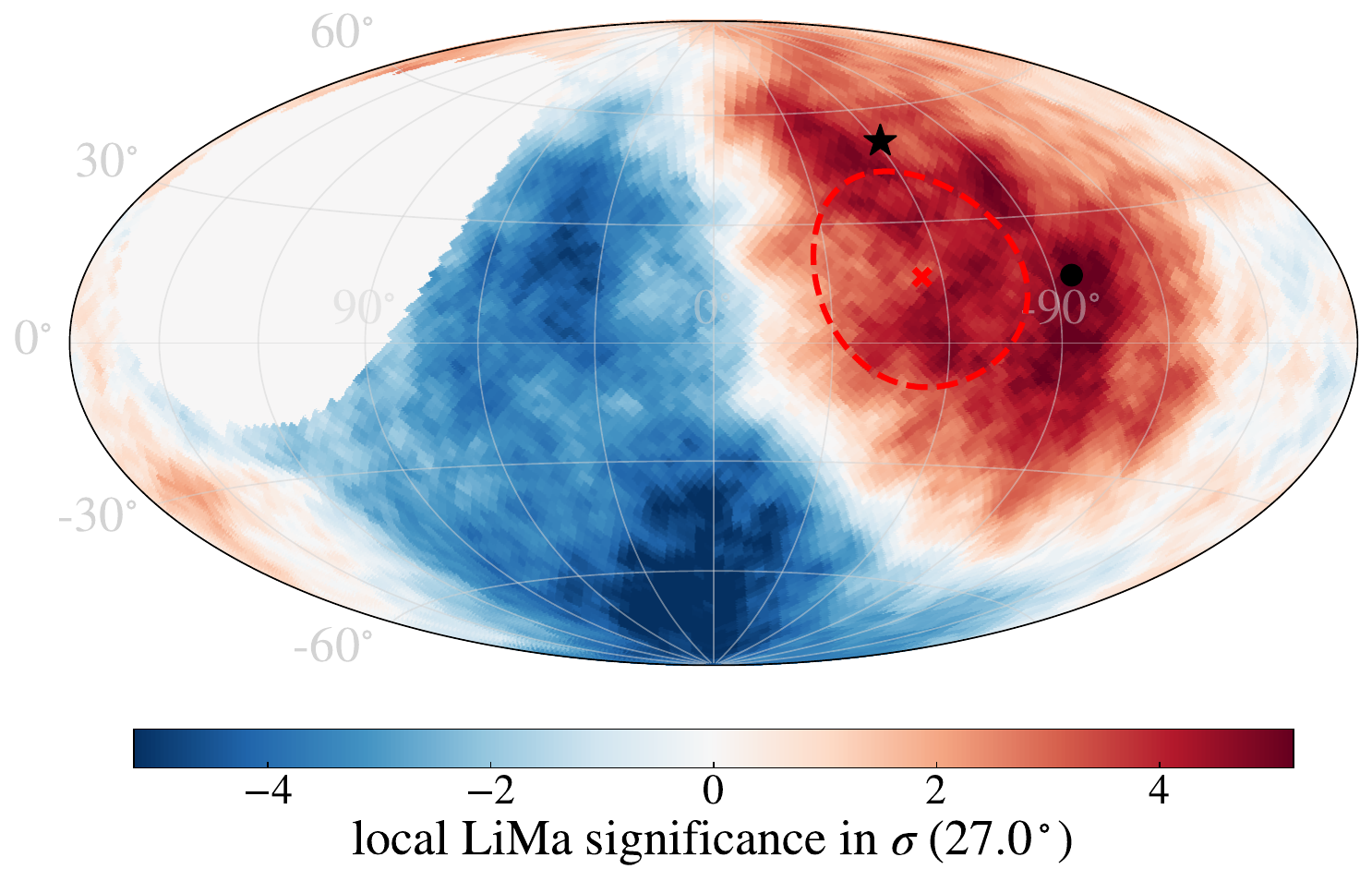}
\includegraphics[width=0.24\textwidth]{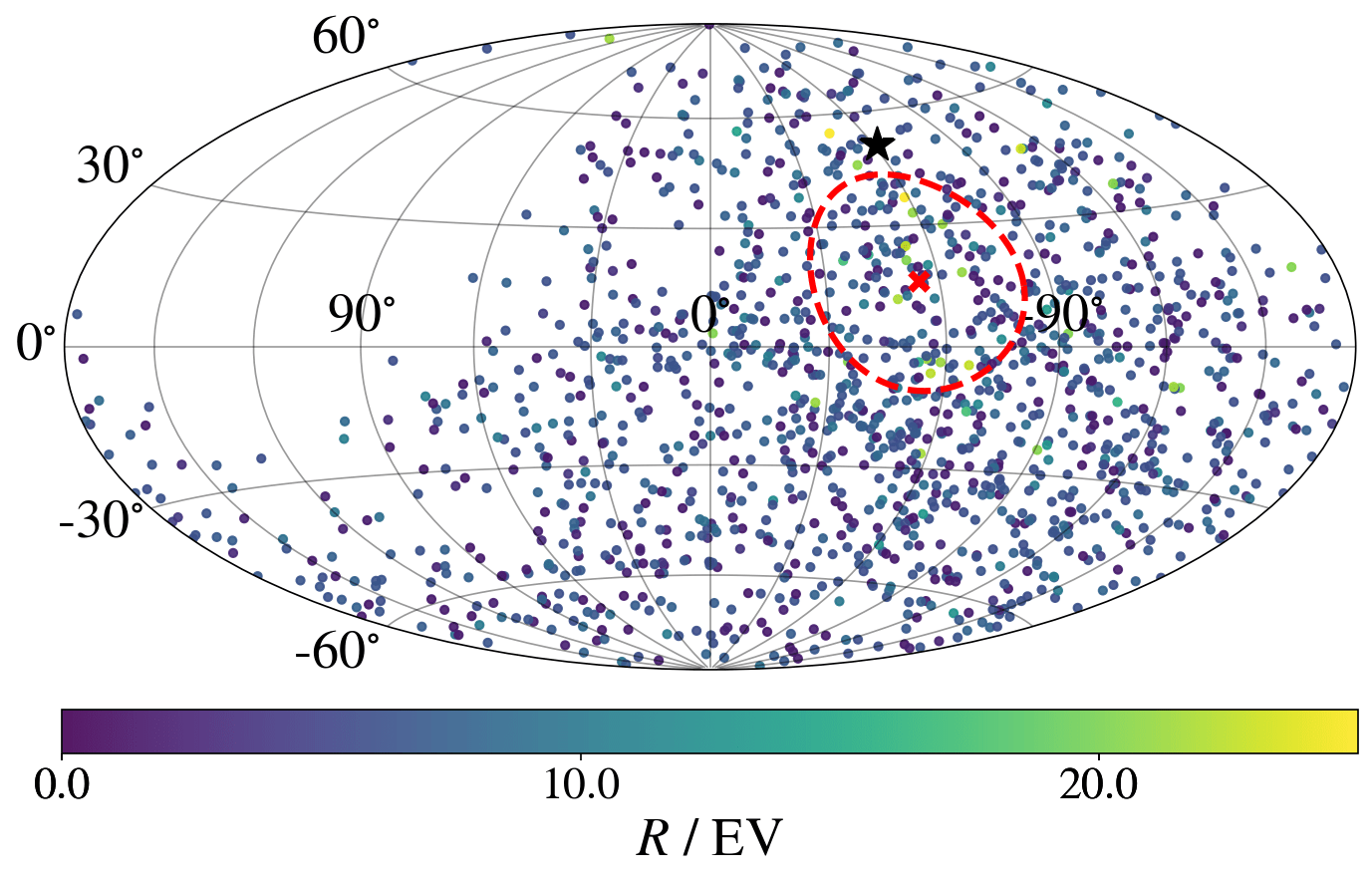}
\includegraphics[width=0.24\textwidth]{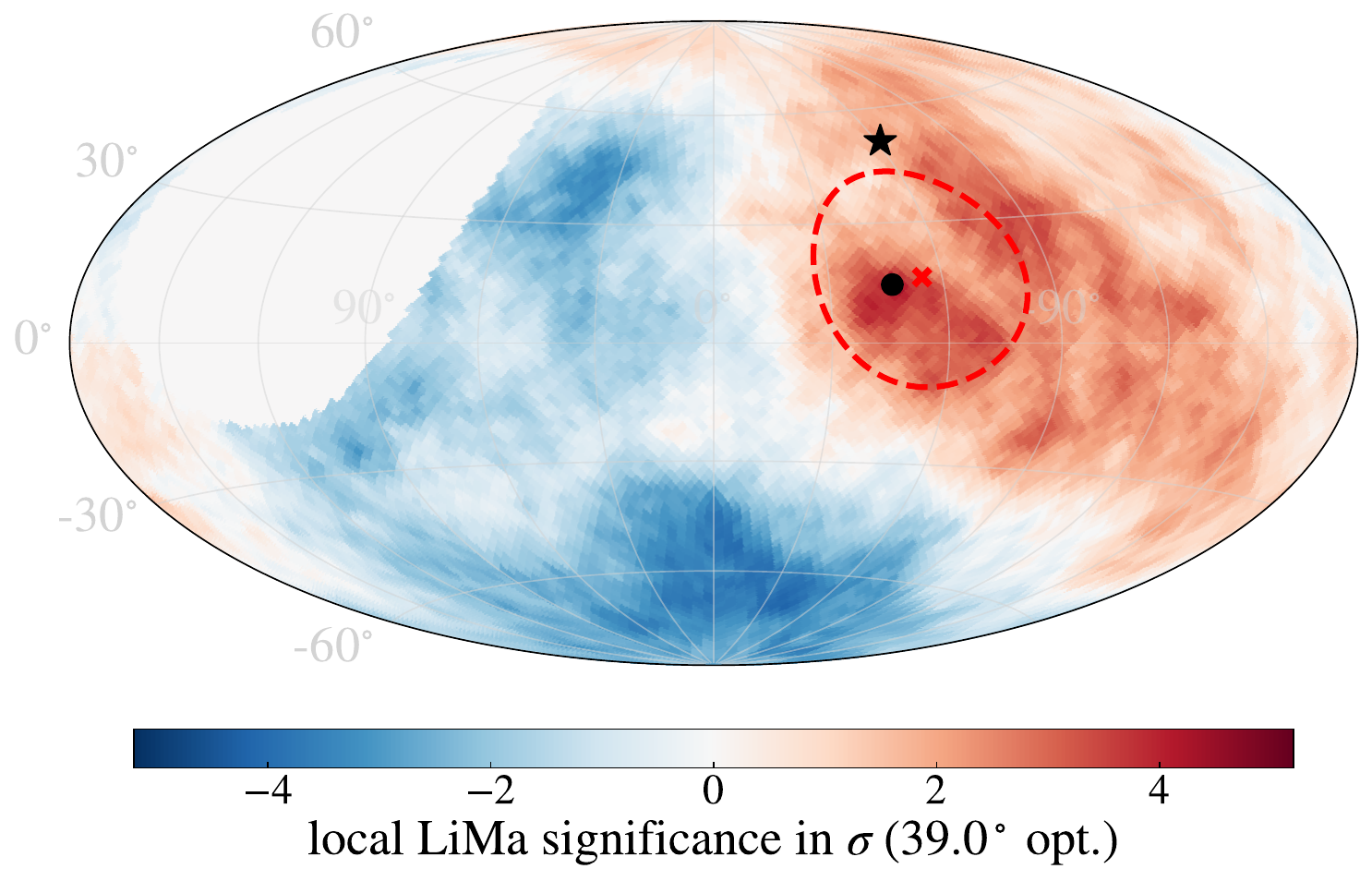}
\caption{Two example simulations with the Sombrero galaxy as the only source (indicated by the black star), with $f=1$, mixed composition, $\beta_\mathrm{EGMF}=16$ and using the \texttt{UF23-base-Pl} (\textit{upper row}) and the \texttt{UF23-twistX-Pl} (\textit{lower row}) GMF models. The left two figures are for $E_\mathrm{min}=20\,\mathrm{EeV}$ and the right two for $E_\mathrm{min}=40\,\mathrm{EeV}$. All skymaps are in Galactic coordinates. See Fig.~\ref{fig:cena_AD} for more details.}
\label{fig:sombrero_f1}
\end{figure}



\bibliographystyle{elsarticle-num} 
\bibliography{references2}

\end{document}